%% file: main-icfp23-pearl.tex
\documentclass[acmsmall,screen]{acmart}
\pdfoutput=1
\usepackage{agda}
\usepackage{stmaryrd}
\usepackage{listings}

\setcopyright{acmcopyright}
\copyrightyear{2023}
\acmYear{2023}
\acmDOI{XXXXXXX.XXXXXXX}

\input{unicodeletters}
\input{agdamacros}

\begin{document}

\title{Intrinsically Typed Sessions With Callbacks}
\subtitle{Functional Pearl}

\author{Peter Thiemann}
\email{thiemann@acm.org}
\orcid{0000-0002-9000-1239}
\affiliation{%
  \institution{University of Freiburg}
  \country{Germany}
}


\begin{abstract}
All formalizations of session types rely on linear types for soundness as
session-typed communication channels must change their type at every
operation. Embedded language implementations of session types follow
suit. They either rely on clever typing constructions to guarantee
linearity statically, or on run-time checks that approximate
linearity.

We present a new language embedded implementation of session types,
which is inspired by the inversion of control design principle. With
our approach, all application programs are intrinsically session typed and
unable to break linearity by construction. Linearity remains a proof
obligation for a tiny encapsulated library that can be discharged
once and for all when the library is built.

We demonstrate that our proposed design extends to a wide range of
features of session type systems: branching, recursion, multichannel
and higher-order session, as well as context-free sessions. The
multichannel extension provides an embedded implementation of
session types which guarantees deadlock freedom by construction.

The development reported in this paper is fully backed by
type-checked Agda code.
\end{abstract}



\keywords{session types, domain specific languages, dependent types, Agda}


\maketitle

\section{Introduction}
\label{sec:introduction}

\input{latex/ST-finite-nonbranching.tex}

Session types provide a type discipline for protocols
in concurrent programming systems. Each session type operator
describes the direction of communication and the type of the
payload (the transmitted value). Moreover, there are ways to specify
control structures for protocols like sequencing, branching, and
looping.

Honda and others
\cite{DBLP:conf/concur/Honda93,DBLP:conf/parle/TakeuchiHK94,DBLP:conf/esop/HondaVK98}
started the investigation of session types originally in the context
of the pi-calculus \cite{DBLP:journals/iandc/MilnerPW92a}. Later the concepts were adapted to other programming
paradigms like functional programming
\cite{DBLP:journals/tcs/VasconcelosGR06,DBLP:journals/jfp/GayV10} and
object-oriented programming
\cite{DBLP:conf/europar/RavaraV97,DBLP:journals/corr/abs-1205-5344}. 

The early works concentrated on binary session types, i.e., protocols
with two parties. Subsequently, multi-party session types were
proposed \cite{DBLP:journals/jacm/HondaYC16} that relax the
restriction to two parties. The Betty project resulted in several survey articles
and books that provide a good overview of the field
\cite{gay17:_behav_types,DBLP:journals/csur/HuttelLVCCDMPRT16,DBLP:journals/jlp/BartolettiCDDGP15,DBLP:journals/ftpl/AnconaBB0CDGGGH16}. 
For this paper, we mostly focus the discussion on the binary case, but draw some
inspiration from multi-party session types.

Session types are inextricably connected to linear types via the
Curry-Howard correspondence. This connection has been discovered and
investigated in quite some depth
\cite{DBLP:conf/concur/CairesP10,DBLP:journals/mscs/CairesPT16,DBLP:journals/jfp/Wadler14}
along with many ramifications \cite{DBLP:journals/pacmpl/BalzerP17}. 
Moreover, this connection is not just of theoretical interest, but it
has affected the implementation and (one could argue) hampered widespread
adoption. Languages constructed around session types are usually
special purpose languages that embrace linearity so that their type
checker rejects violations thereof. Examples are plenty: Links \cite{lindley17:_light_funct_session_types},
Sepi \cite{DBLP:conf/sefm/FrancoV13,DBLP:conf/sfm/Vasconcelos09}, Sill
\cite{DBLP:conf/esop/ToninhoCP13}, C0
\cite{DBLP:journals/corr/WillseyPP17}, and so on.
While these languages and their implementation have fostered research
and encouraged experimentation, they are not widely used.

To boost the use of session types, 
a lot of work has been dedicated to embedding session types in mainstream
languages, most of which do not have native support for linearity. 
There are plenty of examples for such embeddings for functional languages
like
Haskell \cite{DBLP:conf/padl/NeubauerT04,SackmanE08,
  DBLP:conf/haskell/PucellaT08} and
OCaml \cite{DBLP:journals/jfp/Padovani17,DBLP:journals/scp/ImaiYY19},
as well as object-oriented languages like
C\# \cite{DBLP:journals/corr/abs-2004-01325},
Scala \cite{DBLP:conf/ecoop/ScalasY16},
Java \cite{DBLP:conf/ecoop/HuYH08}. Most of these approaches ignore
the issue of linearity at compile time; some ignore it entirely. Some (e.g., \cite{DBLP:journals/jfp/Padovani17}) rely
on run-time checks, others rely on encodings of linearity using lenses
\cite{DBLP:journals/jip/ImaiG19} or monads \cite{DBLP:conf/haskell/PucellaT08}.
There are also extension languages with a separate checker that add
sessions \cite{DBLP:conf/coordination/NgYPHK11} or, more generally,
typestate \cite{DBLP:journals/scp/KouzapasDPG18} to an underlying Java program.
We comment on some recent implementation that rely on Rust in the
related work (Section~\ref{sec:related-work}).

Recent work on multi-party session types
\cite{DBLP:conf/cc/Miu0Y021,DBLP:journals/pacmpl/00020HNY20} suggests
an alternative approach that does not rely on linearity. It is
inspired by the design principle \emph{inversion of control} which is
familiar to programmers from GUI programming. The systems described in
those works translate a description of a multi-party session type into
a library that encapsulates the implementation of all
communication. For each communication action, the library provides an
interface where the programmer specifies a callback function for this
particular action.

To clarify this idea, we give a very simple example, continued and
extended in Section~\ref{sec:finite-nonbr-simple}, in the context of a
functional language. Unlike the cited
work, our work as well as this example rely on \emph{binary} session
types. We start with the following grammar for types $T$ and session
types $S$.
\begin{align*}
  T &::= \Aint \mid \Abool  & S &::= \Atsend{T}{S} \mid \Atrecv{T}{S} \mid \Atend
\end{align*}
The session type $\Atsend{T}{S}$ ($\Atrecv{T}{S}$) describes a channel that is ready to send (receive)
a value of payload type $T$ and the continue as $S$. The session type
$\Atend$ describes a channel that can only be closed.

\begin{description}
\item[Traditional setting] (cf.\ \cite{DBLP:journals/jfp/GayV10}) The
  traditional interface to session-typed communication consists of primitive
  operations like
  \begin{align*}
  \mathtt{send} & : \Atsend{T}{S} \otimes T \multimap S &
                                                   \mathtt{recv} &:
                                                                   \Atrecv{T}{S}
                                                                   \multimap
                                                                   (T
                                                                   \otimes
                                                                   S)
    & \mathtt{close} &: \Atend \multimap ()
  \end{align*}
  that send on a channel, receive from a channel, and close a channel.
  The crucial observation is that the type system must treat channels
  linearly to ensure protocol fidelity.

  Programs typically look like this:
\begin{lstlisting}
negp-server : ?int.!int.end -o ()
negp-server  c0 =                    -- c0 : ?int.!int.end
  let (x, c1) = recv c0 in           -- c1 :      !int.end
  let c2 = send (c1, -x) in          -- c2 :           end
  close c2
\end{lstlisting}
By linearity, the \texttt{recv} operation consumes \texttt{c0};
otherwise, another \texttt{recv} could be applied could \texttt{c0},
thus breaking the protocol.
Analogous arguments apply to \texttt{c1} and \texttt{c2}, e.g., once
the channel \texttt{c2} is closed, it cannot be closed again.
\item[Callback approach] The callback interface to session-typed
  communication proposed in this work consists of two items.
  \begin{enumerate}
  \item A datatype of commands, {\ACommand}, indexed by an application
    state $A$ and a session type. This datatype constitutes an
    intrinsically session-typed encoding of communicating functional programs.
  \item An encapsulated interpreter {\Aexecutor} to execute
    commands.\footnote{The function {\stTypeInterpretationSignature}
      maps type syntax to its interpretation as an Agda type.
      For further
      details see Section~\ref{sec:finite-nonbr-simple}.}
  \end{enumerate}
\stCommand

  The {\ACSEND} command has a callback to obtain the value to be sent
  from the application state. Similarly, the {\ACRECV} command has a
  callback to inject the received value in the application state. Both
  take a continuation command of type {\ACommand~$A$~$S$} that deals
  with the continuation session $S$. The {\ACCLOSE} command signifies
  the end of the protocol.

  A program is expressed as a value of type {\ACommand}. It looks similar to the traditional one where we choose
  {\AZ}, the integers, as the application state. The encoding relies
  strongly on the core idea of functional programming: functions
  (callbacks) as first class values.\footnote{The operator {\Adollar}
    stands for infix function application. It associates to the
    right. (The second {\Adollar} could be omitted.)}
\stNegpCommand

  The least sophisticated interpreter takes a command, a suitable
  initial application state, an untyped channel, and results in an IO
  action that produces the final application state.
\stExecutorSignature

  This interpreter is implemented once and for all in an encapsulated
  library. In a sense, it forms the trusted computing base of our
  approach, as we have the obligation to prove that it performs the
  commands on the channel according to the session type index of the
  channel. 
\end{description}

\subsection{Contributions}
\label{sec:contributions}

\begin{itemize}
\item We introduce the callback approach to binary session
  types in the context of dependently-typed functional
  programming. We deploy it as a proof-of-concept specification in the
  language Agda, but we expect our development to be transferrable to
  Haskell, either via compilation or via translation 
  \cite{DBLP:conf/haskell/CockxME0N22}.
\item Linearity of session handling is ensured by verifying linear
  handling of command execution in a small interpreter that forms the
  trusted computing base of our approach. There is no need for linear
  types in the type system of the host language, nor is there a need
  for clever type constructions to simulate linearity.
\item We demonstrate that the approach extends to most familiar
  session type constructions: branching
  (Section~\ref{sec:select-choice}), recursion
  (Section~\ref{sec:going-circles}), multichannel and
  higher-order sessions (Section~\ref{sec:handl-mult-chann}). In
  Section~\ref{sec:what-about-client} we offer a novel and significant
  improvement of the API-based treatment of recursion.
\item The extension to operate on multiple channels is significant and
  mostly orthogonal to the other features. Our approach is inspired by
  Wadler's GV calculus \cite{DBLP:journals/jfp/Wadler14} and thus
  yields deadlock-free programs by construction. 
\item We propose a new dynamic selection operation in the context of
  branching session types (Section~\ref{sec:select-choice}).
\item We extend the callback approach to context-free session types
  (with branching and recursion), which in turn requires a more
  sophisticated, dependently-typed encoding of commands than regular
  session types (Section~\ref{sec:context-free-session}).
\item For monad lovers Section~\ref{sec:going-monadic} describes a
  version with a monadic encoding of callbacks. 
\end{itemize}

The source of this document includes a number of literate Agda
scripts which will be submitted as an anonymized supplement (to be
turned into an artifact). Every line of code that is typeset in color
has been checked by Agda. At present, the interpreters are implemented
against a small API of monadic IO operations to manipulate untyped
channels. This API can be implemented in Haskell using Agda's foreign
function interface.\footnote{An as-yet unfinished programming exercise
  that we plan to complete for artifact submission.}

As a functional pearl, this paper concentrates on the library design, it contains no
formal proofs of the proof obligations on the library interpreter
(i.e., linearity and freedom of deadlock for the multichannel
case). The discussion in Section~\ref{sec:discussion} contains some
suggestions how this task may be approached.

Working knowledge of Agda is not a hard requirement for understanding the
paper. We strive to make the code accessible to readers who are
knowledgable in Haskell by explaining features specific
to Agda as they are encountered.

\section{Finite non-branching session types}
\label{sec:finite-nonbr-simple}

Let's start straight away with the simplest instance, finite
non-branching simple session types, to convey the
gist of the approach. Subsequent sections show how to add most of the
usual features of session types.

A binary session type describes a bidirectional communication between two peers,
let's call them server and client. The session type is attached to the
type of the communication channel.\footnote{Agda supports a mixfix
  syntax where underlines in the identifier indicate the position of
  the arguments. For example,  $\Atsend{\_}{\_}$ and $\Atrecv{\_}{\_}$ are operators with two
  arguments. We declare these operators to associate to the right to
  save parentheses.}
\stFiniteType
\stFiniteSession
These Agda types correspond to the standard grammar of session types, where
$T$ is the type of payload values that can be transmitted and $S$ is the
type of sessions. 
\begin{align*}
  T &::= \Aint \mid \Abool  & S &::= \Atsend{T}{S} \mid \Atrecv{T}{S} \mid \Atend
\end{align*}
The session type $\Atsend{T}{S}$ ($\Atrecv{T}{S}$) describes a channel that is ready to send (receive)
a value of payload type $T$ and then continue as $S$. The session type
$\Atend$ describes a channel that can only be closed.

Here are two examples for session types: the types of the server for a
binary operation and a unary operation, respectively.
\stExampleBinpUnP

In GV, a widely studied functional session type theory \cite{DBLP:journals/jfp/GayV10}, there are primitives to
send and receive values and to close a channel with types like this:
\begin{align*}
  \mathtt{send} & : (\Atsend{T}{S} \otimes T) \multimap S &
                                                   \mathtt{recv} &:
                                                                   \Atrecv{T}{S}
                                                                   \multimap
                                                                   (T
                                                                   \otimes
                                                                   S)
  & \mathtt{close} &: \Atend \multimap ()
\end{align*}
The types indicate that we must treat channel values of session type
\emph{linearly}: the \texttt{send} operation \emph{consumes} a
channel, which is ready to send, paired with the payload and returns it in a state described
by $S$; the \texttt{recv} operation \emph{consumes} a channel, which
is ready to receive, and returns a pair with the received value and
the updated channel; the \texttt{close} operation \emph{consumes} the
channel and returns a unit value. Enforcing this linearity is required
for soundness.

In this work, we take a different approach inspired by callback
programming. Instead of providing \texttt{send} and \texttt{recv} 
primitives to the programmer, we ask the programmer to define the
``application logic'' by implementing a command value whose type {\ACommand} is indexed by
a session type. This definition relies on an interpretation of
types as Agda types.
\stTypeInterpretation
\stCommand
In this definition, the type parameter $A$ embodies the application state. 
Each $\ACSEND$ command takes a state transformer that extracts the value to
send from the current application state; each $\ACRECV$ command takes
a state transformer that is indexed by the received value; the $\ACCLOSE$
command terminates the session. In fact, we could provide the
application logic by actions in a state monad over the application
state $A$. We defer the shift to a monadic interface to Section~\ref{sec:going-monadic},
when we have the full picture.

Continuing our example, we define commands that implement a server
for the protocols {\Aunaryp} and {\Abinaryp} with the operation instantiated to
negation and addition, respectively.
\stNegpCommand
\stAddpCommand  
 We use explicit lambda abstraction instead of fancy abbreviations and library combinators
for clarity; the variable $a$ stands for the application state and $x$ and $y$ for
the respective values received from the channel. For connoisseurs of pointfree definitions, we give a
more concise version of the addition command using functions from the
standard library:
\stAddpCommandAlternative

To execute commands, we write an
interpreter that relies on primitive operations provided
in the $\AIO$ monad. The point of our approach is that this
interpreter is the single definition where we are obliged to prove
that it handles channels in a linear fashion. 
\stPostulates
This API should be self-explanatory.\footnote{The type $\top$ is like
  Haskell's unit type with single element {\Att}.} It declares an abstract type of
untyped, raw channels with operations to accept a connection, close a channel, as
well as send and receive a value over the channel. It glosses over issues like
serialization, which can be addressed using type class constraints like
{\ASerialize~$A$} (in Haskell) on {\AprimSend} and {\AprimRecv}.

The interpreter itself is defined by induction on the type
{\ACommand}. The actual computation takes place in the {\AIO} monad and is expressed using the {\Ado} notation, both
familiar from Haskell. 
\stExecutorSignature\vspace{-1.5\baselineskip}
\stExecutor
To actually run a server, it remains to provide a wrapper that accepts
a connection and invokes the interpreter.
\stAcceptor
Examining the interpreter, we finally see the full monadic
structure. We need a stack of monad transformers starting with a state
monad for the application state on top of a reader monad providing the
channel on top of the IO monad.

Our proof obligation for the interpreter boils down to verifying that
the interpretation of each command executes the single communication
action designated by the corresponding session type operator. The
correct sequencing according to the session type is imposed by the
sequencing constraint underlying the {\AIO} monad. Indeed, this
observation was the reason to employ monads for APIs to state-based
operations in pure functional languages \cite{DBLP:conf/popl/JonesW93}.

\section{Selection and choice}
\label{sec:select-choice}

Adding branching to our development is straightforward. The standard 
theory of session types allows branching on a finite set of labels using this syntax:
\begin{align*}
  S & ::= \dots \mid \oplus\{ \ell : S_\ell \mid \ell \in L \} \mid
      \&\{\ell: S_\ell \mid \ell \in L\}
\end{align*}
Here $L$ is a finite, non-empty set of labels, which can be chosen differently at
every use of the type operator. 
The type constructor $\oplus$ corresponds to an \emph{internal choice} of the
program.  The \texttt{select} primitive sends one of the labels, say $\ell \in L$, available in the
type and continues according to $S_\ell$:
\begin{align*}
  \mathtt{select}\ \ell &: \oplus\{ \ell : S_\ell \mid \ell \in L \}
                          \multimap S_\ell
\end{align*}
The type constructor $\&$ corresponds to an \emph{external choice}. The
primitive \texttt{branch} receives one of the labels 
mentioned in the type and chooses a continuation according to the
label. In the presence of sum types and linearity, the primitive can be typed as
follows \cite{DBLP:journals/toplas/Padovani19}.
\begin{align*}
  \mathtt{branch} &: \&\{\ell: S_\ell \mid \ell \in L\} \multimap +\{\ell: S_\ell \mid \ell \in L\}
\end{align*}
Our modeling in Agda extends the definitions of
{\ASession}, {\ACommand}, and {\Aexecutor} from
Section~\ref{sec:finite-nonbr-simple}. A label set of size $k$ is
modeled by the type $\AFin~k = \{ 0, \dots, k-1\}$ and the alternative continuation sessions by
functions from labels to {\ASession} (isomorphic to vectors of sessions, cf.\  Section~\ref{sec:select-choice-with}).
\stBranchingType
\stBranchingCommand
The commands for {\ACSELECT} and {\ACCHOICE} only differ in the
placement of the parentheses: {\ACSELECT} takes a label and a command
corresponding to this label, whereas {\ACCHOICE} takes a function that
maps a label to its command. Neither command requires the application
state: the selection is static and does not depend on the current
state; once the external choice has been taken, the chosen command can
reflect that choice in the application state.
\stExecutorSignature\vspace{-1.5\baselineskip}
\stBranchingExecutor

Extending our running example, we define the type of an arithmetic server,
which gives a choice between a binary operation and a unary one. 
This definition uses a smart constructor {\Aamp} for the external choice that
takes a vector of session types and transforms it into the
corresponding function.\footnote{This transformation relies on the
  isomorphism between vectors of size $k$ and non-dependent functions with domain {\AFin~$k$}.}
\stExampleArithP
The command for the server extends in the obvious way. The vector
trick does not work for the {\ACommand} constructors {\ACSELECT} and
{\ACCHOICE} because their arguments are dependent functions that
generally return different types for different arguments.
\stArithpCommand

Finally, we observe that the proposed interface enables a more dynamic
selection operator than in standard session types. While the standard
select operator is indexed by a label that is fixed at compile time,
we can supply a dynamic selector command where the label is computed
by a callback \AgdaFunction{getl} at run time:
\stDynamicBranchingCommand
Extending {\Aexecutor} to this command is straightforward.
There is still room for improvement in the type of this command. We
come back to this issue in Section~\ref{sec:context-free-session}.

\section{Going in circles}
\label{sec:going-circles}
\input{latex/ST-recursive.tex}

Recursive types are a common feature of session types. They are required
to model protocols for servers that perform the same functionality
over and over again. Our running example will be a server that allows
a client to repeatedly perform a unary operation until the client
quits.

The pen-and-paper syntax of recursive types relies on type variables
and a $\mu$ operator like so:\footnote{We gloss over the issue of
  guardedness (or contractiveness) for recursive types to avoid further complexity in the
  types.}
\begin{align*}
  S &::= \Atsend{T}{S} \mid \Atrecv{T}{S} \mid \Atend \mid
      \Amu~X. S
      \mid X
      \mid \dots
\end{align*}
The intended semantics of the $\Amu$ operator is that $\Amu~X.S$ is
equivalent to $S[X \mapsto \Amu~X.S]$, the unfolding, where we
substitute the recursive type itself for the variable $X$ in its body.

For the Agda formalization we choose the standard de Bruijn encoding
of bound variables. The parameter $n$ of
the {\ASession} type denotes the number of type variables in
scope. The {\Amu} operator opens a new scope and {\Aback~$i$}
references the $i$th variable, the innermost binding being $0$.
\rstSession

Further extending our running example, we redefine the protocol {\Aunaryp} as a function
that takes the rest of the protocol and wraps it into a recursive
type. The session type {\Amanyunaryp} is a recursive type that either
runs a unary function and recurses or just ends the protocol. We
define {\Aamp} as a smart constructor as in Section~\ref{sec:select-choice}.
\rstExampleManyUnaryp

\subsection{Commands}
\label{sec:commands}

The {\ACommand} type obtains a new
parameter $n$ to match the parameter of the session type used as an
index. We only show the two new cases.
\rstCommand

With this type we are ready to implement a
service that repeatedly adds numbers as they are received and sends the partial
sum as a response each time.
\rstSumupCommand

\subsection{Interpretation}
\label{sec:interpretaion}

Executing a command with recursion requires a new component in the
executor. Whenever execution enters a {\AMU} command, this component
saves the entire loop on a stack. When we encounter a {\ACONTINUE~$i$}
command, we grab the corresponding loop from the stack and
continue with it. There are a few complications in setting up this new
component that we call {\ACommandStore}.

\begin{figure}[tp]
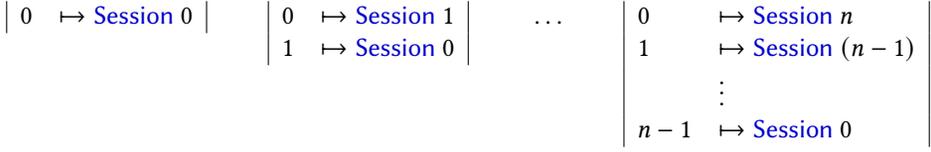

  \begin{align*}
    &
      \begin{array}[t]{|ll|}
        0 &\mapsto \ASession~0
      \end{array}
    &&
      \begin{array}[t]{|ll|}
        0 &\mapsto \ASession~1 \\
        1 &\mapsto \ASession~0
      \end{array}
          &&
             \dots
    &&
       \begin{array}[t]{|ll|}
         0 & \mapsto \ASession~n \\
         1 & \mapsto \ASession~(n-1) \\
           & \vdots \\
         n-1 & \mapsto \ASession~0
       \end{array}
  \end{align*}
  \caption{The session type indices in {\ACommandStore} after 1, 2, and $n$ loop entries}
  \label{fig:commandstore}
\end{figure}
Its type requires some thought.  The top-level type is a
{\ASession~0}. When we enter the first loop, its type is {\ASession~0}. Hence, the {\ACommandStore} contains
one entry (left column in Figure~\ref{fig:commandstore}).
Entering the next loop pushes a loop of type {\ASession~1}
(middle column). After entering the $n$-th loop, we obtain the
picture on the right of Figure~\ref{fig:commandstore}. Here is a
suitable type definition.
\rstCommandStore

For some $(i : \AFin~(\Asuc~ n))$, the function
$\AgdaFunction{opposite}~i$ returns $n-i$ (as an element of the same
type) and the function {\AtoN} injects an element of the finite number
type into the natural numbers (it is an identity function up to the type). The tricky part is implemented in the
typed push and pop operations on that structure.
\rstPops

The \AgdaFunction{push} function takes a {\ACommandStore} and a
suitable {\ACommand} and returns the store extended by this command
(at position {\Azero}). 
The \AgdaFunction{pop1} function pops the first entry off the
{\ACommandStore}. It gets used in defining the inductive step of the
function \AgdaFunction{pop}, which pops any (legal) number $i$ of entries
from the stack. The definitions are simple but omitted from the text
as they require invoking some technical lemmas about injections (i.e.,
identity functions) from $\AFin~n$ to $\AFin~(\Asuc~n)$.

\rstExecutorSignature\vspace{-1.5\baselineskip}
\rstExecutor

The execution of {\AMU} just pushes the loop onto the
{\ACommandStore} and executes the body.
The case for {\ACONTINUE~$i$} jumps to the selected loop and pops
all intervening continuations from the store. It also reveals another complication. The recursive
call to {\Aexecutor} happens on a command drawn from the {\ACommandStore}. This
command is unrelated to the current command, which upsets Agda's
termination checker. To console it we introduce a \AgdaFunction{Gas}
argument, which we decrement at each {\ACONTINUE}.

The question is, what to do if we run out of gas?
There are at least three possibilities.
\begin{enumerate}
\item We regard the current formalization as a proof of concept and
  refer the reader to a ``real'' implementation in Haskell, where the
  \AgdaFunction{Gas} argument is not required. We see no
  problem in creating such an implementation, but it would be more
  involved as dependent types are not yet fully integrated in Haskell.
\item We turn the bug into a feature and require that every
  {\ACONTINUE} communicates with the other end of the channel to
  exchange their respective gas levels and quit the loop if one end
  runs out of gas. This choice implies that we would have to a
  continuation argument to  {\ACONTINUE} that acts like a
  finally-clause for exception and finishes the protocol in an orderly
  way.
\item We can modify {\Aexecutor} so that it always takes a single step
  and returns either a continuation and an intermediate state or a
  final result of type $A$. This modification is straightforward in 
  the setting with loops as a value \AgdaFunction{cms} of type {\ACommandStore~$(\Asuc~n)$~$A$}
  can serve as a continuation. To restart a continuation, we simply invoke
  {\Aexecutor} on \AgdaFunction{cms}~\Azero. For concreteness, we show
  the continuation type and an implementation of the restart function.
  \rstAlternativeExecutorRestart
  With this approach, the management of {\Arestart}  has to be
  included in the trusted computing base as it also has to guarantee linear treatment of
  the channel.
\end{enumerate}
The current formalization leaves that question open and just breaks
the protocol. Our preferred solution is a Haskell implementation,
which omits the gas and thus sticks to the protocol.
u
\subsection{What about the client?}
\label{sec:what-about-client}

The commands we presented so far for recursive sessions are fine for
servers that repeatedly perform the same action. However, a client
might want to perform different actions on each iteration. While such
a behavior may be encoded in the application state, it would not be an
enjoyable experience for the programmer.

Hence, we propose an {\AUNROLL} command that
enables the specification of a command for one loop iteration at a time.
\rstCommandUNROLL
It specifies one command for the session type's loop body and a
continuation command for the whole type to cover subsequent
iterations. 
Executing this command means to execute its body and push its
continuation on the stack.
\rstExecutorUNROLL
As an example, we write a client for the {\Amanyunaryp} protocol that
iterates the protocol two times before it terminates. In each round,
it sends an integer and ignores the response.
\rstClientExample

\section{Going monadic}
\label{sec:going-monadic}
\input{latex/ST-monadic.tex}

We already remarked on the monadic structure apparent in the
callbacks and in the implementation of the {\Aexecutor}
function. Indeed, moving on to a monadic interface might make our session
programs more concise.

We start with a slightly refactored type of commands. More precisely,
we introduce a new {\ACSKIP} command that only performs a state 
transformation via its single callback.
Second, we restate the types of the callbacks for {\ACSEND} and
{\ACRECV} in terms of the state monad.

\mstCommand

Here are the specifics of the types of the callbacks. Each callback
runs in a state monad that handles the application state $A$ on top of
another monad $M$. This construction is expressed in terms of the
monad transformer {\AStateT} applied to the application state
$A$ and the underlying monad $M$. The type of the callback does not give away more
than that.

But wait, using the Agda standard library, we have to state that $M$
has a type that fits a monad and that it implements the
interface {\ARawMonad} (a record that contains the basic monadic
operations).
Fortunately, we can abstract from these issues and adopt a
Haskell-inspired syntax with a straightforward
Agda definition.\footnote{Agda's \AgdaKeyword{syntax} command defines
  a macro that enables abstraction over binders. The newly introduced
  syntax, the definiendum, is \emph{on the right} of the equals
  sign. Putting the record {\ARawMonad~$M$} in double braces enables overloading of the
  monadic operators \cite{DBLP:conf/icfp/DevrieseP11}.}
\mstMonadic

Our running examples become (even more?) concise:
\mstExampleServers
Here, {\Aput}, {\Amodify}, and {\Aget} are monadic functions defined
in the state monad transformer. Their full types are fairly
unreadable, but intuitively we can treat them like this:
\begin{align*}
  \Aput &: A \to ST~A~\top
  & \Amodify &: (A \to A)\to ST~A~A
  & \Aget &: ST~A~A
\end{align*}
{\Aput} overwrites the state with its argument; {\Amodify~$h$} applies 
function $h$ to the state and returns the new state; and {$\Aget = \Amodify~\Aid$} returns the current state.

While we are at it, we also define the {\Aexecutor} function in
monadic style. At this point, the full type of the execution monad
emerges. We already encountered the state transformer on top for
handling the application state. Below we find a reader monad (again
implemented using a monad transformer) that handles access to the
channel. At the bottom, we have the IO monad as before.
\mstExecutor

The most remarkable aspect of this implementation is how unremarkable
it is. However, bear in mind that monads have been used for a long
time to contain linear resources, most notably with the inception of
monadic IO \cite{DBLP:conf/popl/JonesW93}. That is, our proof
obligation of linear handling for the channel becomes even simpler in
this monadic setting.

We conclude this section with the definition of the monadic
acceptor. It essentially invokes the runners of the reader and state
monads and executes the underlying IO action.
\mstAcceptor

\section{Context-free session types}
\label{sec:context-free-session}
\input{latex/ST-indexed-contextfree.tex}

Context-free session types
\cite{DBLP:journals/iandc/AlmeidaMTV22,DBLP:journals/toplas/Padovani19,DBLP:conf/icfp/ThiemannV16}
have been conceived to liberate session types from the restriction to
tail recursion. Alleviating this restriction makes session-typed
programming more compositional and enables low-level programming tasks
like the serialization of tree structures.

The basic idea \cite{DBLP:conf/icfp/ThiemannV16} is to reorganize the
type language of session types as follows.
\begin{align*}
  S &::= \Atcfsend{T} \mid \Atcfrecv{T} \mid \Atcfcomp{S}{S} \mid \Atcfskip
\end{align*}
Now $\Atcfsend{T}$ ($\Atcfrecv{T}$) describes just the act of sending
(receiving) a value of type $T$. To combine two session types, we have
to use sequential composition $\Atcfcomp{S}{S}$ with unit
$\Atcfskip$. The branching types and recursion are as before, so we do
not repeat them here.

The Agda encoding of this structure combines straightforwardly with
the accumulated work of the previous sections.
\cstSession

The revised command structure has a few catches that require
explanation.
\cstCmd

The {\ACommand} datatype now carries four type-related parameters on
top of the session type. A command of type {$\ACommand~A~B~V~W~S$} is
firstly an action that takes an input of type $A$, yields an output of
type $B$, and executes a session according to $S$. The additional
parameters, $V$ and $W$, are vectors of types that control the typing
of the currently pending loops, which are explained with the commands
{\AMU} and {\ACONTINUE}. For now, we can think of them as stack of
the input and output types of those pending loops.

The {\ACSKIP} command is associated with a {\Atcfskip} in the session
type. It comes with a function that transforms $A$s into $B$s.

The {\ACSEND} and {\ACRECV} commands no longer take a command parameter
to process the continuation session. This functionality is now
provided once and for all by the composition operator.

The command {\ACSELECT} prescribes dynamic selection. It takes a
callback that maps a value of input type $A$ to a dependent pair of a
label $i$ and a value of type $F~i$.\footnote{The negative occurrence
  of the $F$ parameter is not permitted in the constructor of a
  {\ASet} datatype. Its presence forces the command type one level up
  into {\ASetOne}.} The continuation takes advantage of this
fine-grained type information by associating the label $i$ with the
input type $F~i$ as produced by the callback.

The {\ACCHOICE} command is as before up to the additional type parameters.

The composition operator for two commands has three additional
callback arguments, which we call {\Asplit}, {\Across}, and
{\Ajoin}. The first callback {\Asplit} splits the input type into one part that gets
consumed by the first command and another that bypasses it.
The second callback {\Across} joins the output type and the bypass into the
input type for the second command.
The third callback {\Ajoin} combines the outputs of the two commands.

In the context-free case, the behavior of the {\AMU} command is more
general than the simple tail-recursive iteration in
Section~\ref{sec:going-circles}. It takes a loop body that transforms
$A$s into $B$s with these same types pushed on the stacks.

The {$\ACONTINUE~i$} command invokes the $i$th pending loop. To do so,
the current type must match the typing of the loop, which we find by
lookup up the input and output type on the stacks at position $i$.

\subsection{Examples}
\label{sec:examples}

Before we delve into the interpreter, let's take stock what we
achieved by reviewing our old examples as well as a new one that
demonstrates the additional expressivity of context-free sessions.
\cstBinaryp

Compare these types with the corresponding ones from
Section~\ref{sec:going-circles}, where the protocol fragments are
metafunctions that take a continuation session as a parameter. No such
parameterization is needed with context-free session types. They are
intrinsically compositional and modular.

The final example {\Amanyunaryp} contains a tail recursive part. The
second component of the choice must make use of the {\Atcfskip} type,
because the arm of a choice cannot be empty.

Let's consider servers that implement those protocols.
\cstCmdExamples

The code of the servers does not look that different from
before. However, unlike before, each server is now reusable as part of
the implementation of a larger protocol. One indication is the use of
the parameters $V$ and $W$ in the types, another is the lack of the
{\ACCLOSE} constructor which prescribes closing the connection.

As these protocols are still tail-recursive, their implementation uses
a binary composition operator for commands that is tailored for this
use case. The split function feeds everything to the first command; the
cross operation ignores the bypassed value, passing only the output of
the first command to the second; and the join operation passes only
the output of the second command.
\cstTailComposition

To further pinpoint the difference between context-free sessions and
regular session we finally present a protocol that
exploits the power of context-free session types by going beyond tail
recursion: servers that receive and send binary trees. 
\cstTreep

The session types {\Aleafp} and {\Abranchp} encode receiving a leaf and
a branch of the binary tree type {\AIntTree}. The session type
{\Atreep} provides the enclosing recursion and choice between the leaf
and branch protocols. As {\Abranchp} contains two recursive calls,
the protocol {\Atreep} is no longer tail recursive.
\cstRecvTree
The receiver for the leaf (choice {\Azero}) wraps the received value in the {\ACLeaf}
constructor and returns it.
The receiver for the branch (choice {\Asuc~\Azero}) obtains the left and right subtrees by
recursive calls {\ACONTINUE~\Azero} and combines the outcomes using
the {\ACBranch} constructor. The {\Asplit} and {\Across} callbacks
just manipulate values of unit type.

The sender of a tree makes full use of the callbacks for command
composition. 
\cstSendTree
The function {\AsplitTree} implements essentially (one half of) the
isomorphism that unfolds the recursion in type {\AIntTree};
function {\AIntTreeF} helps construct its type. It is straightforward
to construct the sender of a tree using these helpers. Using
{\AsplitTree} as the callback for {\ACSELECT}, we find that the
{\Azero} branch is invoked with an integer and the {\Asuc~\Azero}
branch is invoked with a pair of the two subtrees. The {\Asplit}
callback is just the identity, so that the left subtree is passed to
the first recursive call and the right one to the second.

\subsection{Interpretation}
\label{sec:interpretation}

We show the full {\Aexecutor} function and finish with a rationale why
the callback interface in this section is not monadic.
\cstExec

The implementation of {\ACSKIP} just executes the action.
Sending  {\ACSEND}, receiving  {\ACRECV}, and choice {\ACCHOICE} are as usual.
The dynamic selection is improved with respect to
Section~\ref{sec:select-choice}. Previously, there was no connection
between the selected label and the executed continuation, so that a
bug in the interpreter might introduce a mismatch. In the present
interpreter, no such mismatch is possible because it would be a type
error to invoke any other continuation than $\Acont~i$ with $ai$.

Composition first splits the input using the {\Asplit} callback. It performs the left command, obtains its final state in
$b_1$, combines it with the bypass value $a'$ using {\Across}, and
then performs the right command. It obtains its final state in
$b_2$ and returns $\Ajoin~a_1~a_2$.
The implementations for {\AMU} and {\ACONTINUE} are as before in
Section~\ref{sec:going-circles}. The difference is that the
{\ACONTINUE} operation may now appear in the context of a composition
which provides a nontrivial continuation. In
Section~\ref{sec:going-circles}, the function {\Aexecutor} is tail
recursive, but here it is not!

Now we turn to the question of the monadic callback interface. 
Examination of the code reveals that the state monad is no longer the most appropriate model of a
callback. On first sight, the type $A \to 
M~B$ is the same as $\AReaderT~A~M~B $. However, the case 
for composition shows that a reader does not fit because
the type $A$ of the reader's source changes between recursive calls to
the interpreter. Moreover, the callbacks for composition do not fit
the pattern of the reader monad at all.
Thus, we refrain from the monadic interface.

\section{Handling multiple channels}
\label{sec:handl-mult-chann}
\input{latex/ST-multichannel.tex}

We have to amend some final elements to fully encompass traditional binary
session types: a thread can open and manipulate many channels at a
time and channels can be delegated. So far, our interfaces were
restricted to single channels and to transmitting pure data (i.e., no channels). We
now turn to lifting these restrictions.

To concentrate on the new issues, 
we start out by slightly rephrasing session types as we
know them from Section~\ref{sec:select-choice}. In particular, we
restrict to binary branching and leave the extension
to finitary branching as well as the addition of recursion as an
exercise to the reader. Generally, these types describe 
the communication on a single channel, as before. There are two novel
aspects: we factorize the specification of the direction of
communication and we add a special type for channel delegation, i.e.,
sending or receiving a channel.

\multiSession

An actual multichannel session type describes the interleaved
communication on all channels at once. It comes with features and
restrictions very similar to multiparty session types \cite{DBLP:journals/jacm/HondaYC16}.

\multiMSesson

A multichannel session type {\AMSession~n}, ranged over by $M$, is indexed by the number $n$ of
channels that it governs. Multichannel session types are loosely based on Wadler's GV
calculus \cite{DBLP:journals/jfp/Wadler14}. In particular, the
connection topology of multichannel programs is restricted in the same
way as in Wadler's GV, with the pleasing
consequence that they are guaranteed to be free of deadlocks.

The names of the channels are represented by
de Bruijn indices. Unlike in other embedded implementations, a
multichannel type covers the full choreography of a communicating
application: connecting (forking) to a new process is combined with channel creation, branching and transmitting values as usual, closing a
channel, delegation, and terminating the application after all
protocols are finished and their respective channels closed.

Each multichannel type, except {\ACconnect} and {\ACterminate}, takes a parameter
$c$ that identifies the channel on which the command operates. 
The type to transmit a value (\ACtransmit) is as before, save the
direction and channel parameters. Branching (\ACbranch) is as before
except the extra function {\ACausality}: it ensures that a branch is
reasonable by enforcing that the session on every channel except $c$
is not affected by the branch. This restriction is called
``Causality'' in the context of multiparty session types. It is needed
because only the party on other end of channel $c$ knows about the
branching, but the others do not.

The {\ACclose} type has a continuation parameter because a channel
can be closed in the middle of a choreography without terminating the
multichannel protocol. Once all channels are closed, the choreography
can be concluded using the type {\ACterminate}.

The type {\ACconnect} indicates creation of a new thread which runs protocol $M_1$
whereas the current thread continues with $M_2$. The currently open
channels are distributed among the threads according to the {\ASplit}
parameter.
Communication with the new thread is established by 
creating a new channel, which is mapped to address {\Azero} in both
threads. Moreover, the {\ACheckDual} predicate makes sure that the
session types of the two ends of this channel are dual to one
another.\footnote{We elide details about the {\ASplit} type as well as
  duality from the paper for space reasons. See the supplement.}

The delegation types deal with sending (receiving) a channel over
another channel. Their types are sufficiently different so as not to
factor out the direction. To send a channel $j$ using {\ACdelegateOUT}, we need to know that
the transmitted channel is not the same channel we are sending on, its simple session type,
and that the continuation command has one fewer channel.
To receive a channel using {\ACdelegateIN} we just map it to address
{\Azero} in the continuation session.

The function {\Aproject} is similar to the projection from global
types to local types in multiparty session types. We make use of
auxiliary functions that manipulate de Bruijn indices:
{\AlocateSplit~sp-c~c} determines the target thread of channel $c$ and
its local index; {\Aadjust} changes the index to account for a channel
that is removed from choreography. {\ACausality} and {\ACheckDual} are
implemented using the obvious formulas.
\multiProjection

Given the preceding discussion as well as the discussion in previous
sections, the types of commands follow directly.

\multiCmd

Likewise, the interpreter {\Aexec} is an easy exercise.

\multiExec

\section{Discussion}
\label{sec:discussion}



\subsection{Selection and choice with vectors}
\label{sec:select-choice-with}

The reader may wonder why we do not define the constructors for
selection and choice using vectors of length $n$, rather than
functions from $\AFin~n \to \ASession$ (which is isomorphic). Using a function extends
directly to the definition of the {\ACommand} type where the {\ASession} index of the
continuation command depends on the function argument $i : \AFin~n$.
We would have to define a special dependent vector type to achieve similar
expressivity. 

\subsection{Alternative representations}
\label{sec:code-generation}

Instead of interpreting a command value, we could compile it to a
custom library, following the lead of the related work
\cite{DBLP:journals/pacmpl/00020HNY20,DBLP:conf/cc/Miu0Y021}.
Such a compiler can be obtained specializing the command interpreter
with respect to single commands.

The resulting \emph{denotational implementation} corresponds to a library implementation that exposes
the commands, which are reified in a datatype in our approach, as
functions. This approach has been pioneered by Reynolds \cite{Reynolds1994} and
subsequently applied, e.g., in the context of partial evaluation and
program analysis
\cite{DBLP:journals/jfp/Thiemann99,DBLP:journals/jfp/CaretteKS09}.

Here is an excerpt of such an implementation derived from the
interpreter in Section~\ref{sec:finite-nonbr-simple}.
\stCombinators
The interpreters from other sections can be rephrased analogously
without sacrificing the advantages our approach. However, for program
development the command-based approach is more attractive because the
Agda interactive programming support features autocompletion for commands, but not
for combinators.

\subsection{Multiparty session types}
\label{sec:mult-sess-types}

We see no issues in extending the approach presented in this work to
protocols with more than two participants. We refrained from doing so
in this work to avoid the extra
complication. Section~\ref{sec:handl-mult-chann} already gives some
insight into the requirements for a full multiparty version.

\subsection{Verification}
\label{sec:verification}

A fly in the ointment of our approach is the proof obligation that
the interpreter does not break linearity. As the interpreters are very
simple, it is tempting to rely on manual inspection.

To obtain a formal and possible mechanized proof, we envision a
semantics in terms of a free monad
\cite{DBLP:journals/jfp/Swierstra08,DBLP:conf/haskell/KiselyovI15}
with uninterpreted occurrences of 
the operations of our primitive channel IO API. The actual proof might be
conducted using interaction trees
\cite{DBLP:journals/pacmpl/XiaZHHMPZ20}, a mechanized framework for
representing about recursive and impure programs and  reasoning about them.

\section{Related work}
\label{sec:related-work}

In this section, we review how different library implementations of
session types deal with linearity. Specifically, we do not consider
dedicated language implementations like
SILL \cite{DBLP:conf/esop/ToninhoCP13},
SEPI \cite{DBLP:conf/sefm/FrancoV13},
FreeST \cite{DBLP:journals/iandc/AlmeidaMTV22},
or Links \cite{lindley17:_light_funct_session_types}. These
implementations come with dedicated type checker that properly treat
linearity at compile time.

There are libraries with fully dynamic enforcement of a session type
discipline
\cite{DBLP:conf/rv/HuNYDH13,DBLP:journals/pacmpl/MelgrattiP17}. They
suffer from run-time overhead as they have to check every
communication operation and they lack guarantees as they terminate the
protocol when an error is detected at some peer. 

Several libraries perform
linearity checks at run time. While a check for at most one (affine)
use of a resource can be performed with a single bit
\cite{DBLP:conf/esop/TovP10},  checks for linearity are more
expensive, but still deemed lightweight
\cite{DBLP:conf/fase/HuY16}.

Several libraries statically enforce linearity by encoding it
using parameterized monads
\cite{SackmanE08,DBLP:conf/haskell/PucellaT08},  polymorphism
\cite{DBLP:journals/scp/ImaiYY19}, or higher-order 
abstract syntax \cite{DBLP:conf/haskell/LindleyM16}. While these
encodings cleverly exploit the facilities of the host language and
support type inference, they
are nontrivial to explain and yield types that are not easy on the
programmer. No such cleverness is needed in our approach; types are
human-readable and the interactive Agda system helps with constructing
types and programs, but session types are not inferred.

None of the existing embeddings offers a feature like our dynamic
selection. However, dynamic selection can be viewed as a special case
of label-dependent session types
\cite{DBLP:journals/pacmpl/ThiemannV20}, so that our  approach
implements part of that theory, too.

While the dynamic and object-oriented libraries support session
subtyping, our approach currently does not support subtyping.

The most closely related work is by Miu, Ferreira, Yoshida, and Zhou
\cite{DBLP:conf/cc/Miu0Y021}. They develop an
implementation of multiparty session types in TypeScript by generation
of custom libraries from a protocol specification. Their
implementation guarantees freedom from
communication errors, including deadlocks, communication
mismatches, channel usage violation or cancellation errors. They
generate TypeScript APIs in callback style, where the finite state
machine underlying the communication endpoints is reified in terms of
interfaces. Sending and receiving are both encoded via callbacks 
into the library or into the user program, as appropriate. The
generated APIs encapsulate all primitive communication operations.

\subsection{Haskell}
\label{sec:hask-impl}

The earliest Haskell implementation by Neubauer \cite{DBLP:conf/padl/NeubauerT04}
emphasizes the modeling of the type of a single session using phantom
types. Duality is implemented using type classes with functional
dependencies. Linearity is not considered.

Sackman \cite{SackmanE08} and Tov \cite{DBLP:conf/popl/TovP11} model
multiple channels using a parameterized monad that is indexed by a mapping
from channel names to their current session types. The mapping changes
with each operation to keep track of the current state of all
channels. The monadic interface guarantees linearity.

This approach leaves more freedom to the programmer than our proposal
in Section~\ref{sec:handl-mult-chann}. While individual channel types
are independent in their approach, our multi-channel session types
impose a choreography on all connections, which is closer in spirit to
multiparty session types. 

Lindley and Morris \cite{DBLP:conf/haskell/LindleyM16} extend a HOAS
encoding of linear lambda calculus with monads and session primitives
in the style of GV \cite{DBLP:journals/jfp/GayV10}. They call their approach
``parameterized tagless'' \cite{DBLP:journals/jfp/CaretteKS09}, which means that they encode the syntax of
lambda calculus and the session primitives in term of parameterized
functions in a type class with implementations provided subsequently.

Orchard and Yoshida \cite{DBLP:conf/popl/OrchardY16} discuss
connections between session types and effect systems. They present an
implementation of session types in Haskell via an effect system
encoding based on graded monads. Here the ``grading'' keeps track of
the currently active channels and their session types whereas the
monadic structure provides proper sequencing.

\subsection{OCaml}
\label{sec:ocaml}

The implementation of FuSe \cite{DBLP:journals/jfp/Padovani17}
consists of a typed layer on top of untyped channels (just like our
approach).  It provides an API inspired by GV, supports
type inference, and checks linearity at run time. The approach is
extensible to context-free session types
\cite{DBLP:journals/toplas/Padovani19} at the expense of some user
annotations (while keeping type inference).

Session OCaml \cite{DBLP:journals/scp/ImaiYY19} enforces linearity by
parametric polymorphism based on a technique by
Garrigue\footnote{See \url{https://github.com/garrigue/safeio}.} with significant
extensions to deal with session types.
It can handle a fixed number of channels at the same time with
globally defined channel names (slot names). The technique hinges on
the polymorphic types of the global slot names.
Session types are inferred from programs that can be written in a notation
similar to FuSe and GV.

\subsection{Java}
\label{sec:java}

The early interface proposed in ``Session-based distributed programming in Java''
\cite{DBLP:conf/ecoop/HuYH08} relied on special syntax for session
communication and a preprocessor to limit aliasing.
Mungo \cite{DBLP:journals/scp/KouzapasDPG18} and Bica
\cite{DBLP:journals/corr/abs-1205-5344} use similar ideas to implement
typestate and sessions.

Hu and Yoshida \cite{DBLP:conf/fase/HuY16} pioneered code generating
approaches for implementing multiparty session types. They generate protocol-specific
endpoint APIs from multiparty session  
types for Java, but claim generality of their approach for mainstream
languages. They start from the observation that the behavior of an
endpoint of a communication can be represented by a finite state
machine. Each of these states is reified as a state channel type with
methods that imply communication operations corresponding to state
transitions. As Java places no restrictions on object uses, they
deploy ``very light run-time checks in the 
generated API that enforce a linear usage discipline on instances of
the channel types.'' Their abstract I/O state interfaces are closest
to the facilities that we provide.

While their approach reifies the different possible states of a
communication endpoint, these states are implicit in our
approach. Moreover, while an API based on finite state machines is
entirely appropriate for servers, we can provide additional
flexibility for implementing clients of recursive protocols though
commands like {\AUNROLL} described in
Section~\ref{sec:what-about-client}.

Scalas work for Scala \cite{DBLP:conf/ecoop/ScalasY16} follows similar
ideas with run-time checks for linearity. More recent work
\cite{DBLP:conf/ecoop/CledouEJP22} improves on the flexibility by
relying on advanced typing features in Scala~3.

\subsection{Rust}
\label{sec:rust}

Several recent implementations of session types \cite{DBLP:conf/ecoop/ChenBT22,DBLP:conf/ecoop/LagaillardieNY22,DBLP:conf/ppopp/CutnerYV22,DBLP:conf/coordination/CutnerY21,DBLP:conf/coordination/LagaillardieNY20,DBLP:journals/corr/abs-1909-05970,DBLP:conf/icfp/JespersenML15} rely on Rust, a
language with uniqueness types and ownership, all checked at compile
time. While uniqueness types are quite similar to affine types
(describing values that can be used at most once), they do not quite
get to the level needed for sessions: having an affine session type
for a channel means that an agent can drop the connection anytime
without finishing the protocol, which leads to deadlock at the other
end of the connection. With a proper linear session type, every agent has to
fulfil the protocol up to the closing of the connection.


\section{Conclusions}
\label{sec:conclusions}

Our paper demonstrates that a callback-based approach to
implementation session-typed programs is a perfect fit with (mildly)
dependently-typed functional programming. The intrinsically
session-typed way of writing program guarantees protocol fidelity and,
in some instances, deadlock freedom 
by construction. In connection with the callback approach, the host
language does not have to support linearity, so that programs are
statically safe, once the encapsulated library is verified.

Interestingly, the changed angle of attack revealed the possibility of
and the need for two novel constructions, the dynamic selection and
the {\AUNROLL} command for recursive sessions. The first one is
facilitated by our dependently-typed host language. The second one
arose from the need to write client programs for recursive protocols. 

The discussion in Section~\ref{sec:discussion} contains several
pointers for future work. It would also be interesting to investigate
ways to integrate subtyping (an obvious one being the insistence on
explicit coercions).


\bibliographystyle{ACM-Reference-Format}
\bibliography{references}


\end{document}

%% file: unicodeletters.tex
\usepackage{dsfont}
\usepackage{newunicodechar}
\newunicodechar{λ}{\ensuremath{\mathnormal\lambda}}
\newunicodechar{σ}{\ensuremath{\mathnormal\sigma}}
\newunicodechar{τ}{\ensuremath{\mathnormal\tau}}
\newunicodechar{π}{\ensuremath{\mathnormal\pi}}
\newunicodechar{ℕ}{\ensuremath{\mathbb{N}}}
\newunicodechar{∷}{\ensuremath{::}}
\newunicodechar{≡}{\ensuremath{\equiv}}
\newunicodechar{∀}{\ensuremath{\forall}}
\newunicodechar{ᴸ}{\ensuremath{^L}}
\newunicodechar{ᴿ}{\ensuremath{^R}}
\newunicodechar{ʳ}{\ensuremath{^r}}
\newunicodechar{ⱽ}{\ensuremath{^V}}
\newunicodechar{⟧}{\ensuremath{\rrbracket}}
\newunicodechar{⟦}{\ensuremath{\llbracket}}
\newunicodechar{⊤}{\ensuremath{\top}}
\newunicodechar{⊥}{\ensuremath{\bot}}
\newunicodechar{₁}{\ensuremath{_1}}
\newunicodechar{₂}{\ensuremath{_2}}
\newunicodechar{∈}{\ensuremath{\in}}
\newunicodechar{₀}{\ensuremath{_0}}
\newunicodechar{′}{\ensuremath{'}}
\newunicodechar{ˢ}{\ensuremath{^S}}
\newunicodechar{ᴬ}{\ensuremath{^A}}
\newunicodechar{∘}{\ensuremath{\circ}}
\newunicodechar{𝟙}{\ensuremath{\mathds{1}}}  
\newunicodechar{𝟘}{\ensuremath{\mathds{O}}}
\newunicodechar{ᴾ}{\ensuremath{^P}}
\newunicodechar{ᵀ}{\ensuremath{^T}}
\newunicodechar{⊎}{\ensuremath{\uplus}}
\newunicodechar{ι}{\ensuremath{\iota}}
\newunicodechar{⇐}{\ensuremath{\Leftarrow}}
\newunicodechar{⇒}{\ensuremath{\Rightarrow}}
\newunicodechar{∎}{\ensuremath{\mathnormal\blacksquare}}
\newunicodechar{➙}{\ensuremath{\to^P}}
\newunicodechar{Δ}{\ensuremath{\Delta}}
\newunicodechar{∅}{\ensuremath{\emptyset}}
\newunicodechar{⁺}{\ensuremath{^+}}
\newunicodechar{𝕏}{\ensuremath{\mathbb{X}}}
\newunicodechar{∙}{\ensuremath{\cdot}}
\newunicodechar{⁇}{\ensuremath{?}}
\newunicodechar{‼}{\ensuremath{!}}
\newunicodechar{⊕}{\ensuremath{\oplus}}
\newunicodechar{ℤ}{\ensuremath{\mathbb{Z}}}
\newunicodechar{μ}{\ensuremath{\mu}}
\newunicodechar{∃}{\ensuremath{\exists}}
\newunicodechar{⨟}{\ensuremath{\fatsemi}}
\newunicodechar{Σ}{\ensuremath{\Sigma}}
\newunicodechar{ᵣ}{\ensuremath{_r}}
\newunicodechar{≢}{\ensuremath{\nequiv}}
\newunicodechar{≟}{\ensuremath{\stackrel{{\tiny?}}{=}}}

%% file: agdamacros.tex
\newcommand\Aamp{\AgdaFunction{\ensuremath{\&}}}
\newcommand\Att{\AgdaInductiveConstructor{tt}}
\newcommand\Ado{\AgdaKeyword{do}}
\newcommand\AZ{\AgdaDatatype{ℤ}}
\newcommand\Asuc{\AgdaInductiveConstructor{suc}}
\newcommand\Azero{\AgdaInductiveConstructor{zero}}

\newcommand\Aint{\AgdaInductiveConstructor{int}}
\newcommand\Abool{\AgdaInductiveConstructor{bool}}
\newcommand\Atend{\AgdaInductiveConstructor{end}}
\newcommand\Atsend[2]{\AgdaInductiveConstructor{‼{\textcolor{black}{\ensuremath{#1}}}∙{\textcolor{black}{\ensuremath{#2}}}}}
\newcommand\Atrecv[2]{\AgdaInductiveConstructor{⁇{\textcolor{black}{\ensuremath{#1}}}∙{\textcolor{black}{\ensuremath{#2}}}}}
\newcommand\Atcfsend[1]{\AgdaInductiveConstructor{‼{\textcolor{black}{\ensuremath{#1}}}}}
\newcommand\Atcfrecv[1]{\AgdaInductiveConstructor{⁇{\textcolor{black}{\ensuremath{#1}}}}}
\newcommand\Atcfcomp[2]{\AgdaInductiveConstructor{{\textcolor{black}{\ensuremath{#1}}}⨟{\textcolor{black}{\ensuremath{#2}}}}}
\newcommand\Atcfskip{\AgdaInductiveConstructor{skip}}
\newcommand\ACSKIP{\AgdaInductiveConstructor{SKIP}}

\newcommand\ACCLOSE{\AgdaInductiveConstructor{CLOSE}}

\newcommand\ACconnect{\AgdaInductiveConstructor{connect}}
\newcommand\ACterminate{\AgdaInductiveConstructor{terminate}}
\newcommand\ACdelegateIN{\AgdaInductiveConstructor{delegateIN}}
\newcommand\ACdelegateOUT{\AgdaInductiveConstructor{delegateOUT}}
\newcommand\ACtransmit{\AgdaInductiveConstructor{transmit}}
\newcommand\ACbranch{\AgdaInductiveConstructor{branch}}
\newcommand\ACclose{\AgdaInductiveConstructor{close}}
\newcommand\ACSEND{\AgdaInductiveConstructor{SEND}}
\newcommand\ACRECV{\AgdaInductiveConstructor{RECV}}
\newcommand\ACSELECT{\AgdaInductiveConstructor{SELECT}}
\newcommand\ACCHOICE{\AgdaInductiveConstructor{CHOICE}}
\newcommand\AFin{\AgdaDatatype{Fin}}
\newcommand\ACommand{\AgdaDatatype{Cmd}}
\newcommand\ACommandStore{\AgdaDatatype{CmdStore}}
\newcommand\ASession{\AgdaDatatype{Session}}
\newcommand\ASplit{\AgdaDatatype{Split}}
\newcommand\AMSession{\AgdaDatatype{MSession}}
\newcommand\ASet{\AgdaDatatype{Set}}
\newcommand\ASetOne{\AgdaDatatype{Set$_1$}}
\newcommand\Abinaryp{\AgdaFunction{binaryp}}
\newcommand\Aunaryp{\AgdaFunction{unaryp}}

\newcommand\ACausality{\AgdaFunction{Causality}}
\newcommand\ACheckDual{\AgdaFunction{CheckDual0}}
\newcommand\Aexecutor{\AgdaFunction{exec}}
\newcommand\Aexec{\AgdaFunction{exec}}
\newcommand\AIO{\AgdaFunction{IO}}
\newcommand\Amu{\AgdaInductiveConstructor{\ensuremath{\mu}}}
\newcommand\AMU{\AgdaInductiveConstructor{LOOP}}
\newcommand\AUNROLL{\AgdaInductiveConstructor{UNROLL}}
\newcommand\ACONTINUE{\AgdaInductiveConstructor{CONTINUE}}
\newcommand\Aback{\AgdaInductiveConstructor{\ensuremath{`}}}
\newcommand\Amanyunaryp{\AgdaFunction{many-unaryp}}
\newcommand\Arestart{\AgdaFunction{restart}}
\newcommand\ASerialize{\AgdaRecord{Serialize}}
\newcommand\ARawMonad{\AgdaRecord{RawMonad}}
\newcommand\AReaderT{\AgdaRecord{ReaderT}}
\newcommand\AStateT{\AgdaRecord{StateT}}
\newcommand\Aput{\AgdaFunction{put}}
\newcommand\Amodify{\AgdaFunction{modify}}
\newcommand\Aget{\AgdaFunction{get}}
\newcommand\Aproject{\AgdaFunction{project}}
\newcommand\AlocateSplit{\AgdaFunction{locate-split}}
\newcommand\Aadjust{\AgdaFunction{adjust}}
\newcommand\Aid{\AgdaFunction{id}}
\newcommand\AprimSend{\AgdaFunction{primSend}}
\newcommand\AprimRecv{\AgdaFunction{primRecv}}
\newcommand\Aleafp{\AgdaFunction{leafp}}
\newcommand\Abranchp{\AgdaFunction{branchp}}
\newcommand\Atreep{\AgdaFunction{treep}}
\newcommand\AIntTree{\AgdaDatatype{IntTree}}
\newcommand\AIntTreeF{\AgdaFunction{IntTreeF}}
\newcommand\ACLeaf{\AgdaInductiveConstructor{Leaf}}
\newcommand\ACBranch{\AgdaInductiveConstructor{Branch}}
\newcommand\AtoN{\AgdaFunction{toℕ}}
\newcommand\Asplit{\AgdaFunction{split}}
\newcommand\Across{\AgdaFunction{cross}}
\newcommand\Ajoin{\AgdaFunction{join}}
\newcommand\AsplitTree{\AgdaFunction{splitTree}}
\newcommand\Acont{\AgdaFunction{cont}}
\newcommand\Adollar{\AgdaOperator{\AgdaFunction{\AgdaUnderscore{}\$\AgdaUnderscore{}}}}


%% file: latex/ST-finite-nonbranching.tex
\begin{code}[hide]%
\>[0]\AgdaSymbol{\{-\#}\AgdaSpace{}%
\AgdaKeyword{OPTIONS}\AgdaSpace{}%
\AgdaPragma{--guardedness}\AgdaSpace{}%
\AgdaSymbol{\#-\}}\AgdaSpace{}%
\AgdaComment{\{-\ required\ for\ IO\ -\}}\<%
\\
\>[0]\AgdaKeyword{module}\AgdaSpace{}%
\AgdaModule{ST-finite-nonbranching}\AgdaSpace{}%
\AgdaKeyword{where}\<%
\\
\\[\AgdaEmptyExtraSkip]%
\>[0]\AgdaKeyword{open}\AgdaSpace{}%
\AgdaKeyword{import}\AgdaSpace{}%
\AgdaModule{Data.Bool}\AgdaSpace{}%
\AgdaKeyword{using}\AgdaSpace{}%
\AgdaSymbol{(}\AgdaDatatype{Bool}\AgdaSymbol{;}\AgdaSpace{}%
\AgdaInductiveConstructor{true}\AgdaSymbol{;}\AgdaSpace{}%
\AgdaInductiveConstructor{false}\AgdaSymbol{)}\<%
\\
\>[0]\AgdaKeyword{open}\AgdaSpace{}%
\AgdaKeyword{import}\AgdaSpace{}%
\AgdaModule{Data.Fin}\AgdaSpace{}%
\AgdaKeyword{using}\AgdaSpace{}%
\AgdaSymbol{(}\AgdaDatatype{Fin}\AgdaSymbol{;}\AgdaSpace{}%
\AgdaInductiveConstructor{zero}\AgdaSymbol{;}\AgdaSpace{}%
\AgdaInductiveConstructor{suc}\AgdaSymbol{)}\<%
\\
\>[0]\AgdaKeyword{open}\AgdaSpace{}%
\AgdaKeyword{import}\AgdaSpace{}%
\AgdaModule{Data.Integer}\AgdaSpace{}%
\AgdaKeyword{using}\AgdaSpace{}%
\AgdaSymbol{(}\AgdaDatatype{ℤ}\AgdaSymbol{;}\AgdaSpace{}%
\AgdaFunction{0ℤ}\AgdaSymbol{;}\AgdaSpace{}%
\AgdaOperator{\AgdaFunction{\AgdaUnderscore{}+\AgdaUnderscore{}}}\AgdaSymbol{;}\AgdaSpace{}%
\AgdaOperator{\AgdaFunction{-\AgdaUnderscore{}}}\AgdaSymbol{)}\<%
\\
\>[0]\AgdaKeyword{open}\AgdaSpace{}%
\AgdaKeyword{import}\AgdaSpace{}%
\AgdaModule{Data.Nat}\AgdaSpace{}%
\AgdaKeyword{using}\AgdaSpace{}%
\AgdaSymbol{(}\AgdaDatatype{ℕ}\AgdaSymbol{;}\AgdaSpace{}%
\AgdaInductiveConstructor{zero}\AgdaSymbol{;}\AgdaSpace{}%
\AgdaInductiveConstructor{suc}\AgdaSymbol{)}\<%
\\
\>[0]\AgdaKeyword{open}\AgdaSpace{}%
\AgdaKeyword{import}\AgdaSpace{}%
\AgdaModule{Data.Product}\AgdaSpace{}%
\AgdaKeyword{using}\AgdaSpace{}%
\AgdaSymbol{(}\AgdaOperator{\AgdaFunction{\AgdaUnderscore{}×\AgdaUnderscore{}}}\AgdaSymbol{;}\AgdaSpace{}%
\AgdaRecord{Σ}\AgdaSymbol{;}\AgdaSpace{}%
\AgdaField{proj₁}\AgdaSymbol{;}\AgdaSpace{}%
\AgdaField{proj₂}\AgdaSymbol{;}\AgdaSpace{}%
\AgdaOperator{\AgdaFunction{<\AgdaUnderscore{},\AgdaUnderscore{}>}}\AgdaSymbol{)}\AgdaSpace{}%
\AgdaKeyword{renaming}\AgdaSpace{}%
\AgdaSymbol{(}\AgdaOperator{\AgdaInductiveConstructor{\AgdaUnderscore{},\AgdaUnderscore{}}}\AgdaSpace{}%
\AgdaSymbol{to}\AgdaSpace{}%
\AgdaOperator{\AgdaInductiveConstructor{⟨\AgdaUnderscore{},\AgdaUnderscore{}⟩}}\AgdaSymbol{)}\<%
\\
\>[0]\AgdaKeyword{open}\AgdaSpace{}%
\AgdaKeyword{import}\AgdaSpace{}%
\AgdaModule{Data.Sum}\AgdaSpace{}%
\AgdaKeyword{using}\AgdaSpace{}%
\AgdaSymbol{(}\AgdaOperator{\AgdaDatatype{\AgdaUnderscore{}⊎\AgdaUnderscore{}}}\AgdaSymbol{;}\AgdaSpace{}%
\AgdaInductiveConstructor{inj₁}\AgdaSymbol{;}\AgdaSpace{}%
\AgdaInductiveConstructor{inj₂}\AgdaSymbol{)}\<%
\\
\>[0]\AgdaKeyword{open}\AgdaSpace{}%
\AgdaKeyword{import}\AgdaSpace{}%
\AgdaModule{Data.Vec}\AgdaSpace{}%
\AgdaKeyword{using}\AgdaSpace{}%
\AgdaSymbol{(}\AgdaDatatype{Vec}\AgdaSymbol{;}\AgdaSpace{}%
\AgdaInductiveConstructor{[]}\AgdaSymbol{;}\AgdaSpace{}%
\AgdaOperator{\AgdaInductiveConstructor{\AgdaUnderscore{}∷\AgdaUnderscore{}}}\AgdaSymbol{)}\<%
\\
\\[\AgdaEmptyExtraSkip]%
\>[0]\AgdaKeyword{open}\AgdaSpace{}%
\AgdaKeyword{import}\AgdaSpace{}%
\AgdaModule{Data.Unit}\AgdaSpace{}%
\AgdaKeyword{using}\AgdaSpace{}%
\AgdaSymbol{(}\AgdaRecord{⊤}\AgdaSymbol{;}\AgdaSpace{}%
\AgdaInductiveConstructor{tt}\AgdaSymbol{)}\<%
\\
\\[\AgdaEmptyExtraSkip]%
\>[0]\AgdaKeyword{open}\AgdaSpace{}%
\AgdaKeyword{import}\AgdaSpace{}%
\AgdaModule{Function.Base}\AgdaSpace{}%
\AgdaKeyword{using}\AgdaSpace{}%
\AgdaSymbol{(}\AgdaOperator{\AgdaFunction{case\AgdaUnderscore{}of\AgdaUnderscore{}}}\AgdaSymbol{;}\AgdaSpace{}%
\AgdaOperator{\AgdaFunction{\AgdaUnderscore{}∘\AgdaUnderscore{}}}\AgdaSymbol{;}\AgdaSpace{}%
\AgdaFunction{const}\AgdaSymbol{;}\AgdaSpace{}%
\AgdaOperator{\AgdaFunction{\AgdaUnderscore{}\$\AgdaUnderscore{}}}\AgdaSymbol{;}\AgdaSpace{}%
\AgdaFunction{id}\AgdaSymbol{)}\<%
\\
\\[\AgdaEmptyExtraSkip]%
\>[0]\AgdaKeyword{open}\AgdaSpace{}%
\AgdaKeyword{import}\AgdaSpace{}%
\AgdaModule{IO}\<%
\\
\\[\AgdaEmptyExtraSkip]%
\>[0]\AgdaKeyword{open}\AgdaSpace{}%
\AgdaKeyword{import}\AgdaSpace{}%
\AgdaModule{Utils}\<%
\\
\\[\AgdaEmptyExtraSkip]%
\\[\AgdaEmptyExtraSkip]%
\>[0]\AgdaKeyword{pattern}\AgdaSpace{}%
\AgdaOperator{\AgdaInductiveConstructor{[\AgdaUnderscore{}]}}\AgdaSpace{}%
\AgdaBound{x}\AgdaSpace{}%
\AgdaSymbol{=}\AgdaSpace{}%
\AgdaBound{x}\AgdaSpace{}%
\AgdaOperator{\AgdaInductiveConstructor{∷}}\AgdaSpace{}%
\AgdaInductiveConstructor{[]}\<%
\\
\>[0]\AgdaKeyword{pattern}\AgdaSpace{}%
\AgdaOperator{\AgdaInductiveConstructor{[\AgdaUnderscore{},\AgdaUnderscore{}]}}\AgdaSpace{}%
\AgdaBound{x}\AgdaSpace{}%
\AgdaBound{y}\AgdaSpace{}%
\AgdaSymbol{=}\AgdaSpace{}%
\AgdaBound{x}\AgdaSpace{}%
\AgdaOperator{\AgdaInductiveConstructor{∷}}\AgdaSpace{}%
\AgdaBound{y}\AgdaSpace{}%
\AgdaOperator{\AgdaInductiveConstructor{∷}}\AgdaSpace{}%
\AgdaInductiveConstructor{[]}\<%
\\
\>[0]\AgdaKeyword{pattern}\AgdaSpace{}%
\AgdaOperator{\AgdaInductiveConstructor{[\AgdaUnderscore{},\AgdaUnderscore{},\AgdaUnderscore{}]}}\AgdaSpace{}%
\AgdaBound{x}\AgdaSpace{}%
\AgdaBound{y}\AgdaSpace{}%
\AgdaBound{z}\AgdaSpace{}%
\AgdaSymbol{=}\AgdaSpace{}%
\AgdaBound{x}\AgdaSpace{}%
\AgdaOperator{\AgdaInductiveConstructor{∷}}\AgdaSpace{}%
\AgdaBound{y}\AgdaSpace{}%
\AgdaOperator{\AgdaInductiveConstructor{∷}}\AgdaSpace{}%
\AgdaBound{z}\AgdaSpace{}%
\AgdaOperator{\AgdaInductiveConstructor{∷}}\AgdaSpace{}%
\AgdaInductiveConstructor{[]}\<%
\\
\\[\AgdaEmptyExtraSkip]%
\\[\AgdaEmptyExtraSkip]%
\>[0]\AgdaKeyword{variable}\<%
\\
\>[0][@{}l@{\AgdaIndent{0}}]%
\>[2]\AgdaGeneralizable{n}\AgdaSpace{}%
\AgdaGeneralizable{k}\AgdaSpace{}%
\AgdaSymbol{:}\AgdaSpace{}%
\AgdaDatatype{ℕ}\<%
\end{code}
\newcommand\stFiniteType{%
\begin{code}%
\>[0]\AgdaKeyword{data}\AgdaSpace{}%
\AgdaDatatype{Type}\AgdaSpace{}%
\AgdaSymbol{:}\AgdaSpace{}%
\AgdaPrimitive{Set}\AgdaSpace{}%
\AgdaKeyword{where}\<%
\\
\>[0][@{}l@{\AgdaIndent{0}}]%
\>[2]\AgdaInductiveConstructor{int}\AgdaSpace{}%
\AgdaSymbol{:}\AgdaSpace{}%
\AgdaDatatype{Type}\<%
\\
\>[2]\AgdaInductiveConstructor{bool}\AgdaSpace{}%
\AgdaSymbol{:}\AgdaSpace{}%
\AgdaDatatype{Type}\<%
\end{code}}
\begin{code}[hide]%
\>[2]\AgdaInductiveConstructor{nat}\AgdaSpace{}%
\AgdaSymbol{:}\AgdaSpace{}%
\AgdaDatatype{Type}\<%
\\
\>[2]\AgdaInductiveConstructor{fin}\AgdaSpace{}%
\AgdaSymbol{:}\AgdaSpace{}%
\AgdaDatatype{ℕ}\AgdaSpace{}%
\AgdaSymbol{→}\AgdaSpace{}%
\AgdaDatatype{Type}\<%
\\
\\[\AgdaEmptyExtraSkip]%
\>[0]\AgdaKeyword{module}\AgdaSpace{}%
\AgdaModule{formatting1}\AgdaSpace{}%
\AgdaKeyword{where}\<%
\end{code}
\newcommand\stBranchingType{%
\begin{code}%
\>[0][@{}l@{\AgdaIndent{1}}]%
\>[2]\AgdaKeyword{data}\AgdaSpace{}%
\AgdaDatatype{Session}\AgdaSpace{}%
\AgdaSymbol{:}\AgdaSpace{}%
\AgdaPrimitive{Set}\AgdaSpace{}%
\AgdaKeyword{where}\<%
\\
\>[2][@{}l@{\AgdaIndent{0}}]%
\>[4]\AgdaInductiveConstructor{⊕′}\AgdaSpace{}%
\AgdaSymbol{:}\AgdaSpace{}%
\AgdaSymbol{(}\AgdaBound{Si}\AgdaSpace{}%
\AgdaSymbol{:}\AgdaSpace{}%
\AgdaSymbol{(}\AgdaBound{i}\AgdaSpace{}%
\AgdaSymbol{:}\AgdaSpace{}%
\AgdaDatatype{Fin}\AgdaSpace{}%
\AgdaGeneralizable{k}\AgdaSymbol{)}\AgdaSpace{}%
\AgdaSymbol{→}\AgdaSpace{}%
\AgdaDatatype{Session}\AgdaSymbol{)}\AgdaSpace{}%
\AgdaSymbol{→}\AgdaSpace{}%
\AgdaDatatype{Session}\<%
\\
\>[4]\AgdaInductiveConstructor{\&′}\AgdaSpace{}%
\AgdaSymbol{:}\AgdaSpace{}%
\AgdaSymbol{(}\AgdaBound{Si}\AgdaSpace{}%
\AgdaSymbol{:}\AgdaSpace{}%
\AgdaSymbol{(}\AgdaBound{i}\AgdaSpace{}%
\AgdaSymbol{:}\AgdaSpace{}%
\AgdaDatatype{Fin}\AgdaSpace{}%
\AgdaGeneralizable{k}\AgdaSymbol{)}\AgdaSpace{}%
\AgdaSymbol{→}\AgdaSpace{}%
\AgdaDatatype{Session}\AgdaSymbol{)}\AgdaSpace{}%
\AgdaSymbol{→}\AgdaSpace{}%
\AgdaDatatype{Session}\<%
\end{code}}
\newcommand\stFiniteSession{%
\begin{code}%
\>[0]\AgdaKeyword{data}\AgdaSpace{}%
\AgdaDatatype{Session}\AgdaSpace{}%
\AgdaSymbol{:}\AgdaSpace{}%
\AgdaPrimitive{Set}\AgdaSpace{}%
\AgdaKeyword{where}\<%
\\
\>[0][@{}l@{\AgdaIndent{0}}]%
\>[2]\AgdaOperator{\AgdaInductiveConstructor{‼\AgdaUnderscore{}∙\AgdaUnderscore{}}}\AgdaSpace{}%
\AgdaSymbol{:}\AgdaSpace{}%
\AgdaDatatype{Type}\AgdaSpace{}%
\AgdaSymbol{→}\AgdaSpace{}%
\AgdaDatatype{Session}\AgdaSpace{}%
\AgdaSymbol{→}\AgdaSpace{}%
\AgdaDatatype{Session}\<%
\\
\>[2]\AgdaOperator{\AgdaInductiveConstructor{⁇\AgdaUnderscore{}∙\AgdaUnderscore{}}}\AgdaSpace{}%
\AgdaSymbol{:}\AgdaSpace{}%
\AgdaDatatype{Type}\AgdaSpace{}%
\AgdaSymbol{→}\AgdaSpace{}%
\AgdaDatatype{Session}\AgdaSpace{}%
\AgdaSymbol{→}\AgdaSpace{}%
\AgdaDatatype{Session}\<%
\\
\>[2]\AgdaInductiveConstructor{end}\AgdaSpace{}%
\AgdaSymbol{:}\AgdaSpace{}%
\AgdaDatatype{Session}\<%
\end{code}}
\begin{code}[hide]%
\>[2]\AgdaInductiveConstructor{⊕′}\AgdaSpace{}%
\AgdaSymbol{:}\AgdaSpace{}%
\AgdaSymbol{∀}\AgdaSpace{}%
\AgdaSymbol{\{}\AgdaBound{k}\AgdaSymbol{\}}\AgdaSpace{}%
\AgdaSymbol{→}\AgdaSpace{}%
\AgdaSymbol{(}\AgdaBound{Si}\AgdaSpace{}%
\AgdaSymbol{:}\AgdaSpace{}%
\AgdaSymbol{(}\AgdaBound{i}\AgdaSpace{}%
\AgdaSymbol{:}\AgdaSpace{}%
\AgdaDatatype{Fin}\AgdaSpace{}%
\AgdaBound{k}\AgdaSymbol{)}\AgdaSpace{}%
\AgdaSymbol{→}\AgdaSpace{}%
\AgdaDatatype{Session}\AgdaSymbol{)}\AgdaSpace{}%
\AgdaSymbol{→}\AgdaSpace{}%
\AgdaDatatype{Session}\<%
\\
\>[2]\AgdaInductiveConstructor{\&′}\AgdaSpace{}%
\AgdaSymbol{:}\AgdaSpace{}%
\AgdaSymbol{∀}\AgdaSpace{}%
\AgdaSymbol{\{}\AgdaBound{k}\AgdaSymbol{\}}\AgdaSpace{}%
\AgdaSymbol{→}\AgdaSpace{}%
\AgdaSymbol{(}\AgdaBound{Si}\AgdaSpace{}%
\AgdaSymbol{:}\AgdaSpace{}%
\AgdaSymbol{(}\AgdaBound{i}\AgdaSpace{}%
\AgdaSymbol{:}\AgdaSpace{}%
\AgdaDatatype{Fin}\AgdaSpace{}%
\AgdaBound{k}\AgdaSymbol{)}\AgdaSpace{}%
\AgdaSymbol{→}\AgdaSpace{}%
\AgdaDatatype{Session}\AgdaSymbol{)}\AgdaSpace{}%
\AgdaSymbol{→}\AgdaSpace{}%
\AgdaDatatype{Session}\<%
\end{code}
\begin{code}[hide]%
\>[2]\AgdaInductiveConstructor{sel}\AgdaSpace{}%
\AgdaInductiveConstructor{chc}\AgdaSpace{}%
\AgdaSymbol{:}\AgdaSpace{}%
\AgdaSymbol{∀\{}\AgdaBound{k}\AgdaSymbol{\}}\AgdaSpace{}%
\AgdaSymbol{→}\AgdaSpace{}%
\AgdaDatatype{Vec}\AgdaSpace{}%
\AgdaDatatype{Session}\AgdaSpace{}%
\AgdaBound{k}\AgdaSpace{}%
\AgdaSymbol{→}\AgdaSpace{}%
\AgdaDatatype{Session}\<%
\\
\>[2]\AgdaInductiveConstructor{select}\AgdaSpace{}%
\AgdaInductiveConstructor{choice}\AgdaSpace{}%
\AgdaSymbol{:}\AgdaSpace{}%
\AgdaDatatype{Session}\AgdaSpace{}%
\AgdaSymbol{→}\AgdaSpace{}%
\AgdaDatatype{Session}\AgdaSpace{}%
\AgdaSymbol{→}\AgdaSpace{}%
\AgdaDatatype{Session}\<%
\\
\\[\AgdaEmptyExtraSkip]%
\>[0]\AgdaKeyword{pattern}\AgdaSpace{}%
\AgdaInductiveConstructor{recv}\AgdaSpace{}%
\AgdaBound{t}\AgdaSpace{}%
\AgdaBound{s}\AgdaSpace{}%
\AgdaSymbol{=}\AgdaSpace{}%
\AgdaOperator{\AgdaInductiveConstructor{⁇}}\AgdaSpace{}%
\AgdaBound{t}\AgdaSpace{}%
\AgdaOperator{\AgdaInductiveConstructor{∙}}\AgdaSpace{}%
\AgdaBound{s}\<%
\\
\>[0]\AgdaKeyword{pattern}\AgdaSpace{}%
\AgdaInductiveConstructor{send}\AgdaSpace{}%
\AgdaBound{t}\AgdaSpace{}%
\AgdaBound{s}\AgdaSpace{}%
\AgdaSymbol{=}\AgdaSpace{}%
\AgdaOperator{\AgdaInductiveConstructor{‼}}\AgdaSpace{}%
\AgdaBound{t}\AgdaSpace{}%
\AgdaOperator{\AgdaInductiveConstructor{∙}}\AgdaSpace{}%
\AgdaBound{s}\<%
\\
\\[\AgdaEmptyExtraSkip]%
\>[0]\AgdaKeyword{infixr}\AgdaSpace{}%
\AgdaNumber{20}\AgdaSpace{}%
\AgdaOperator{\AgdaInductiveConstructor{‼\AgdaUnderscore{}∙\AgdaUnderscore{}}}\AgdaSpace{}%
\AgdaOperator{\AgdaInductiveConstructor{⁇\AgdaUnderscore{}∙\AgdaUnderscore{}}}\<%
\\
\\[\AgdaEmptyExtraSkip]%
\>[0]\AgdaFunction{⊕}\AgdaSpace{}%
\AgdaSymbol{:}\AgdaSpace{}%
\AgdaDatatype{Vec}\AgdaSpace{}%
\AgdaDatatype{Session}\AgdaSpace{}%
\AgdaGeneralizable{n}\AgdaSpace{}%
\AgdaSymbol{→}\AgdaSpace{}%
\AgdaDatatype{Session}\<%
\\
\>[0]\AgdaFunction{⊕}\AgdaSpace{}%
\AgdaSymbol{=}\AgdaSpace{}%
\AgdaInductiveConstructor{⊕′}\AgdaSpace{}%
\AgdaOperator{\AgdaFunction{∘}}\AgdaSpace{}%
\AgdaFunction{vec→fin}\<%
\\
\\[\AgdaEmptyExtraSkip]%
\>[0]\AgdaFunction{\&}\AgdaSpace{}%
\AgdaSymbol{:}\AgdaSpace{}%
\AgdaDatatype{Vec}\AgdaSpace{}%
\AgdaDatatype{Session}\AgdaSpace{}%
\AgdaGeneralizable{n}\AgdaSpace{}%
\AgdaSymbol{→}\AgdaSpace{}%
\AgdaDatatype{Session}\<%
\\
\>[0]\AgdaFunction{\&}\AgdaSpace{}%
\AgdaSymbol{=}\AgdaSpace{}%
\AgdaInductiveConstructor{\&′}\AgdaSpace{}%
\AgdaOperator{\AgdaFunction{∘}}\AgdaSpace{}%
\AgdaFunction{vec→fin}\<%
\\
\\[\AgdaEmptyExtraSkip]%
\>[0]\AgdaComment{--\ service\ protocol\ for\ a\ binary\ function}\<%
\\
\>[0]\AgdaFunction{binaryp}\AgdaSpace{}%
\AgdaSymbol{:}\AgdaSpace{}%
\AgdaDatatype{Session}\<%
\\
\>[0]\AgdaComment{--\ service\ protocol\ for\ a\ unary\ function}\<%
\\
\>[0]\AgdaFunction{unaryp}\AgdaSpace{}%
\AgdaSymbol{:}\AgdaSpace{}%
\AgdaDatatype{Session}\<%
\\
\>[0]\AgdaComment{--\ service\ protocol\ for\ choosing\ between\ a\ binary\ and\ a\ unary\ function}\<%
\\
\>[0]\AgdaFunction{arithp}\AgdaSpace{}%
\AgdaSymbol{:}\AgdaSpace{}%
\AgdaDatatype{Session}\<%
\end{code}
\newcommand\stExampleBinpUnP{%
\begin{code}%
\>[0]\AgdaFunction{binaryp}\AgdaSpace{}%
\AgdaSymbol{=}\AgdaSpace{}%
\AgdaOperator{\AgdaInductiveConstructor{⁇}}\AgdaSpace{}%
\AgdaInductiveConstructor{int}\AgdaSpace{}%
\AgdaOperator{\AgdaInductiveConstructor{∙}}\AgdaSpace{}%
\AgdaOperator{\AgdaInductiveConstructor{⁇}}\AgdaSpace{}%
\AgdaInductiveConstructor{int}\AgdaSpace{}%
\AgdaOperator{\AgdaInductiveConstructor{∙}}\AgdaSpace{}%
\AgdaOperator{\AgdaInductiveConstructor{‼}}\AgdaSpace{}%
\AgdaInductiveConstructor{int}\AgdaSpace{}%
\AgdaOperator{\AgdaInductiveConstructor{∙}}\AgdaSpace{}%
\AgdaInductiveConstructor{end}\<%
\\
\>[0]\AgdaFunction{unaryp}\AgdaSpace{}%
\AgdaSymbol{=}\AgdaSpace{}%
\AgdaOperator{\AgdaInductiveConstructor{⁇}}\AgdaSpace{}%
\AgdaInductiveConstructor{int}\AgdaSpace{}%
\AgdaOperator{\AgdaInductiveConstructor{∙}}\AgdaSpace{}%
\AgdaOperator{\AgdaInductiveConstructor{‼}}\AgdaSpace{}%
\AgdaInductiveConstructor{int}\AgdaSpace{}%
\AgdaOperator{\AgdaInductiveConstructor{∙}}\AgdaSpace{}%
\AgdaInductiveConstructor{end}\<%
\end{code}}
\newcommand\stExampleArithP{%
\begin{code}%
\>[0]\AgdaFunction{arithp}\AgdaSpace{}%
\AgdaSymbol{=}\AgdaSpace{}%
\AgdaFunction{\&}\AgdaSpace{}%
\AgdaOperator{\AgdaInductiveConstructor{[}}\AgdaSpace{}%
\AgdaFunction{binaryp}\AgdaSpace{}%
\AgdaOperator{\AgdaInductiveConstructor{,}}\AgdaSpace{}%
\AgdaFunction{unaryp}\AgdaSpace{}%
\AgdaOperator{\AgdaInductiveConstructor{]}}\<%
\end{code}}
\begin{code}[hide]%
\>[0]\AgdaFunction{arithp-raw}\AgdaSpace{}%
\AgdaSymbol{=}\AgdaSpace{}%
\AgdaInductiveConstructor{\&′}\AgdaSpace{}%
\AgdaSymbol{\{}\AgdaNumber{2}\AgdaSymbol{\}}\AgdaSpace{}%
\AgdaSymbol{(λ\{}\AgdaSpace{}%
\AgdaInductiveConstructor{zero}\AgdaSpace{}%
\AgdaSymbol{→}\AgdaSpace{}%
\AgdaFunction{binaryp}\AgdaSpace{}%
\AgdaSymbol{;}\AgdaSpace{}%
\AgdaSymbol{(}\AgdaInductiveConstructor{suc}\AgdaSpace{}%
\AgdaInductiveConstructor{zero}\AgdaSymbol{)}\AgdaSpace{}%
\AgdaSymbol{→}\AgdaSpace{}%
\AgdaFunction{unaryp}\AgdaSymbol{\})}\<%
\\
\\[\AgdaEmptyExtraSkip]%
\>[0]\AgdaFunction{arithpv}\AgdaSpace{}%
\AgdaSymbol{:}\AgdaSpace{}%
\AgdaDatatype{Session}\<%
\\
\>[0]\AgdaFunction{arithpv}\AgdaSpace{}%
\AgdaSymbol{=}\AgdaSpace{}%
\AgdaInductiveConstructor{chc}\AgdaSpace{}%
\AgdaSymbol{(}\AgdaFunction{binaryp}\AgdaSpace{}%
\AgdaOperator{\AgdaInductiveConstructor{∷}}\AgdaSpace{}%
\AgdaSymbol{(}\AgdaFunction{unaryp}\AgdaSpace{}%
\AgdaOperator{\AgdaInductiveConstructor{∷}}\AgdaSpace{}%
\AgdaInductiveConstructor{[]}\AgdaSymbol{))}\<%
\\
\\[\AgdaEmptyExtraSkip]%
\>[0]\AgdaKeyword{variable}\<%
\\
\>[0][@{}l@{\AgdaIndent{0}}]%
\>[2]\AgdaGeneralizable{A}\AgdaSpace{}%
\AgdaGeneralizable{A′}\AgdaSpace{}%
\AgdaGeneralizable{A″}\AgdaSpace{}%
\AgdaGeneralizable{A₁}\AgdaSpace{}%
\AgdaGeneralizable{A₂}\AgdaSpace{}%
\AgdaSymbol{:}\AgdaSpace{}%
\AgdaPrimitive{Set}\<%
\\
\>[2]\AgdaGeneralizable{T}\AgdaSpace{}%
\AgdaGeneralizable{t}\AgdaSpace{}%
\AgdaSymbol{:}\AgdaSpace{}%
\AgdaDatatype{Type}\<%
\\
\>[2]\AgdaGeneralizable{S}\AgdaSpace{}%
\AgdaGeneralizable{s}\AgdaSpace{}%
\AgdaGeneralizable{s₁}\AgdaSpace{}%
\AgdaGeneralizable{s₂}\AgdaSpace{}%
\AgdaSymbol{:}\AgdaSpace{}%
\AgdaDatatype{Session}\<%
\\
\\[\AgdaEmptyExtraSkip]%
\>[0]\AgdaKeyword{module}\AgdaSpace{}%
\AgdaModule{type-formatting}\AgdaSpace{}%
\AgdaKeyword{where}\<%
\\
\>[0][@{}l@{\AgdaIndent{0}}]%
\>[2]\AgdaKeyword{postulate}\<%
\end{code}
\newcommand\stTypeInterpretationSignature{%
\begin{code}[inline]%
\>[2][@{}l@{\AgdaIndent{1}}]%
\>[4]\AgdaOperator{\AgdaPostulate{T⟦\AgdaUnderscore{}⟧}}\AgdaSpace{}%
\AgdaSymbol{:}\AgdaSpace{}%
\AgdaDatatype{Type}\AgdaSpace{}%
\AgdaSymbol{→}\AgdaSpace{}%
\AgdaPrimitive{Set}\<%
\end{code}}
\newcommand\stTypeInterpretation{%
\begin{code}%
\>[0]\AgdaOperator{\AgdaFunction{T⟦\AgdaUnderscore{}⟧}}\AgdaSpace{}%
\AgdaSymbol{:}\AgdaSpace{}%
\AgdaDatatype{Type}\AgdaSpace{}%
\AgdaSymbol{→}\AgdaSpace{}%
\AgdaPrimitive{Set}\<%
\\
\>[0]\AgdaOperator{\AgdaFunction{T⟦}}\AgdaSpace{}%
\AgdaInductiveConstructor{int}\AgdaSpace{}%
\AgdaOperator{\AgdaFunction{⟧}}\AgdaSpace{}%
\AgdaSymbol{=}\AgdaSpace{}%
\AgdaDatatype{ℤ}\<%
\\
\>[0]\AgdaOperator{\AgdaFunction{T⟦}}\AgdaSpace{}%
\AgdaInductiveConstructor{bool}\AgdaSpace{}%
\AgdaOperator{\AgdaFunction{⟧}}\AgdaSpace{}%
\AgdaSymbol{=}\AgdaSpace{}%
\AgdaDatatype{Bool}\<%
\end{code}}
\begin{code}[hide]%
\>[0]\AgdaOperator{\AgdaFunction{T⟦}}\AgdaSpace{}%
\AgdaInductiveConstructor{nat}\AgdaSpace{}%
\AgdaOperator{\AgdaFunction{⟧}}\AgdaSpace{}%
\AgdaSymbol{=}\AgdaSpace{}%
\AgdaDatatype{ℕ}\<%
\\
\>[0]\AgdaOperator{\AgdaFunction{T⟦}}\AgdaSpace{}%
\AgdaInductiveConstructor{fin}\AgdaSpace{}%
\AgdaBound{k}\AgdaSpace{}%
\AgdaOperator{\AgdaFunction{⟧}}\AgdaSpace{}%
\AgdaSymbol{=}\AgdaSpace{}%
\AgdaDatatype{Fin}\AgdaSpace{}%
\AgdaBound{k}\<%
\\
\\[\AgdaEmptyExtraSkip]%
\>[0]\AgdaKeyword{module}\AgdaSpace{}%
\AgdaModule{formatting2}\AgdaSpace{}%
\AgdaKeyword{where}\<%
\\
\>[0]\<%
\end{code}
\newcommand\stBranchingCommand{%
\begin{code}%
\>[0][@{}l@{\AgdaIndent{1}}]%
\>[2]\AgdaKeyword{data}\AgdaSpace{}%
\AgdaDatatype{Cmd}\AgdaSpace{}%
\AgdaSymbol{(}\AgdaBound{A}\AgdaSpace{}%
\AgdaSymbol{:}\AgdaSpace{}%
\AgdaPrimitive{Set}\AgdaSymbol{)}\AgdaSpace{}%
\AgdaSymbol{:}\AgdaSpace{}%
\AgdaDatatype{Session}\AgdaSpace{}%
\AgdaSymbol{→}\AgdaSpace{}%
\AgdaPrimitive{Set}\AgdaSpace{}%
\AgdaKeyword{where}\<%
\\
\>[2][@{}l@{\AgdaIndent{0}}]%
\>[4]\AgdaInductiveConstructor{SELECT}\AgdaSpace{}%
\AgdaSymbol{:}\AgdaSpace{}%
\AgdaSymbol{∀}\AgdaSpace{}%
\AgdaSymbol{\{}\AgdaBound{Si}\AgdaSymbol{\}}\AgdaSpace{}%
\AgdaSymbol{→}\AgdaSpace{}%
\AgdaSymbol{(}\AgdaBound{i}\AgdaSpace{}%
\AgdaSymbol{:}\AgdaSpace{}%
\AgdaDatatype{Fin}\AgdaSpace{}%
\AgdaGeneralizable{k}\AgdaSymbol{)}\AgdaSpace{}%
\AgdaSymbol{→}\AgdaSpace{}%
\AgdaDatatype{Cmd}\AgdaSpace{}%
\AgdaBound{A}\AgdaSpace{}%
\AgdaSymbol{(}\AgdaBound{Si}\AgdaSpace{}%
\AgdaBound{i}\AgdaSymbol{)}\AgdaSpace{}%
\AgdaSymbol{→}\AgdaSpace{}%
\AgdaDatatype{Cmd}\AgdaSpace{}%
\AgdaBound{A}\AgdaSpace{}%
\AgdaSymbol{(}\AgdaInductiveConstructor{⊕′}\AgdaSpace{}%
\AgdaBound{Si}\AgdaSymbol{)}\<%
\\
\>[4]\AgdaInductiveConstructor{CHOICE}\AgdaSpace{}%
\AgdaSymbol{:}\AgdaSpace{}%
\AgdaSymbol{∀}\AgdaSpace{}%
\AgdaSymbol{\{}\AgdaBound{Si}\AgdaSymbol{\}}\AgdaSpace{}%
\AgdaSymbol{→}\AgdaSpace{}%
\AgdaSymbol{((}\AgdaBound{i}\AgdaSpace{}%
\AgdaSymbol{:}\AgdaSpace{}%
\AgdaDatatype{Fin}\AgdaSpace{}%
\AgdaGeneralizable{k}\AgdaSymbol{)}\AgdaSpace{}%
\AgdaSymbol{→}\AgdaSpace{}%
\AgdaDatatype{Cmd}\AgdaSpace{}%
\AgdaBound{A}\AgdaSpace{}%
\AgdaSymbol{(}\AgdaBound{Si}\AgdaSpace{}%
\AgdaBound{i}\AgdaSymbol{))}\AgdaSpace{}%
\AgdaSymbol{→}\AgdaSpace{}%
\AgdaDatatype{Cmd}\AgdaSpace{}%
\AgdaBound{A}\AgdaSpace{}%
\AgdaSymbol{(}\AgdaInductiveConstructor{\&′}\AgdaSpace{}%
\AgdaBound{Si}\AgdaSymbol{)}\<%
\end{code}}
\begin{code}[hide]%
\>[0]\AgdaKeyword{module}\AgdaSpace{}%
\AgdaModule{formatting-deselect}\AgdaSpace{}%
\AgdaKeyword{where}\<%
\end{code}
\newcommand\stDynamicBranchingCommand{%
\begin{code}%
\>[0][@{}l@{\AgdaIndent{1}}]%
\>[2]\AgdaKeyword{data}\AgdaSpace{}%
\AgdaDatatype{Cmd}\AgdaSpace{}%
\AgdaSymbol{(}\AgdaBound{A}\AgdaSpace{}%
\AgdaSymbol{:}\AgdaSpace{}%
\AgdaPrimitive{Set}\AgdaSymbol{)}\AgdaSpace{}%
\AgdaSymbol{:}\AgdaSpace{}%
\AgdaDatatype{Session}\AgdaSpace{}%
\AgdaSymbol{→}\AgdaSpace{}%
\AgdaPrimitive{Set}\AgdaSpace{}%
\AgdaKeyword{where}\<%
\\
\>[2][@{}l@{\AgdaIndent{0}}]%
\>[4]\AgdaInductiveConstructor{DSELECT}\AgdaSpace{}%
\AgdaSymbol{:}\AgdaSpace{}%
\AgdaSymbol{∀}\AgdaSpace{}%
\AgdaSymbol{\{}\AgdaBound{Si}\AgdaSymbol{\}}%
\>[401I]\AgdaSymbol{→}\AgdaSpace{}%
\AgdaSymbol{(}\AgdaBound{getl}\AgdaSpace{}%
\AgdaSymbol{:}\AgdaSpace{}%
\AgdaBound{A}\AgdaSpace{}%
\AgdaSymbol{→}\AgdaSpace{}%
\AgdaBound{A}\AgdaSpace{}%
\AgdaOperator{\AgdaFunction{×}}\AgdaSpace{}%
\AgdaDatatype{Fin}\AgdaSpace{}%
\AgdaGeneralizable{k}\AgdaSymbol{)}\<%
\\
\>[.][@{}l@{}]\<[401I]%
\>[21]\AgdaSymbol{→}\AgdaSpace{}%
\AgdaSymbol{((}\AgdaBound{i}\AgdaSpace{}%
\AgdaSymbol{:}\AgdaSpace{}%
\AgdaDatatype{Fin}\AgdaSpace{}%
\AgdaGeneralizable{k}\AgdaSymbol{)}\AgdaSpace{}%
\AgdaSymbol{→}\AgdaSpace{}%
\AgdaDatatype{Cmd}\AgdaSpace{}%
\AgdaBound{A}\AgdaSpace{}%
\AgdaSymbol{(}\AgdaBound{Si}\AgdaSpace{}%
\AgdaBound{i}\AgdaSymbol{))}\<%
\\
\>[21]\AgdaSymbol{→}\AgdaSpace{}%
\AgdaDatatype{Cmd}\AgdaSpace{}%
\AgdaBound{A}\AgdaSpace{}%
\AgdaSymbol{(}\AgdaInductiveConstructor{⊕′}\AgdaSpace{}%
\AgdaBound{Si}\AgdaSymbol{)}\<%
\end{code}}
\newcommand\stCommand{%
\begin{code}%
\>[0]\AgdaKeyword{data}\AgdaSpace{}%
\AgdaDatatype{Cmd}\AgdaSpace{}%
\AgdaSymbol{(}\AgdaBound{A}\AgdaSpace{}%
\AgdaSymbol{:}\AgdaSpace{}%
\AgdaPrimitive{Set}\AgdaSymbol{)}\AgdaSpace{}%
\AgdaSymbol{:}\AgdaSpace{}%
\AgdaDatatype{Session}\AgdaSpace{}%
\AgdaSymbol{→}\AgdaSpace{}%
\AgdaPrimitive{Set}\AgdaSpace{}%
\AgdaKeyword{where}\<%
\\
\>[0][@{}l@{\AgdaIndent{0}}]%
\>[2]\AgdaInductiveConstructor{CLOSE}%
\>[9]\AgdaSymbol{:}\AgdaSpace{}%
\AgdaDatatype{Cmd}\AgdaSpace{}%
\AgdaBound{A}\AgdaSpace{}%
\AgdaInductiveConstructor{end}\<%
\\
\>[2]\AgdaInductiveConstructor{SEND}%
\>[9]\AgdaSymbol{:}\AgdaSpace{}%
\AgdaSymbol{(}\AgdaBound{A}\AgdaSpace{}%
\AgdaSymbol{→}\AgdaSpace{}%
\AgdaBound{A}\AgdaSpace{}%
\AgdaOperator{\AgdaFunction{×}}\AgdaSpace{}%
\AgdaOperator{\AgdaFunction{T⟦}}\AgdaSpace{}%
\AgdaGeneralizable{T}\AgdaSpace{}%
\AgdaOperator{\AgdaFunction{⟧}}\AgdaSymbol{)}\AgdaSpace{}%
\AgdaSymbol{→}\AgdaSpace{}%
\AgdaDatatype{Cmd}\AgdaSpace{}%
\AgdaBound{A}\AgdaSpace{}%
\AgdaGeneralizable{S}\AgdaSpace{}%
\AgdaSymbol{→}\AgdaSpace{}%
\AgdaDatatype{Cmd}\AgdaSpace{}%
\AgdaBound{A}\AgdaSpace{}%
\AgdaSymbol{(}\AgdaOperator{\AgdaInductiveConstructor{‼}}\AgdaSpace{}%
\AgdaGeneralizable{T}\AgdaSpace{}%
\AgdaOperator{\AgdaInductiveConstructor{∙}}\AgdaSpace{}%
\AgdaGeneralizable{S}\AgdaSymbol{)}\<%
\\
\>[2]\AgdaInductiveConstructor{RECV}%
\>[9]\AgdaSymbol{:}\AgdaSpace{}%
\AgdaSymbol{(}\AgdaOperator{\AgdaFunction{T⟦}}\AgdaSpace{}%
\AgdaGeneralizable{T}\AgdaSpace{}%
\AgdaOperator{\AgdaFunction{⟧}}\AgdaSpace{}%
\AgdaSymbol{→}\AgdaSpace{}%
\AgdaBound{A}\AgdaSpace{}%
\AgdaSymbol{→}\AgdaSpace{}%
\AgdaBound{A}\AgdaSymbol{)}\AgdaSpace{}%
\AgdaSymbol{→}\AgdaSpace{}%
\AgdaDatatype{Cmd}\AgdaSpace{}%
\AgdaBound{A}\AgdaSpace{}%
\AgdaGeneralizable{S}\AgdaSpace{}%
\AgdaSymbol{→}\AgdaSpace{}%
\AgdaDatatype{Cmd}\AgdaSpace{}%
\AgdaBound{A}\AgdaSpace{}%
\AgdaSymbol{(}\AgdaOperator{\AgdaInductiveConstructor{⁇}}\AgdaSpace{}%
\AgdaGeneralizable{T}\AgdaSpace{}%
\AgdaOperator{\AgdaInductiveConstructor{∙}}\AgdaSpace{}%
\AgdaGeneralizable{S}\AgdaSymbol{)}\<%
\end{code}}
\begin{code}[hide]%
\>[2]\AgdaInductiveConstructor{SELECT}\AgdaSpace{}%
\AgdaSymbol{:}\AgdaSpace{}%
\AgdaSymbol{∀}\AgdaSpace{}%
\AgdaSymbol{\{}\AgdaBound{Si}\AgdaSymbol{\}}\AgdaSpace{}%
\AgdaSymbol{→}\AgdaSpace{}%
\AgdaSymbol{(}\AgdaBound{i}\AgdaSpace{}%
\AgdaSymbol{:}\AgdaSpace{}%
\AgdaDatatype{Fin}\AgdaSpace{}%
\AgdaGeneralizable{k}\AgdaSymbol{)}\AgdaSpace{}%
\AgdaSymbol{→}\AgdaSpace{}%
\AgdaDatatype{Cmd}\AgdaSpace{}%
\AgdaBound{A}\AgdaSpace{}%
\AgdaSymbol{(}\AgdaBound{Si}\AgdaSpace{}%
\AgdaBound{i}\AgdaSymbol{)}\AgdaSpace{}%
\AgdaSymbol{→}\AgdaSpace{}%
\AgdaDatatype{Cmd}\AgdaSpace{}%
\AgdaBound{A}\AgdaSpace{}%
\AgdaSymbol{(}\AgdaInductiveConstructor{⊕′}\AgdaSpace{}%
\AgdaBound{Si}\AgdaSymbol{)}\<%
\\
\>[2]\AgdaInductiveConstructor{CHOICE}\AgdaSpace{}%
\AgdaSymbol{:}\AgdaSpace{}%
\AgdaSymbol{∀}\AgdaSpace{}%
\AgdaSymbol{\{}\AgdaBound{Si}\AgdaSymbol{\}}\AgdaSpace{}%
\AgdaSymbol{→}\AgdaSpace{}%
\AgdaSymbol{((}\AgdaBound{i}\AgdaSpace{}%
\AgdaSymbol{:}\AgdaSpace{}%
\AgdaDatatype{Fin}\AgdaSpace{}%
\AgdaGeneralizable{k}\AgdaSymbol{)}\AgdaSpace{}%
\AgdaSymbol{→}\AgdaSpace{}%
\AgdaDatatype{Cmd}\AgdaSpace{}%
\AgdaBound{A}\AgdaSpace{}%
\AgdaSymbol{(}\AgdaBound{Si}\AgdaSpace{}%
\AgdaBound{i}\AgdaSymbol{))}\AgdaSpace{}%
\AgdaSymbol{→}\AgdaSpace{}%
\AgdaDatatype{Cmd}\AgdaSpace{}%
\AgdaBound{A}\AgdaSpace{}%
\AgdaSymbol{(}\AgdaInductiveConstructor{\&′}\AgdaSpace{}%
\AgdaBound{Si}\AgdaSymbol{)}\<%
\end{code}
\begin{code}[hide]%
\>[2]\AgdaInductiveConstructor{SELECT2}\AgdaSpace{}%
\AgdaSymbol{:}\AgdaSpace{}%
\AgdaSymbol{(}\AgdaBound{A}\AgdaSpace{}%
\AgdaSymbol{→}\AgdaSpace{}%
\AgdaDatatype{Bool}\AgdaSpace{}%
\AgdaOperator{\AgdaFunction{×}}\AgdaSpace{}%
\AgdaBound{A}\AgdaSymbol{)}\AgdaSpace{}%
\AgdaSymbol{→}\AgdaSpace{}%
\AgdaDatatype{Cmd}\AgdaSpace{}%
\AgdaBound{A}\AgdaSpace{}%
\AgdaGeneralizable{s₁}\AgdaSpace{}%
\AgdaSymbol{→}\AgdaSpace{}%
\AgdaDatatype{Cmd}\AgdaSpace{}%
\AgdaBound{A}\AgdaSpace{}%
\AgdaGeneralizable{s₂}\AgdaSpace{}%
\AgdaSymbol{→}\AgdaSpace{}%
\AgdaDatatype{Cmd}\AgdaSpace{}%
\AgdaBound{A}\AgdaSpace{}%
\AgdaSymbol{(}\AgdaInductiveConstructor{select}\AgdaSpace{}%
\AgdaGeneralizable{s₁}\AgdaSpace{}%
\AgdaGeneralizable{s₂}\AgdaSymbol{)}\<%
\\
\>[2]\AgdaInductiveConstructor{CHOICE2}\AgdaSpace{}%
\AgdaSymbol{:}\AgdaSpace{}%
\AgdaSymbol{(}\AgdaDatatype{Bool}\AgdaSpace{}%
\AgdaSymbol{→}\AgdaSpace{}%
\AgdaBound{A}\AgdaSpace{}%
\AgdaSymbol{→}\AgdaSpace{}%
\AgdaRecord{⊤}\AgdaSpace{}%
\AgdaOperator{\AgdaFunction{×}}\AgdaSpace{}%
\AgdaBound{A}\AgdaSymbol{)}\AgdaSpace{}%
\AgdaSymbol{→}\AgdaSpace{}%
\AgdaDatatype{Cmd}\AgdaSpace{}%
\AgdaBound{A}\AgdaSpace{}%
\AgdaGeneralizable{s₁}\AgdaSpace{}%
\AgdaSymbol{→}\AgdaSpace{}%
\AgdaDatatype{Cmd}\AgdaSpace{}%
\AgdaBound{A}\AgdaSpace{}%
\AgdaGeneralizable{s₂}\AgdaSpace{}%
\AgdaSymbol{→}\AgdaSpace{}%
\AgdaDatatype{Cmd}\AgdaSpace{}%
\AgdaBound{A}\AgdaSpace{}%
\AgdaSymbol{(}\AgdaInductiveConstructor{choice}\AgdaSpace{}%
\AgdaGeneralizable{s₁}\AgdaSpace{}%
\AgdaGeneralizable{s₂}\AgdaSymbol{)}\<%
\\
\>[0]\<%
\end{code}
\newcommand\stAddpCommand{%
\begin{code}%
\>[0]\AgdaFunction{addp-command}\AgdaSpace{}%
\AgdaSymbol{:}\AgdaSpace{}%
\AgdaDatatype{Cmd}\AgdaSpace{}%
\AgdaDatatype{ℤ}\AgdaSpace{}%
\AgdaFunction{binaryp}\<%
\\
\>[0]\AgdaFunction{addp-command}\AgdaSpace{}%
\AgdaSymbol{=}\AgdaSpace{}%
\AgdaInductiveConstructor{RECV}\AgdaSpace{}%
\AgdaSymbol{(λ}\AgdaSpace{}%
\AgdaBound{x}\AgdaSpace{}%
\AgdaBound{a}\AgdaSpace{}%
\AgdaSymbol{→}\AgdaSpace{}%
\AgdaBound{x}\AgdaSymbol{)}\AgdaSpace{}%
\AgdaOperator{\AgdaFunction{\$}}\AgdaSpace{}%
\AgdaInductiveConstructor{RECV}\AgdaSpace{}%
\AgdaSymbol{(λ}\AgdaSpace{}%
\AgdaBound{y}\AgdaSpace{}%
\AgdaBound{a}\AgdaSpace{}%
\AgdaSymbol{→}\AgdaSpace{}%
\AgdaBound{y}\AgdaSpace{}%
\AgdaOperator{\AgdaFunction{+}}\AgdaSpace{}%
\AgdaBound{a}\AgdaSymbol{)}\AgdaSpace{}%
\AgdaOperator{\AgdaFunction{\$}}\AgdaSpace{}%
\AgdaInductiveConstructor{SEND}\AgdaSpace{}%
\AgdaSymbol{(λ}\AgdaSpace{}%
\AgdaBound{a}\AgdaSpace{}%
\AgdaSymbol{→}\AgdaSpace{}%
\AgdaOperator{\AgdaInductiveConstructor{⟨}}\AgdaSpace{}%
\AgdaBound{a}\AgdaSpace{}%
\AgdaOperator{\AgdaInductiveConstructor{,}}\AgdaSpace{}%
\AgdaBound{a}\AgdaSpace{}%
\AgdaOperator{\AgdaInductiveConstructor{⟩}}\AgdaSymbol{)}\AgdaSpace{}%
\AgdaOperator{\AgdaFunction{\$}}\AgdaSpace{}%
\AgdaInductiveConstructor{CLOSE}\<%
\end{code}}
\newcommand\stAddpCommandAlternative{%
\begin{code}%
\>[0]\AgdaFunction{addp-command′}\AgdaSpace{}%
\AgdaSymbol{:}\AgdaSpace{}%
\AgdaDatatype{Cmd}\AgdaSpace{}%
\AgdaDatatype{ℤ}\AgdaSpace{}%
\AgdaFunction{binaryp}\<%
\\
\>[0]\AgdaFunction{addp-command′}\AgdaSpace{}%
\AgdaSymbol{=}\AgdaSpace{}%
\AgdaInductiveConstructor{RECV}\AgdaSpace{}%
\AgdaFunction{const}\AgdaSpace{}%
\AgdaOperator{\AgdaFunction{\$}}\AgdaSpace{}%
\AgdaInductiveConstructor{RECV}\AgdaSpace{}%
\AgdaOperator{\AgdaFunction{\AgdaUnderscore{}+\AgdaUnderscore{}}}\AgdaSpace{}%
\AgdaOperator{\AgdaFunction{\$}}\AgdaSpace{}%
\AgdaInductiveConstructor{SEND}\AgdaSpace{}%
\AgdaOperator{\AgdaFunction{<}}\AgdaSpace{}%
\AgdaFunction{id}\AgdaSpace{}%
\AgdaOperator{\AgdaFunction{,}}\AgdaSpace{}%
\AgdaFunction{id}\AgdaSpace{}%
\AgdaOperator{\AgdaFunction{>}}\AgdaSpace{}%
\AgdaOperator{\AgdaFunction{\$}}\AgdaSpace{}%
\AgdaInductiveConstructor{CLOSE}\<%
\end{code}}
\newcommand\stNegpCommand{%
\begin{code}%
\>[0]\AgdaFunction{negp-command}\AgdaSpace{}%
\AgdaSymbol{:}\AgdaSpace{}%
\AgdaDatatype{Cmd}\AgdaSpace{}%
\AgdaDatatype{ℤ}\AgdaSpace{}%
\AgdaSymbol{(}\AgdaOperator{\AgdaInductiveConstructor{⁇}}\AgdaSpace{}%
\AgdaInductiveConstructor{int}\AgdaSpace{}%
\AgdaOperator{\AgdaInductiveConstructor{∙}}\AgdaSpace{}%
\AgdaOperator{\AgdaInductiveConstructor{‼}}\AgdaSpace{}%
\AgdaInductiveConstructor{int}\AgdaSpace{}%
\AgdaOperator{\AgdaInductiveConstructor{∙}}\AgdaSpace{}%
\AgdaInductiveConstructor{end}\AgdaSymbol{)}\<%
\\
\>[0]\AgdaFunction{negp-command}\AgdaSpace{}%
\AgdaSymbol{=}\AgdaSpace{}%
\AgdaInductiveConstructor{RECV}\AgdaSpace{}%
\AgdaSymbol{(λ}\AgdaSpace{}%
\AgdaBound{x}\AgdaSpace{}%
\AgdaBound{a}\AgdaSpace{}%
\AgdaSymbol{→}\AgdaSpace{}%
\AgdaBound{x}\AgdaSymbol{)}\AgdaSpace{}%
\AgdaOperator{\AgdaFunction{\$}}\AgdaSpace{}%
\AgdaInductiveConstructor{SEND}\AgdaSpace{}%
\AgdaSymbol{(λ}\AgdaSpace{}%
\AgdaBound{a}\AgdaSpace{}%
\AgdaSymbol{→}\AgdaSpace{}%
\AgdaOperator{\AgdaInductiveConstructor{⟨}}\AgdaSpace{}%
\AgdaBound{a}\AgdaSpace{}%
\AgdaOperator{\AgdaInductiveConstructor{,}}\AgdaSpace{}%
\AgdaOperator{\AgdaFunction{-}}\AgdaSpace{}%
\AgdaBound{a}\AgdaSpace{}%
\AgdaOperator{\AgdaInductiveConstructor{⟩}}\AgdaSymbol{)}\AgdaSpace{}%
\AgdaOperator{\AgdaFunction{\$}}\AgdaSpace{}%
\AgdaInductiveConstructor{CLOSE}\<%
\end{code}}
\newcommand\stNegpCommandAlternative{%
\begin{code}%
\>[0]\AgdaFunction{negp-command′}\AgdaSpace{}%
\AgdaSymbol{:}\AgdaSpace{}%
\AgdaDatatype{Cmd}\AgdaSpace{}%
\AgdaDatatype{ℤ}\AgdaSpace{}%
\AgdaSymbol{(}\AgdaOperator{\AgdaInductiveConstructor{⁇}}\AgdaSpace{}%
\AgdaInductiveConstructor{int}\AgdaSpace{}%
\AgdaOperator{\AgdaInductiveConstructor{∙}}\AgdaSpace{}%
\AgdaOperator{\AgdaInductiveConstructor{‼}}\AgdaSpace{}%
\AgdaInductiveConstructor{int}\AgdaSpace{}%
\AgdaOperator{\AgdaInductiveConstructor{∙}}\AgdaSpace{}%
\AgdaInductiveConstructor{end}\AgdaSymbol{)}\<%
\\
\>[0]\AgdaFunction{negp-command′}\AgdaSpace{}%
\AgdaSymbol{=}\AgdaSpace{}%
\AgdaInductiveConstructor{RECV}\AgdaSpace{}%
\AgdaFunction{const}\AgdaSpace{}%
\AgdaOperator{\AgdaFunction{\$}}\AgdaSpace{}%
\AgdaInductiveConstructor{SEND}\AgdaSpace{}%
\AgdaSymbol{(λ}\AgdaSpace{}%
\AgdaBound{a}\AgdaSpace{}%
\AgdaSymbol{→}\AgdaSpace{}%
\AgdaOperator{\AgdaInductiveConstructor{⟨}}\AgdaSpace{}%
\AgdaBound{a}\AgdaSpace{}%
\AgdaOperator{\AgdaInductiveConstructor{,}}\AgdaSpace{}%
\AgdaOperator{\AgdaFunction{-}}\AgdaSpace{}%
\AgdaBound{a}\AgdaSpace{}%
\AgdaOperator{\AgdaInductiveConstructor{⟩}}\AgdaSymbol{)}\AgdaSpace{}%
\AgdaOperator{\AgdaFunction{\$}}\AgdaSpace{}%
\AgdaInductiveConstructor{CLOSE}\<%
\end{code}}
\newcommand\stArithpCommand{%
\begin{code}%
\>[0]\AgdaFunction{arithp-command}\AgdaSpace{}%
\AgdaSymbol{:}\AgdaSpace{}%
\AgdaDatatype{Cmd}\AgdaSpace{}%
\AgdaDatatype{ℤ}\AgdaSpace{}%
\AgdaFunction{arithp}\<%
\\
\>[0]\AgdaFunction{arithp-command}\AgdaSpace{}%
\AgdaSymbol{=}\AgdaSpace{}%
\AgdaInductiveConstructor{CHOICE}\AgdaSpace{}%
\AgdaSymbol{λ}\AgdaSpace{}%
\AgdaKeyword{where}\<%
\\
\>[0][@{}l@{\AgdaIndent{0}}]%
\>[2]\AgdaInductiveConstructor{zero}\AgdaSpace{}%
\AgdaSymbol{→}\AgdaSpace{}%
\AgdaFunction{addp-command}\<%
\\
\>[2]\AgdaSymbol{(}\AgdaInductiveConstructor{suc}\AgdaSpace{}%
\AgdaInductiveConstructor{zero}\AgdaSymbol{)}\AgdaSpace{}%
\AgdaSymbol{→}\AgdaSpace{}%
\AgdaFunction{negp-command}\<%
\end{code}}
\newcommand\stPostulates{%
\begin{code}%
\>[0]\AgdaKeyword{postulate}\<%
\\
\>[0][@{}l@{\AgdaIndent{0}}]%
\>[2]\AgdaPostulate{Channel}\AgdaSpace{}%
\AgdaSymbol{:}\AgdaSpace{}%
\AgdaPrimitive{Set}\<%
\\
\>[2]\AgdaPostulate{primAccept}\AgdaSpace{}%
\AgdaSymbol{:}\AgdaSpace{}%
\AgdaDatatype{IO}\AgdaSpace{}%
\AgdaPostulate{Channel}%
\>[34]\AgdaComment{--\ accept\ a\ connection,\ return\ a\ new\ channel}\<%
\\
\>[2]\AgdaPostulate{primClose}%
\>[13]\AgdaSymbol{:}\AgdaSpace{}%
\AgdaPostulate{Channel}\AgdaSpace{}%
\AgdaSymbol{→}\AgdaSpace{}%
\AgdaDatatype{IO}\AgdaSpace{}%
\AgdaRecord{⊤}%
\>[34]\AgdaComment{--\ close\ a\ connection}\<%
\\
\>[2]\AgdaPostulate{primSend}%
\>[13]\AgdaSymbol{:}\AgdaSpace{}%
\AgdaGeneralizable{A}\AgdaSpace{}%
\AgdaSymbol{→}\AgdaSpace{}%
\AgdaPostulate{Channel}\AgdaSpace{}%
\AgdaSymbol{→}\AgdaSpace{}%
\AgdaDatatype{IO}\AgdaSpace{}%
\AgdaRecord{⊤}\AgdaSpace{}%
\AgdaComment{--\ send\ value\ of\ type\ A}\<%
\\
\>[2]\AgdaPostulate{primRecv}%
\>[13]\AgdaSymbol{:}\AgdaSpace{}%
\AgdaPostulate{Channel}\AgdaSpace{}%
\AgdaSymbol{→}\AgdaSpace{}%
\AgdaDatatype{IO}\AgdaSpace{}%
\AgdaGeneralizable{A}%
\>[34]\AgdaComment{--\ receive\ value\ of\ type\ A}\<%
\end{code}}
\newcommand\stExecutorSignature{%
\begin{code}%
\>[0]\AgdaFunction{exec}\AgdaSpace{}%
\AgdaSymbol{:}\AgdaSpace{}%
\AgdaDatatype{Cmd}\AgdaSpace{}%
\AgdaGeneralizable{A}\AgdaSpace{}%
\AgdaGeneralizable{S}\AgdaSpace{}%
\AgdaSymbol{→}\AgdaSpace{}%
\AgdaGeneralizable{A}\AgdaSpace{}%
\AgdaSymbol{→}\AgdaSpace{}%
\AgdaPostulate{Channel}\AgdaSpace{}%
\AgdaSymbol{→}\AgdaSpace{}%
\AgdaDatatype{IO}\AgdaSpace{}%
\AgdaGeneralizable{A}\<%
\end{code}}
\newcommand\stExecutor{%
\begin{code}%
\>[0]\AgdaFunction{exec}\AgdaSpace{}%
\AgdaInductiveConstructor{CLOSE}\AgdaSpace{}%
\AgdaBound{state}\AgdaSpace{}%
\AgdaBound{ch}\AgdaSpace{}%
\AgdaSymbol{=}\AgdaSpace{}%
\AgdaKeyword{do}\<%
\\
\>[0][@{}l@{\AgdaIndent{0}}]%
\>[2]\AgdaPostulate{primClose}\AgdaSpace{}%
\AgdaBound{ch}\<%
\\
\>[2]\AgdaInductiveConstructor{pure}\AgdaSpace{}%
\AgdaBound{state}\<%
\\
\>[0]\AgdaFunction{exec}\AgdaSpace{}%
\AgdaSymbol{(}\AgdaInductiveConstructor{SEND}\AgdaSpace{}%
\AgdaBound{getx}\AgdaSpace{}%
\AgdaBound{cmd}\AgdaSymbol{)}\AgdaSpace{}%
\AgdaBound{state}\AgdaSpace{}%
\AgdaBound{ch}\AgdaSpace{}%
\AgdaSymbol{=}\AgdaSpace{}%
\AgdaKeyword{do}\<%
\\
\>[0][@{}l@{\AgdaIndent{0}}]%
\>[2]\AgdaKeyword{let}\AgdaSpace{}%
\AgdaOperator{\AgdaInductiveConstructor{⟨}}\AgdaSpace{}%
\AgdaBound{state′}\AgdaSpace{}%
\AgdaOperator{\AgdaInductiveConstructor{,}}\AgdaSpace{}%
\AgdaBound{x}\AgdaSpace{}%
\AgdaOperator{\AgdaInductiveConstructor{⟩}}\AgdaSpace{}%
\AgdaSymbol{=}\AgdaSpace{}%
\AgdaBound{getx}\AgdaSpace{}%
\AgdaBound{state}\<%
\\
\>[2]\AgdaPostulate{primSend}\AgdaSpace{}%
\AgdaBound{x}\AgdaSpace{}%
\AgdaBound{ch}\<%
\\
\>[2]\AgdaFunction{exec}\AgdaSpace{}%
\AgdaBound{cmd}\AgdaSpace{}%
\AgdaBound{state′}\AgdaSpace{}%
\AgdaBound{ch}\<%
\\
\>[0]\AgdaFunction{exec}\AgdaSpace{}%
\AgdaSymbol{(}\AgdaInductiveConstructor{RECV}\AgdaSpace{}%
\AgdaBound{putx}\AgdaSpace{}%
\AgdaBound{cmd}\AgdaSymbol{)}\AgdaSpace{}%
\AgdaBound{state}\AgdaSpace{}%
\AgdaBound{ch}\AgdaSpace{}%
\AgdaSymbol{=}\AgdaSpace{}%
\AgdaKeyword{do}\<%
\\
\>[0][@{}l@{\AgdaIndent{0}}]%
\>[2]\AgdaBound{x}\AgdaSpace{}%
\AgdaOperator{\AgdaFunction{←}}\AgdaSpace{}%
\AgdaPostulate{primRecv}\AgdaSpace{}%
\AgdaBound{ch}\<%
\\
\>[2]\AgdaKeyword{let}\AgdaSpace{}%
\AgdaBound{state′}\AgdaSpace{}%
\AgdaSymbol{=}\AgdaSpace{}%
\AgdaBound{putx}\AgdaSpace{}%
\AgdaBound{x}\AgdaSpace{}%
\AgdaBound{state}\<%
\\
\>[2]\AgdaFunction{exec}\AgdaSpace{}%
\AgdaBound{cmd}\AgdaSpace{}%
\AgdaBound{state′}\AgdaSpace{}%
\AgdaBound{ch}\<%
\end{code}}
\newcommand\stBranchingExecutor{%
\begin{code}%
\>[0]\AgdaFunction{exec}\AgdaSpace{}%
\AgdaSymbol{(}\AgdaInductiveConstructor{SELECT}\AgdaSpace{}%
\AgdaBound{i}\AgdaSpace{}%
\AgdaBound{cmd}\AgdaSymbol{)}\AgdaSpace{}%
\AgdaBound{state}\AgdaSpace{}%
\AgdaBound{ch}\AgdaSpace{}%
\AgdaSymbol{=}\AgdaSpace{}%
\AgdaKeyword{do}\<%
\\
\>[0][@{}l@{\AgdaIndent{0}}]%
\>[2]\AgdaPostulate{primSend}\AgdaSpace{}%
\AgdaBound{i}\AgdaSpace{}%
\AgdaBound{ch}\<%
\\
\>[2]\AgdaFunction{exec}\AgdaSpace{}%
\AgdaBound{cmd}\AgdaSpace{}%
\AgdaBound{state}\AgdaSpace{}%
\AgdaBound{ch}\<%
\\
\\[\AgdaEmptyExtraSkip]%
\>[0]\AgdaFunction{exec}\AgdaSpace{}%
\AgdaSymbol{(}\AgdaInductiveConstructor{CHOICE}\AgdaSpace{}%
\AgdaBound{cont}\AgdaSymbol{)}\AgdaSpace{}%
\AgdaBound{state}\AgdaSpace{}%
\AgdaBound{ch}\AgdaSpace{}%
\AgdaSymbol{=}\AgdaSpace{}%
\AgdaKeyword{do}\<%
\\
\>[0][@{}l@{\AgdaIndent{0}}]%
\>[2]\AgdaBound{x}\AgdaSpace{}%
\AgdaOperator{\AgdaFunction{←}}\AgdaSpace{}%
\AgdaPostulate{primRecv}\AgdaSpace{}%
\AgdaBound{ch}\<%
\\
\>[2]\AgdaFunction{exec}\AgdaSpace{}%
\AgdaSymbol{(}\AgdaBound{cont}\AgdaSpace{}%
\AgdaBound{x}\AgdaSymbol{)}\AgdaSpace{}%
\AgdaBound{state}\AgdaSpace{}%
\AgdaBound{ch}\<%
\end{code}}
\begin{code}[hide]%
\>[0]\AgdaFunction{exec}\AgdaSpace{}%
\AgdaSymbol{(}\AgdaInductiveConstructor{SELECT2}\AgdaSpace{}%
\AgdaBound{getx}\AgdaSpace{}%
\AgdaBound{cmd₁}\AgdaSpace{}%
\AgdaBound{cmd₂}\AgdaSymbol{)}\AgdaSpace{}%
\AgdaBound{state}\AgdaSpace{}%
\AgdaBound{ch}\AgdaSpace{}%
\AgdaSymbol{=}\AgdaSpace{}%
\AgdaKeyword{do}\<%
\\
\>[0][@{}l@{\AgdaIndent{0}}]%
\>[2]\AgdaKeyword{let}\AgdaSpace{}%
\AgdaOperator{\AgdaInductiveConstructor{⟨}}\AgdaSpace{}%
\AgdaBound{x}\AgdaSpace{}%
\AgdaOperator{\AgdaInductiveConstructor{,}}\AgdaSpace{}%
\AgdaBound{state′}\AgdaSpace{}%
\AgdaOperator{\AgdaInductiveConstructor{⟩}}\AgdaSpace{}%
\AgdaSymbol{=}\AgdaSpace{}%
\AgdaBound{getx}\AgdaSpace{}%
\AgdaBound{state}\<%
\\
\>[2]\AgdaPostulate{primSend}\AgdaSpace{}%
\AgdaSymbol{\{}\AgdaDatatype{Bool}\AgdaSymbol{\}}\AgdaSpace{}%
\AgdaBound{x}\AgdaSpace{}%
\AgdaBound{ch}\<%
\\
\>[2]\AgdaSymbol{(}\AgdaOperator{\AgdaFunction{case}}\AgdaSpace{}%
\AgdaBound{x}\AgdaSpace{}%
\AgdaOperator{\AgdaFunction{of}}\AgdaSpace{}%
\AgdaSymbol{(λ\{}\AgdaSpace{}%
\AgdaInductiveConstructor{false}\AgdaSpace{}%
\AgdaSymbol{→}\AgdaSpace{}%
\AgdaFunction{exec}\AgdaSpace{}%
\AgdaBound{cmd₁}\AgdaSpace{}%
\AgdaBound{state′}\AgdaSpace{}%
\AgdaBound{ch}\AgdaSpace{}%
\AgdaSymbol{;}\AgdaSpace{}%
\AgdaInductiveConstructor{true}\AgdaSpace{}%
\AgdaSymbol{→}\AgdaSpace{}%
\AgdaFunction{exec}\AgdaSpace{}%
\AgdaBound{cmd₂}\AgdaSpace{}%
\AgdaBound{state′}\AgdaSpace{}%
\AgdaBound{ch}\AgdaSymbol{\}))}\<%
\\
\>[0]\AgdaFunction{exec}\AgdaSpace{}%
\AgdaSymbol{(}\AgdaInductiveConstructor{CHOICE2}\AgdaSpace{}%
\AgdaBound{putx}\AgdaSpace{}%
\AgdaBound{cmd₁}\AgdaSpace{}%
\AgdaBound{cmd₂}\AgdaSymbol{)}\AgdaSpace{}%
\AgdaBound{state}\AgdaSpace{}%
\AgdaBound{ch}\AgdaSpace{}%
\AgdaSymbol{=}\AgdaSpace{}%
\AgdaKeyword{do}\<%
\\
\>[0][@{}l@{\AgdaIndent{0}}]%
\>[2]\AgdaBound{x}\AgdaSpace{}%
\AgdaOperator{\AgdaFunction{←}}\AgdaSpace{}%
\AgdaPostulate{primRecv}\AgdaSpace{}%
\AgdaSymbol{\{}\AgdaDatatype{Bool}\AgdaSymbol{\}}\AgdaSpace{}%
\AgdaBound{ch}\<%
\\
\>[2]\AgdaKeyword{let}\AgdaSpace{}%
\AgdaOperator{\AgdaInductiveConstructor{⟨}}\AgdaSpace{}%
\AgdaSymbol{\AgdaUnderscore{}}\AgdaSpace{}%
\AgdaOperator{\AgdaInductiveConstructor{,}}\AgdaSpace{}%
\AgdaBound{state′}\AgdaSpace{}%
\AgdaOperator{\AgdaInductiveConstructor{⟩}}\AgdaSpace{}%
\AgdaSymbol{=}\AgdaSpace{}%
\AgdaBound{putx}\AgdaSpace{}%
\AgdaBound{x}\AgdaSpace{}%
\AgdaBound{state}\<%
\\
\>[2]\AgdaSymbol{(}\AgdaOperator{\AgdaFunction{case}}\AgdaSpace{}%
\AgdaBound{x}\AgdaSpace{}%
\AgdaOperator{\AgdaFunction{of}}\AgdaSpace{}%
\AgdaSymbol{(λ\{}\AgdaSpace{}%
\AgdaInductiveConstructor{false}\AgdaSpace{}%
\AgdaSymbol{→}\AgdaSpace{}%
\AgdaFunction{exec}\AgdaSpace{}%
\AgdaBound{cmd₁}\AgdaSpace{}%
\AgdaBound{state′}\AgdaSpace{}%
\AgdaBound{ch}\AgdaSpace{}%
\AgdaSymbol{;}\AgdaSpace{}%
\AgdaInductiveConstructor{true}\AgdaSpace{}%
\AgdaSymbol{→}\AgdaSpace{}%
\AgdaFunction{exec}\AgdaSpace{}%
\AgdaBound{cmd₂}\AgdaSpace{}%
\AgdaBound{state′}\AgdaSpace{}%
\AgdaBound{ch}\AgdaSymbol{\}))}\<%
\end{code}
\newcommand\stAcceptor{%
\begin{code}%
\>[0]\AgdaKeyword{record}\AgdaSpace{}%
\AgdaRecord{Accepting}\AgdaSpace{}%
\AgdaBound{A}\AgdaSpace{}%
\AgdaBound{S}\AgdaSpace{}%
\AgdaSymbol{:}\AgdaSpace{}%
\AgdaPrimitive{Set}\AgdaSpace{}%
\AgdaKeyword{where}\<%
\\
\>[0][@{}l@{\AgdaIndent{0}}]%
\>[2]\AgdaKeyword{constructor}\AgdaSpace{}%
\AgdaInductiveConstructor{ACC}\<%
\\
\>[2]\AgdaKeyword{field}\AgdaSpace{}%
\AgdaField{cmd}\AgdaSpace{}%
\AgdaSymbol{:}\AgdaSpace{}%
\AgdaDatatype{Cmd}\AgdaSpace{}%
\AgdaBound{A}\AgdaSpace{}%
\AgdaBound{S}\<%
\\
\\[\AgdaEmptyExtraSkip]%
\>[0]\AgdaFunction{acceptor}\AgdaSpace{}%
\AgdaSymbol{:}\AgdaSpace{}%
\AgdaRecord{Accepting}\AgdaSpace{}%
\AgdaGeneralizable{A}\AgdaSpace{}%
\AgdaGeneralizable{S}\AgdaSpace{}%
\AgdaSymbol{→}\AgdaSpace{}%
\AgdaGeneralizable{A}\AgdaSpace{}%
\AgdaSymbol{→}\AgdaSpace{}%
\AgdaDatatype{IO}\AgdaSpace{}%
\AgdaGeneralizable{A}\<%
\\
\>[0]\AgdaFunction{acceptor}\AgdaSpace{}%
\AgdaSymbol{(}\AgdaInductiveConstructor{ACC}\AgdaSpace{}%
\AgdaBound{cmd}\AgdaSymbol{)}\AgdaSpace{}%
\AgdaBound{a}\AgdaSpace{}%
\AgdaSymbol{=}\AgdaSpace{}%
\AgdaPostulate{primAccept}\AgdaSpace{}%
\AgdaOperator{\AgdaFunction{>>=}}\AgdaSpace{}%
\AgdaFunction{exec}\AgdaSpace{}%
\AgdaBound{cmd}\AgdaSpace{}%
\AgdaBound{a}\<%
\end{code}}
\newcommand\stCombinators{%
\begin{code}%
\>[0]\AgdaFunction{XCmd}\AgdaSpace{}%
\AgdaSymbol{:}\AgdaSpace{}%
\AgdaPrimitive{Set}\AgdaSpace{}%
\AgdaSymbol{→}\AgdaSpace{}%
\AgdaDatatype{Session}\AgdaSpace{}%
\AgdaSymbol{→}\AgdaSpace{}%
\AgdaPrimitive{Set₁}\<%
\\
\>[0]\AgdaFunction{XCmd}\AgdaSpace{}%
\AgdaBound{A}\AgdaSpace{}%
\AgdaBound{s}\AgdaSpace{}%
\AgdaSymbol{=}\AgdaSpace{}%
\AgdaBound{A}\AgdaSpace{}%
\AgdaSymbol{→}\AgdaSpace{}%
\AgdaPostulate{Channel}\AgdaSpace{}%
\AgdaSymbol{→}\AgdaSpace{}%
\AgdaDatatype{IO}\AgdaSpace{}%
\AgdaBound{A}\<%
\\
\\[\AgdaEmptyExtraSkip]%
\>[0]\AgdaFunction{xclose}\AgdaSpace{}%
\AgdaSymbol{:}\AgdaSpace{}%
\AgdaFunction{XCmd}\AgdaSpace{}%
\AgdaGeneralizable{A}\AgdaSpace{}%
\AgdaInductiveConstructor{end}\<%
\\
\>[0]\AgdaFunction{xclose}\AgdaSpace{}%
\AgdaBound{state}\AgdaSpace{}%
\AgdaBound{ch}\AgdaSpace{}%
\AgdaSymbol{=}\AgdaSpace{}%
\AgdaKeyword{do}\<%
\\
\>[0][@{}l@{\AgdaIndent{0}}]%
\>[2]\AgdaPostulate{primClose}\AgdaSpace{}%
\AgdaBound{ch}\<%
\\
\>[2]\AgdaInductiveConstructor{pure}\AgdaSpace{}%
\AgdaBound{state}\<%
\\
\\[\AgdaEmptyExtraSkip]%
\>[0]\AgdaFunction{xsend}\AgdaSpace{}%
\AgdaSymbol{:}\AgdaSpace{}%
\AgdaSymbol{(}\AgdaGeneralizable{A}\AgdaSpace{}%
\AgdaSymbol{→}\AgdaSpace{}%
\AgdaGeneralizable{A}\AgdaSpace{}%
\AgdaOperator{\AgdaFunction{×}}\AgdaSpace{}%
\AgdaOperator{\AgdaFunction{T⟦}}\AgdaSpace{}%
\AgdaGeneralizable{T}\AgdaSpace{}%
\AgdaOperator{\AgdaFunction{⟧}}\AgdaSymbol{)}\AgdaSpace{}%
\AgdaSymbol{→}\AgdaSpace{}%
\AgdaFunction{XCmd}\AgdaSpace{}%
\AgdaGeneralizable{A}\AgdaSpace{}%
\AgdaGeneralizable{S}\AgdaSpace{}%
\AgdaSymbol{→}\AgdaSpace{}%
\AgdaFunction{XCmd}\AgdaSpace{}%
\AgdaGeneralizable{A}\AgdaSpace{}%
\AgdaSymbol{(}\AgdaOperator{\AgdaInductiveConstructor{‼}}\AgdaSpace{}%
\AgdaGeneralizable{T}\AgdaSpace{}%
\AgdaOperator{\AgdaInductiveConstructor{∙}}\AgdaSpace{}%
\AgdaGeneralizable{S}\AgdaSymbol{)}\<%
\\
\>[0]\AgdaFunction{xsend}\AgdaSpace{}%
\AgdaBound{f}\AgdaSpace{}%
\AgdaBound{xcmd}\AgdaSpace{}%
\AgdaSymbol{=}\AgdaSpace{}%
\AgdaSymbol{λ}\AgdaSpace{}%
\AgdaBound{state}\AgdaSpace{}%
\AgdaBound{ch}\AgdaSpace{}%
\AgdaSymbol{→}\AgdaSpace{}%
\AgdaKeyword{do}\<%
\\
\>[0][@{}l@{\AgdaIndent{0}}]%
\>[2]\AgdaKeyword{let}\AgdaSpace{}%
\AgdaOperator{\AgdaInductiveConstructor{⟨}}\AgdaSpace{}%
\AgdaBound{state′}\AgdaSpace{}%
\AgdaOperator{\AgdaInductiveConstructor{,}}\AgdaSpace{}%
\AgdaBound{x}\AgdaSpace{}%
\AgdaOperator{\AgdaInductiveConstructor{⟩}}\AgdaSpace{}%
\AgdaSymbol{=}\AgdaSpace{}%
\AgdaBound{f}\AgdaSpace{}%
\AgdaBound{state}\<%
\\
\>[2]\AgdaPostulate{primSend}\AgdaSpace{}%
\AgdaBound{x}\AgdaSpace{}%
\AgdaBound{ch}\<%
\\
\>[2]\AgdaBound{xcmd}\AgdaSpace{}%
\AgdaBound{state′}\AgdaSpace{}%
\AgdaBound{ch}\<%
\end{code}}
\begin{code}[hide]%
\>[0]\AgdaFunction{xrecv}\AgdaSpace{}%
\AgdaSymbol{:}\AgdaSpace{}%
\AgdaSymbol{(}\AgdaOperator{\AgdaFunction{T⟦}}\AgdaSpace{}%
\AgdaGeneralizable{T}\AgdaSpace{}%
\AgdaOperator{\AgdaFunction{⟧}}\AgdaSpace{}%
\AgdaSymbol{→}\AgdaSpace{}%
\AgdaGeneralizable{A}\AgdaSpace{}%
\AgdaSymbol{→}\AgdaSpace{}%
\AgdaGeneralizable{A}\AgdaSymbol{)}\AgdaSpace{}%
\AgdaSymbol{→}\AgdaSpace{}%
\AgdaFunction{XCmd}\AgdaSpace{}%
\AgdaGeneralizable{A}\AgdaSpace{}%
\AgdaGeneralizable{S}\AgdaSpace{}%
\AgdaSymbol{→}\AgdaSpace{}%
\AgdaFunction{XCmd}\AgdaSpace{}%
\AgdaGeneralizable{A}\AgdaSpace{}%
\AgdaSymbol{(}\AgdaOperator{\AgdaInductiveConstructor{⁇}}\AgdaSpace{}%
\AgdaGeneralizable{T}\AgdaSpace{}%
\AgdaOperator{\AgdaInductiveConstructor{∙}}\AgdaSpace{}%
\AgdaGeneralizable{S}\AgdaSymbol{)}\<%
\\
\>[0]\AgdaFunction{xrecv}\AgdaSpace{}%
\AgdaBound{f}\AgdaSpace{}%
\AgdaBound{xcmd}\AgdaSpace{}%
\AgdaSymbol{=}\AgdaSpace{}%
\AgdaSymbol{λ}\AgdaSpace{}%
\AgdaBound{state}\AgdaSpace{}%
\AgdaBound{ch}\AgdaSpace{}%
\AgdaSymbol{→}\AgdaSpace{}%
\AgdaKeyword{do}\<%
\\
\>[0][@{}l@{\AgdaIndent{0}}]%
\>[2]\AgdaBound{x}\AgdaSpace{}%
\AgdaOperator{\AgdaFunction{←}}\AgdaSpace{}%
\AgdaPostulate{primRecv}\AgdaSpace{}%
\AgdaBound{ch}\<%
\\
\>[2]\AgdaKeyword{let}\AgdaSpace{}%
\AgdaBound{state′}\AgdaSpace{}%
\AgdaSymbol{=}\AgdaSpace{}%
\AgdaBound{f}\AgdaSpace{}%
\AgdaBound{x}\AgdaSpace{}%
\AgdaBound{state}\<%
\\
\>[2]\AgdaBound{xcmd}\AgdaSpace{}%
\AgdaBound{state′}\AgdaSpace{}%
\AgdaBound{ch}\<%
\\
\\[\AgdaEmptyExtraSkip]%
\>[0]\AgdaFunction{xselect}\AgdaSpace{}%
\AgdaSymbol{:}\AgdaSpace{}%
\AgdaSymbol{∀}\AgdaSpace{}%
\AgdaSymbol{\{}\AgdaBound{Si}\AgdaSymbol{\}}\AgdaSpace{}%
\AgdaSymbol{→}\AgdaSpace{}%
\AgdaSymbol{(}\AgdaBound{i}\AgdaSpace{}%
\AgdaSymbol{:}\AgdaSpace{}%
\AgdaDatatype{Fin}\AgdaSpace{}%
\AgdaGeneralizable{k}\AgdaSymbol{)}\AgdaSpace{}%
\AgdaSymbol{→}\AgdaSpace{}%
\AgdaFunction{XCmd}\AgdaSpace{}%
\AgdaGeneralizable{A}\AgdaSpace{}%
\AgdaSymbol{(}\AgdaBound{Si}\AgdaSpace{}%
\AgdaBound{i}\AgdaSymbol{)}\AgdaSpace{}%
\AgdaSymbol{→}\AgdaSpace{}%
\AgdaFunction{XCmd}\AgdaSpace{}%
\AgdaGeneralizable{A}\AgdaSpace{}%
\AgdaSymbol{(}\AgdaInductiveConstructor{⊕′}\AgdaSpace{}%
\AgdaBound{Si}\AgdaSymbol{)}\<%
\\
\>[0]\AgdaFunction{xselect}\AgdaSpace{}%
\AgdaBound{i}\AgdaSpace{}%
\AgdaBound{xcmd}\AgdaSpace{}%
\AgdaSymbol{=}\AgdaSpace{}%
\AgdaSymbol{λ}\AgdaSpace{}%
\AgdaBound{state}\AgdaSpace{}%
\AgdaBound{ch}\AgdaSpace{}%
\AgdaSymbol{→}\AgdaSpace{}%
\AgdaKeyword{do}\<%
\\
\>[0][@{}l@{\AgdaIndent{0}}]%
\>[2]\AgdaPostulate{primSend}\AgdaSpace{}%
\AgdaBound{i}\AgdaSpace{}%
\AgdaBound{ch}\<%
\\
\>[2]\AgdaBound{xcmd}\AgdaSpace{}%
\AgdaBound{state}\AgdaSpace{}%
\AgdaBound{ch}\<%
\\
\\[\AgdaEmptyExtraSkip]%
\>[0]\AgdaFunction{xchoice}\AgdaSpace{}%
\AgdaSymbol{:}\AgdaSpace{}%
\AgdaSymbol{∀}\AgdaSpace{}%
\AgdaSymbol{\{}\AgdaBound{Si}\AgdaSymbol{\}}\AgdaSpace{}%
\AgdaSymbol{→}\AgdaSpace{}%
\AgdaSymbol{((}\AgdaBound{i}\AgdaSpace{}%
\AgdaSymbol{:}\AgdaSpace{}%
\AgdaDatatype{Fin}\AgdaSpace{}%
\AgdaGeneralizable{k}\AgdaSymbol{)}\AgdaSpace{}%
\AgdaSymbol{→}\AgdaSpace{}%
\AgdaFunction{XCmd}\AgdaSpace{}%
\AgdaGeneralizable{A}\AgdaSpace{}%
\AgdaSymbol{(}\AgdaBound{Si}\AgdaSpace{}%
\AgdaBound{i}\AgdaSymbol{))}\AgdaSpace{}%
\AgdaSymbol{→}\AgdaSpace{}%
\AgdaFunction{XCmd}\AgdaSpace{}%
\AgdaGeneralizable{A}\AgdaSpace{}%
\AgdaSymbol{(}\AgdaInductiveConstructor{\&′}\AgdaSpace{}%
\AgdaBound{Si}\AgdaSymbol{)}\<%
\\
\>[0]\AgdaFunction{xchoice}\AgdaSpace{}%
\AgdaBound{f-xcont}\AgdaSpace{}%
\AgdaSymbol{=}\AgdaSpace{}%
\AgdaSymbol{λ}\AgdaSpace{}%
\AgdaBound{state}\AgdaSpace{}%
\AgdaBound{ch}\AgdaSpace{}%
\AgdaSymbol{→}\AgdaSpace{}%
\AgdaKeyword{do}\<%
\\
\>[0][@{}l@{\AgdaIndent{0}}]%
\>[2]\AgdaBound{x}\AgdaSpace{}%
\AgdaOperator{\AgdaFunction{←}}\AgdaSpace{}%
\AgdaPostulate{primRecv}\AgdaSpace{}%
\AgdaBound{ch}\<%
\\
\>[2]\AgdaBound{f-xcont}\AgdaSpace{}%
\AgdaBound{x}\AgdaSpace{}%
\AgdaBound{state}\AgdaSpace{}%
\AgdaBound{ch}\<%
\end{code}
\begin{code}[hide]%
\>[0]\AgdaComment{----------------------------------------------------------------------}\<%
\\
\>[0]\AgdaComment{--\ a\ Σ\ type\ isomorphic\ to\ A₁\ ⊎\ A₂}\<%
\\
\\[\AgdaEmptyExtraSkip]%
\>[0]\AgdaFunction{ifb}\AgdaSpace{}%
\AgdaSymbol{:}\AgdaSpace{}%
\AgdaPrimitive{Set}\AgdaSpace{}%
\AgdaSymbol{→}\AgdaSpace{}%
\AgdaPrimitive{Set}\AgdaSpace{}%
\AgdaSymbol{→}\AgdaSpace{}%
\AgdaDatatype{Bool}\AgdaSpace{}%
\AgdaSymbol{→}\AgdaSpace{}%
\AgdaPrimitive{Set}\<%
\\
\>[0]\AgdaFunction{ifb}\AgdaSpace{}%
\AgdaBound{A₁}\AgdaSpace{}%
\AgdaBound{A₂}\AgdaSpace{}%
\AgdaInductiveConstructor{false}\AgdaSpace{}%
\AgdaSymbol{=}\AgdaSpace{}%
\AgdaBound{A₁}\<%
\\
\>[0]\AgdaFunction{ifb}\AgdaSpace{}%
\AgdaBound{A₁}\AgdaSpace{}%
\AgdaBound{A₂}\AgdaSpace{}%
\AgdaInductiveConstructor{true}\AgdaSpace{}%
\AgdaSymbol{=}\AgdaSpace{}%
\AgdaBound{A₂}\<%
\\
\\[\AgdaEmptyExtraSkip]%
\>[0]\AgdaFunction{zzfalse}\AgdaSpace{}%
\AgdaSymbol{:}\AgdaSpace{}%
\AgdaRecord{Σ}\AgdaSpace{}%
\AgdaSymbol{\AgdaUnderscore{}}\AgdaSpace{}%
\AgdaSymbol{(}\AgdaFunction{ifb}\AgdaSpace{}%
\AgdaDatatype{Bool}\AgdaSpace{}%
\AgdaDatatype{ℕ}\AgdaSymbol{)}\<%
\\
\>[0]\AgdaFunction{zzfalse}\AgdaSpace{}%
\AgdaSymbol{=}\AgdaSpace{}%
\AgdaOperator{\AgdaInductiveConstructor{⟨}}\AgdaSpace{}%
\AgdaInductiveConstructor{false}\AgdaSpace{}%
\AgdaOperator{\AgdaInductiveConstructor{,}}\AgdaSpace{}%
\AgdaInductiveConstructor{false}\AgdaSpace{}%
\AgdaOperator{\AgdaInductiveConstructor{⟩}}\<%
\\
\\[\AgdaEmptyExtraSkip]%
\>[0]\AgdaFunction{zztrue}\AgdaSpace{}%
\AgdaSymbol{:}%
\>[10]\AgdaRecord{Σ}\AgdaSpace{}%
\AgdaSymbol{\AgdaUnderscore{}}\AgdaSpace{}%
\AgdaSymbol{(}\AgdaFunction{ifb}\AgdaSpace{}%
\AgdaDatatype{Bool}\AgdaSpace{}%
\AgdaDatatype{ℕ}\AgdaSymbol{)}\<%
\\
\>[0]\AgdaFunction{zztrue}\AgdaSpace{}%
\AgdaSymbol{=}%
\>[10]\AgdaOperator{\AgdaInductiveConstructor{⟨}}\AgdaSpace{}%
\AgdaInductiveConstructor{true}\AgdaSpace{}%
\AgdaOperator{\AgdaInductiveConstructor{,}}\AgdaSpace{}%
\AgdaNumber{42}\AgdaSpace{}%
\AgdaOperator{\AgdaInductiveConstructor{⟩}}\<%
\\
\\[\AgdaEmptyExtraSkip]%
\>[0]\AgdaFunction{fffun}%
\>[7]\AgdaSymbol{:}\AgdaSpace{}%
\AgdaSymbol{(}\AgdaBound{x}\AgdaSpace{}%
\AgdaSymbol{:}\AgdaSpace{}%
\AgdaDatatype{Bool}\AgdaSymbol{)}\AgdaSpace{}%
\AgdaSymbol{→}\AgdaSpace{}%
\AgdaFunction{ifb}\AgdaSpace{}%
\AgdaDatatype{Bool}\AgdaSpace{}%
\AgdaDatatype{ℕ}\AgdaSpace{}%
\AgdaBound{x}\<%
\\
\>[0]\AgdaFunction{fffun}\AgdaSpace{}%
\AgdaInductiveConstructor{false}\AgdaSpace{}%
\AgdaSymbol{=}\AgdaSpace{}%
\AgdaInductiveConstructor{false}\<%
\\
\>[0]\AgdaFunction{fffun}\AgdaSpace{}%
\AgdaInductiveConstructor{true}\AgdaSpace{}%
\AgdaSymbol{=}\AgdaSpace{}%
\AgdaNumber{42}\<%
\\
\\[\AgdaEmptyExtraSkip]%
\>[0]\AgdaFunction{ΣB}\AgdaSpace{}%
\AgdaSymbol{:}\AgdaSpace{}%
\AgdaPrimitive{Set}\AgdaSpace{}%
\AgdaSymbol{→}\AgdaSpace{}%
\AgdaPrimitive{Set}\AgdaSpace{}%
\AgdaSymbol{→}\AgdaSpace{}%
\AgdaPrimitive{Set}\<%
\\
\>[0]\AgdaFunction{ΣB}\AgdaSpace{}%
\AgdaBound{A₁}\AgdaSpace{}%
\AgdaBound{A₂}\AgdaSpace{}%
\AgdaSymbol{=}\AgdaSpace{}%
\AgdaRecord{Σ}\AgdaSpace{}%
\AgdaSymbol{\AgdaUnderscore{}}\AgdaSpace{}%
\AgdaSymbol{(}\AgdaFunction{ifb}\AgdaSpace{}%
\AgdaBound{A₁}\AgdaSpace{}%
\AgdaBound{A₂}\AgdaSymbol{)}\<%
\\
\\[\AgdaEmptyExtraSkip]%
\\[\AgdaEmptyExtraSkip]%
\>[0]\AgdaKeyword{data}\AgdaSpace{}%
\AgdaDatatype{Cmd′}\AgdaSpace{}%
\AgdaSymbol{(}\AgdaBound{A}\AgdaSpace{}%
\AgdaSymbol{:}\AgdaSpace{}%
\AgdaPrimitive{Set}\AgdaSymbol{)}\AgdaSpace{}%
\AgdaSymbol{:}\AgdaSpace{}%
\AgdaPrimitive{Set}\AgdaSpace{}%
\AgdaSymbol{→}\AgdaSpace{}%
\AgdaDatatype{Session}\AgdaSpace{}%
\AgdaSymbol{→}\AgdaSpace{}%
\AgdaPrimitive{Set₁}\AgdaSpace{}%
\AgdaKeyword{where}\<%
\\
\>[0][@{}l@{\AgdaIndent{0}}]%
\>[2]\AgdaInductiveConstructor{CLOSE}%
\>[9]\AgdaSymbol{:}\AgdaSpace{}%
\AgdaDatatype{Cmd′}\AgdaSpace{}%
\AgdaBound{A}\AgdaSpace{}%
\AgdaBound{A}\AgdaSpace{}%
\AgdaInductiveConstructor{end}\<%
\\
\>[2]\AgdaInductiveConstructor{SEND}%
\>[9]\AgdaSymbol{:}\AgdaSpace{}%
\AgdaSymbol{(}\AgdaBound{A}\AgdaSpace{}%
\AgdaSymbol{→}\AgdaSpace{}%
\AgdaOperator{\AgdaFunction{T⟦}}\AgdaSpace{}%
\AgdaGeneralizable{T}\AgdaSpace{}%
\AgdaOperator{\AgdaFunction{⟧}}\AgdaSpace{}%
\AgdaOperator{\AgdaFunction{×}}\AgdaSpace{}%
\AgdaGeneralizable{A′}\AgdaSymbol{)}\AgdaSpace{}%
\AgdaSymbol{→}\AgdaSpace{}%
\AgdaDatatype{Cmd′}\AgdaSpace{}%
\AgdaGeneralizable{A′}\AgdaSpace{}%
\AgdaGeneralizable{A″}\AgdaSpace{}%
\AgdaGeneralizable{S}\AgdaSpace{}%
\AgdaSymbol{→}\AgdaSpace{}%
\AgdaDatatype{Cmd′}\AgdaSpace{}%
\AgdaBound{A}\AgdaSpace{}%
\AgdaGeneralizable{A″}\AgdaSpace{}%
\AgdaSymbol{(}\AgdaInductiveConstructor{send}\AgdaSpace{}%
\AgdaGeneralizable{T}\AgdaSpace{}%
\AgdaGeneralizable{S}\AgdaSymbol{)}\<%
\\
\>[2]\AgdaInductiveConstructor{RECV}%
\>[9]\AgdaSymbol{:}\AgdaSpace{}%
\AgdaSymbol{(}\AgdaOperator{\AgdaFunction{T⟦}}\AgdaSpace{}%
\AgdaGeneralizable{T}\AgdaSpace{}%
\AgdaOperator{\AgdaFunction{⟧}}\AgdaSpace{}%
\AgdaSymbol{→}\AgdaSpace{}%
\AgdaBound{A}\AgdaSpace{}%
\AgdaSymbol{→}\AgdaSpace{}%
\AgdaGeneralizable{A′}\AgdaSymbol{)}\AgdaSpace{}%
\AgdaSymbol{→}\AgdaSpace{}%
\AgdaDatatype{Cmd′}\AgdaSpace{}%
\AgdaGeneralizable{A′}\AgdaSpace{}%
\AgdaGeneralizable{A″}\AgdaSpace{}%
\AgdaGeneralizable{S}\AgdaSpace{}%
\AgdaSymbol{→}\AgdaSpace{}%
\AgdaDatatype{Cmd′}\AgdaSpace{}%
\AgdaBound{A}\AgdaSpace{}%
\AgdaGeneralizable{A″}\AgdaSpace{}%
\AgdaSymbol{(}\AgdaInductiveConstructor{recv}\AgdaSpace{}%
\AgdaGeneralizable{T}\AgdaSpace{}%
\AgdaGeneralizable{S}\AgdaSymbol{)}\<%
\\
\>[2]\AgdaInductiveConstructor{SELECT21}\AgdaSpace{}%
\AgdaSymbol{:}\AgdaSpace{}%
\AgdaSymbol{(}\AgdaBound{A}\AgdaSpace{}%
\AgdaSymbol{→}\AgdaSpace{}%
\AgdaGeneralizable{A₁}\AgdaSpace{}%
\AgdaOperator{\AgdaDatatype{⊎}}\AgdaSpace{}%
\AgdaGeneralizable{A₂}\AgdaSymbol{)}\AgdaSpace{}%
\AgdaSymbol{→}\AgdaSpace{}%
\AgdaDatatype{Cmd′}\AgdaSpace{}%
\AgdaGeneralizable{A₁}\AgdaSpace{}%
\AgdaGeneralizable{A″}\AgdaSpace{}%
\AgdaGeneralizable{s₁}\AgdaSpace{}%
\AgdaSymbol{→}\AgdaSpace{}%
\AgdaDatatype{Cmd′}\AgdaSpace{}%
\AgdaGeneralizable{A₂}\AgdaSpace{}%
\AgdaGeneralizable{A″}\AgdaSpace{}%
\AgdaGeneralizable{s₂}\AgdaSpace{}%
\AgdaSymbol{→}\AgdaSpace{}%
\AgdaDatatype{Cmd′}\AgdaSpace{}%
\AgdaBound{A}\AgdaSpace{}%
\AgdaGeneralizable{A″}\AgdaSpace{}%
\AgdaSymbol{(}\AgdaInductiveConstructor{select}\AgdaSpace{}%
\AgdaGeneralizable{s₁}\AgdaSpace{}%
\AgdaGeneralizable{s₂}\AgdaSymbol{)}\<%
\\
\>[2]\AgdaInductiveConstructor{CHOICE21}\AgdaSpace{}%
\AgdaSymbol{:}\AgdaSpace{}%
\AgdaSymbol{((}\AgdaBound{x}\AgdaSpace{}%
\AgdaSymbol{:}\AgdaSpace{}%
\AgdaDatatype{Bool}\AgdaSymbol{)}\AgdaSpace{}%
\AgdaSymbol{→}\AgdaSpace{}%
\AgdaBound{A}\AgdaSpace{}%
\AgdaSymbol{→}\AgdaSpace{}%
\AgdaSymbol{(}\AgdaOperator{\AgdaFunction{case}}\AgdaSpace{}%
\AgdaBound{x}\AgdaSpace{}%
\AgdaOperator{\AgdaFunction{of}}\AgdaSpace{}%
\AgdaSymbol{λ\{}\AgdaInductiveConstructor{false}\AgdaSpace{}%
\AgdaSymbol{→}\AgdaSpace{}%
\AgdaGeneralizable{A₁}\AgdaSymbol{;}\AgdaSpace{}%
\AgdaInductiveConstructor{true}\AgdaSpace{}%
\AgdaSymbol{→}\AgdaSpace{}%
\AgdaGeneralizable{A₂}\AgdaSymbol{\}))}\AgdaSpace{}%
\AgdaSymbol{→}\AgdaSpace{}%
\AgdaDatatype{Cmd′}\AgdaSpace{}%
\AgdaGeneralizable{A₁}\AgdaSpace{}%
\AgdaGeneralizable{A″}\AgdaSpace{}%
\AgdaGeneralizable{s₁}\AgdaSpace{}%
\AgdaSymbol{→}\AgdaSpace{}%
\AgdaDatatype{Cmd′}\AgdaSpace{}%
\AgdaGeneralizable{A₂}\AgdaSpace{}%
\AgdaGeneralizable{A″}\AgdaSpace{}%
\AgdaGeneralizable{s₂}\AgdaSpace{}%
\AgdaSymbol{→}\AgdaSpace{}%
\AgdaDatatype{Cmd′}\AgdaSpace{}%
\AgdaBound{A}\AgdaSpace{}%
\AgdaGeneralizable{A″}\AgdaSpace{}%
\AgdaSymbol{(}\AgdaInductiveConstructor{choice}\AgdaSpace{}%
\AgdaGeneralizable{s₁}\AgdaSpace{}%
\AgdaGeneralizable{s₂}\AgdaSymbol{)}\<%
\\
\\[\AgdaEmptyExtraSkip]%
\>[2]\AgdaInductiveConstructor{SELECT22}\AgdaSpace{}%
\AgdaSymbol{:}\AgdaSpace{}%
\AgdaSymbol{(}\AgdaBound{A}\AgdaSpace{}%
\AgdaSymbol{→}\AgdaSpace{}%
\AgdaFunction{ΣB}\AgdaSpace{}%
\AgdaGeneralizable{A₁}\AgdaSpace{}%
\AgdaGeneralizable{A₂}\AgdaSymbol{)}\AgdaSpace{}%
\AgdaSymbol{→}\AgdaSpace{}%
\AgdaDatatype{Cmd′}\AgdaSpace{}%
\AgdaGeneralizable{A₁}\AgdaSpace{}%
\AgdaGeneralizable{A″}\AgdaSpace{}%
\AgdaGeneralizable{s₁}\AgdaSpace{}%
\AgdaSymbol{→}\AgdaSpace{}%
\AgdaDatatype{Cmd′}\AgdaSpace{}%
\AgdaGeneralizable{A₂}\AgdaSpace{}%
\AgdaGeneralizable{A″}\AgdaSpace{}%
\AgdaGeneralizable{s₂}\AgdaSpace{}%
\AgdaSymbol{→}\AgdaSpace{}%
\AgdaDatatype{Cmd′}\AgdaSpace{}%
\AgdaBound{A}\AgdaSpace{}%
\AgdaGeneralizable{A″}\AgdaSpace{}%
\AgdaSymbol{(}\AgdaInductiveConstructor{select}\AgdaSpace{}%
\AgdaGeneralizable{s₁}\AgdaSpace{}%
\AgdaGeneralizable{s₂}\AgdaSymbol{)}\<%
\\
\\[\AgdaEmptyExtraSkip]%
\>[0]\AgdaFunction{exec′}\AgdaSpace{}%
\AgdaSymbol{:}\AgdaSpace{}%
\AgdaSymbol{\{}\AgdaBound{s}\AgdaSpace{}%
\AgdaSymbol{:}\AgdaSpace{}%
\AgdaDatatype{Session}\AgdaSymbol{\}}\AgdaSpace{}%
\AgdaSymbol{→}\AgdaSpace{}%
\AgdaDatatype{Cmd′}\AgdaSpace{}%
\AgdaGeneralizable{A}\AgdaSpace{}%
\AgdaGeneralizable{A″}\AgdaSpace{}%
\AgdaBound{s}\AgdaSpace{}%
\AgdaSymbol{→}\AgdaSpace{}%
\AgdaSymbol{(}\AgdaBound{init}\AgdaSpace{}%
\AgdaSymbol{:}\AgdaSpace{}%
\AgdaGeneralizable{A}\AgdaSymbol{)}\AgdaSpace{}%
\AgdaSymbol{→}\AgdaSpace{}%
\AgdaPostulate{Channel}\AgdaSpace{}%
\AgdaSymbol{→}\AgdaSpace{}%
\AgdaDatatype{IO}\AgdaSpace{}%
\AgdaGeneralizable{A″}\<%
\\
\>[0]\AgdaFunction{exec′}\AgdaSpace{}%
\AgdaInductiveConstructor{CLOSE}\AgdaSpace{}%
\AgdaBound{state}\AgdaSpace{}%
\AgdaBound{ch}\AgdaSpace{}%
\AgdaSymbol{=}\AgdaSpace{}%
\AgdaKeyword{do}\<%
\\
\>[0][@{}l@{\AgdaIndent{0}}]%
\>[2]\AgdaPostulate{primClose}\AgdaSpace{}%
\AgdaBound{ch}\<%
\\
\>[2]\AgdaInductiveConstructor{pure}\AgdaSpace{}%
\AgdaBound{state}\<%
\\
\>[0]\AgdaFunction{exec′}\AgdaSpace{}%
\AgdaSymbol{(}\AgdaInductiveConstructor{SEND}\AgdaSpace{}%
\AgdaBound{getx}\AgdaSpace{}%
\AgdaBound{cmd}\AgdaSymbol{)}\AgdaSpace{}%
\AgdaBound{state}\AgdaSpace{}%
\AgdaBound{ch}\AgdaSpace{}%
\AgdaSymbol{=}\AgdaSpace{}%
\AgdaKeyword{do}\<%
\\
\>[0][@{}l@{\AgdaIndent{0}}]%
\>[2]\AgdaKeyword{let}\AgdaSpace{}%
\AgdaOperator{\AgdaInductiveConstructor{⟨}}\AgdaSpace{}%
\AgdaBound{x}\AgdaSpace{}%
\AgdaOperator{\AgdaInductiveConstructor{,}}\AgdaSpace{}%
\AgdaBound{state′}\AgdaSpace{}%
\AgdaOperator{\AgdaInductiveConstructor{⟩}}\AgdaSpace{}%
\AgdaSymbol{=}\AgdaSpace{}%
\AgdaBound{getx}\AgdaSpace{}%
\AgdaBound{state}\<%
\\
\>[2]\AgdaPostulate{primSend}\AgdaSpace{}%
\AgdaBound{x}\AgdaSpace{}%
\AgdaBound{ch}\<%
\\
\>[2]\AgdaFunction{exec′}\AgdaSpace{}%
\AgdaBound{cmd}\AgdaSpace{}%
\AgdaBound{state′}\AgdaSpace{}%
\AgdaBound{ch}\<%
\\
\>[0]\AgdaFunction{exec′}\AgdaSpace{}%
\AgdaSymbol{(}\AgdaInductiveConstructor{RECV}\AgdaSpace{}%
\AgdaBound{putx}\AgdaSpace{}%
\AgdaBound{cmd}\AgdaSymbol{)}\AgdaSpace{}%
\AgdaBound{state}\AgdaSpace{}%
\AgdaBound{ch}\AgdaSpace{}%
\AgdaSymbol{=}\AgdaSpace{}%
\AgdaKeyword{do}\<%
\\
\>[0][@{}l@{\AgdaIndent{0}}]%
\>[2]\AgdaBound{x}\AgdaSpace{}%
\AgdaOperator{\AgdaFunction{←}}\AgdaSpace{}%
\AgdaPostulate{primRecv}\AgdaSpace{}%
\AgdaBound{ch}\<%
\\
\>[2]\AgdaKeyword{let}\AgdaSpace{}%
\AgdaBound{state′}\AgdaSpace{}%
\AgdaSymbol{=}\AgdaSpace{}%
\AgdaBound{putx}\AgdaSpace{}%
\AgdaBound{x}\AgdaSpace{}%
\AgdaBound{state}\<%
\\
\>[2]\AgdaFunction{exec′}\AgdaSpace{}%
\AgdaBound{cmd}\AgdaSpace{}%
\AgdaBound{state′}\AgdaSpace{}%
\AgdaBound{ch}\<%
\\
\>[0]\AgdaFunction{exec′}\AgdaSpace{}%
\AgdaSymbol{(}\AgdaInductiveConstructor{SELECT21}\AgdaSpace{}%
\AgdaBound{getx}\AgdaSpace{}%
\AgdaBound{cmd₁}\AgdaSpace{}%
\AgdaBound{cmd₂}\AgdaSymbol{)}\AgdaSpace{}%
\AgdaBound{state}\AgdaSpace{}%
\AgdaBound{ch}\<%
\\
\>[0][@{}l@{\AgdaIndent{0}}]%
\>[2]\AgdaKeyword{with}\AgdaSpace{}%
\AgdaBound{getx}\AgdaSpace{}%
\AgdaBound{state}\<%
\\
\>[0]\AgdaSymbol{...}\AgdaSpace{}%
\AgdaSymbol{|}%
\>[1306I]\AgdaInductiveConstructor{inj₁}\AgdaSpace{}%
\AgdaBound{state₁}\AgdaSpace{}%
\AgdaSymbol{=}\AgdaSpace{}%
\AgdaKeyword{do}\<%
\\
\>[.][@{}l@{}]\<[1306I]%
\>[6]\AgdaPostulate{primSend}\AgdaSpace{}%
\AgdaSymbol{\{}\AgdaDatatype{Bool}\AgdaSymbol{\}}\AgdaSpace{}%
\AgdaInductiveConstructor{true}\AgdaSpace{}%
\AgdaBound{ch}\<%
\\
\>[6]\AgdaFunction{exec′}\AgdaSpace{}%
\AgdaBound{cmd₁}\AgdaSpace{}%
\AgdaBound{state₁}\AgdaSpace{}%
\AgdaBound{ch}\<%
\\
\>[0]\AgdaSymbol{...}\AgdaSpace{}%
\AgdaSymbol{|}%
\>[1317I]\AgdaInductiveConstructor{inj₂}\AgdaSpace{}%
\AgdaBound{state₂}\AgdaSpace{}%
\AgdaSymbol{=}\AgdaSpace{}%
\AgdaKeyword{do}\<%
\\
\>[.][@{}l@{}]\<[1317I]%
\>[6]\AgdaPostulate{primSend}\AgdaSpace{}%
\AgdaSymbol{\{}\AgdaDatatype{Bool}\AgdaSymbol{\}}\AgdaSpace{}%
\AgdaInductiveConstructor{false}\AgdaSpace{}%
\AgdaBound{ch}\<%
\\
\>[6]\AgdaFunction{exec′}\AgdaSpace{}%
\AgdaBound{cmd₂}\AgdaSpace{}%
\AgdaBound{state₂}\AgdaSpace{}%
\AgdaBound{ch}\<%
\\
\>[0]\AgdaFunction{exec′}\AgdaSpace{}%
\AgdaSymbol{(}\AgdaInductiveConstructor{CHOICE21}\AgdaSpace{}%
\AgdaBound{putx}\AgdaSpace{}%
\AgdaBound{cmd₁}\AgdaSpace{}%
\AgdaBound{cmd₂}\AgdaSymbol{)}\AgdaSpace{}%
\AgdaBound{state}\AgdaSpace{}%
\AgdaBound{ch}\AgdaSpace{}%
\AgdaSymbol{=}\AgdaSpace{}%
\AgdaKeyword{do}\<%
\\
\>[0][@{}l@{\AgdaIndent{0}}]%
\>[2]\AgdaInductiveConstructor{false}\AgdaSpace{}%
\AgdaOperator{\AgdaFunction{←}}\AgdaSpace{}%
\AgdaPostulate{primRecv}\AgdaSpace{}%
\AgdaSymbol{\{}\AgdaDatatype{Bool}\AgdaSymbol{\}}\AgdaSpace{}%
\AgdaBound{ch}\<%
\\
\>[2][@{}l@{\AgdaIndent{0}}]%
\>[4]\AgdaKeyword{where}\<%
\\
\>[4][@{}l@{\AgdaIndent{0}}]%
\>[6]\AgdaInductiveConstructor{true}\AgdaSpace{}%
\AgdaSymbol{→}\AgdaSpace{}%
\AgdaKeyword{do}\<%
\\
\>[6][@{}l@{\AgdaIndent{0}}]%
\>[8]\AgdaKeyword{let}\AgdaSpace{}%
\AgdaBound{state′}\AgdaSpace{}%
\AgdaSymbol{=}\AgdaSpace{}%
\AgdaBound{putx}\AgdaSpace{}%
\AgdaInductiveConstructor{true}\AgdaSpace{}%
\AgdaBound{state}\<%
\\
\>[8]\AgdaFunction{exec′}\AgdaSpace{}%
\AgdaBound{cmd₂}\AgdaSpace{}%
\AgdaBound{state′}\AgdaSpace{}%
\AgdaBound{ch}\<%
\\
\>[2]\AgdaKeyword{let}\AgdaSpace{}%
\AgdaBound{state′}\AgdaSpace{}%
\AgdaSymbol{=}\AgdaSpace{}%
\AgdaBound{putx}\AgdaSpace{}%
\AgdaInductiveConstructor{false}\AgdaSpace{}%
\AgdaBound{state}\<%
\\
\>[2]\AgdaFunction{exec′}\AgdaSpace{}%
\AgdaBound{cmd₁}\AgdaSpace{}%
\AgdaBound{state′}\AgdaSpace{}%
\AgdaBound{ch}\<%
\\
\>[0]\AgdaFunction{exec′}\AgdaSpace{}%
\AgdaSymbol{(}\AgdaInductiveConstructor{SELECT22}\AgdaSpace{}%
\AgdaBound{getx}\AgdaSpace{}%
\AgdaBound{cmd₁}\AgdaSpace{}%
\AgdaBound{cmd₂}\AgdaSymbol{)}\AgdaSpace{}%
\AgdaBound{state}\AgdaSpace{}%
\AgdaBound{ch}\AgdaSpace{}%
\AgdaSymbol{=}\AgdaSpace{}%
\AgdaKeyword{do}\<%
\\
\>[0][@{}l@{\AgdaIndent{0}}]%
\>[2]\AgdaKeyword{let}\AgdaSpace{}%
\AgdaBound{bst}\AgdaSpace{}%
\AgdaSymbol{=}\AgdaSpace{}%
\AgdaBound{getx}\AgdaSpace{}%
\AgdaBound{state}\<%
\\
\>[2]\AgdaPostulate{primSend}\AgdaSpace{}%
\AgdaSymbol{\{}\AgdaDatatype{Bool}\AgdaSymbol{\}}\AgdaSpace{}%
\AgdaSymbol{(}\AgdaField{proj₁}\AgdaSpace{}%
\AgdaBound{bst}\AgdaSymbol{)}\AgdaSpace{}%
\AgdaBound{ch}\<%
\\
\>[2]\AgdaOperator{\AgdaInductiveConstructor{⟨}}%
\>[1373I]\AgdaInductiveConstructor{false}\AgdaSpace{}%
\AgdaOperator{\AgdaInductiveConstructor{,}}\AgdaSpace{}%
\AgdaBound{state₁}\AgdaSpace{}%
\AgdaOperator{\AgdaInductiveConstructor{⟩}}\AgdaSpace{}%
\AgdaOperator{\AgdaFunction{←}}\AgdaSpace{}%
\AgdaInductiveConstructor{pure}\AgdaSpace{}%
\AgdaBound{bst}\<%
\\
\>[.][@{}l@{}]\<[1373I]%
\>[4]\AgdaKeyword{where}\<%
\\
\>[4][@{}l@{\AgdaIndent{0}}]%
\>[6]\AgdaOperator{\AgdaInductiveConstructor{⟨}}\AgdaSpace{}%
\AgdaInductiveConstructor{true}\AgdaSpace{}%
\AgdaOperator{\AgdaInductiveConstructor{,}}\AgdaSpace{}%
\AgdaBound{state₂}\AgdaSpace{}%
\AgdaOperator{\AgdaInductiveConstructor{⟩}}\AgdaSpace{}%
\AgdaSymbol{→}\AgdaSpace{}%
\AgdaFunction{exec′}\AgdaSpace{}%
\AgdaBound{cmd₂}\AgdaSpace{}%
\AgdaBound{state₂}\AgdaSpace{}%
\AgdaBound{ch}\<%
\\
\>[2]\AgdaFunction{exec′}\AgdaSpace{}%
\AgdaBound{cmd₁}\AgdaSpace{}%
\AgdaBound{state₁}\AgdaSpace{}%
\AgdaBound{ch}\<%
\\
\>[0]\<%
\end{code}

%% file: latex/ST-recursive.tex
\begin{code}[hide]%
\>[0]\AgdaSymbol{\{-\#}\AgdaSpace{}%
\AgdaKeyword{OPTIONS}\AgdaSpace{}%
\AgdaPragma{--guardedness}\AgdaSpace{}%
\AgdaSymbol{\#-\}}\AgdaSpace{}%
\AgdaComment{\{-\ required\ for\ IO\ -\}}\<%
\\
\>[0]\AgdaKeyword{module}\AgdaSpace{}%
\AgdaModule{ST-recursive}\AgdaSpace{}%
\AgdaKeyword{where}\<%
\\
\\[\AgdaEmptyExtraSkip]%
\>[0]\AgdaKeyword{open}\AgdaSpace{}%
\AgdaKeyword{import}\AgdaSpace{}%
\AgdaModule{Data.Bool}\AgdaSpace{}%
\AgdaKeyword{using}\AgdaSpace{}%
\AgdaSymbol{(}\AgdaDatatype{Bool}\AgdaSymbol{;}\AgdaSpace{}%
\AgdaInductiveConstructor{true}\AgdaSymbol{;}\AgdaSpace{}%
\AgdaInductiveConstructor{false}\AgdaSymbol{)}\<%
\\
\>[0]\AgdaKeyword{open}\AgdaSpace{}%
\AgdaKeyword{import}\AgdaSpace{}%
\AgdaModule{Data.Fin}\AgdaSpace{}%
\AgdaKeyword{using}\AgdaSpace{}%
\AgdaSymbol{(}\AgdaDatatype{Fin}\AgdaSymbol{;}\AgdaSpace{}%
\AgdaInductiveConstructor{zero}\AgdaSymbol{;}\AgdaSpace{}%
\AgdaInductiveConstructor{suc}\AgdaSymbol{;}\AgdaSpace{}%
\AgdaFunction{toℕ}\AgdaSymbol{;}\AgdaSpace{}%
\AgdaFunction{opposite}\AgdaSymbol{)}\<%
\\
\>[0]\AgdaKeyword{open}\AgdaSpace{}%
\AgdaKeyword{import}\AgdaSpace{}%
\AgdaModule{Data.Fin.Properties}\AgdaSpace{}%
\AgdaKeyword{using}\AgdaSpace{}%
\AgdaSymbol{(}\AgdaFunction{toℕ-fromℕ}\AgdaSymbol{;}\AgdaSpace{}%
\AgdaFunction{toℕ-inject₁}\AgdaSymbol{)}\<%
\\
\>[0]\AgdaKeyword{open}\AgdaSpace{}%
\AgdaKeyword{import}\AgdaSpace{}%
\AgdaModule{Data.Integer}\AgdaSpace{}%
\AgdaKeyword{using}\AgdaSpace{}%
\AgdaSymbol{(}\AgdaDatatype{ℤ}\AgdaSymbol{;}\AgdaSpace{}%
\AgdaFunction{0ℤ}\AgdaSymbol{;}\AgdaSpace{}%
\AgdaOperator{\AgdaFunction{\AgdaUnderscore{}+\AgdaUnderscore{}}}\AgdaSymbol{;}\AgdaSpace{}%
\AgdaOperator{\AgdaFunction{-\AgdaUnderscore{}}}\AgdaSymbol{;}\AgdaSpace{}%
\AgdaInductiveConstructor{+\AgdaUnderscore{}}\AgdaSymbol{)}\<%
\\
\>[0]\AgdaKeyword{open}\AgdaSpace{}%
\AgdaKeyword{import}\AgdaSpace{}%
\AgdaModule{Data.Nat}\AgdaSpace{}%
\AgdaKeyword{using}\AgdaSpace{}%
\AgdaSymbol{(}\AgdaDatatype{ℕ}\AgdaSymbol{;}\AgdaSpace{}%
\AgdaInductiveConstructor{zero}\AgdaSymbol{;}\AgdaSpace{}%
\AgdaInductiveConstructor{suc}\AgdaSymbol{)}\<%
\\
\>[0]\AgdaKeyword{open}\AgdaSpace{}%
\AgdaKeyword{import}\AgdaSpace{}%
\AgdaModule{Data.Product}\AgdaSpace{}%
\AgdaKeyword{using}\AgdaSpace{}%
\AgdaSymbol{(}\AgdaOperator{\AgdaFunction{\AgdaUnderscore{}×\AgdaUnderscore{}}}\AgdaSymbol{;}\AgdaSpace{}%
\AgdaRecord{Σ}\AgdaSymbol{;}\AgdaSpace{}%
\AgdaField{proj₁}\AgdaSymbol{;}\AgdaSpace{}%
\AgdaField{proj₂}\AgdaSymbol{;}\AgdaSpace{}%
\AgdaFunction{∃-syntax}\AgdaSymbol{)}\AgdaSpace{}%
\AgdaKeyword{renaming}\AgdaSpace{}%
\AgdaSymbol{(}\AgdaOperator{\AgdaInductiveConstructor{\AgdaUnderscore{},\AgdaUnderscore{}}}\AgdaSpace{}%
\AgdaSymbol{to}\AgdaSpace{}%
\AgdaOperator{\AgdaInductiveConstructor{⟨\AgdaUnderscore{},\AgdaUnderscore{}⟩}}\AgdaSymbol{)}\<%
\\
\>[0]\AgdaKeyword{open}\AgdaSpace{}%
\AgdaKeyword{import}\AgdaSpace{}%
\AgdaModule{Data.Sum}\AgdaSpace{}%
\AgdaKeyword{using}\AgdaSpace{}%
\AgdaSymbol{(}\AgdaOperator{\AgdaDatatype{\AgdaUnderscore{}⊎\AgdaUnderscore{}}}\AgdaSymbol{;}\AgdaSpace{}%
\AgdaInductiveConstructor{inj₁}\AgdaSymbol{;}\AgdaSpace{}%
\AgdaInductiveConstructor{inj₂}\AgdaSymbol{)}\<%
\\
\>[0]\AgdaKeyword{open}\AgdaSpace{}%
\AgdaKeyword{import}\AgdaSpace{}%
\AgdaModule{Data.Vec}\AgdaSpace{}%
\AgdaKeyword{using}\AgdaSpace{}%
\AgdaSymbol{(}\AgdaDatatype{Vec}\AgdaSymbol{;}\AgdaSpace{}%
\AgdaInductiveConstructor{[]}\AgdaSymbol{;}\AgdaSpace{}%
\AgdaOperator{\AgdaInductiveConstructor{\AgdaUnderscore{}∷\AgdaUnderscore{}}}\AgdaSymbol{)}\<%
\\
\\[\AgdaEmptyExtraSkip]%
\>[0]\AgdaKeyword{open}\AgdaSpace{}%
\AgdaKeyword{import}\AgdaSpace{}%
\AgdaModule{Data.Unit}\AgdaSpace{}%
\AgdaKeyword{using}\AgdaSpace{}%
\AgdaSymbol{(}\AgdaRecord{⊤}\AgdaSymbol{;}\AgdaSpace{}%
\AgdaInductiveConstructor{tt}\AgdaSymbol{)}\<%
\\
\\[\AgdaEmptyExtraSkip]%
\>[0]\AgdaKeyword{open}\AgdaSpace{}%
\AgdaKeyword{import}\AgdaSpace{}%
\AgdaModule{Function.Base}\AgdaSpace{}%
\AgdaKeyword{using}\AgdaSpace{}%
\AgdaSymbol{(}\AgdaOperator{\AgdaFunction{case\AgdaUnderscore{}of\AgdaUnderscore{}}}\AgdaSymbol{;}\AgdaSpace{}%
\AgdaOperator{\AgdaFunction{\AgdaUnderscore{}∘\AgdaUnderscore{}}}\AgdaSymbol{;}\AgdaSpace{}%
\AgdaFunction{const}\AgdaSymbol{;}\AgdaSpace{}%
\AgdaFunction{constᵣ}\AgdaSymbol{;}\AgdaSpace{}%
\AgdaOperator{\AgdaFunction{\AgdaUnderscore{}\$\AgdaUnderscore{}}}\AgdaSymbol{;}\AgdaSpace{}%
\AgdaFunction{id}\AgdaSymbol{)}\<%
\\
\\[\AgdaEmptyExtraSkip]%
\>[0]\AgdaKeyword{open}\AgdaSpace{}%
\AgdaKeyword{import}\AgdaSpace{}%
\AgdaModule{Relation.Binary.PropositionalEquality}\AgdaSpace{}%
\AgdaKeyword{using}\AgdaSpace{}%
\AgdaSymbol{(}\AgdaOperator{\AgdaDatatype{\AgdaUnderscore{}≡\AgdaUnderscore{}}}\AgdaSymbol{;}\AgdaSpace{}%
\AgdaInductiveConstructor{refl}\AgdaSymbol{;}\AgdaSpace{}%
\AgdaFunction{subst}\AgdaSymbol{;}\AgdaSpace{}%
\AgdaFunction{sym}\AgdaSymbol{;}\AgdaSpace{}%
\AgdaFunction{cong}\AgdaSymbol{;}\AgdaSpace{}%
\AgdaFunction{cong₂}\AgdaSymbol{;}\AgdaSpace{}%
\AgdaFunction{trans}\AgdaSymbol{;}\AgdaSpace{}%
\AgdaKeyword{module}\AgdaSpace{}%
\AgdaModule{≡-Reasoning}\AgdaSymbol{)}\<%
\\
\\[\AgdaEmptyExtraSkip]%
\>[0]\AgdaKeyword{open}\AgdaSpace{}%
\AgdaKeyword{import}\AgdaSpace{}%
\AgdaModule{IO}\<%
\\
\\[\AgdaEmptyExtraSkip]%
\>[0]\AgdaKeyword{open}\AgdaSpace{}%
\AgdaKeyword{import}\AgdaSpace{}%
\AgdaModule{Utils}\<%
\\
\\[\AgdaEmptyExtraSkip]%
\\[\AgdaEmptyExtraSkip]%
\>[0]\AgdaKeyword{pattern}\AgdaSpace{}%
\AgdaOperator{\AgdaInductiveConstructor{[\AgdaUnderscore{}]}}\AgdaSpace{}%
\AgdaBound{x}\AgdaSpace{}%
\AgdaSymbol{=}\AgdaSpace{}%
\AgdaBound{x}\AgdaSpace{}%
\AgdaOperator{\AgdaInductiveConstructor{∷}}\AgdaSpace{}%
\AgdaInductiveConstructor{[]}\<%
\\
\>[0]\AgdaKeyword{pattern}\AgdaSpace{}%
\AgdaOperator{\AgdaInductiveConstructor{[\AgdaUnderscore{},\AgdaUnderscore{}]}}\AgdaSpace{}%
\AgdaBound{x}\AgdaSpace{}%
\AgdaBound{y}\AgdaSpace{}%
\AgdaSymbol{=}\AgdaSpace{}%
\AgdaBound{x}\AgdaSpace{}%
\AgdaOperator{\AgdaInductiveConstructor{∷}}\AgdaSpace{}%
\AgdaBound{y}\AgdaSpace{}%
\AgdaOperator{\AgdaInductiveConstructor{∷}}\AgdaSpace{}%
\AgdaInductiveConstructor{[]}\<%
\\
\>[0]\AgdaKeyword{pattern}\AgdaSpace{}%
\AgdaOperator{\AgdaInductiveConstructor{[\AgdaUnderscore{},\AgdaUnderscore{},\AgdaUnderscore{}]}}\AgdaSpace{}%
\AgdaBound{x}\AgdaSpace{}%
\AgdaBound{y}\AgdaSpace{}%
\AgdaBound{z}\AgdaSpace{}%
\AgdaSymbol{=}\AgdaSpace{}%
\AgdaBound{x}\AgdaSpace{}%
\AgdaOperator{\AgdaInductiveConstructor{∷}}\AgdaSpace{}%
\AgdaBound{y}\AgdaSpace{}%
\AgdaOperator{\AgdaInductiveConstructor{∷}}\AgdaSpace{}%
\AgdaBound{z}\AgdaSpace{}%
\AgdaOperator{\AgdaInductiveConstructor{∷}}\AgdaSpace{}%
\AgdaInductiveConstructor{[]}\<%
\\
\\[\AgdaEmptyExtraSkip]%
\\[\AgdaEmptyExtraSkip]%
\>[0]\AgdaKeyword{variable}\<%
\\
\>[0][@{}l@{\AgdaIndent{0}}]%
\>[2]\AgdaGeneralizable{n}\AgdaSpace{}%
\AgdaGeneralizable{k}\AgdaSpace{}%
\AgdaSymbol{:}\AgdaSpace{}%
\AgdaDatatype{ℕ}\<%
\end{code}
\newcommand\rstFiniteType{%
\begin{code}%
\>[0]\AgdaKeyword{data}\AgdaSpace{}%
\AgdaDatatype{Type}\AgdaSpace{}%
\AgdaSymbol{:}\AgdaSpace{}%
\AgdaPrimitive{Set}\AgdaSpace{}%
\AgdaKeyword{where}\<%
\\
\>[0][@{}l@{\AgdaIndent{0}}]%
\>[2]\AgdaInductiveConstructor{int}\AgdaSpace{}%
\AgdaInductiveConstructor{bool}\AgdaSpace{}%
\AgdaSymbol{:}\AgdaSpace{}%
\AgdaDatatype{Type}\<%
\end{code}}
\begin{code}[hide]%
\>[2]\AgdaInductiveConstructor{nat}\AgdaSpace{}%
\AgdaSymbol{:}\AgdaSpace{}%
\AgdaDatatype{Type}\<%
\\
\>[2]\AgdaInductiveConstructor{fin}\AgdaSpace{}%
\AgdaSymbol{:}\AgdaSpace{}%
\AgdaDatatype{ℕ}\AgdaSpace{}%
\AgdaSymbol{→}\AgdaSpace{}%
\AgdaDatatype{Type}\<%
\\
\\[\AgdaEmptyExtraSkip]%
\>[0]\AgdaKeyword{module}\AgdaSpace{}%
\AgdaModule{formatting1}\AgdaSpace{}%
\AgdaKeyword{where}\<%
\end{code}
\newcommand\rstBranchingType{%
\begin{code}%
\>[0][@{}l@{\AgdaIndent{1}}]%
\>[2]\AgdaKeyword{data}\AgdaSpace{}%
\AgdaDatatype{Session}\AgdaSpace{}%
\AgdaSymbol{(}\AgdaBound{n}\AgdaSpace{}%
\AgdaSymbol{:}\AgdaSpace{}%
\AgdaDatatype{ℕ}\AgdaSymbol{)}\AgdaSpace{}%
\AgdaSymbol{:}\AgdaSpace{}%
\AgdaPrimitive{Set}\AgdaSpace{}%
\AgdaKeyword{where}\<%
\\
\>[2][@{}l@{\AgdaIndent{0}}]%
\>[4]\AgdaInductiveConstructor{⊕′}\AgdaSpace{}%
\AgdaSymbol{:}\AgdaSpace{}%
\AgdaSymbol{(}\AgdaBound{Si}\AgdaSpace{}%
\AgdaSymbol{:}\AgdaSpace{}%
\AgdaSymbol{(}\AgdaBound{i}\AgdaSpace{}%
\AgdaSymbol{:}\AgdaSpace{}%
\AgdaDatatype{Fin}\AgdaSpace{}%
\AgdaGeneralizable{k}\AgdaSymbol{)}\AgdaSpace{}%
\AgdaSymbol{→}\AgdaSpace{}%
\AgdaDatatype{Session}\AgdaSpace{}%
\AgdaBound{n}\AgdaSymbol{)}\AgdaSpace{}%
\AgdaSymbol{→}\AgdaSpace{}%
\AgdaDatatype{Session}\AgdaSpace{}%
\AgdaBound{n}\<%
\\
\>[4]\AgdaInductiveConstructor{\&′}\AgdaSpace{}%
\AgdaSymbol{:}\AgdaSpace{}%
\AgdaSymbol{(}\AgdaBound{Si}\AgdaSpace{}%
\AgdaSymbol{:}\AgdaSpace{}%
\AgdaSymbol{(}\AgdaBound{i}\AgdaSpace{}%
\AgdaSymbol{:}\AgdaSpace{}%
\AgdaDatatype{Fin}\AgdaSpace{}%
\AgdaGeneralizable{k}\AgdaSymbol{)}\AgdaSpace{}%
\AgdaSymbol{→}\AgdaSpace{}%
\AgdaDatatype{Session}\AgdaSpace{}%
\AgdaBound{n}\AgdaSymbol{)}\AgdaSpace{}%
\AgdaSymbol{→}\AgdaSpace{}%
\AgdaDatatype{Session}\AgdaSpace{}%
\AgdaBound{n}\<%
\end{code}}
\newcommand\rstSession{%
\begin{code}%
\>[0]\AgdaKeyword{data}\AgdaSpace{}%
\AgdaDatatype{Session}\AgdaSpace{}%
\AgdaSymbol{(}\AgdaBound{n}\AgdaSpace{}%
\AgdaSymbol{:}\AgdaSpace{}%
\AgdaDatatype{ℕ}\AgdaSymbol{)}\AgdaSpace{}%
\AgdaSymbol{:}\AgdaSpace{}%
\AgdaPrimitive{Set}\AgdaSpace{}%
\AgdaKeyword{where}\<%
\\
\>[0][@{}l@{\AgdaIndent{0}}]%
\>[2]\AgdaOperator{\AgdaInductiveConstructor{‼\AgdaUnderscore{}∙\AgdaUnderscore{}}}\AgdaSpace{}%
\AgdaSymbol{:}\AgdaSpace{}%
\AgdaDatatype{Type}\AgdaSpace{}%
\AgdaSymbol{→}\AgdaSpace{}%
\AgdaDatatype{Session}\AgdaSpace{}%
\AgdaBound{n}\AgdaSpace{}%
\AgdaSymbol{→}\AgdaSpace{}%
\AgdaDatatype{Session}\AgdaSpace{}%
\AgdaBound{n}\<%
\\
\>[2]\AgdaOperator{\AgdaInductiveConstructor{⁇\AgdaUnderscore{}∙\AgdaUnderscore{}}}\AgdaSpace{}%
\AgdaSymbol{:}\AgdaSpace{}%
\AgdaDatatype{Type}\AgdaSpace{}%
\AgdaSymbol{→}\AgdaSpace{}%
\AgdaDatatype{Session}\AgdaSpace{}%
\AgdaBound{n}\AgdaSpace{}%
\AgdaSymbol{→}\AgdaSpace{}%
\AgdaDatatype{Session}\AgdaSpace{}%
\AgdaBound{n}\<%
\\
\>[2]\AgdaInductiveConstructor{end}\AgdaSpace{}%
\AgdaSymbol{:}\AgdaSpace{}%
\AgdaDatatype{Session}\AgdaSpace{}%
\AgdaBound{n}\<%
\\
\>[2]\AgdaInductiveConstructor{⊕′}\AgdaSpace{}%
\AgdaSymbol{:}\AgdaSpace{}%
\AgdaSymbol{(}\AgdaBound{Si}\AgdaSpace{}%
\AgdaSymbol{:}\AgdaSpace{}%
\AgdaSymbol{(}\AgdaBound{i}\AgdaSpace{}%
\AgdaSymbol{:}\AgdaSpace{}%
\AgdaDatatype{Fin}\AgdaSpace{}%
\AgdaGeneralizable{k}\AgdaSymbol{)}\AgdaSpace{}%
\AgdaSymbol{→}\AgdaSpace{}%
\AgdaDatatype{Session}\AgdaSpace{}%
\AgdaBound{n}\AgdaSymbol{)}\AgdaSpace{}%
\AgdaSymbol{→}\AgdaSpace{}%
\AgdaDatatype{Session}\AgdaSpace{}%
\AgdaBound{n}\<%
\\
\>[2]\AgdaInductiveConstructor{\&′}\AgdaSpace{}%
\AgdaSymbol{:}\AgdaSpace{}%
\AgdaSymbol{(}\AgdaBound{Si}\AgdaSpace{}%
\AgdaSymbol{:}\AgdaSpace{}%
\AgdaSymbol{(}\AgdaBound{i}\AgdaSpace{}%
\AgdaSymbol{:}\AgdaSpace{}%
\AgdaDatatype{Fin}\AgdaSpace{}%
\AgdaGeneralizable{k}\AgdaSymbol{)}\AgdaSpace{}%
\AgdaSymbol{→}\AgdaSpace{}%
\AgdaDatatype{Session}\AgdaSpace{}%
\AgdaBound{n}\AgdaSymbol{)}\AgdaSpace{}%
\AgdaSymbol{→}\AgdaSpace{}%
\AgdaDatatype{Session}\AgdaSpace{}%
\AgdaBound{n}\<%
\\
\>[2]\AgdaOperator{\AgdaInductiveConstructor{μ\AgdaUnderscore{}}}\AgdaSpace{}%
\AgdaSymbol{:}\AgdaSpace{}%
\AgdaDatatype{Session}\AgdaSpace{}%
\AgdaSymbol{(}\AgdaInductiveConstructor{suc}\AgdaSpace{}%
\AgdaBound{n}\AgdaSymbol{)}\AgdaSpace{}%
\AgdaSymbol{→}\AgdaSpace{}%
\AgdaDatatype{Session}\AgdaSpace{}%
\AgdaBound{n}\<%
\\
\>[2]\AgdaOperator{\AgdaInductiveConstructor{`\AgdaUnderscore{}}}\AgdaSpace{}%
\AgdaSymbol{:}\AgdaSpace{}%
\AgdaDatatype{Fin}\AgdaSpace{}%
\AgdaBound{n}\AgdaSpace{}%
\AgdaSymbol{→}\AgdaSpace{}%
\AgdaDatatype{Session}\AgdaSpace{}%
\AgdaBound{n}\<%
\end{code}}
\begin{code}[hide]%
\>[0]\<%
\\
\>[0]\AgdaKeyword{pattern}\AgdaSpace{}%
\AgdaInductiveConstructor{recv}\AgdaSpace{}%
\AgdaBound{t}\AgdaSpace{}%
\AgdaBound{s}\AgdaSpace{}%
\AgdaSymbol{=}\AgdaSpace{}%
\AgdaOperator{\AgdaInductiveConstructor{⁇}}\AgdaSpace{}%
\AgdaBound{t}\AgdaSpace{}%
\AgdaOperator{\AgdaInductiveConstructor{∙}}\AgdaSpace{}%
\AgdaBound{s}\<%
\\
\>[0]\AgdaKeyword{pattern}\AgdaSpace{}%
\AgdaInductiveConstructor{send}\AgdaSpace{}%
\AgdaBound{t}\AgdaSpace{}%
\AgdaBound{s}\AgdaSpace{}%
\AgdaSymbol{=}\AgdaSpace{}%
\AgdaOperator{\AgdaInductiveConstructor{‼}}\AgdaSpace{}%
\AgdaBound{t}\AgdaSpace{}%
\AgdaOperator{\AgdaInductiveConstructor{∙}}\AgdaSpace{}%
\AgdaBound{s}\<%
\\
\\[\AgdaEmptyExtraSkip]%
\>[0]\AgdaKeyword{infixr}\AgdaSpace{}%
\AgdaNumber{20}\AgdaSpace{}%
\AgdaOperator{\AgdaInductiveConstructor{‼\AgdaUnderscore{}∙\AgdaUnderscore{}}}\AgdaSpace{}%
\AgdaOperator{\AgdaInductiveConstructor{⁇\AgdaUnderscore{}∙\AgdaUnderscore{}}}\<%
\\
\>[0]\AgdaKeyword{infixr}\AgdaSpace{}%
\AgdaNumber{20}\AgdaSpace{}%
\AgdaOperator{\AgdaInductiveConstructor{μ\AgdaUnderscore{}}}\AgdaSpace{}%
\AgdaOperator{\AgdaInductiveConstructor{`\AgdaUnderscore{}}}\<%
\\
\\[\AgdaEmptyExtraSkip]%
\>[0]\AgdaComment{--\ duality}\<%
\\
\>[0]\AgdaFunction{dual}\AgdaSpace{}%
\AgdaSymbol{:}\AgdaSpace{}%
\AgdaDatatype{Session}\AgdaSpace{}%
\AgdaGeneralizable{n}\AgdaSpace{}%
\AgdaSymbol{→}\AgdaSpace{}%
\AgdaDatatype{Session}\AgdaSpace{}%
\AgdaGeneralizable{n}\<%
\\
\>[0]\AgdaFunction{dual}\AgdaSpace{}%
\AgdaSymbol{(}\AgdaInductiveConstructor{send}\AgdaSpace{}%
\AgdaBound{T}\AgdaSpace{}%
\AgdaBound{S}\AgdaSymbol{)}\AgdaSpace{}%
\AgdaSymbol{=}\AgdaSpace{}%
\AgdaInductiveConstructor{recv}\AgdaSpace{}%
\AgdaBound{T}\AgdaSpace{}%
\AgdaSymbol{(}\AgdaFunction{dual}\AgdaSpace{}%
\AgdaBound{S}\AgdaSymbol{)}\<%
\\
\>[0]\AgdaFunction{dual}\AgdaSpace{}%
\AgdaSymbol{(}\AgdaInductiveConstructor{recv}\AgdaSpace{}%
\AgdaBound{T}\AgdaSpace{}%
\AgdaBound{S}\AgdaSymbol{)}\AgdaSpace{}%
\AgdaSymbol{=}\AgdaSpace{}%
\AgdaInductiveConstructor{send}\AgdaSpace{}%
\AgdaBound{T}\AgdaSpace{}%
\AgdaSymbol{(}\AgdaFunction{dual}\AgdaSpace{}%
\AgdaBound{S}\AgdaSymbol{)}\<%
\\
\>[0]\AgdaFunction{dual}\AgdaSpace{}%
\AgdaInductiveConstructor{end}\AgdaSpace{}%
\AgdaSymbol{=}\AgdaSpace{}%
\AgdaInductiveConstructor{end}\<%
\\
\>[0]\AgdaFunction{dual}\AgdaSpace{}%
\AgdaSymbol{(}\AgdaInductiveConstructor{⊕′}\AgdaSpace{}%
\AgdaBound{Si}\AgdaSymbol{)}\AgdaSpace{}%
\AgdaSymbol{=}\AgdaSpace{}%
\AgdaInductiveConstructor{\&′}\AgdaSpace{}%
\AgdaSymbol{(}\AgdaFunction{dual}\AgdaSpace{}%
\AgdaOperator{\AgdaFunction{∘}}\AgdaSpace{}%
\AgdaBound{Si}\AgdaSymbol{)}\<%
\\
\>[0]\AgdaFunction{dual}\AgdaSpace{}%
\AgdaSymbol{(}\AgdaInductiveConstructor{\&′}\AgdaSpace{}%
\AgdaBound{Si}\AgdaSymbol{)}\AgdaSpace{}%
\AgdaSymbol{=}\AgdaSpace{}%
\AgdaInductiveConstructor{⊕′}\AgdaSpace{}%
\AgdaSymbol{(}\AgdaFunction{dual}\AgdaSpace{}%
\AgdaOperator{\AgdaFunction{∘}}\AgdaSpace{}%
\AgdaBound{Si}\AgdaSymbol{)}\<%
\\
\>[0]\AgdaFunction{dual}\AgdaSpace{}%
\AgdaSymbol{(}\AgdaOperator{\AgdaInductiveConstructor{μ}}\AgdaSpace{}%
\AgdaBound{S}\AgdaSymbol{)}\AgdaSpace{}%
\AgdaSymbol{=}\AgdaSpace{}%
\AgdaOperator{\AgdaInductiveConstructor{μ}}\AgdaSpace{}%
\AgdaSymbol{(}\AgdaFunction{dual}\AgdaSpace{}%
\AgdaBound{S}\AgdaSymbol{)}\<%
\\
\>[0]\AgdaFunction{dual}\AgdaSpace{}%
\AgdaSymbol{(}\AgdaOperator{\AgdaInductiveConstructor{`}}\AgdaSpace{}%
\AgdaBound{x}\AgdaSymbol{)}\AgdaSpace{}%
\AgdaSymbol{=}\AgdaSpace{}%
\AgdaOperator{\AgdaInductiveConstructor{`}}\AgdaSpace{}%
\AgdaBound{x}\<%
\\
\\[\AgdaEmptyExtraSkip]%
\\[\AgdaEmptyExtraSkip]%
\>[0]\AgdaFunction{⊕}\AgdaSpace{}%
\AgdaSymbol{:}\AgdaSpace{}%
\AgdaDatatype{Vec}\AgdaSpace{}%
\AgdaSymbol{(}\AgdaDatatype{Session}\AgdaSpace{}%
\AgdaGeneralizable{n}\AgdaSymbol{)}\AgdaSpace{}%
\AgdaGeneralizable{k}\AgdaSpace{}%
\AgdaSymbol{→}\AgdaSpace{}%
\AgdaDatatype{Session}\AgdaSpace{}%
\AgdaGeneralizable{n}\<%
\\
\>[0]\AgdaFunction{⊕}\AgdaSpace{}%
\AgdaSymbol{=}\AgdaSpace{}%
\AgdaInductiveConstructor{⊕′}\AgdaSpace{}%
\AgdaOperator{\AgdaFunction{∘}}\AgdaSpace{}%
\AgdaFunction{vec→fin}\<%
\\
\\[\AgdaEmptyExtraSkip]%
\>[0]\AgdaFunction{\&}\AgdaSpace{}%
\AgdaSymbol{:}\AgdaSpace{}%
\AgdaDatatype{Vec}\AgdaSpace{}%
\AgdaSymbol{(}\AgdaDatatype{Session}\AgdaSpace{}%
\AgdaGeneralizable{n}\AgdaSymbol{)}\AgdaSpace{}%
\AgdaGeneralizable{k}\AgdaSpace{}%
\AgdaSymbol{→}\AgdaSpace{}%
\AgdaDatatype{Session}\AgdaSpace{}%
\AgdaGeneralizable{n}\<%
\\
\>[0]\AgdaFunction{\&}\AgdaSpace{}%
\AgdaSymbol{=}\AgdaSpace{}%
\AgdaInductiveConstructor{\&′}\AgdaSpace{}%
\AgdaOperator{\AgdaFunction{∘}}\AgdaSpace{}%
\AgdaFunction{vec→fin}\<%
\\
\>[0]\<%
\\
\\[\AgdaEmptyExtraSkip]%
\>[0]\AgdaComment{--\ service\ protocol\ for\ a\ binary\ function}\<%
\\
\>[0]\AgdaFunction{binaryp}\AgdaSpace{}%
\AgdaSymbol{:}\AgdaSpace{}%
\AgdaDatatype{Session}\AgdaSpace{}%
\AgdaGeneralizable{n}\AgdaSpace{}%
\AgdaSymbol{→}\AgdaSpace{}%
\AgdaDatatype{Session}\AgdaSpace{}%
\AgdaGeneralizable{n}\<%
\\
\>[0]\AgdaFunction{binaryp}\AgdaSpace{}%
\AgdaBound{S}\AgdaSpace{}%
\AgdaSymbol{=}\AgdaSpace{}%
\AgdaOperator{\AgdaInductiveConstructor{⁇}}\AgdaSpace{}%
\AgdaInductiveConstructor{int}\AgdaSpace{}%
\AgdaOperator{\AgdaInductiveConstructor{∙}}\AgdaSpace{}%
\AgdaOperator{\AgdaInductiveConstructor{⁇}}\AgdaSpace{}%
\AgdaInductiveConstructor{int}\AgdaSpace{}%
\AgdaOperator{\AgdaInductiveConstructor{∙}}\AgdaSpace{}%
\AgdaOperator{\AgdaInductiveConstructor{‼}}\AgdaSpace{}%
\AgdaInductiveConstructor{int}\AgdaSpace{}%
\AgdaOperator{\AgdaInductiveConstructor{∙}}\AgdaSpace{}%
\AgdaBound{S}\<%
\\
\>[0]\AgdaComment{--\ service\ protocol\ for\ a\ unary\ function}\<%
\\
\>[0]\AgdaComment{--\ service\ protocol\ for\ choosing\ between\ a\ binary\ and\ a\ unary\ function}\<%
\\
\>[0]\AgdaFunction{arithp}\AgdaSpace{}%
\AgdaSymbol{:}\AgdaSpace{}%
\AgdaDatatype{Session}\AgdaSpace{}%
\AgdaGeneralizable{n}\AgdaSpace{}%
\AgdaSymbol{→}\AgdaSpace{}%
\AgdaDatatype{Session}\AgdaSpace{}%
\AgdaGeneralizable{n}\<%
\\
\>[0]\AgdaComment{--\ service\ protocol\ for\ many\ unary\ ops}\<%
\end{code}
\newcommand\rstExampleManyUnaryp{%
\begin{code}%
\>[0]\AgdaFunction{unaryp}\AgdaSpace{}%
\AgdaSymbol{:}\AgdaSpace{}%
\AgdaDatatype{Session}\AgdaSpace{}%
\AgdaGeneralizable{n}\AgdaSpace{}%
\AgdaSymbol{→}\AgdaSpace{}%
\AgdaDatatype{Session}\AgdaSpace{}%
\AgdaGeneralizable{n}\<%
\\
\>[0]\AgdaFunction{unaryp}\AgdaSpace{}%
\AgdaBound{S}\AgdaSpace{}%
\AgdaSymbol{=}\AgdaSpace{}%
\AgdaOperator{\AgdaInductiveConstructor{⁇}}\AgdaSpace{}%
\AgdaInductiveConstructor{int}\AgdaSpace{}%
\AgdaOperator{\AgdaInductiveConstructor{∙}}\AgdaSpace{}%
\AgdaOperator{\AgdaInductiveConstructor{‼}}\AgdaSpace{}%
\AgdaInductiveConstructor{int}\AgdaSpace{}%
\AgdaOperator{\AgdaInductiveConstructor{∙}}\AgdaSpace{}%
\AgdaBound{S}\<%
\\
\\[\AgdaEmptyExtraSkip]%
\>[0]\AgdaFunction{many-unaryp}\AgdaSpace{}%
\AgdaSymbol{:}\AgdaSpace{}%
\AgdaDatatype{Session}\AgdaSpace{}%
\AgdaNumber{0}\<%
\\
\>[0]\AgdaFunction{many-unaryp}\AgdaSpace{}%
\AgdaSymbol{=}\AgdaSpace{}%
\AgdaOperator{\AgdaInductiveConstructor{μ}}\AgdaSpace{}%
\AgdaFunction{\&}\AgdaSpace{}%
\AgdaOperator{\AgdaInductiveConstructor{[}}\AgdaSpace{}%
\AgdaFunction{unaryp}\AgdaSpace{}%
\AgdaSymbol{(}\AgdaOperator{\AgdaInductiveConstructor{`}}\AgdaSpace{}%
\AgdaInductiveConstructor{zero}\AgdaSymbol{)}\AgdaSpace{}%
\AgdaOperator{\AgdaInductiveConstructor{,}}\AgdaSpace{}%
\AgdaInductiveConstructor{end}\AgdaSpace{}%
\AgdaOperator{\AgdaInductiveConstructor{]}}\<%
\end{code}}
\newcommand\rstExampleArithP{%
\begin{code}%
\>[0]\AgdaFunction{arithp}\AgdaSpace{}%
\AgdaBound{S}\AgdaSpace{}%
\AgdaSymbol{=}\AgdaSpace{}%
\AgdaFunction{\&}\AgdaSpace{}%
\AgdaOperator{\AgdaInductiveConstructor{[}}\AgdaSpace{}%
\AgdaFunction{binaryp}\AgdaSpace{}%
\AgdaBound{S}\AgdaSpace{}%
\AgdaOperator{\AgdaInductiveConstructor{,}}\AgdaSpace{}%
\AgdaFunction{unaryp}\AgdaSpace{}%
\AgdaBound{S}\AgdaSpace{}%
\AgdaOperator{\AgdaInductiveConstructor{]}}\<%
\end{code}}
\begin{code}[hide]%
\>[0]\AgdaFunction{arithp-raw}\AgdaSpace{}%
\AgdaSymbol{:}\AgdaSpace{}%
\AgdaDatatype{Session}\AgdaSpace{}%
\AgdaGeneralizable{n}\AgdaSpace{}%
\AgdaSymbol{→}\AgdaSpace{}%
\AgdaDatatype{Session}\AgdaSpace{}%
\AgdaGeneralizable{n}\<%
\\
\>[0]\AgdaFunction{arithp-raw}\AgdaSpace{}%
\AgdaBound{S}\AgdaSpace{}%
\AgdaSymbol{=}\AgdaSpace{}%
\AgdaInductiveConstructor{\&′}\AgdaSpace{}%
\AgdaSymbol{\{}\AgdaArgument{k}\AgdaSpace{}%
\AgdaSymbol{=}\AgdaSpace{}%
\AgdaNumber{2}\AgdaSymbol{\}}\AgdaSpace{}%
\AgdaSymbol{(λ\{}\AgdaSpace{}%
\AgdaInductiveConstructor{zero}\AgdaSpace{}%
\AgdaSymbol{→}\AgdaSpace{}%
\AgdaFunction{binaryp}\AgdaSpace{}%
\AgdaBound{S}\AgdaSpace{}%
\AgdaSymbol{;}\AgdaSpace{}%
\AgdaSymbol{(}\AgdaInductiveConstructor{suc}\AgdaSpace{}%
\AgdaInductiveConstructor{zero}\AgdaSymbol{)}\AgdaSpace{}%
\AgdaSymbol{→}\AgdaSpace{}%
\AgdaFunction{unaryp}\AgdaSpace{}%
\AgdaBound{S}\AgdaSymbol{\})}\<%
\\
\\[\AgdaEmptyExtraSkip]%
\\[\AgdaEmptyExtraSkip]%
\>[0]\AgdaKeyword{variable}\<%
\\
\>[0][@{}l@{\AgdaIndent{0}}]%
\>[2]\AgdaGeneralizable{A}\AgdaSpace{}%
\AgdaGeneralizable{A′}\AgdaSpace{}%
\AgdaGeneralizable{A″}\AgdaSpace{}%
\AgdaGeneralizable{A₁}\AgdaSpace{}%
\AgdaGeneralizable{A₂}\AgdaSpace{}%
\AgdaSymbol{:}\AgdaSpace{}%
\AgdaPrimitive{Set}\<%
\\
\>[2]\AgdaGeneralizable{T}\AgdaSpace{}%
\AgdaGeneralizable{t}\AgdaSpace{}%
\AgdaSymbol{:}\AgdaSpace{}%
\AgdaDatatype{Type}\<%
\\
\>[2]\AgdaGeneralizable{S}\AgdaSpace{}%
\AgdaGeneralizable{s}\AgdaSpace{}%
\AgdaGeneralizable{s₁}\AgdaSpace{}%
\AgdaGeneralizable{s₂}\AgdaSpace{}%
\AgdaSymbol{:}\AgdaSpace{}%
\AgdaDatatype{Session}\AgdaSpace{}%
\AgdaGeneralizable{n}\<%
\end{code}
\newcommand\rstTypeInterpretation{%
\begin{code}%
\>[0]\AgdaOperator{\AgdaFunction{T⟦\AgdaUnderscore{}⟧}}\AgdaSpace{}%
\AgdaSymbol{:}\AgdaSpace{}%
\AgdaDatatype{Type}\AgdaSpace{}%
\AgdaSymbol{→}\AgdaSpace{}%
\AgdaPrimitive{Set}\<%
\\
\>[0]\AgdaOperator{\AgdaFunction{T⟦}}\AgdaSpace{}%
\AgdaInductiveConstructor{int}\AgdaSpace{}%
\AgdaOperator{\AgdaFunction{⟧}}\AgdaSpace{}%
\AgdaSymbol{=}\AgdaSpace{}%
\AgdaDatatype{ℤ}\<%
\\
\>[0]\AgdaOperator{\AgdaFunction{T⟦}}\AgdaSpace{}%
\AgdaInductiveConstructor{bool}\AgdaSpace{}%
\AgdaOperator{\AgdaFunction{⟧}}\AgdaSpace{}%
\AgdaSymbol{=}\AgdaSpace{}%
\AgdaDatatype{Bool}\<%
\end{code}}
\begin{code}[hide]%
\>[0]\AgdaOperator{\AgdaFunction{T⟦}}\AgdaSpace{}%
\AgdaInductiveConstructor{nat}\AgdaSpace{}%
\AgdaOperator{\AgdaFunction{⟧}}\AgdaSpace{}%
\AgdaSymbol{=}\AgdaSpace{}%
\AgdaDatatype{ℕ}\<%
\\
\>[0]\AgdaOperator{\AgdaFunction{T⟦}}\AgdaSpace{}%
\AgdaInductiveConstructor{fin}\AgdaSpace{}%
\AgdaBound{k}\AgdaSpace{}%
\AgdaOperator{\AgdaFunction{⟧}}\AgdaSpace{}%
\AgdaSymbol{=}\AgdaSpace{}%
\AgdaDatatype{Fin}\AgdaSpace{}%
\AgdaBound{k}\<%
\\
\\[\AgdaEmptyExtraSkip]%
\>[0]\AgdaKeyword{module}\AgdaSpace{}%
\AgdaModule{formatting2}\AgdaSpace{}%
\AgdaKeyword{where}\<%
\\
\>[0]\<%
\end{code}
\newcommand\rstBranchingCommand{%
\begin{code}%
\>[0][@{}l@{\AgdaIndent{1}}]%
\>[2]\AgdaComment{--\ data\ Cmd\ (A\ :\ Set)\ :\ Session\ n\ →\ Set\ where}\<%
\\
\>[2]\AgdaComment{--\ \ \ SELECT\ :\ ∀\ \{si\}\ →\ (setl\ :\ A\ →\ Fin\ k\ ×\ A)}\<%
\\
\>[2]\AgdaComment{--\ \ \ \ \ \ \ \ \ \ \ \ \ \ \ \ \ \ \ →\ ((i\ :\ Fin\ k)\ →\ Cmd\ A\ (si\ i))}\<%
\\
\>[2]\AgdaComment{--\ \ \ \ \ \ \ \ \ \ \ \ \ \ \ \ \ \ \ →\ Cmd\ A\ (⊕\ si)}\<%
\\
\>[2]\AgdaComment{--\ \ \ CHOICE\ :\ ∀\ \{si\}\ →\ (getl\ :\ Fin\ k\ →\ A\ →\ A)}\<%
\\
\>[2]\AgdaComment{--\ \ \ \ \ \ \ \ \ \ \ \ \ \ \ \ \ \ \ →\ ((i\ :\ Fin\ k)\ →\ Cmd\ A\ (si\ i))}\<%
\\
\>[2]\AgdaComment{--\ \ \ \ \ \ \ \ \ \ \ \ \ \ \ \ \ \ \ →\ Cmd\ A\ (\&\ si)}\<%
\end{code}}
\newcommand\rstCommand{%
\begin{code}%
\>[0]\AgdaKeyword{data}\AgdaSpace{}%
\AgdaDatatype{Cmd}\AgdaSpace{}%
\AgdaSymbol{(}\AgdaBound{n}\AgdaSpace{}%
\AgdaSymbol{:}\AgdaSpace{}%
\AgdaDatatype{ℕ}\AgdaSymbol{)}\AgdaSpace{}%
\AgdaSymbol{(}\AgdaBound{A}\AgdaSpace{}%
\AgdaSymbol{:}\AgdaSpace{}%
\AgdaPrimitive{Set}\AgdaSymbol{)}\AgdaSpace{}%
\AgdaSymbol{:}\AgdaSpace{}%
\AgdaDatatype{Session}\AgdaSpace{}%
\AgdaBound{n}\AgdaSpace{}%
\AgdaSymbol{→}\AgdaSpace{}%
\AgdaPrimitive{Set}\AgdaSpace{}%
\AgdaKeyword{where}\<%
\\
\>[0][@{}l@{\AgdaIndent{0}}]%
\>[2]\AgdaInductiveConstructor{LOOP}%
\>[11]\AgdaSymbol{:}\AgdaSpace{}%
\AgdaDatatype{Cmd}\AgdaSpace{}%
\AgdaSymbol{(}\AgdaInductiveConstructor{suc}\AgdaSpace{}%
\AgdaBound{n}\AgdaSymbol{)}\AgdaSpace{}%
\AgdaBound{A}\AgdaSpace{}%
\AgdaGeneralizable{S}\AgdaSpace{}%
\AgdaSymbol{→}\AgdaSpace{}%
\AgdaDatatype{Cmd}\AgdaSpace{}%
\AgdaBound{n}\AgdaSpace{}%
\AgdaBound{A}\AgdaSpace{}%
\AgdaSymbol{(}\AgdaOperator{\AgdaInductiveConstructor{μ}}\AgdaSpace{}%
\AgdaGeneralizable{S}\AgdaSymbol{)}\<%
\\
\>[2]\AgdaInductiveConstructor{CONTINUE}\AgdaSpace{}%
\AgdaSymbol{:}\AgdaSpace{}%
\AgdaSymbol{(}\AgdaBound{i}\AgdaSpace{}%
\AgdaSymbol{:}\AgdaSpace{}%
\AgdaDatatype{Fin}\AgdaSpace{}%
\AgdaBound{n}\AgdaSymbol{)}\AgdaSpace{}%
\AgdaSymbol{→}\AgdaSpace{}%
\AgdaDatatype{Cmd}\AgdaSpace{}%
\AgdaBound{n}\AgdaSpace{}%
\AgdaBound{A}\AgdaSpace{}%
\AgdaSymbol{(}\AgdaOperator{\AgdaInductiveConstructor{`}}\AgdaSpace{}%
\AgdaBound{i}\AgdaSymbol{)}\<%
\end{code}}
\newcommand\rstCommandUNROLL{%
\begin{code}%
\>[2]\AgdaInductiveConstructor{UNROLL}%
\>[11]\AgdaSymbol{:}\AgdaSpace{}%
\AgdaDatatype{Cmd}\AgdaSpace{}%
\AgdaSymbol{(}\AgdaInductiveConstructor{suc}\AgdaSpace{}%
\AgdaBound{n}\AgdaSymbol{)}\AgdaSpace{}%
\AgdaBound{A}\AgdaSpace{}%
\AgdaGeneralizable{S}\AgdaSpace{}%
\AgdaSymbol{→}\AgdaSpace{}%
\AgdaDatatype{Cmd}\AgdaSpace{}%
\AgdaBound{n}\AgdaSpace{}%
\AgdaBound{A}\AgdaSpace{}%
\AgdaSymbol{(}\AgdaOperator{\AgdaInductiveConstructor{μ}}\AgdaSpace{}%
\AgdaGeneralizable{S}\AgdaSymbol{)}\AgdaSpace{}%
\AgdaSymbol{→}\AgdaSpace{}%
\AgdaDatatype{Cmd}\AgdaSpace{}%
\AgdaBound{n}\AgdaSpace{}%
\AgdaBound{A}\AgdaSpace{}%
\AgdaSymbol{(}\AgdaOperator{\AgdaInductiveConstructor{μ}}\AgdaSpace{}%
\AgdaGeneralizable{S}\AgdaSymbol{)}\<%
\end{code}}
\begin{code}[hide]%
\>[2]\AgdaInductiveConstructor{CLOSE}%
\>[9]\AgdaSymbol{:}\AgdaSpace{}%
\AgdaDatatype{Cmd}\AgdaSpace{}%
\AgdaBound{n}\AgdaSpace{}%
\AgdaBound{A}\AgdaSpace{}%
\AgdaInductiveConstructor{end}\<%
\\
\>[2]\AgdaInductiveConstructor{SEND}%
\>[9]\AgdaSymbol{:}\AgdaSpace{}%
\AgdaSymbol{(}\AgdaBound{A}\AgdaSpace{}%
\AgdaSymbol{→}\AgdaSpace{}%
\AgdaBound{A}\AgdaSpace{}%
\AgdaOperator{\AgdaFunction{×}}\AgdaSpace{}%
\AgdaOperator{\AgdaFunction{T⟦}}\AgdaSpace{}%
\AgdaGeneralizable{T}\AgdaSpace{}%
\AgdaOperator{\AgdaFunction{⟧}}\AgdaSymbol{)}\AgdaSpace{}%
\AgdaSymbol{→}\AgdaSpace{}%
\AgdaDatatype{Cmd}\AgdaSpace{}%
\AgdaBound{n}\AgdaSpace{}%
\AgdaBound{A}\AgdaSpace{}%
\AgdaGeneralizable{S}\AgdaSpace{}%
\AgdaSymbol{→}\AgdaSpace{}%
\AgdaDatatype{Cmd}\AgdaSpace{}%
\AgdaBound{n}\AgdaSpace{}%
\AgdaBound{A}\AgdaSpace{}%
\AgdaSymbol{(}\AgdaOperator{\AgdaInductiveConstructor{‼}}\AgdaSpace{}%
\AgdaGeneralizable{T}\AgdaSpace{}%
\AgdaOperator{\AgdaInductiveConstructor{∙}}\AgdaSpace{}%
\AgdaGeneralizable{S}\AgdaSymbol{)}\<%
\\
\>[2]\AgdaInductiveConstructor{RECV}%
\>[9]\AgdaSymbol{:}\AgdaSpace{}%
\AgdaSymbol{(}\AgdaOperator{\AgdaFunction{T⟦}}\AgdaSpace{}%
\AgdaGeneralizable{T}\AgdaSpace{}%
\AgdaOperator{\AgdaFunction{⟧}}\AgdaSpace{}%
\AgdaSymbol{→}\AgdaSpace{}%
\AgdaBound{A}\AgdaSpace{}%
\AgdaSymbol{→}\AgdaSpace{}%
\AgdaBound{A}\AgdaSymbol{)}\AgdaSpace{}%
\AgdaSymbol{→}\AgdaSpace{}%
\AgdaDatatype{Cmd}\AgdaSpace{}%
\AgdaBound{n}\AgdaSpace{}%
\AgdaBound{A}\AgdaSpace{}%
\AgdaGeneralizable{S}\AgdaSpace{}%
\AgdaSymbol{→}\AgdaSpace{}%
\AgdaDatatype{Cmd}\AgdaSpace{}%
\AgdaBound{n}\AgdaSpace{}%
\AgdaBound{A}\AgdaSpace{}%
\AgdaSymbol{(}\AgdaOperator{\AgdaInductiveConstructor{⁇}}\AgdaSpace{}%
\AgdaGeneralizable{T}\AgdaSpace{}%
\AgdaOperator{\AgdaInductiveConstructor{∙}}\AgdaSpace{}%
\AgdaGeneralizable{S}\AgdaSymbol{)}\<%
\\
\>[2]\AgdaInductiveConstructor{SELECT}\AgdaSpace{}%
\AgdaSymbol{:}\AgdaSpace{}%
\AgdaSymbol{∀}\AgdaSpace{}%
\AgdaSymbol{\{}\AgdaBound{Si}\AgdaSymbol{\}}\AgdaSpace{}%
\AgdaSymbol{→}\AgdaSpace{}%
\AgdaSymbol{(}\AgdaBound{i}\AgdaSpace{}%
\AgdaSymbol{:}\AgdaSpace{}%
\AgdaDatatype{Fin}\AgdaSpace{}%
\AgdaGeneralizable{k}\AgdaSymbol{)}\AgdaSpace{}%
\AgdaSymbol{→}\AgdaSpace{}%
\AgdaDatatype{Cmd}\AgdaSpace{}%
\AgdaBound{n}\AgdaSpace{}%
\AgdaBound{A}\AgdaSpace{}%
\AgdaSymbol{(}\AgdaBound{Si}\AgdaSpace{}%
\AgdaBound{i}\AgdaSymbol{)}\AgdaSpace{}%
\AgdaSymbol{→}\AgdaSpace{}%
\AgdaDatatype{Cmd}\AgdaSpace{}%
\AgdaBound{n}\AgdaSpace{}%
\AgdaBound{A}\AgdaSpace{}%
\AgdaSymbol{(}\AgdaInductiveConstructor{⊕′}\AgdaSpace{}%
\AgdaBound{Si}\AgdaSymbol{)}\<%
\\
\>[2]\AgdaInductiveConstructor{CHOICE}\AgdaSpace{}%
\AgdaSymbol{:}\AgdaSpace{}%
\AgdaSymbol{∀}\AgdaSpace{}%
\AgdaSymbol{\{}\AgdaBound{Si}\AgdaSymbol{\}}\AgdaSpace{}%
\AgdaSymbol{→}\AgdaSpace{}%
\AgdaSymbol{((}\AgdaBound{i}\AgdaSpace{}%
\AgdaSymbol{:}\AgdaSpace{}%
\AgdaDatatype{Fin}\AgdaSpace{}%
\AgdaGeneralizable{k}\AgdaSymbol{)}\AgdaSpace{}%
\AgdaSymbol{→}\AgdaSpace{}%
\AgdaDatatype{Cmd}\AgdaSpace{}%
\AgdaBound{n}\AgdaSpace{}%
\AgdaBound{A}\AgdaSpace{}%
\AgdaSymbol{(}\AgdaBound{Si}\AgdaSpace{}%
\AgdaBound{i}\AgdaSymbol{))}\AgdaSpace{}%
\AgdaSymbol{→}\AgdaSpace{}%
\AgdaDatatype{Cmd}\AgdaSpace{}%
\AgdaBound{n}\AgdaSpace{}%
\AgdaBound{A}\AgdaSpace{}%
\AgdaSymbol{(}\AgdaInductiveConstructor{\&′}\AgdaSpace{}%
\AgdaBound{Si}\AgdaSymbol{)}\<%
\end{code}
\newcommand\rstAddpCommand{%
\begin{code}%
\>[0]\AgdaFunction{addp-command}\AgdaSpace{}%
\AgdaSymbol{:}\AgdaSpace{}%
\AgdaDatatype{Cmd}\AgdaSpace{}%
\AgdaGeneralizable{n}\AgdaSpace{}%
\AgdaDatatype{ℤ}\AgdaSpace{}%
\AgdaGeneralizable{S}\AgdaSpace{}%
\AgdaSymbol{→}\AgdaSpace{}%
\AgdaDatatype{Cmd}\AgdaSpace{}%
\AgdaGeneralizable{n}\AgdaSpace{}%
\AgdaDatatype{ℤ}\AgdaSpace{}%
\AgdaSymbol{(}\AgdaFunction{binaryp}\AgdaSpace{}%
\AgdaGeneralizable{S}\AgdaSymbol{)}\<%
\\
\>[0]\AgdaFunction{addp-command}\AgdaSpace{}%
\AgdaBound{cmd}\AgdaSpace{}%
\AgdaSymbol{=}\AgdaSpace{}%
\AgdaInductiveConstructor{RECV}\AgdaSpace{}%
\AgdaSymbol{(λ}\AgdaSpace{}%
\AgdaBound{x}\AgdaSpace{}%
\AgdaBound{a}\AgdaSpace{}%
\AgdaSymbol{→}\AgdaSpace{}%
\AgdaBound{x}\AgdaSymbol{)}\AgdaSpace{}%
\AgdaOperator{\AgdaFunction{\$}}\AgdaSpace{}%
\AgdaInductiveConstructor{RECV}\AgdaSpace{}%
\AgdaSymbol{(λ}\AgdaSpace{}%
\AgdaBound{y}\AgdaSpace{}%
\AgdaBound{a}\AgdaSpace{}%
\AgdaSymbol{→}\AgdaSpace{}%
\AgdaBound{y}\AgdaSpace{}%
\AgdaOperator{\AgdaFunction{+}}\AgdaSpace{}%
\AgdaBound{a}\AgdaSymbol{)}\AgdaSpace{}%
\AgdaOperator{\AgdaFunction{\$}}\AgdaSpace{}%
\AgdaInductiveConstructor{SEND}\AgdaSpace{}%
\AgdaSymbol{(λ}\AgdaSpace{}%
\AgdaBound{a}\AgdaSpace{}%
\AgdaSymbol{→}\AgdaSpace{}%
\AgdaOperator{\AgdaInductiveConstructor{⟨}}\AgdaSpace{}%
\AgdaBound{a}\AgdaSpace{}%
\AgdaOperator{\AgdaInductiveConstructor{,}}\AgdaSpace{}%
\AgdaBound{a}\AgdaSpace{}%
\AgdaOperator{\AgdaInductiveConstructor{⟩}}\AgdaSymbol{)}\AgdaSpace{}%
\AgdaOperator{\AgdaFunction{\$}}\AgdaSpace{}%
\AgdaBound{cmd}\<%
\\
\>[0]\<%
\end{code}}
\newcommand\rstSumupCommand{%
\begin{code}%
\>[0]\AgdaFunction{addup-command}\AgdaSpace{}%
\AgdaSymbol{:}\AgdaSpace{}%
\AgdaDatatype{Cmd}\AgdaSpace{}%
\AgdaGeneralizable{n}\AgdaSpace{}%
\AgdaDatatype{ℤ}\AgdaSpace{}%
\AgdaGeneralizable{S}\AgdaSpace{}%
\AgdaSymbol{→}\AgdaSpace{}%
\AgdaDatatype{Cmd}\AgdaSpace{}%
\AgdaGeneralizable{n}\AgdaSpace{}%
\AgdaDatatype{ℤ}\AgdaSpace{}%
\AgdaSymbol{(}\AgdaFunction{unaryp}\AgdaSpace{}%
\AgdaGeneralizable{S}\AgdaSymbol{)}\<%
\\
\>[0]\AgdaFunction{addup-command}\AgdaSpace{}%
\AgdaBound{cmd}\AgdaSpace{}%
\AgdaSymbol{=}\AgdaSpace{}%
\AgdaInductiveConstructor{RECV}\AgdaSpace{}%
\AgdaSymbol{(λ}\AgdaSpace{}%
\AgdaBound{x}\AgdaSpace{}%
\AgdaBound{a}\AgdaSpace{}%
\AgdaSymbol{→}\AgdaSpace{}%
\AgdaBound{x}\AgdaSpace{}%
\AgdaOperator{\AgdaFunction{+}}\AgdaSpace{}%
\AgdaBound{a}\AgdaSymbol{)}\AgdaSpace{}%
\AgdaOperator{\AgdaFunction{\$}}\AgdaSpace{}%
\AgdaInductiveConstructor{SEND}\AgdaSpace{}%
\AgdaSymbol{(λ}\AgdaSpace{}%
\AgdaBound{a}\AgdaSpace{}%
\AgdaSymbol{→}\AgdaSpace{}%
\AgdaOperator{\AgdaInductiveConstructor{⟨}}\AgdaSpace{}%
\AgdaBound{a}\AgdaSpace{}%
\AgdaOperator{\AgdaInductiveConstructor{,}}\AgdaSpace{}%
\AgdaBound{a}\AgdaSpace{}%
\AgdaOperator{\AgdaInductiveConstructor{⟩}}\AgdaSymbol{)}\AgdaSpace{}%
\AgdaOperator{\AgdaFunction{\$}}\AgdaSpace{}%
\AgdaBound{cmd}\<%
\\
\\[\AgdaEmptyExtraSkip]%
\>[0]\AgdaFunction{runningsum-command}\AgdaSpace{}%
\AgdaSymbol{:}\AgdaSpace{}%
\AgdaDatatype{Cmd}\AgdaSpace{}%
\AgdaNumber{0}\AgdaSpace{}%
\AgdaDatatype{ℤ}\AgdaSpace{}%
\AgdaFunction{many-unaryp}\<%
\\
\>[0]\AgdaFunction{runningsum-command}\AgdaSpace{}%
\AgdaSymbol{=}\AgdaSpace{}%
\AgdaInductiveConstructor{LOOP}\AgdaSpace{}%
\AgdaOperator{\AgdaFunction{\$}}\AgdaSpace{}%
\AgdaInductiveConstructor{CHOICE}\AgdaSpace{}%
\AgdaSymbol{λ}\AgdaSpace{}%
\AgdaKeyword{where}\<%
\\
\>[0][@{}l@{\AgdaIndent{0}}]%
\>[2]\AgdaInductiveConstructor{zero}\AgdaSpace{}%
\AgdaSymbol{→}\AgdaSpace{}%
\AgdaFunction{addup-command}\AgdaSpace{}%
\AgdaSymbol{(}\AgdaInductiveConstructor{CONTINUE}\AgdaSpace{}%
\AgdaInductiveConstructor{zero}\AgdaSymbol{)}\<%
\\
\>[2]\AgdaSymbol{(}\AgdaInductiveConstructor{suc}\AgdaSpace{}%
\AgdaInductiveConstructor{zero}\AgdaSymbol{)}\AgdaSpace{}%
\AgdaSymbol{→}\AgdaSpace{}%
\AgdaInductiveConstructor{CLOSE}\<%
\end{code}}
\newcommand\rstPostulates{%
\begin{code}%
\>[0]\AgdaKeyword{postulate}\<%
\\
\>[0][@{}l@{\AgdaIndent{0}}]%
\>[2]\AgdaPostulate{Channel}\AgdaSpace{}%
\AgdaSymbol{:}\AgdaSpace{}%
\AgdaPrimitive{Set}\<%
\\
\>[2]\AgdaPostulate{primAccept}\AgdaSpace{}%
\AgdaSymbol{:}\AgdaSpace{}%
\AgdaDatatype{IO}\AgdaSpace{}%
\AgdaPostulate{Channel}\<%
\\
\>[2]\AgdaPostulate{primClose}%
\>[13]\AgdaSymbol{:}\AgdaSpace{}%
\AgdaPostulate{Channel}\AgdaSpace{}%
\AgdaSymbol{→}\AgdaSpace{}%
\AgdaDatatype{IO}\AgdaSpace{}%
\AgdaRecord{⊤}\<%
\\
\>[2]\AgdaPostulate{primSend}%
\>[13]\AgdaSymbol{:}\AgdaSpace{}%
\AgdaGeneralizable{A}\AgdaSpace{}%
\AgdaSymbol{→}\AgdaSpace{}%
\AgdaPostulate{Channel}\AgdaSpace{}%
\AgdaSymbol{→}\AgdaSpace{}%
\AgdaDatatype{IO}\AgdaSpace{}%
\AgdaRecord{⊤}\<%
\\
\>[2]\AgdaPostulate{primRecv}%
\>[13]\AgdaSymbol{:}\AgdaSpace{}%
\AgdaPostulate{Channel}\AgdaSpace{}%
\AgdaSymbol{→}\AgdaSpace{}%
\AgdaDatatype{IO}\AgdaSpace{}%
\AgdaGeneralizable{A}\<%
\end{code}}
\newcommand\rstCommandStore{%
\begin{code}%
\>[0]\AgdaFunction{CmdStore}\AgdaSpace{}%
\AgdaSymbol{:}\AgdaSpace{}%
\AgdaDatatype{ℕ}\AgdaSpace{}%
\AgdaSymbol{→}\AgdaSpace{}%
\AgdaPrimitive{Set}\AgdaSpace{}%
\AgdaSymbol{→}\AgdaSpace{}%
\AgdaPrimitive{Set}\<%
\\
\>[0]\AgdaFunction{CmdStore}\AgdaSpace{}%
\AgdaBound{n}\AgdaSpace{}%
\AgdaBound{A}\AgdaSpace{}%
\AgdaSymbol{=}\AgdaSpace{}%
\AgdaSymbol{(}\AgdaBound{i}\AgdaSpace{}%
\AgdaSymbol{:}\AgdaSpace{}%
\AgdaDatatype{Fin}\AgdaSpace{}%
\AgdaBound{n}\AgdaSymbol{)}\AgdaSpace{}%
\AgdaSymbol{→}\AgdaSpace{}%
\AgdaFunction{∃[}\AgdaSpace{}%
\AgdaBound{S}\AgdaSpace{}%
\AgdaFunction{]}\AgdaSpace{}%
\AgdaSymbol{(}\AgdaDatatype{Cmd}\AgdaSpace{}%
\AgdaSymbol{(}\AgdaFunction{toℕ}\AgdaSpace{}%
\AgdaSymbol{(}\AgdaFunction{opposite}\AgdaSpace{}%
\AgdaBound{i}\AgdaSymbol{))}\AgdaSpace{}%
\AgdaBound{A}\AgdaSpace{}%
\AgdaBound{S}\AgdaSymbol{)}\<%
\end{code}}
\newcommand\rstPops{%
\begin{code}%
\>[0]\AgdaFunction{push}\AgdaSpace{}%
\AgdaSymbol{:}\AgdaSpace{}%
\AgdaFunction{CmdStore}\AgdaSpace{}%
\AgdaGeneralizable{n}\AgdaSpace{}%
\AgdaGeneralizable{A}\AgdaSpace{}%
\AgdaSymbol{→}\AgdaSpace{}%
\AgdaDatatype{Cmd}\AgdaSpace{}%
\AgdaGeneralizable{n}\AgdaSpace{}%
\AgdaGeneralizable{A}\AgdaSpace{}%
\AgdaGeneralizable{S}\AgdaSpace{}%
\AgdaSymbol{→}\AgdaSpace{}%
\AgdaFunction{CmdStore}\AgdaSpace{}%
\AgdaSymbol{(}\AgdaInductiveConstructor{suc}\AgdaSpace{}%
\AgdaGeneralizable{n}\AgdaSymbol{)}\AgdaSpace{}%
\AgdaGeneralizable{A}\<%
\\
\>[0]\AgdaFunction{pop1}\AgdaSpace{}%
\AgdaSymbol{:}\AgdaSpace{}%
\AgdaFunction{CmdStore}\AgdaSpace{}%
\AgdaSymbol{(}\AgdaInductiveConstructor{suc}\AgdaSpace{}%
\AgdaGeneralizable{n}\AgdaSymbol{)}\AgdaSpace{}%
\AgdaGeneralizable{A}\AgdaSpace{}%
\AgdaSymbol{→}\AgdaSpace{}%
\AgdaFunction{CmdStore}\AgdaSpace{}%
\AgdaGeneralizable{n}\AgdaSpace{}%
\AgdaGeneralizable{A}\<%
\\
\>[0]\AgdaFunction{pop}%
\>[5]\AgdaSymbol{:}\AgdaSpace{}%
\AgdaFunction{CmdStore}\AgdaSpace{}%
\AgdaSymbol{(}\AgdaInductiveConstructor{suc}\AgdaSpace{}%
\AgdaGeneralizable{n}\AgdaSymbol{)}\AgdaSpace{}%
\AgdaGeneralizable{A}\AgdaSpace{}%
\AgdaSymbol{→}\AgdaSpace{}%
\AgdaSymbol{(}\AgdaBound{i}\AgdaSpace{}%
\AgdaSymbol{:}\AgdaSpace{}%
\AgdaDatatype{Fin}\AgdaSpace{}%
\AgdaSymbol{(}\AgdaInductiveConstructor{suc}\AgdaSpace{}%
\AgdaGeneralizable{n}\AgdaSymbol{))}\AgdaSpace{}%
\AgdaSymbol{→}\AgdaSpace{}%
\AgdaFunction{CmdStore}\AgdaSpace{}%
\AgdaSymbol{(}\AgdaInductiveConstructor{suc}\AgdaSpace{}%
\AgdaSymbol{(}\AgdaFunction{toℕ}\AgdaSpace{}%
\AgdaSymbol{(}\AgdaFunction{opposite}\AgdaSpace{}%
\AgdaBound{i}\AgdaSymbol{)))}\AgdaSpace{}%
\AgdaGeneralizable{A}\<%
\end{code}}
\begin{code}[hide]%
\>[0]\AgdaFunction{push}\AgdaSpace{}%
\AgdaSymbol{\{}\AgdaBound{n}\AgdaSymbol{\}\{}\AgdaArgument{S}\AgdaSpace{}%
\AgdaSymbol{=}\AgdaSpace{}%
\AgdaBound{S}\AgdaSymbol{\}}\AgdaSpace{}%
\AgdaBound{cms}\AgdaSpace{}%
\AgdaBound{cmd}\AgdaSpace{}%
\AgdaInductiveConstructor{zero}%
\>[32]\AgdaKeyword{rewrite}\AgdaSpace{}%
\AgdaFunction{toℕ-fromℕ}\AgdaSpace{}%
\AgdaBound{n}\AgdaSpace{}%
\AgdaSymbol{=}\AgdaSpace{}%
\AgdaOperator{\AgdaInductiveConstructor{⟨}}\AgdaSpace{}%
\AgdaBound{S}\AgdaSpace{}%
\AgdaOperator{\AgdaInductiveConstructor{,}}\AgdaSpace{}%
\AgdaBound{cmd}\AgdaSpace{}%
\AgdaOperator{\AgdaInductiveConstructor{⟩}}\<%
\\
\>[0]\AgdaFunction{push}\AgdaSpace{}%
\AgdaBound{cms}\AgdaSpace{}%
\AgdaBound{cmd}\AgdaSpace{}%
\AgdaSymbol{(}\AgdaInductiveConstructor{suc}\AgdaSpace{}%
\AgdaBound{i}\AgdaSymbol{)}\AgdaSpace{}%
\AgdaKeyword{rewrite}\AgdaSpace{}%
\AgdaFunction{toℕ-inject₁}\AgdaSpace{}%
\AgdaSymbol{(}\AgdaFunction{opposite}\AgdaSpace{}%
\AgdaBound{i}\AgdaSymbol{)}\AgdaSpace{}%
\AgdaSymbol{=}\AgdaSpace{}%
\AgdaBound{cms}\AgdaSpace{}%
\AgdaBound{i}\<%
\\
\\[\AgdaEmptyExtraSkip]%
\>[0]\AgdaFunction{pop1}\AgdaSpace{}%
\AgdaBound{cms}\AgdaSpace{}%
\AgdaBound{i}\AgdaSpace{}%
\AgdaKeyword{with}\AgdaSpace{}%
\AgdaBound{cms}\AgdaSpace{}%
\AgdaSymbol{(}\AgdaInductiveConstructor{suc}\AgdaSpace{}%
\AgdaBound{i}\AgdaSymbol{)}\<%
\\
\>[0]\AgdaSymbol{...}\AgdaSpace{}%
\AgdaSymbol{|}\AgdaSpace{}%
\AgdaBound{cms₁}\AgdaSpace{}%
\AgdaKeyword{rewrite}\AgdaSpace{}%
\AgdaFunction{toℕ-inject₁}\AgdaSpace{}%
\AgdaSymbol{(}\AgdaFunction{opposite}\AgdaSpace{}%
\AgdaBound{i}\AgdaSymbol{)}\AgdaSpace{}%
\AgdaSymbol{=}\AgdaSpace{}%
\AgdaBound{cms₁}\<%
\\
\\[\AgdaEmptyExtraSkip]%
\>[0]\AgdaFunction{pop}\AgdaSpace{}%
\AgdaSymbol{\{}\AgdaBound{n}\AgdaSymbol{\}}\AgdaSpace{}%
\AgdaBound{cms}\AgdaSpace{}%
\AgdaInductiveConstructor{zero}\AgdaSpace{}%
\AgdaKeyword{rewrite}\AgdaSpace{}%
\AgdaFunction{toℕ-fromℕ}\AgdaSpace{}%
\AgdaBound{n}\AgdaSpace{}%
\AgdaSymbol{=}\AgdaSpace{}%
\AgdaBound{cms}\<%
\\
\>[0]\AgdaFunction{pop}\AgdaSpace{}%
\AgdaSymbol{\{}\AgdaInductiveConstructor{suc}\AgdaSpace{}%
\AgdaBound{n}\AgdaSymbol{\}}\AgdaSpace{}%
\AgdaBound{cms}\AgdaSpace{}%
\AgdaSymbol{(}\AgdaInductiveConstructor{suc}\AgdaSpace{}%
\AgdaBound{i}\AgdaSymbol{)}\AgdaSpace{}%
\AgdaSymbol{=}\AgdaSpace{}%
\AgdaFunction{subst}\AgdaSpace{}%
\AgdaSymbol{(λ}\AgdaSpace{}%
\AgdaBound{H}\AgdaSpace{}%
\AgdaSymbol{→}\AgdaSpace{}%
\AgdaFunction{CmdStore}\AgdaSpace{}%
\AgdaSymbol{(}\AgdaInductiveConstructor{suc}\AgdaSpace{}%
\AgdaBound{H}\AgdaSymbol{)}\AgdaSpace{}%
\AgdaSymbol{\AgdaUnderscore{})}\AgdaSpace{}%
\AgdaSymbol{(}\AgdaFunction{sym}\AgdaSpace{}%
\AgdaSymbol{(}\AgdaFunction{toℕ-inject₁}\AgdaSpace{}%
\AgdaSymbol{(}\AgdaFunction{opposite}\AgdaSpace{}%
\AgdaBound{i}\AgdaSymbol{)))}\AgdaSpace{}%
\AgdaSymbol{(}\AgdaFunction{pop}\AgdaSpace{}%
\AgdaSymbol{(}\AgdaFunction{pop1}\AgdaSpace{}%
\AgdaBound{cms}\AgdaSymbol{)}\AgdaSpace{}%
\AgdaBound{i}\AgdaSymbol{)}\<%
\end{code}
\begin{code}[hide]%
\>[0]\AgdaKeyword{module}\AgdaSpace{}%
\AgdaModule{alternative-executor}\AgdaSpace{}%
\AgdaKeyword{where}\<%
\\
\>[0][@{}l@{\AgdaIndent{0}}]%
\>[2]\AgdaFunction{exec}\AgdaSpace{}%
\AgdaSymbol{:}\AgdaSpace{}%
\AgdaDatatype{Cmd}\AgdaSpace{}%
\AgdaGeneralizable{n}\AgdaSpace{}%
\AgdaGeneralizable{A}\AgdaSpace{}%
\AgdaGeneralizable{S}\AgdaSpace{}%
\AgdaSymbol{→}\AgdaSpace{}%
\AgdaFunction{CmdStore}\AgdaSpace{}%
\AgdaGeneralizable{n}\AgdaSpace{}%
\AgdaGeneralizable{A}\AgdaSpace{}%
\AgdaSymbol{→}\AgdaSpace{}%
\AgdaSymbol{(}\AgdaBound{init}\AgdaSpace{}%
\AgdaSymbol{:}\AgdaSpace{}%
\AgdaGeneralizable{A}\AgdaSymbol{)}\AgdaSpace{}%
\AgdaSymbol{→}\AgdaSpace{}%
\AgdaPostulate{Channel}\<%
\\
\>[2][@{}l@{\AgdaIndent{0}}]%
\>[4]\AgdaSymbol{→}\AgdaSpace{}%
\AgdaDatatype{IO}\AgdaSpace{}%
\AgdaSymbol{(}\AgdaFunction{∃[}\AgdaSpace{}%
\AgdaBound{n}\AgdaSpace{}%
\AgdaFunction{]}\AgdaSpace{}%
\AgdaSymbol{(}\AgdaFunction{CmdStore}\AgdaSpace{}%
\AgdaSymbol{(}\AgdaInductiveConstructor{suc}\AgdaSpace{}%
\AgdaBound{n}\AgdaSymbol{)}\AgdaSpace{}%
\AgdaGeneralizable{A}\AgdaSpace{}%
\AgdaOperator{\AgdaFunction{×}}\AgdaSpace{}%
\AgdaGeneralizable{A}\AgdaSymbol{)}\AgdaSpace{}%
\AgdaOperator{\AgdaDatatype{⊎}}\AgdaSpace{}%
\AgdaGeneralizable{A}\AgdaSymbol{)}\<%
\\
\>[2]\AgdaFunction{exec}\AgdaSpace{}%
\AgdaSymbol{(}\AgdaInductiveConstructor{UNROLL}\AgdaSpace{}%
\AgdaBound{body-cmd}\AgdaSpace{}%
\AgdaBound{next-cmd}\AgdaSymbol{)}\AgdaSpace{}%
\AgdaBound{cms}\AgdaSpace{}%
\AgdaBound{st}\AgdaSpace{}%
\AgdaBound{ch}\AgdaSpace{}%
\AgdaSymbol{=}\AgdaSpace{}%
\AgdaFunction{exec}\AgdaSpace{}%
\AgdaBound{body-cmd}\AgdaSpace{}%
\AgdaSymbol{(}\AgdaFunction{push}\AgdaSpace{}%
\AgdaBound{cms}\AgdaSpace{}%
\AgdaBound{next-cmd}\AgdaSymbol{)}\AgdaSpace{}%
\AgdaBound{st}\AgdaSpace{}%
\AgdaBound{ch}\<%
\\
\>[2]\AgdaFunction{exec}\AgdaSpace{}%
\AgdaSymbol{(}\AgdaInductiveConstructor{LOOP}\AgdaSpace{}%
\AgdaBound{cmd}\AgdaSymbol{)}\AgdaSpace{}%
\AgdaBound{cms}\AgdaSpace{}%
\AgdaBound{st}\AgdaSpace{}%
\AgdaBound{ch}\AgdaSpace{}%
\AgdaSymbol{=}\AgdaSpace{}%
\AgdaFunction{exec}\AgdaSpace{}%
\AgdaBound{cmd}\AgdaSpace{}%
\AgdaSymbol{(}\AgdaFunction{push}\AgdaSpace{}%
\AgdaBound{cms}\AgdaSpace{}%
\AgdaSymbol{(}\AgdaInductiveConstructor{LOOP}\AgdaSpace{}%
\AgdaBound{cmd}\AgdaSymbol{))}\AgdaSpace{}%
\AgdaBound{st}\AgdaSpace{}%
\AgdaBound{ch}\<%
\\
\>[2]\AgdaFunction{exec}\AgdaSpace{}%
\AgdaSymbol{\{}\AgdaArgument{n}\AgdaSpace{}%
\AgdaSymbol{=}\AgdaSpace{}%
\AgdaInductiveConstructor{suc}\AgdaSpace{}%
\AgdaBound{n}\AgdaSymbol{\}}\AgdaSpace{}%
\AgdaSymbol{(}\AgdaInductiveConstructor{CONTINUE}\AgdaSpace{}%
\AgdaBound{i}\AgdaSymbol{)}\AgdaSpace{}%
\AgdaBound{cms}\AgdaSpace{}%
\AgdaBound{st}\AgdaSpace{}%
\AgdaBound{ch}\AgdaSpace{}%
\AgdaSymbol{=}\AgdaSpace{}%
\AgdaInductiveConstructor{pure}\AgdaSpace{}%
\AgdaSymbol{(}\AgdaInductiveConstructor{inj₁}\AgdaSpace{}%
\AgdaOperator{\AgdaInductiveConstructor{⟨}}\AgdaSpace{}%
\AgdaSymbol{\AgdaUnderscore{}}\AgdaSpace{}%
\AgdaOperator{\AgdaInductiveConstructor{,}}\AgdaSpace{}%
\AgdaOperator{\AgdaInductiveConstructor{⟨}}\AgdaSpace{}%
\AgdaFunction{pop}\AgdaSpace{}%
\AgdaBound{cms}\AgdaSpace{}%
\AgdaBound{i}\AgdaSpace{}%
\AgdaOperator{\AgdaInductiveConstructor{,}}\AgdaSpace{}%
\AgdaBound{st}\AgdaSpace{}%
\AgdaOperator{\AgdaInductiveConstructor{⟩}}\AgdaSpace{}%
\AgdaOperator{\AgdaInductiveConstructor{⟩}}\AgdaSymbol{)}\<%
\\
\>[2]\AgdaFunction{exec}\AgdaSpace{}%
\AgdaInductiveConstructor{CLOSE}\AgdaSpace{}%
\AgdaBound{cms}\AgdaSpace{}%
\AgdaBound{st}\AgdaSpace{}%
\AgdaBound{ch}\AgdaSpace{}%
\AgdaSymbol{=}\AgdaSpace{}%
\AgdaKeyword{do}\<%
\\
\>[2][@{}l@{\AgdaIndent{0}}]%
\>[4]\AgdaPostulate{primClose}\AgdaSpace{}%
\AgdaBound{ch}\<%
\\
\>[4]\AgdaInductiveConstructor{pure}\AgdaSpace{}%
\AgdaSymbol{(}\AgdaInductiveConstructor{inj₂}\AgdaSpace{}%
\AgdaBound{st}\AgdaSymbol{)}\<%
\\
\>[2]\AgdaFunction{exec}\AgdaSpace{}%
\AgdaSymbol{(}\AgdaInductiveConstructor{SEND}\AgdaSpace{}%
\AgdaBound{getx}\AgdaSpace{}%
\AgdaBound{cmd}\AgdaSymbol{)}\AgdaSpace{}%
\AgdaBound{cms}\AgdaSpace{}%
\AgdaBound{st}\AgdaSpace{}%
\AgdaBound{ch}\AgdaSpace{}%
\AgdaSymbol{=}\AgdaSpace{}%
\AgdaKeyword{do}\<%
\\
\>[2][@{}l@{\AgdaIndent{0}}]%
\>[4]\AgdaKeyword{let}\AgdaSpace{}%
\AgdaOperator{\AgdaInductiveConstructor{⟨}}\AgdaSpace{}%
\AgdaBound{st′}\AgdaSpace{}%
\AgdaOperator{\AgdaInductiveConstructor{,}}\AgdaSpace{}%
\AgdaBound{x}\AgdaSpace{}%
\AgdaOperator{\AgdaInductiveConstructor{⟩}}\AgdaSpace{}%
\AgdaSymbol{=}\AgdaSpace{}%
\AgdaBound{getx}\AgdaSpace{}%
\AgdaBound{st}\<%
\\
\>[4]\AgdaPostulate{primSend}\AgdaSpace{}%
\AgdaBound{x}\AgdaSpace{}%
\AgdaBound{ch}\<%
\\
\>[4]\AgdaFunction{exec}\AgdaSpace{}%
\AgdaBound{cmd}\AgdaSpace{}%
\AgdaBound{cms}\AgdaSpace{}%
\AgdaBound{st′}\AgdaSpace{}%
\AgdaBound{ch}\<%
\\
\>[2]\AgdaFunction{exec}\AgdaSpace{}%
\AgdaSymbol{(}\AgdaInductiveConstructor{RECV}\AgdaSpace{}%
\AgdaBound{putx}\AgdaSpace{}%
\AgdaBound{cmd}\AgdaSymbol{)}\AgdaSpace{}%
\AgdaBound{cms}\AgdaSpace{}%
\AgdaBound{st}\AgdaSpace{}%
\AgdaBound{ch}\AgdaSpace{}%
\AgdaSymbol{=}\AgdaSpace{}%
\AgdaKeyword{do}\<%
\\
\>[2][@{}l@{\AgdaIndent{0}}]%
\>[4]\AgdaBound{x}\AgdaSpace{}%
\AgdaOperator{\AgdaFunction{←}}\AgdaSpace{}%
\AgdaPostulate{primRecv}\AgdaSpace{}%
\AgdaBound{ch}\<%
\\
\>[4]\AgdaKeyword{let}\AgdaSpace{}%
\AgdaBound{st′}\AgdaSpace{}%
\AgdaSymbol{=}\AgdaSpace{}%
\AgdaBound{putx}\AgdaSpace{}%
\AgdaBound{x}\AgdaSpace{}%
\AgdaBound{st}\<%
\\
\>[4]\AgdaFunction{exec}\AgdaSpace{}%
\AgdaBound{cmd}\AgdaSpace{}%
\AgdaBound{cms}\AgdaSpace{}%
\AgdaBound{st′}\AgdaSpace{}%
\AgdaBound{ch}\<%
\\
\>[2]\AgdaFunction{exec}\AgdaSpace{}%
\AgdaSymbol{(}\AgdaInductiveConstructor{SELECT}\AgdaSpace{}%
\AgdaBound{i}\AgdaSpace{}%
\AgdaBound{cmd}\AgdaSymbol{)}\AgdaSpace{}%
\AgdaBound{cms}\AgdaSpace{}%
\AgdaBound{st}\AgdaSpace{}%
\AgdaBound{ch}\AgdaSpace{}%
\AgdaSymbol{=}\AgdaSpace{}%
\AgdaKeyword{do}\<%
\\
\>[2][@{}l@{\AgdaIndent{0}}]%
\>[4]\AgdaPostulate{primSend}\AgdaSpace{}%
\AgdaBound{i}\AgdaSpace{}%
\AgdaBound{ch}\<%
\\
\>[4]\AgdaFunction{exec}\AgdaSpace{}%
\AgdaBound{cmd}\AgdaSpace{}%
\AgdaBound{cms}\AgdaSpace{}%
\AgdaBound{st}\AgdaSpace{}%
\AgdaBound{ch}\<%
\\
\>[2]\AgdaFunction{exec}\AgdaSpace{}%
\AgdaSymbol{(}\AgdaInductiveConstructor{CHOICE}\AgdaSpace{}%
\AgdaBound{f-cmd}\AgdaSymbol{)}\AgdaSpace{}%
\AgdaBound{cms}\AgdaSpace{}%
\AgdaBound{st}\AgdaSpace{}%
\AgdaBound{ch}\AgdaSpace{}%
\AgdaSymbol{=}\AgdaSpace{}%
\AgdaKeyword{do}\<%
\\
\>[2][@{}l@{\AgdaIndent{0}}]%
\>[4]\AgdaBound{x}\AgdaSpace{}%
\AgdaOperator{\AgdaFunction{←}}\AgdaSpace{}%
\AgdaPostulate{primRecv}\AgdaSpace{}%
\AgdaBound{ch}\<%
\\
\>[4]\AgdaFunction{exec}\AgdaSpace{}%
\AgdaSymbol{(}\AgdaBound{f-cmd}\AgdaSpace{}%
\AgdaBound{x}\AgdaSymbol{)}\AgdaSpace{}%
\AgdaBound{cms}\AgdaSpace{}%
\AgdaBound{st}\AgdaSpace{}%
\AgdaBound{ch}\<%
\end{code}
\newcommand\rstAlternativeExecutorRestart{%
\begin{code}%
\>[2]\AgdaFunction{CmdCont}\AgdaSpace{}%
\AgdaSymbol{:}\AgdaSpace{}%
\AgdaPrimitive{Set}\AgdaSpace{}%
\AgdaSymbol{→}\AgdaSpace{}%
\AgdaPrimitive{Set}\<%
\\
\>[2]\AgdaFunction{CmdCont}\AgdaSpace{}%
\AgdaBound{A}\AgdaSpace{}%
\AgdaSymbol{=}\AgdaSpace{}%
\AgdaFunction{∃[}\AgdaSpace{}%
\AgdaBound{n}\AgdaSpace{}%
\AgdaFunction{]}\AgdaSpace{}%
\AgdaSymbol{(}\AgdaFunction{CmdStore}\AgdaSpace{}%
\AgdaSymbol{(}\AgdaInductiveConstructor{suc}\AgdaSpace{}%
\AgdaBound{n}\AgdaSymbol{)}\AgdaSpace{}%
\AgdaBound{A}\AgdaSpace{}%
\AgdaOperator{\AgdaFunction{×}}\AgdaSpace{}%
\AgdaBound{A}\AgdaSymbol{)}\<%
\\
\\[\AgdaEmptyExtraSkip]%
\>[2]\AgdaFunction{restart}\AgdaSpace{}%
\AgdaSymbol{:}\AgdaSpace{}%
\AgdaFunction{CmdCont}\AgdaSpace{}%
\AgdaGeneralizable{A}\AgdaSpace{}%
\AgdaSymbol{→}\AgdaSpace{}%
\AgdaPostulate{Channel}\AgdaSpace{}%
\AgdaSymbol{→}\AgdaSpace{}%
\AgdaDatatype{IO}\AgdaSpace{}%
\AgdaSymbol{(}\AgdaFunction{CmdCont}\AgdaSpace{}%
\AgdaGeneralizable{A}\AgdaSpace{}%
\AgdaOperator{\AgdaDatatype{⊎}}\AgdaSpace{}%
\AgdaGeneralizable{A}\AgdaSymbol{)}\<%
\\
\>[2]\AgdaFunction{restart}\AgdaSpace{}%
\AgdaOperator{\AgdaInductiveConstructor{⟨}}\AgdaSpace{}%
\AgdaBound{n}\AgdaSpace{}%
\AgdaOperator{\AgdaInductiveConstructor{,}}\AgdaSpace{}%
\AgdaOperator{\AgdaInductiveConstructor{⟨}}\AgdaSpace{}%
\AgdaBound{cms}\AgdaSpace{}%
\AgdaOperator{\AgdaInductiveConstructor{,}}\AgdaSpace{}%
\AgdaBound{st}\AgdaSpace{}%
\AgdaOperator{\AgdaInductiveConstructor{⟩}}\AgdaSpace{}%
\AgdaOperator{\AgdaInductiveConstructor{⟩}}\AgdaSpace{}%
\AgdaBound{ch}\<%
\\
\>[2][@{}l@{\AgdaIndent{0}}]%
\>[4]\AgdaKeyword{with}\AgdaSpace{}%
\AgdaBound{cms}\AgdaSpace{}%
\AgdaInductiveConstructor{zero}\<%
\\
\>[2]\AgdaSymbol{...}\AgdaSpace{}%
\AgdaSymbol{|}\AgdaSpace{}%
\AgdaOperator{\AgdaInductiveConstructor{⟨}}\AgdaSpace{}%
\AgdaBound{s₀}\AgdaSpace{}%
\AgdaOperator{\AgdaInductiveConstructor{,}}\AgdaSpace{}%
\AgdaBound{cmd₀}\AgdaSpace{}%
\AgdaOperator{\AgdaInductiveConstructor{⟩}}\AgdaSpace{}%
\AgdaKeyword{rewrite}\AgdaSpace{}%
\AgdaFunction{toℕ-fromℕ}\AgdaSpace{}%
\AgdaBound{n}\AgdaSpace{}%
\AgdaSymbol{=}\AgdaSpace{}%
\AgdaFunction{exec}\AgdaSpace{}%
\AgdaBound{cmd₀}\AgdaSpace{}%
\AgdaSymbol{(}\AgdaFunction{pop1}\AgdaSpace{}%
\AgdaBound{cms}\AgdaSymbol{)}\AgdaSpace{}%
\AgdaBound{st}\AgdaSpace{}%
\AgdaBound{ch}\<%
\end{code}}
\newcommand\rstExecutorSignature{%
\begin{code}%
\>[0]\AgdaFunction{Gas}\AgdaSpace{}%
\AgdaSymbol{=}\AgdaSpace{}%
\AgdaDatatype{ℕ}\<%
\\
\>[0]\AgdaFunction{exec}\AgdaSpace{}%
\AgdaSymbol{:}\AgdaSpace{}%
\AgdaFunction{Gas}\AgdaSpace{}%
\AgdaSymbol{→}\AgdaSpace{}%
\AgdaDatatype{Cmd}\AgdaSpace{}%
\AgdaGeneralizable{n}\AgdaSpace{}%
\AgdaGeneralizable{A}\AgdaSpace{}%
\AgdaGeneralizable{S}\AgdaSpace{}%
\AgdaSymbol{→}\AgdaSpace{}%
\AgdaFunction{CmdStore}\AgdaSpace{}%
\AgdaGeneralizable{n}\AgdaSpace{}%
\AgdaGeneralizable{A}\AgdaSpace{}%
\AgdaSymbol{→}\AgdaSpace{}%
\AgdaSymbol{(}\AgdaBound{init}\AgdaSpace{}%
\AgdaSymbol{:}\AgdaSpace{}%
\AgdaGeneralizable{A}\AgdaSymbol{)}\AgdaSpace{}%
\AgdaSymbol{→}\AgdaSpace{}%
\AgdaPostulate{Channel}\AgdaSpace{}%
\AgdaSymbol{→}\AgdaSpace{}%
\AgdaDatatype{IO}\AgdaSpace{}%
\AgdaGeneralizable{A}\<%
\end{code}}
\begin{code}[hide]%
\>[0]\AgdaFunction{exec}\AgdaSpace{}%
\AgdaBound{k}\AgdaSpace{}%
\AgdaInductiveConstructor{CLOSE}\AgdaSpace{}%
\AgdaBound{cms}\AgdaSpace{}%
\AgdaBound{state}\AgdaSpace{}%
\AgdaBound{ch}\AgdaSpace{}%
\AgdaSymbol{=}\AgdaSpace{}%
\AgdaKeyword{do}\<%
\\
\>[0][@{}l@{\AgdaIndent{0}}]%
\>[2]\AgdaPostulate{primClose}\AgdaSpace{}%
\AgdaBound{ch}\<%
\\
\>[2]\AgdaInductiveConstructor{pure}\AgdaSpace{}%
\AgdaBound{state}\<%
\\
\>[0]\AgdaFunction{exec}\AgdaSpace{}%
\AgdaBound{k}\AgdaSpace{}%
\AgdaSymbol{(}\AgdaInductiveConstructor{SEND}\AgdaSpace{}%
\AgdaBound{getx}\AgdaSpace{}%
\AgdaBound{cmd}\AgdaSymbol{)}\AgdaSpace{}%
\AgdaBound{cms}\AgdaSpace{}%
\AgdaBound{state}\AgdaSpace{}%
\AgdaBound{ch}\AgdaSpace{}%
\AgdaSymbol{=}\AgdaSpace{}%
\AgdaKeyword{do}\<%
\\
\>[0][@{}l@{\AgdaIndent{0}}]%
\>[2]\AgdaKeyword{let}\AgdaSpace{}%
\AgdaOperator{\AgdaInductiveConstructor{⟨}}\AgdaSpace{}%
\AgdaBound{state′}\AgdaSpace{}%
\AgdaOperator{\AgdaInductiveConstructor{,}}\AgdaSpace{}%
\AgdaBound{x}\AgdaSpace{}%
\AgdaOperator{\AgdaInductiveConstructor{⟩}}\AgdaSpace{}%
\AgdaSymbol{=}\AgdaSpace{}%
\AgdaBound{getx}\AgdaSpace{}%
\AgdaBound{state}\<%
\\
\>[2]\AgdaPostulate{primSend}\AgdaSpace{}%
\AgdaBound{x}\AgdaSpace{}%
\AgdaBound{ch}\<%
\\
\>[2]\AgdaFunction{exec}\AgdaSpace{}%
\AgdaBound{k}\AgdaSpace{}%
\AgdaBound{cmd}\AgdaSpace{}%
\AgdaBound{cms}\AgdaSpace{}%
\AgdaBound{state′}\AgdaSpace{}%
\AgdaBound{ch}\<%
\\
\>[0]\AgdaFunction{exec}\AgdaSpace{}%
\AgdaBound{k}\AgdaSpace{}%
\AgdaSymbol{(}\AgdaInductiveConstructor{RECV}\AgdaSpace{}%
\AgdaBound{putx}\AgdaSpace{}%
\AgdaBound{cmd}\AgdaSymbol{)}\AgdaSpace{}%
\AgdaBound{cms}\AgdaSpace{}%
\AgdaBound{state}\AgdaSpace{}%
\AgdaBound{ch}\AgdaSpace{}%
\AgdaSymbol{=}\AgdaSpace{}%
\AgdaKeyword{do}\<%
\\
\>[0][@{}l@{\AgdaIndent{0}}]%
\>[2]\AgdaBound{x}\AgdaSpace{}%
\AgdaOperator{\AgdaFunction{←}}\AgdaSpace{}%
\AgdaPostulate{primRecv}\AgdaSpace{}%
\AgdaBound{ch}\<%
\\
\>[2]\AgdaKeyword{let}\AgdaSpace{}%
\AgdaBound{state′}\AgdaSpace{}%
\AgdaSymbol{=}\AgdaSpace{}%
\AgdaBound{putx}\AgdaSpace{}%
\AgdaBound{x}\AgdaSpace{}%
\AgdaBound{state}\<%
\\
\>[2]\AgdaFunction{exec}\AgdaSpace{}%
\AgdaBound{k}\AgdaSpace{}%
\AgdaBound{cmd}\AgdaSpace{}%
\AgdaBound{cms}\AgdaSpace{}%
\AgdaBound{state′}\AgdaSpace{}%
\AgdaBound{ch}\<%
\\
\>[0]\AgdaFunction{exec}\AgdaSpace{}%
\AgdaBound{k}\AgdaSpace{}%
\AgdaSymbol{(}\AgdaInductiveConstructor{SELECT}\AgdaSpace{}%
\AgdaBound{i}\AgdaSpace{}%
\AgdaBound{cmd}\AgdaSymbol{)}\AgdaSpace{}%
\AgdaBound{cms}\AgdaSpace{}%
\AgdaBound{state}\AgdaSpace{}%
\AgdaBound{ch}\AgdaSpace{}%
\AgdaSymbol{=}\AgdaSpace{}%
\AgdaKeyword{do}\<%
\\
\>[0][@{}l@{\AgdaIndent{0}}]%
\>[2]\AgdaPostulate{primSend}\AgdaSpace{}%
\AgdaBound{i}\AgdaSpace{}%
\AgdaBound{ch}\<%
\\
\>[2]\AgdaFunction{exec}\AgdaSpace{}%
\AgdaBound{k}\AgdaSpace{}%
\AgdaBound{cmd}\AgdaSpace{}%
\AgdaBound{cms}\AgdaSpace{}%
\AgdaBound{state}\AgdaSpace{}%
\AgdaBound{ch}\<%
\\
\>[0]\AgdaFunction{exec}\AgdaSpace{}%
\AgdaBound{k}\AgdaSpace{}%
\AgdaSymbol{(}\AgdaInductiveConstructor{CHOICE}\AgdaSpace{}%
\AgdaBound{f-cmd}\AgdaSymbol{)}\AgdaSpace{}%
\AgdaBound{cms}\AgdaSpace{}%
\AgdaBound{state}\AgdaSpace{}%
\AgdaBound{ch}\AgdaSpace{}%
\AgdaSymbol{=}\AgdaSpace{}%
\AgdaKeyword{do}\<%
\\
\>[0][@{}l@{\AgdaIndent{0}}]%
\>[2]\AgdaBound{x}\AgdaSpace{}%
\AgdaOperator{\AgdaFunction{←}}\AgdaSpace{}%
\AgdaPostulate{primRecv}\AgdaSpace{}%
\AgdaBound{ch}\<%
\\
\>[2]\AgdaFunction{exec}\AgdaSpace{}%
\AgdaBound{k}\AgdaSpace{}%
\AgdaSymbol{(}\AgdaBound{f-cmd}\AgdaSpace{}%
\AgdaBound{x}\AgdaSymbol{)}\AgdaSpace{}%
\AgdaBound{cms}\AgdaSpace{}%
\AgdaBound{state}\AgdaSpace{}%
\AgdaBound{ch}\<%
\end{code}
\newcommand\rstExecutor{%
\begin{code}%
\>[0]\AgdaFunction{exec}\AgdaSpace{}%
\AgdaBound{g}\AgdaSpace{}%
\AgdaSymbol{(}\AgdaInductiveConstructor{LOOP}\AgdaSpace{}%
\AgdaBound{cmd}\AgdaSymbol{)}\AgdaSpace{}%
\AgdaBound{cms}\AgdaSpace{}%
\AgdaBound{state}\AgdaSpace{}%
\AgdaBound{ch}\AgdaSpace{}%
\AgdaSymbol{=}\AgdaSpace{}%
\AgdaFunction{exec}\AgdaSpace{}%
\AgdaBound{g}\AgdaSpace{}%
\AgdaBound{cmd}\AgdaSpace{}%
\AgdaSymbol{(}\AgdaFunction{push}\AgdaSpace{}%
\AgdaBound{cms}\AgdaSpace{}%
\AgdaSymbol{(}\AgdaInductiveConstructor{LOOP}\AgdaSpace{}%
\AgdaBound{cmd}\AgdaSymbol{))}\AgdaSpace{}%
\AgdaBound{state}\AgdaSpace{}%
\AgdaBound{ch}\<%
\\
\>[0]\AgdaFunction{exec}\AgdaSpace{}%
\AgdaSymbol{\{}\AgdaInductiveConstructor{suc}\AgdaSpace{}%
\AgdaBound{n}\AgdaSymbol{\}}\AgdaSpace{}%
\AgdaSymbol{\{}\AgdaBound{A}\AgdaSymbol{\}}\AgdaSpace{}%
\AgdaInductiveConstructor{zero}\AgdaSpace{}%
\AgdaSymbol{(}\AgdaInductiveConstructor{CONTINUE}\AgdaSpace{}%
\AgdaBound{i}\AgdaSymbol{)}\AgdaSpace{}%
\AgdaBound{cms}\AgdaSpace{}%
\AgdaBound{state}\AgdaSpace{}%
\AgdaBound{ch}\AgdaSpace{}%
\AgdaSymbol{=}\AgdaSpace{}%
\AgdaInductiveConstructor{pure}\AgdaSpace{}%
\AgdaBound{state}\AgdaSpace{}%
\AgdaComment{--\ hack\ alert!}\<%
\\
\>[0]\AgdaFunction{exec}\AgdaSpace{}%
\AgdaSymbol{\{}\AgdaInductiveConstructor{suc}\AgdaSpace{}%
\AgdaBound{n}\AgdaSymbol{\}}\AgdaSpace{}%
\AgdaSymbol{\{}\AgdaBound{A}\AgdaSymbol{\}}\AgdaSpace{}%
\AgdaSymbol{(}\AgdaInductiveConstructor{suc}\AgdaSpace{}%
\AgdaBound{g}\AgdaSymbol{)}\AgdaSpace{}%
\AgdaSymbol{(}\AgdaInductiveConstructor{CONTINUE}\AgdaSpace{}%
\AgdaBound{i}\AgdaSymbol{)}\AgdaSpace{}%
\AgdaBound{cms}\AgdaSpace{}%
\AgdaBound{state}\AgdaSpace{}%
\AgdaBound{ch}\<%
\\
\>[0][@{}l@{\AgdaIndent{0}}]%
\>[2]\AgdaKeyword{with}\AgdaSpace{}%
\AgdaBound{cms}\AgdaSpace{}%
\AgdaBound{i}\<%
\\
\>[0]\AgdaSymbol{...}\AgdaSpace{}%
\AgdaSymbol{|}\AgdaSpace{}%
\AgdaOperator{\AgdaInductiveConstructor{⟨}}\AgdaSpace{}%
\AgdaSymbol{\AgdaUnderscore{}}\AgdaSpace{}%
\AgdaOperator{\AgdaInductiveConstructor{,}}\AgdaSpace{}%
\AgdaBound{cmd-i}\AgdaSpace{}%
\AgdaOperator{\AgdaInductiveConstructor{⟩}}\AgdaSpace{}%
\AgdaSymbol{=}\AgdaSpace{}%
\AgdaFunction{exec}\AgdaSpace{}%
\AgdaBound{g}\AgdaSpace{}%
\AgdaBound{cmd-i}\AgdaSpace{}%
\AgdaSymbol{(}\AgdaFunction{pop1}\AgdaSpace{}%
\AgdaSymbol{(}\AgdaFunction{pop}\AgdaSpace{}%
\AgdaBound{cms}\AgdaSpace{}%
\AgdaBound{i}\AgdaSymbol{))}\AgdaSpace{}%
\AgdaBound{state}\AgdaSpace{}%
\AgdaBound{ch}\<%
\end{code}}
\newcommand\rstExecutorUNROLL{%
\begin{code}%
\>[0]\AgdaFunction{exec}\AgdaSpace{}%
\AgdaBound{g}\AgdaSpace{}%
\AgdaSymbol{(}\AgdaInductiveConstructor{UNROLL}\AgdaSpace{}%
\AgdaBound{body-cmd}\AgdaSpace{}%
\AgdaBound{next-cmd}\AgdaSymbol{)}\AgdaSpace{}%
\AgdaBound{cms}\AgdaSpace{}%
\AgdaBound{st}\AgdaSpace{}%
\AgdaBound{ch}\AgdaSpace{}%
\AgdaSymbol{=}\AgdaSpace{}%
\AgdaFunction{exec}\AgdaSpace{}%
\AgdaBound{g}\AgdaSpace{}%
\AgdaBound{body-cmd}\AgdaSpace{}%
\AgdaSymbol{(}\AgdaFunction{push}\AgdaSpace{}%
\AgdaBound{cms}\AgdaSpace{}%
\AgdaBound{next-cmd}\AgdaSymbol{)}\AgdaSpace{}%
\AgdaBound{st}\AgdaSpace{}%
\AgdaBound{ch}\<%
\end{code}}
\newcommand\rstAcceptor{%
\begin{code}%
\>[0]\AgdaKeyword{record}\AgdaSpace{}%
\AgdaRecord{Accepting}\AgdaSpace{}%
\AgdaSymbol{\{}\AgdaBound{n}\AgdaSymbol{\}}\AgdaSpace{}%
\AgdaBound{A}\AgdaSpace{}%
\AgdaBound{S}\AgdaSpace{}%
\AgdaSymbol{:}\AgdaSpace{}%
\AgdaPrimitive{Set}\AgdaSpace{}%
\AgdaKeyword{where}\<%
\\
\>[0][@{}l@{\AgdaIndent{0}}]%
\>[2]\AgdaKeyword{constructor}\AgdaSpace{}%
\AgdaInductiveConstructor{ACC}\<%
\\
\>[2]\AgdaKeyword{field}\AgdaSpace{}%
\AgdaField{cmd}\AgdaSpace{}%
\AgdaSymbol{:}\AgdaSpace{}%
\AgdaDatatype{Cmd}\AgdaSpace{}%
\AgdaBound{n}\AgdaSpace{}%
\AgdaBound{A}\AgdaSpace{}%
\AgdaBound{S}\<%
\\
\\[\AgdaEmptyExtraSkip]%
\>[0]\AgdaFunction{acceptor}\AgdaSpace{}%
\AgdaSymbol{:}\AgdaSpace{}%
\AgdaSymbol{\{}\AgdaBound{S}\AgdaSpace{}%
\AgdaSymbol{:}\AgdaSpace{}%
\AgdaDatatype{Session}\AgdaSpace{}%
\AgdaNumber{0}\AgdaSymbol{\}}\AgdaSpace{}%
\AgdaSymbol{→}\AgdaSpace{}%
\AgdaFunction{Gas}\AgdaSpace{}%
\AgdaSymbol{→}\AgdaSpace{}%
\AgdaRecord{Accepting}\AgdaSpace{}%
\AgdaGeneralizable{A}\AgdaSpace{}%
\AgdaBound{S}\AgdaSpace{}%
\AgdaSymbol{→}\AgdaSpace{}%
\AgdaGeneralizable{A}\AgdaSpace{}%
\AgdaSymbol{→}\AgdaSpace{}%
\AgdaDatatype{IO}\AgdaSpace{}%
\AgdaGeneralizable{A}\<%
\\
\>[0]\AgdaFunction{acceptor}\AgdaSpace{}%
\AgdaBound{k}\AgdaSpace{}%
\AgdaSymbol{(}\AgdaInductiveConstructor{ACC}\AgdaSpace{}%
\AgdaBound{cmd}\AgdaSymbol{)}\AgdaSpace{}%
\AgdaBound{a}\AgdaSpace{}%
\AgdaSymbol{=}\AgdaSpace{}%
\AgdaPostulate{primAccept}\AgdaSpace{}%
\AgdaOperator{\AgdaFunction{>>=}}\AgdaSpace{}%
\AgdaFunction{exec}\AgdaSpace{}%
\AgdaBound{k}\AgdaSpace{}%
\AgdaBound{cmd}\AgdaSpace{}%
\AgdaSymbol{(λ())}\AgdaSpace{}%
\AgdaBound{a}\<%
\end{code}}
\newcommand\rstClientExample{%
\begin{code}%
\>[0]\AgdaFunction{runningsum-client}\AgdaSpace{}%
\AgdaSymbol{:}\AgdaSpace{}%
\AgdaDatatype{Cmd}\AgdaSpace{}%
\AgdaNumber{0}\AgdaSpace{}%
\AgdaRecord{⊤}\AgdaSpace{}%
\AgdaSymbol{(}\AgdaFunction{dual}\AgdaSpace{}%
\AgdaFunction{many-unaryp}\AgdaSymbol{)}\<%
\\
\>[0]\AgdaFunction{runningsum-client}\AgdaSpace{}%
\AgdaSymbol{=}\<%
\\
\>[0][@{}l@{\AgdaIndent{0}}]%
\>[2]\AgdaInductiveConstructor{UNROLL}\AgdaSpace{}%
\AgdaSymbol{(}\AgdaInductiveConstructor{SELECT}\AgdaSpace{}%
\AgdaInductiveConstructor{zero}\AgdaSpace{}%
\AgdaOperator{\AgdaFunction{\$}}\AgdaSpace{}%
\AgdaInductiveConstructor{SEND}\AgdaSpace{}%
\AgdaSymbol{(λ}\AgdaSpace{}%
\AgdaBound{x}\AgdaSpace{}%
\AgdaSymbol{→}\AgdaSpace{}%
\AgdaOperator{\AgdaInductiveConstructor{⟨}}\AgdaSpace{}%
\AgdaInductiveConstructor{tt}\AgdaSpace{}%
\AgdaOperator{\AgdaInductiveConstructor{,}}\AgdaSpace{}%
\AgdaOperator{\AgdaInductiveConstructor{+}}\AgdaSpace{}%
\AgdaNumber{17}\AgdaSpace{}%
\AgdaOperator{\AgdaInductiveConstructor{⟩}}\AgdaSymbol{)}\AgdaSpace{}%
\AgdaOperator{\AgdaFunction{\$}}\AgdaSpace{}%
\AgdaInductiveConstructor{RECV}\AgdaSpace{}%
\AgdaFunction{constᵣ}\AgdaSpace{}%
\AgdaSymbol{(}\AgdaInductiveConstructor{CONTINUE}\AgdaSpace{}%
\AgdaInductiveConstructor{zero}\AgdaSymbol{))}\AgdaSpace{}%
\AgdaOperator{\AgdaFunction{\$}}\<%
\\
\>[2]\AgdaInductiveConstructor{UNROLL}\AgdaSpace{}%
\AgdaSymbol{(}\AgdaInductiveConstructor{SELECT}\AgdaSpace{}%
\AgdaInductiveConstructor{zero}\AgdaSpace{}%
\AgdaOperator{\AgdaFunction{\$}}\AgdaSpace{}%
\AgdaInductiveConstructor{SEND}\AgdaSpace{}%
\AgdaSymbol{(λ}\AgdaSpace{}%
\AgdaBound{x}\AgdaSpace{}%
\AgdaSymbol{→}\AgdaSpace{}%
\AgdaOperator{\AgdaInductiveConstructor{⟨}}\AgdaSpace{}%
\AgdaInductiveConstructor{tt}\AgdaSpace{}%
\AgdaOperator{\AgdaInductiveConstructor{,}}\AgdaSpace{}%
\AgdaOperator{\AgdaInductiveConstructor{+}}\AgdaSpace{}%
\AgdaNumber{4}\AgdaSpace{}%
\AgdaOperator{\AgdaInductiveConstructor{⟩}}\AgdaSymbol{)}%
\>[51]\AgdaOperator{\AgdaFunction{\$}}\AgdaSpace{}%
\AgdaInductiveConstructor{RECV}\AgdaSpace{}%
\AgdaFunction{constᵣ}\AgdaSpace{}%
\AgdaSymbol{(}\AgdaInductiveConstructor{CONTINUE}\AgdaSpace{}%
\AgdaInductiveConstructor{zero}\AgdaSymbol{))}\AgdaSpace{}%
\AgdaOperator{\AgdaFunction{\$}}\<%
\\
\>[2]\AgdaInductiveConstructor{LOOP}\AgdaSpace{}%
\AgdaSymbol{(}\AgdaInductiveConstructor{SELECT}\AgdaSpace{}%
\AgdaSymbol{(}\AgdaInductiveConstructor{suc}\AgdaSpace{}%
\AgdaInductiveConstructor{zero}\AgdaSymbol{)}\AgdaSpace{}%
\AgdaInductiveConstructor{CLOSE}\AgdaSymbol{)}\<%
\end{code}}

%% file: latex/ST-monadic.tex
\begin{code}[hide]%
\>[0]\AgdaSymbol{\{-\#}\AgdaSpace{}%
\AgdaKeyword{OPTIONS}\AgdaSpace{}%
\AgdaPragma{--guardedness}\AgdaSpace{}%
\AgdaSymbol{\#-\}}\AgdaSpace{}%
\AgdaComment{\{-\ required\ for\ IO\ -\}}\<%
\\
\>[0]\AgdaKeyword{open}\AgdaSpace{}%
\AgdaKeyword{import}\AgdaSpace{}%
\AgdaModule{Level}\AgdaSpace{}%
\AgdaKeyword{using}\AgdaSpace{}%
\AgdaSymbol{(}\AgdaPostulate{Level}\AgdaSymbol{)}\AgdaSpace{}%
\AgdaKeyword{renaming}\AgdaSpace{}%
\AgdaSymbol{(}\AgdaPrimitive{zero}\AgdaSpace{}%
\AgdaSymbol{to}\AgdaSpace{}%
\AgdaPrimitive{lzero}\AgdaSymbol{)}\<%
\\
\\[\AgdaEmptyExtraSkip]%
\>[0]\AgdaKeyword{module}\AgdaSpace{}%
\AgdaModule{ST-monadic}\AgdaSpace{}%
\AgdaKeyword{where}\<%
\\
\\[\AgdaEmptyExtraSkip]%
\>[0]\AgdaKeyword{open}\AgdaSpace{}%
\AgdaKeyword{import}\AgdaSpace{}%
\AgdaModule{Data.Bool}\AgdaSpace{}%
\AgdaKeyword{using}\AgdaSpace{}%
\AgdaSymbol{(}\AgdaDatatype{Bool}\AgdaSymbol{;}\AgdaSpace{}%
\AgdaInductiveConstructor{true}\AgdaSymbol{;}\AgdaSpace{}%
\AgdaInductiveConstructor{false}\AgdaSymbol{;}\AgdaSpace{}%
\AgdaOperator{\AgdaFunction{if\AgdaUnderscore{}then\AgdaUnderscore{}else\AgdaUnderscore{}}}\AgdaSymbol{)}\<%
\\
\>[0]\AgdaKeyword{open}\AgdaSpace{}%
\AgdaKeyword{import}\AgdaSpace{}%
\AgdaModule{Data.Fin}\AgdaSpace{}%
\AgdaKeyword{using}\AgdaSpace{}%
\AgdaSymbol{(}\AgdaDatatype{Fin}\AgdaSymbol{;}\AgdaSpace{}%
\AgdaInductiveConstructor{zero}\AgdaSymbol{;}\AgdaSpace{}%
\AgdaInductiveConstructor{suc}\AgdaSymbol{)}\<%
\\
\>[0]\AgdaKeyword{open}\AgdaSpace{}%
\AgdaKeyword{import}\AgdaSpace{}%
\AgdaModule{Data.Nat}\AgdaSpace{}%
\AgdaKeyword{using}\AgdaSpace{}%
\AgdaSymbol{(}\AgdaDatatype{ℕ}\AgdaSymbol{;}\AgdaSpace{}%
\AgdaInductiveConstructor{zero}\AgdaSymbol{;}\AgdaSpace{}%
\AgdaInductiveConstructor{suc}\AgdaSymbol{)}\<%
\\
\>[0]\AgdaKeyword{open}\AgdaSpace{}%
\AgdaKeyword{import}\AgdaSpace{}%
\AgdaModule{Data.Integer}\AgdaSpace{}%
\AgdaKeyword{using}\AgdaSpace{}%
\AgdaSymbol{(}\AgdaDatatype{ℤ}\AgdaSymbol{;}\AgdaSpace{}%
\AgdaOperator{\AgdaFunction{\AgdaUnderscore{}+\AgdaUnderscore{}}}\AgdaSymbol{;}\AgdaSpace{}%
\AgdaFunction{0ℤ}\AgdaSymbol{;}\AgdaSpace{}%
\AgdaOperator{\AgdaFunction{-\AgdaUnderscore{}}}\AgdaSymbol{)}\<%
\\
\>[0]\AgdaKeyword{open}\AgdaSpace{}%
\AgdaKeyword{import}\AgdaSpace{}%
\AgdaModule{Data.Product}\AgdaSpace{}%
\AgdaKeyword{using}\AgdaSpace{}%
\AgdaSymbol{(}\AgdaOperator{\AgdaFunction{\AgdaUnderscore{}×\AgdaUnderscore{}}}\AgdaSymbol{;}\AgdaSpace{}%
\AgdaRecord{Σ}\AgdaSymbol{;}\AgdaSpace{}%
\AgdaField{proj₁}\AgdaSymbol{;}\AgdaSpace{}%
\AgdaField{proj₂}\AgdaSymbol{)}\AgdaSpace{}%
\AgdaKeyword{renaming}\AgdaSpace{}%
\AgdaSymbol{(}\AgdaOperator{\AgdaInductiveConstructor{\AgdaUnderscore{},\AgdaUnderscore{}}}\AgdaSpace{}%
\AgdaSymbol{to}\AgdaSpace{}%
\AgdaOperator{\AgdaInductiveConstructor{⟨\AgdaUnderscore{},\AgdaUnderscore{}⟩}}\AgdaSymbol{)}\<%
\\
\>[0]\AgdaKeyword{open}\AgdaSpace{}%
\AgdaKeyword{import}\AgdaSpace{}%
\AgdaModule{Data.Sum}\AgdaSpace{}%
\AgdaKeyword{using}\AgdaSpace{}%
\AgdaSymbol{(}\AgdaOperator{\AgdaDatatype{\AgdaUnderscore{}⊎\AgdaUnderscore{}}}\AgdaSymbol{;}\AgdaSpace{}%
\AgdaInductiveConstructor{inj₁}\AgdaSymbol{;}\AgdaSpace{}%
\AgdaInductiveConstructor{inj₂}\AgdaSymbol{)}\<%
\\
\>[0]\AgdaKeyword{open}\AgdaSpace{}%
\AgdaKeyword{import}\AgdaSpace{}%
\AgdaModule{Data.Vec}\AgdaSpace{}%
\AgdaKeyword{using}\AgdaSpace{}%
\AgdaSymbol{(}\AgdaDatatype{Vec}\AgdaSymbol{;}\AgdaSpace{}%
\AgdaInductiveConstructor{[]}\AgdaSymbol{;}\AgdaSpace{}%
\AgdaOperator{\AgdaInductiveConstructor{\AgdaUnderscore{}∷\AgdaUnderscore{}}}\AgdaSymbol{)}\<%
\\
\\[\AgdaEmptyExtraSkip]%
\>[0]\AgdaComment{--\ stdlib\ 2.0!}\<%
\\
\>[0]\AgdaKeyword{open}\AgdaSpace{}%
\AgdaKeyword{import}\AgdaSpace{}%
\AgdaModule{Data.Unit.Polymorphic.Base}\AgdaSpace{}%
\AgdaKeyword{using}\AgdaSpace{}%
\AgdaSymbol{(}\AgdaFunction{⊤}\AgdaSymbol{;}\AgdaSpace{}%
\AgdaFunction{tt}\AgdaSymbol{)}\<%
\\
\\[\AgdaEmptyExtraSkip]%
\>[0]\AgdaKeyword{open}\AgdaSpace{}%
\AgdaKeyword{import}\AgdaSpace{}%
\AgdaModule{Effect.Functor}\<%
\\
\\[\AgdaEmptyExtraSkip]%
\>[0]\AgdaKeyword{open}\AgdaSpace{}%
\AgdaKeyword{import}\AgdaSpace{}%
\AgdaModule{Effect.Monad}\<%
\\
\>[0]\AgdaKeyword{open}\AgdaSpace{}%
\AgdaKeyword{import}\AgdaSpace{}%
\AgdaModule{Effect.Monad.State}\<%
\\
\>[0]\AgdaKeyword{open}\AgdaSpace{}%
\AgdaKeyword{import}\AgdaSpace{}%
\AgdaModule{Effect.Monad.Reader.Transformer}\AgdaSpace{}%
\AgdaSymbol{as}\AgdaSpace{}%
\AgdaModule{Reader}\<%
\\
\>[0]\AgdaKeyword{open}\AgdaSpace{}%
\AgdaKeyword{import}\AgdaSpace{}%
\AgdaModule{Effect.Monad.State.Transformer}\AgdaSpace{}%
\AgdaSymbol{as}\AgdaSpace{}%
\AgdaModule{State}\<%
\\
\>[0]\AgdaKeyword{open}\AgdaSpace{}%
\AgdaKeyword{import}\AgdaSpace{}%
\AgdaModule{Effect.Monad.IO}\<%
\\
\\[\AgdaEmptyExtraSkip]%
\>[0]\AgdaKeyword{open}\AgdaSpace{}%
\AgdaKeyword{import}\AgdaSpace{}%
\AgdaModule{Effect.Monad.Reader.Instances}\<%
\\
\>[0]\AgdaKeyword{open}\AgdaSpace{}%
\AgdaKeyword{import}\AgdaSpace{}%
\AgdaModule{Effect.Monad.State.Instances}\<%
\\
\>[0]\AgdaKeyword{open}\AgdaSpace{}%
\AgdaKeyword{import}\AgdaSpace{}%
\AgdaModule{Effect.Monad.Identity.Instances}\<%
\\
\>[0]\AgdaKeyword{open}\AgdaSpace{}%
\AgdaKeyword{import}\AgdaSpace{}%
\AgdaModule{Effect.Monad.IO.Instances}\<%
\\
\>[0]\AgdaKeyword{open}\AgdaSpace{}%
\AgdaKeyword{import}\AgdaSpace{}%
\AgdaModule{IO.Instances}\<%
\\
\\[\AgdaEmptyExtraSkip]%
\>[0]\AgdaKeyword{open}\AgdaSpace{}%
\AgdaModule{RawMonad}\AgdaSpace{}%
\AgdaSymbol{\{\{...\}\}}\<%
\\
\>[0]\AgdaKeyword{open}\AgdaSpace{}%
\AgdaModule{RawMonadState}\AgdaSpace{}%
\AgdaSymbol{\{\{...\}\}}\<%
\\
\>[0]\AgdaKeyword{open}\AgdaSpace{}%
\AgdaModule{RawMonadReader}\AgdaSpace{}%
\AgdaSymbol{\{\{...\}\}}\<%
\\
\>[0]\AgdaKeyword{open}\AgdaSpace{}%
\AgdaModule{RawMonadIO}\AgdaSpace{}%
\AgdaSymbol{\{\{...\}\}}\<%
\\
\>[0]\AgdaKeyword{open}\AgdaSpace{}%
\AgdaModule{RawFunctor}\AgdaSpace{}%
\AgdaSymbol{\{\{...\}\}}\<%
\\
\\[\AgdaEmptyExtraSkip]%
\>[0]\AgdaKeyword{open}\AgdaSpace{}%
\AgdaKeyword{import}\AgdaSpace{}%
\AgdaModule{Function.Base}\AgdaSpace{}%
\AgdaKeyword{using}\AgdaSpace{}%
\AgdaSymbol{(}\AgdaOperator{\AgdaFunction{case\AgdaUnderscore{}of\AgdaUnderscore{}}}\AgdaSymbol{;}\AgdaSpace{}%
\AgdaOperator{\AgdaFunction{\AgdaUnderscore{}∘\AgdaUnderscore{}}}\AgdaSymbol{;}\AgdaSpace{}%
\AgdaOperator{\AgdaFunction{\AgdaUnderscore{}∘′\AgdaUnderscore{}}}\AgdaSymbol{;}\AgdaSpace{}%
\AgdaFunction{const}\AgdaSymbol{;}\AgdaSpace{}%
\AgdaFunction{id}\AgdaSymbol{;}\AgdaSpace{}%
\AgdaOperator{\AgdaFunction{\AgdaUnderscore{}\$\AgdaUnderscore{}}}\AgdaSymbol{)}\<%
\\
\\[\AgdaEmptyExtraSkip]%
\>[0]\AgdaKeyword{open}\AgdaSpace{}%
\AgdaKeyword{import}\AgdaSpace{}%
\AgdaModule{IO.Base}\AgdaSpace{}%
\AgdaKeyword{using}\AgdaSpace{}%
\AgdaSymbol{(}\AgdaDatatype{IO}\AgdaSymbol{)}\<%
\\
\\[\AgdaEmptyExtraSkip]%
\>[0]\AgdaKeyword{pattern}\AgdaSpace{}%
\AgdaOperator{\AgdaInductiveConstructor{[\AgdaUnderscore{}]}}\AgdaSpace{}%
\AgdaBound{x}\AgdaSpace{}%
\AgdaSymbol{=}\AgdaSpace{}%
\AgdaBound{x}\AgdaSpace{}%
\AgdaOperator{\AgdaInductiveConstructor{∷}}\AgdaSpace{}%
\AgdaInductiveConstructor{[]}\<%
\\
\>[0]\AgdaKeyword{pattern}\AgdaSpace{}%
\AgdaOperator{\AgdaInductiveConstructor{[\AgdaUnderscore{},\AgdaUnderscore{}]}}\AgdaSpace{}%
\AgdaBound{x}\AgdaSpace{}%
\AgdaBound{y}\AgdaSpace{}%
\AgdaSymbol{=}\AgdaSpace{}%
\AgdaBound{x}\AgdaSpace{}%
\AgdaOperator{\AgdaInductiveConstructor{∷}}\AgdaSpace{}%
\AgdaBound{y}\AgdaSpace{}%
\AgdaOperator{\AgdaInductiveConstructor{∷}}\AgdaSpace{}%
\AgdaInductiveConstructor{[]}\<%
\\
\>[0]\AgdaKeyword{pattern}\AgdaSpace{}%
\AgdaOperator{\AgdaInductiveConstructor{[\AgdaUnderscore{},\AgdaUnderscore{},\AgdaUnderscore{}]}}\AgdaSpace{}%
\AgdaBound{x}\AgdaSpace{}%
\AgdaBound{y}\AgdaSpace{}%
\AgdaBound{z}\AgdaSpace{}%
\AgdaSymbol{=}\AgdaSpace{}%
\AgdaBound{x}\AgdaSpace{}%
\AgdaOperator{\AgdaInductiveConstructor{∷}}\AgdaSpace{}%
\AgdaBound{y}\AgdaSpace{}%
\AgdaOperator{\AgdaInductiveConstructor{∷}}\AgdaSpace{}%
\AgdaBound{z}\AgdaSpace{}%
\AgdaOperator{\AgdaInductiveConstructor{∷}}\AgdaSpace{}%
\AgdaInductiveConstructor{[]}\<%
\\
\\[\AgdaEmptyExtraSkip]%
\>[0]\AgdaKeyword{variable}\AgdaSpace{}%
\AgdaGeneralizable{n}\AgdaSpace{}%
\AgdaGeneralizable{k}\AgdaSpace{}%
\AgdaSymbol{:}\AgdaSpace{}%
\AgdaDatatype{ℕ}\<%
\\
\\[\AgdaEmptyExtraSkip]%
\>[0]\AgdaKeyword{postulate}\<%
\\
\>[0][@{}l@{\AgdaIndent{0}}]%
\>[2]\AgdaPostulate{Channel}\AgdaSpace{}%
\AgdaSymbol{:}\AgdaSpace{}%
\AgdaPrimitive{Set}\<%
\\
\>[2]\AgdaPostulate{primAccept}\AgdaSpace{}%
\AgdaSymbol{:}\AgdaSpace{}%
\AgdaDatatype{IO}\AgdaSpace{}%
\AgdaPostulate{Channel}\<%
\\
\>[2]\AgdaPostulate{primClose}\AgdaSpace{}%
\AgdaSymbol{:}\AgdaSpace{}%
\AgdaPostulate{Channel}\AgdaSpace{}%
\AgdaSymbol{→}\AgdaSpace{}%
\AgdaDatatype{IO}\AgdaSpace{}%
\AgdaSymbol{\{}\AgdaPrimitive{lzero}\AgdaSymbol{\}}\AgdaSpace{}%
\AgdaFunction{⊤}\<%
\\
\>[2]\AgdaPostulate{primSend}\AgdaSpace{}%
\AgdaSymbol{:}\AgdaSpace{}%
\AgdaSymbol{∀}\AgdaSpace{}%
\AgdaSymbol{\{}\AgdaBound{A}\AgdaSpace{}%
\AgdaSymbol{:}\AgdaSpace{}%
\AgdaPrimitive{Set}\AgdaSymbol{\}}\AgdaSpace{}%
\AgdaSymbol{→}\AgdaSpace{}%
\AgdaBound{A}\AgdaSpace{}%
\AgdaSymbol{→}\AgdaSpace{}%
\AgdaPostulate{Channel}\AgdaSpace{}%
\AgdaSymbol{→}\AgdaSpace{}%
\AgdaDatatype{IO}\AgdaSpace{}%
\AgdaSymbol{\{}\AgdaPrimitive{lzero}\AgdaSymbol{\}}\AgdaSpace{}%
\AgdaFunction{⊤}\<%
\\
\>[2]\AgdaPostulate{primRecv}\AgdaSpace{}%
\AgdaSymbol{:}\AgdaSpace{}%
\AgdaSymbol{∀}\AgdaSpace{}%
\AgdaSymbol{\{}\AgdaBound{A}\AgdaSpace{}%
\AgdaSymbol{:}\AgdaSpace{}%
\AgdaPrimitive{Set}\AgdaSymbol{\}}\AgdaSpace{}%
\AgdaSymbol{→}\AgdaSpace{}%
\AgdaPostulate{Channel}\AgdaSpace{}%
\AgdaSymbol{→}\AgdaSpace{}%
\AgdaDatatype{IO}\AgdaSpace{}%
\AgdaBound{A}\<%
\\
\\[\AgdaEmptyExtraSkip]%
\\[\AgdaEmptyExtraSkip]%
\>[0]\AgdaKeyword{data}\AgdaSpace{}%
\AgdaDatatype{Type}\AgdaSpace{}%
\AgdaSymbol{:}\AgdaSpace{}%
\AgdaPrimitive{Set}\AgdaSpace{}%
\AgdaKeyword{where}\<%
\\
\>[0][@{}l@{\AgdaIndent{0}}]%
\>[2]\AgdaInductiveConstructor{nat}\AgdaSpace{}%
\AgdaInductiveConstructor{int}\AgdaSpace{}%
\AgdaInductiveConstructor{bool}\AgdaSpace{}%
\AgdaSymbol{:}\AgdaSpace{}%
\AgdaDatatype{Type}\<%
\\
\\[\AgdaEmptyExtraSkip]%
\>[0]\AgdaKeyword{data}\AgdaSpace{}%
\AgdaDatatype{Session}\AgdaSpace{}%
\AgdaSymbol{:}\AgdaSpace{}%
\AgdaPrimitive{Set}\AgdaSpace{}%
\AgdaKeyword{where}\<%
\\
\>[0][@{}l@{\AgdaIndent{0}}]%
\>[2]\AgdaInductiveConstructor{⊕′}\AgdaSpace{}%
\AgdaInductiveConstructor{\&′}\AgdaSpace{}%
\AgdaSymbol{:}\AgdaSpace{}%
\AgdaSymbol{∀}\AgdaSpace{}%
\AgdaSymbol{\{}\AgdaBound{k}\AgdaSymbol{\}}\AgdaSpace{}%
\AgdaSymbol{→}\AgdaSpace{}%
\AgdaSymbol{(}\AgdaBound{si}\AgdaSpace{}%
\AgdaSymbol{:}\AgdaSpace{}%
\AgdaSymbol{(}\AgdaBound{i}\AgdaSpace{}%
\AgdaSymbol{:}\AgdaSpace{}%
\AgdaDatatype{Fin}\AgdaSpace{}%
\AgdaBound{k}\AgdaSymbol{)}\AgdaSpace{}%
\AgdaSymbol{→}\AgdaSpace{}%
\AgdaDatatype{Session}\AgdaSymbol{)}\AgdaSpace{}%
\AgdaSymbol{→}\AgdaSpace{}%
\AgdaDatatype{Session}\<%
\\
\>[2]\AgdaInductiveConstructor{send}\AgdaSpace{}%
\AgdaInductiveConstructor{recv}\AgdaSpace{}%
\AgdaSymbol{:}\AgdaSpace{}%
\AgdaDatatype{Type}\AgdaSpace{}%
\AgdaSymbol{→}\AgdaSpace{}%
\AgdaDatatype{Session}\AgdaSpace{}%
\AgdaSymbol{→}\AgdaSpace{}%
\AgdaDatatype{Session}\<%
\\
\>[2]\AgdaInductiveConstructor{end}\AgdaSpace{}%
\AgdaSymbol{:}\AgdaSpace{}%
\AgdaDatatype{Session}\<%
\\
\\[\AgdaEmptyExtraSkip]%
\>[0]\AgdaKeyword{pattern}\AgdaSpace{}%
\AgdaOperator{\AgdaInductiveConstructor{⁇\AgdaUnderscore{}∙\AgdaUnderscore{}}}\AgdaSpace{}%
\AgdaBound{t}\AgdaSpace{}%
\AgdaBound{s}\AgdaSpace{}%
\AgdaSymbol{=}\AgdaSpace{}%
\AgdaInductiveConstructor{recv}\AgdaSpace{}%
\AgdaBound{t}\AgdaSpace{}%
\AgdaBound{s}\<%
\\
\>[0]\AgdaKeyword{pattern}\AgdaSpace{}%
\AgdaOperator{\AgdaInductiveConstructor{‼\AgdaUnderscore{}∙\AgdaUnderscore{}}}\AgdaSpace{}%
\AgdaBound{t}\AgdaSpace{}%
\AgdaBound{s}\AgdaSpace{}%
\AgdaSymbol{=}\AgdaSpace{}%
\AgdaInductiveConstructor{send}\AgdaSpace{}%
\AgdaBound{t}\AgdaSpace{}%
\AgdaBound{s}\<%
\\
\\[\AgdaEmptyExtraSkip]%
\>[0]\AgdaKeyword{infixr}\AgdaSpace{}%
\AgdaNumber{20}\AgdaSpace{}%
\AgdaOperator{\AgdaInductiveConstructor{‼\AgdaUnderscore{}∙\AgdaUnderscore{}}}\AgdaSpace{}%
\AgdaOperator{\AgdaInductiveConstructor{⁇\AgdaUnderscore{}∙\AgdaUnderscore{}}}\<%
\\
\\[\AgdaEmptyExtraSkip]%
\>[0]\AgdaFunction{vec→fin}\AgdaSpace{}%
\AgdaSymbol{:}\AgdaSpace{}%
\AgdaDatatype{Vec}\AgdaSpace{}%
\AgdaDatatype{Session}\AgdaSpace{}%
\AgdaGeneralizable{k}\AgdaSpace{}%
\AgdaSymbol{→}\AgdaSpace{}%
\AgdaSymbol{(}\AgdaDatatype{Fin}\AgdaSpace{}%
\AgdaGeneralizable{k}\AgdaSpace{}%
\AgdaSymbol{→}\AgdaSpace{}%
\AgdaDatatype{Session}\AgdaSymbol{)}\<%
\\
\>[0]\AgdaFunction{vec→fin}\AgdaSpace{}%
\AgdaSymbol{\{}\AgdaArgument{k}\AgdaSpace{}%
\AgdaSymbol{=}\AgdaSpace{}%
\AgdaInductiveConstructor{zero}\AgdaSymbol{\}}\AgdaSpace{}%
\AgdaInductiveConstructor{[]}\AgdaSpace{}%
\AgdaSymbol{=}\AgdaSpace{}%
\AgdaSymbol{λ()}\<%
\\
\>[0]\AgdaFunction{vec→fin}\AgdaSpace{}%
\AgdaSymbol{\{}\AgdaArgument{k}\AgdaSpace{}%
\AgdaSymbol{=}\AgdaSpace{}%
\AgdaInductiveConstructor{suc}\AgdaSpace{}%
\AgdaBound{k}\AgdaSymbol{\}}\AgdaSpace{}%
\AgdaSymbol{(}\AgdaBound{x}\AgdaSpace{}%
\AgdaOperator{\AgdaInductiveConstructor{∷}}\AgdaSpace{}%
\AgdaBound{v}\AgdaSymbol{)}\AgdaSpace{}%
\AgdaSymbol{=}\AgdaSpace{}%
\AgdaSymbol{λ}\AgdaSpace{}%
\AgdaKeyword{where}\<%
\\
\>[0][@{}l@{\AgdaIndent{0}}]%
\>[2]\AgdaInductiveConstructor{zero}\AgdaSpace{}%
\AgdaSymbol{→}\AgdaSpace{}%
\AgdaBound{x}\<%
\\
\>[2]\AgdaSymbol{(}\AgdaInductiveConstructor{suc}\AgdaSpace{}%
\AgdaBound{i}\AgdaSymbol{)}\AgdaSpace{}%
\AgdaSymbol{→}\AgdaSpace{}%
\AgdaFunction{vec→fin}\AgdaSpace{}%
\AgdaBound{v}\AgdaSpace{}%
\AgdaBound{i}\<%
\\
\\[\AgdaEmptyExtraSkip]%
\>[0]\AgdaFunction{⊕}\AgdaSpace{}%
\AgdaSymbol{:}\AgdaSpace{}%
\AgdaDatatype{Vec}\AgdaSpace{}%
\AgdaDatatype{Session}\AgdaSpace{}%
\AgdaGeneralizable{k}\AgdaSpace{}%
\AgdaSymbol{→}\AgdaSpace{}%
\AgdaDatatype{Session}\<%
\\
\>[0]\AgdaFunction{⊕}\AgdaSpace{}%
\AgdaSymbol{=}\AgdaSpace{}%
\AgdaInductiveConstructor{⊕′}\AgdaSpace{}%
\AgdaOperator{\AgdaFunction{∘}}\AgdaSpace{}%
\AgdaFunction{vec→fin}\<%
\\
\\[\AgdaEmptyExtraSkip]%
\>[0]\AgdaFunction{\&}\AgdaSpace{}%
\AgdaSymbol{:}\AgdaSpace{}%
\AgdaDatatype{Vec}\AgdaSpace{}%
\AgdaDatatype{Session}\AgdaSpace{}%
\AgdaGeneralizable{k}\AgdaSpace{}%
\AgdaSymbol{→}\AgdaSpace{}%
\AgdaDatatype{Session}\<%
\\
\>[0]\AgdaFunction{\&}\AgdaSpace{}%
\AgdaSymbol{=}\AgdaSpace{}%
\AgdaInductiveConstructor{\&′}\AgdaSpace{}%
\AgdaOperator{\AgdaFunction{∘}}\AgdaSpace{}%
\AgdaFunction{vec→fin}\<%
\\
\>[0]\<%
\\
\>[0]\AgdaComment{--\ service\ protocol\ for\ a\ binary\ function}\<%
\\
\>[0]\AgdaFunction{binaryp}\AgdaSpace{}%
\AgdaSymbol{:}\AgdaSpace{}%
\AgdaDatatype{Session}\<%
\\
\>[0]\AgdaFunction{binaryp}\AgdaSpace{}%
\AgdaSymbol{=}\AgdaSpace{}%
\AgdaSymbol{(}\AgdaInductiveConstructor{recv}\AgdaSpace{}%
\AgdaInductiveConstructor{int}\AgdaSpace{}%
\AgdaSymbol{(}\AgdaInductiveConstructor{recv}\AgdaSpace{}%
\AgdaInductiveConstructor{int}\AgdaSpace{}%
\AgdaSymbol{(}\AgdaInductiveConstructor{send}\AgdaSpace{}%
\AgdaInductiveConstructor{int}\AgdaSpace{}%
\AgdaInductiveConstructor{end}\AgdaSymbol{)))}\<%
\\
\\[\AgdaEmptyExtraSkip]%
\>[0]\AgdaComment{--\ service\ protocol\ for\ a\ unary\ function}\<%
\\
\>[0]\AgdaFunction{unaryp}\AgdaSpace{}%
\AgdaSymbol{:}\AgdaSpace{}%
\AgdaDatatype{Session}\<%
\\
\>[0]\AgdaFunction{unaryp}\AgdaSpace{}%
\AgdaSymbol{=}\AgdaSpace{}%
\AgdaSymbol{(}\AgdaInductiveConstructor{recv}\AgdaSpace{}%
\AgdaInductiveConstructor{int}\AgdaSpace{}%
\AgdaSymbol{(}\AgdaInductiveConstructor{send}\AgdaSpace{}%
\AgdaInductiveConstructor{int}\AgdaSpace{}%
\AgdaInductiveConstructor{end}\AgdaSymbol{))}\<%
\\
\\[\AgdaEmptyExtraSkip]%
\>[0]\AgdaComment{--\ service\ protocol\ for\ choosing\ between\ a\ binary\ and\ a\ unary\ function}\<%
\\
\>[0]\AgdaFunction{arithp}\AgdaSpace{}%
\AgdaSymbol{:}\AgdaSpace{}%
\AgdaDatatype{Session}\<%
\\
\>[0]\AgdaFunction{arithp}\AgdaSpace{}%
\AgdaSymbol{=}\AgdaSpace{}%
\AgdaFunction{\&}\AgdaSpace{}%
\AgdaOperator{\AgdaInductiveConstructor{[}}\AgdaSpace{}%
\AgdaFunction{binaryp}\AgdaSpace{}%
\AgdaOperator{\AgdaInductiveConstructor{,}}\AgdaSpace{}%
\AgdaFunction{unaryp}\AgdaSpace{}%
\AgdaOperator{\AgdaInductiveConstructor{]}}\<%
\\
\\[\AgdaEmptyExtraSkip]%
\>[0]\AgdaKeyword{variable}\<%
\\
\>[0][@{}l@{\AgdaIndent{0}}]%
\>[2]\AgdaGeneralizable{a}\AgdaSpace{}%
\AgdaGeneralizable{b}\AgdaSpace{}%
\AgdaSymbol{:}\AgdaSpace{}%
\AgdaPostulate{Level}\<%
\\
\>[2]\AgdaGeneralizable{A}\AgdaSpace{}%
\AgdaGeneralizable{A′}\AgdaSpace{}%
\AgdaGeneralizable{A″}\AgdaSpace{}%
\AgdaGeneralizable{A₁}\AgdaSpace{}%
\AgdaGeneralizable{A₂}\AgdaSpace{}%
\AgdaSymbol{:}\AgdaSpace{}%
\AgdaPrimitive{Set}\<%
\\
\>[2]\AgdaGeneralizable{T}\AgdaSpace{}%
\AgdaSymbol{:}\AgdaSpace{}%
\AgdaDatatype{Type}\<%
\\
\>[2]\AgdaGeneralizable{s}\AgdaSpace{}%
\AgdaGeneralizable{s₁}\AgdaSpace{}%
\AgdaGeneralizable{s₂}\AgdaSpace{}%
\AgdaSymbol{:}\AgdaSpace{}%
\AgdaDatatype{Session}\<%
\\
\>[2]\AgdaGeneralizable{S}\AgdaSpace{}%
\AgdaSymbol{:}\AgdaSpace{}%
\AgdaDatatype{Session}\<%
\\
\>[2]\AgdaGeneralizable{M}\AgdaSpace{}%
\AgdaSymbol{:}\AgdaSpace{}%
\AgdaPrimitive{Set}\AgdaSpace{}%
\AgdaGeneralizable{a}\AgdaSpace{}%
\AgdaSymbol{→}\AgdaSpace{}%
\AgdaPrimitive{Set}\AgdaSpace{}%
\AgdaGeneralizable{b}\<%
\\
\\[\AgdaEmptyExtraSkip]%
\>[0]\AgdaOperator{\AgdaFunction{T⟦\AgdaUnderscore{}⟧}}\AgdaSpace{}%
\AgdaSymbol{:}\AgdaSpace{}%
\AgdaDatatype{Type}\AgdaSpace{}%
\AgdaSymbol{→}\AgdaSpace{}%
\AgdaPrimitive{Set}\<%
\\
\>[0]\AgdaOperator{\AgdaFunction{T⟦}}\AgdaSpace{}%
\AgdaInductiveConstructor{nat}\AgdaSpace{}%
\AgdaOperator{\AgdaFunction{⟧}}\AgdaSpace{}%
\AgdaSymbol{=}\AgdaSpace{}%
\AgdaDatatype{ℕ}\<%
\\
\>[0]\AgdaOperator{\AgdaFunction{T⟦}}\AgdaSpace{}%
\AgdaInductiveConstructor{bool}\AgdaSpace{}%
\AgdaOperator{\AgdaFunction{⟧}}\AgdaSpace{}%
\AgdaSymbol{=}\AgdaSpace{}%
\AgdaDatatype{Bool}\<%
\\
\>[0]\AgdaOperator{\AgdaFunction{T⟦}}\AgdaSpace{}%
\AgdaInductiveConstructor{int}\AgdaSpace{}%
\AgdaOperator{\AgdaFunction{⟧}}\AgdaSpace{}%
\AgdaSymbol{=}\AgdaSpace{}%
\AgdaDatatype{ℤ}\<%
\end{code}
\newcommand\mstMonadic{%
\begin{code}%
\>[0]\AgdaFunction{Monadic}\AgdaSpace{}%
\AgdaSymbol{:}\AgdaSpace{}%
\AgdaSymbol{((}\AgdaPrimitive{Set}\AgdaSpace{}%
\AgdaSymbol{→}\AgdaSpace{}%
\AgdaPrimitive{Set₁}\AgdaSymbol{)}\AgdaSpace{}%
\AgdaSymbol{→}\AgdaSpace{}%
\AgdaPrimitive{Set₁}\AgdaSymbol{)}\AgdaSpace{}%
\AgdaSymbol{→}\AgdaSpace{}%
\AgdaPrimitive{Set₂}\<%
\\
\>[0]\AgdaFunction{Monadic}\AgdaSpace{}%
\AgdaBound{f}\AgdaSpace{}%
\AgdaSymbol{=}\AgdaSpace{}%
\AgdaSymbol{∀}\AgdaSpace{}%
\AgdaSymbol{\{}\AgdaBound{M}\AgdaSpace{}%
\AgdaSymbol{:}\AgdaSpace{}%
\AgdaPrimitive{Set}\AgdaSpace{}%
\AgdaSymbol{→}\AgdaSpace{}%
\AgdaPrimitive{Set₁}\AgdaSymbol{\}}\AgdaSpace{}%
\AgdaSymbol{→}\AgdaSpace{}%
\AgdaSymbol{\{\{}\AgdaRecord{RawMonad}\AgdaSpace{}%
\AgdaBound{M}\AgdaSymbol{\}\}}\AgdaSpace{}%
\AgdaSymbol{→}\AgdaSpace{}%
\AgdaBound{f}\AgdaSpace{}%
\AgdaBound{M}\<%
\\
\\[\AgdaEmptyExtraSkip]%
\>[0]\AgdaKeyword{syntax}\AgdaSpace{}%
\AgdaFunction{Monadic}\AgdaSpace{}%
\AgdaSymbol{(λ}\AgdaSpace{}%
\AgdaBound{M}\AgdaSpace{}%
\AgdaSymbol{→}\AgdaSpace{}%
\AgdaBound{X}\AgdaSymbol{)}\AgdaSpace{}%
\AgdaSymbol{=}\AgdaSpace{}%
\AgdaFunction{Monad}\AgdaSpace{}%
\AgdaBound{M}\AgdaSpace{}%
\AgdaFunction{⇒}\AgdaSpace{}%
\AgdaBound{X}\<%
\end{code}}
\newcommand\mstCommand{%
\begin{code}%
\>[0]\AgdaKeyword{data}\AgdaSpace{}%
\AgdaDatatype{Cmd}\AgdaSpace{}%
\AgdaSymbol{(}\AgdaBound{A}\AgdaSpace{}%
\AgdaSymbol{:}\AgdaSpace{}%
\AgdaPrimitive{Set}\AgdaSymbol{)}\AgdaSpace{}%
\AgdaSymbol{:}\AgdaSpace{}%
\AgdaDatatype{Session}\AgdaSpace{}%
\AgdaSymbol{→}\AgdaSpace{}%
\AgdaPrimitive{Set₂}\AgdaSpace{}%
\AgdaKeyword{where}\<%
\\
\>[0][@{}l@{\AgdaIndent{0}}]%
\>[2]\AgdaInductiveConstructor{CLOSE}%
\>[9]\AgdaSymbol{:}\AgdaSpace{}%
\AgdaDatatype{Cmd}\AgdaSpace{}%
\AgdaBound{A}\AgdaSpace{}%
\AgdaInductiveConstructor{end}\<%
\\
\>[2]\AgdaInductiveConstructor{SKIP}%
\>[9]\AgdaSymbol{:}\AgdaSpace{}%
\AgdaSymbol{(}\AgdaFunction{Monad}\AgdaSpace{}%
\AgdaBound{M}\AgdaSpace{}%
\AgdaFunction{⇒}\AgdaSpace{}%
\AgdaRecord{StateT}\AgdaSpace{}%
\AgdaBound{A}\AgdaSpace{}%
\AgdaBound{M}\AgdaSpace{}%
\AgdaFunction{⊤}\AgdaSymbol{)}\AgdaSpace{}%
\AgdaSymbol{→}\AgdaSpace{}%
\AgdaDatatype{Cmd}\AgdaSpace{}%
\AgdaBound{A}\AgdaSpace{}%
\AgdaGeneralizable{S}\AgdaSpace{}%
\AgdaSymbol{→}\AgdaSpace{}%
\AgdaDatatype{Cmd}\AgdaSpace{}%
\AgdaBound{A}\AgdaSpace{}%
\AgdaGeneralizable{S}\<%
\\
\>[2]\AgdaInductiveConstructor{SEND}%
\>[9]\AgdaSymbol{:}\AgdaSpace{}%
\AgdaSymbol{(}\AgdaFunction{Monad}\AgdaSpace{}%
\AgdaBound{M}\AgdaSpace{}%
\AgdaFunction{⇒}\AgdaSpace{}%
\AgdaRecord{StateT}\AgdaSpace{}%
\AgdaBound{A}\AgdaSpace{}%
\AgdaBound{M}\AgdaSpace{}%
\AgdaOperator{\AgdaFunction{T⟦}}\AgdaSpace{}%
\AgdaGeneralizable{T}\AgdaSpace{}%
\AgdaOperator{\AgdaFunction{⟧}}\AgdaSymbol{)}\AgdaSpace{}%
\AgdaSymbol{→}\AgdaSpace{}%
\AgdaDatatype{Cmd}\AgdaSpace{}%
\AgdaBound{A}\AgdaSpace{}%
\AgdaGeneralizable{S}\AgdaSpace{}%
\AgdaSymbol{→}\AgdaSpace{}%
\AgdaDatatype{Cmd}\AgdaSpace{}%
\AgdaBound{A}\AgdaSpace{}%
\AgdaSymbol{(}\AgdaInductiveConstructor{send}\AgdaSpace{}%
\AgdaGeneralizable{T}\AgdaSpace{}%
\AgdaGeneralizable{S}\AgdaSymbol{)}\<%
\\
\>[2]\AgdaInductiveConstructor{RECV}%
\>[9]\AgdaSymbol{:}\AgdaSpace{}%
\AgdaSymbol{(}\AgdaFunction{Monad}\AgdaSpace{}%
\AgdaBound{M}\AgdaSpace{}%
\AgdaFunction{⇒}\AgdaSpace{}%
\AgdaSymbol{(}\AgdaOperator{\AgdaFunction{T⟦}}\AgdaSpace{}%
\AgdaGeneralizable{T}\AgdaSpace{}%
\AgdaOperator{\AgdaFunction{⟧}}\AgdaSpace{}%
\AgdaSymbol{→}\AgdaSpace{}%
\AgdaRecord{StateT}\AgdaSpace{}%
\AgdaBound{A}\AgdaSpace{}%
\AgdaBound{M}\AgdaSpace{}%
\AgdaFunction{⊤}\AgdaSymbol{))}\AgdaSpace{}%
\AgdaSymbol{→}\AgdaSpace{}%
\AgdaDatatype{Cmd}\AgdaSpace{}%
\AgdaBound{A}\AgdaSpace{}%
\AgdaGeneralizable{S}\AgdaSpace{}%
\AgdaSymbol{→}\AgdaSpace{}%
\AgdaDatatype{Cmd}\AgdaSpace{}%
\AgdaBound{A}\AgdaSpace{}%
\AgdaSymbol{(}\AgdaInductiveConstructor{recv}\AgdaSpace{}%
\AgdaGeneralizable{T}\AgdaSpace{}%
\AgdaGeneralizable{S}\AgdaSymbol{)}\<%
\\
\>[2]\AgdaInductiveConstructor{SELECT}\AgdaSpace{}%
\AgdaSymbol{:}\AgdaSpace{}%
\AgdaSymbol{∀}\AgdaSpace{}%
\AgdaSymbol{\{}\AgdaBound{Si}\AgdaSymbol{\}}\AgdaSpace{}%
\AgdaSymbol{→}\AgdaSpace{}%
\AgdaSymbol{(}\AgdaBound{i}\AgdaSpace{}%
\AgdaSymbol{:}\AgdaSpace{}%
\AgdaDatatype{Fin}\AgdaSpace{}%
\AgdaGeneralizable{k}\AgdaSymbol{)}\AgdaSpace{}%
\AgdaSymbol{→}\AgdaSpace{}%
\AgdaDatatype{Cmd}\AgdaSpace{}%
\AgdaBound{A}\AgdaSpace{}%
\AgdaSymbol{(}\AgdaBound{Si}\AgdaSpace{}%
\AgdaBound{i}\AgdaSymbol{)}\AgdaSpace{}%
\AgdaSymbol{→}\AgdaSpace{}%
\AgdaDatatype{Cmd}\AgdaSpace{}%
\AgdaBound{A}\AgdaSpace{}%
\AgdaSymbol{(}\AgdaInductiveConstructor{⊕′}\AgdaSpace{}%
\AgdaBound{Si}\AgdaSymbol{)}\<%
\\
\>[2]\AgdaInductiveConstructor{CHOICE}\AgdaSpace{}%
\AgdaSymbol{:}\AgdaSpace{}%
\AgdaSymbol{∀}\AgdaSpace{}%
\AgdaSymbol{\{}\AgdaBound{Si}\AgdaSymbol{\}}\AgdaSpace{}%
\AgdaSymbol{→}\AgdaSpace{}%
\AgdaSymbol{((}\AgdaBound{i}\AgdaSpace{}%
\AgdaSymbol{:}\AgdaSpace{}%
\AgdaDatatype{Fin}\AgdaSpace{}%
\AgdaGeneralizable{k}\AgdaSymbol{)}\AgdaSpace{}%
\AgdaSymbol{→}\AgdaSpace{}%
\AgdaDatatype{Cmd}\AgdaSpace{}%
\AgdaBound{A}\AgdaSpace{}%
\AgdaSymbol{(}\AgdaBound{Si}\AgdaSpace{}%
\AgdaBound{i}\AgdaSymbol{))}\AgdaSpace{}%
\AgdaSymbol{→}\AgdaSpace{}%
\AgdaDatatype{Cmd}\AgdaSpace{}%
\AgdaBound{A}\AgdaSpace{}%
\AgdaSymbol{(}\AgdaInductiveConstructor{\&′}\AgdaSpace{}%
\AgdaBound{Si}\AgdaSymbol{)}\<%
\end{code}}
\newcommand\mstExampleServers{%
\begin{code}%
\>[0]\AgdaFunction{addp-command}\AgdaSpace{}%
\AgdaSymbol{:}\AgdaSpace{}%
\AgdaDatatype{Cmd}\AgdaSpace{}%
\AgdaDatatype{ℤ}\AgdaSpace{}%
\AgdaFunction{binaryp}\<%
\\
\>[0]\AgdaFunction{addp-command}\AgdaSpace{}%
\AgdaSymbol{=}\AgdaSpace{}%
\AgdaInductiveConstructor{RECV}\AgdaSpace{}%
\AgdaFunction{put}\AgdaSpace{}%
\AgdaOperator{\AgdaFunction{\$}}\AgdaSpace{}%
\AgdaInductiveConstructor{RECV}\AgdaSpace{}%
\AgdaSymbol{(}\AgdaField{modify}\AgdaSpace{}%
\AgdaOperator{\AgdaFunction{∘′}}\AgdaSpace{}%
\AgdaOperator{\AgdaFunction{\AgdaUnderscore{}+\AgdaUnderscore{}}}\AgdaSymbol{)}\AgdaSpace{}%
\AgdaOperator{\AgdaFunction{\$}}\AgdaSpace{}%
\AgdaInductiveConstructor{SEND}\AgdaSpace{}%
\AgdaFunction{get}\AgdaSpace{}%
\AgdaOperator{\AgdaFunction{\$}}\AgdaSpace{}%
\AgdaInductiveConstructor{CLOSE}\<%
\\
\\[\AgdaEmptyExtraSkip]%
\>[0]\AgdaFunction{negp-command}\AgdaSpace{}%
\AgdaSymbol{:}\AgdaSpace{}%
\AgdaDatatype{Cmd}\AgdaSpace{}%
\AgdaDatatype{ℤ}\AgdaSpace{}%
\AgdaFunction{unaryp}\<%
\\
\>[0]\AgdaFunction{negp-command}\AgdaSpace{}%
\AgdaSymbol{=}\AgdaSpace{}%
\AgdaInductiveConstructor{RECV}\AgdaSpace{}%
\AgdaSymbol{(}\AgdaFunction{put}\AgdaSpace{}%
\AgdaOperator{\AgdaFunction{∘}}\AgdaSpace{}%
\AgdaOperator{\AgdaFunction{-\AgdaUnderscore{}}}\AgdaSymbol{)}\AgdaSpace{}%
\AgdaOperator{\AgdaFunction{\$}}\AgdaSpace{}%
\AgdaInductiveConstructor{SEND}\AgdaSpace{}%
\AgdaFunction{get}\AgdaSpace{}%
\AgdaOperator{\AgdaFunction{\$}}\AgdaSpace{}%
\AgdaInductiveConstructor{CLOSE}\<%
\\
\\[\AgdaEmptyExtraSkip]%
\>[0]\AgdaFunction{arithp-command}\AgdaSpace{}%
\AgdaSymbol{:}\AgdaSpace{}%
\AgdaDatatype{Cmd}\AgdaSpace{}%
\AgdaDatatype{ℤ}\AgdaSpace{}%
\AgdaFunction{arithp}\<%
\\
\>[0]\AgdaFunction{arithp-command}\AgdaSpace{}%
\AgdaSymbol{=}\AgdaSpace{}%
\AgdaInductiveConstructor{CHOICE}\AgdaSpace{}%
\AgdaSymbol{λ}\AgdaSpace{}%
\AgdaKeyword{where}\<%
\\
\>[0][@{}l@{\AgdaIndent{0}}]%
\>[2]\AgdaInductiveConstructor{zero}\AgdaSpace{}%
\AgdaSymbol{→}\AgdaSpace{}%
\AgdaFunction{addp-command}\<%
\\
\>[2]\AgdaSymbol{(}\AgdaInductiveConstructor{suc}\AgdaSpace{}%
\AgdaInductiveConstructor{zero}\AgdaSymbol{)}\AgdaSpace{}%
\AgdaSymbol{→}\AgdaSpace{}%
\AgdaFunction{negp-command}\<%
\end{code}}
\newcommand\mstExecutor{%
\begin{code}%
\>[0]\AgdaFunction{exec}\AgdaSpace{}%
\AgdaSymbol{:}\AgdaSpace{}%
\AgdaDatatype{Cmd}\AgdaSpace{}%
\AgdaGeneralizable{A}\AgdaSpace{}%
\AgdaGeneralizable{S}\AgdaSpace{}%
\AgdaSymbol{→}\AgdaSpace{}%
\AgdaRecord{StateT}\AgdaSpace{}%
\AgdaGeneralizable{A}\AgdaSpace{}%
\AgdaSymbol{(}\AgdaRecord{ReaderT}\AgdaSpace{}%
\AgdaPostulate{Channel}\AgdaSpace{}%
\AgdaDatatype{IO}\AgdaSymbol{)}\AgdaSpace{}%
\AgdaFunction{⊤}\<%
\\
\>[0]\AgdaFunction{exec}\AgdaSpace{}%
\AgdaInductiveConstructor{CLOSE}%
\>[21]\AgdaSymbol{=}\AgdaSpace{}%
\AgdaFunction{ask}\AgdaSpace{}%
\AgdaOperator{\AgdaField{>>=}}\AgdaSpace{}%
\AgdaField{liftIO}\AgdaSpace{}%
\AgdaOperator{\AgdaFunction{∘}}\AgdaSpace{}%
\AgdaPostulate{primClose}\<%
\\
\>[0]\AgdaFunction{exec}\AgdaSpace{}%
\AgdaSymbol{(}\AgdaInductiveConstructor{SKIP}\AgdaSpace{}%
\AgdaBound{act}\AgdaSpace{}%
\AgdaBound{cmd}\AgdaSymbol{)}%
\>[21]\AgdaSymbol{=}\AgdaSpace{}%
\AgdaBound{act}\AgdaSpace{}%
\AgdaOperator{\AgdaFunction{>>}}\AgdaSpace{}%
\AgdaFunction{exec}\AgdaSpace{}%
\AgdaBound{cmd}\<%
\\
\>[0]\AgdaFunction{exec}\AgdaSpace{}%
\AgdaSymbol{(}\AgdaInductiveConstructor{SEND}\AgdaSpace{}%
\AgdaBound{getx}\AgdaSpace{}%
\AgdaBound{cmd}\AgdaSymbol{)}\AgdaSpace{}%
\AgdaSymbol{=}\AgdaSpace{}%
\AgdaBound{getx}\AgdaSpace{}%
\AgdaOperator{\AgdaField{>>=}}\AgdaSpace{}%
\AgdaSymbol{λ}\AgdaSpace{}%
\AgdaBound{x}\AgdaSpace{}%
\AgdaSymbol{→}\AgdaSpace{}%
\AgdaFunction{ask}\AgdaSpace{}%
\AgdaOperator{\AgdaField{>>=}}\AgdaSpace{}%
\AgdaField{liftIO}\AgdaSpace{}%
\AgdaOperator{\AgdaFunction{∘}}\AgdaSpace{}%
\AgdaPostulate{primSend}\AgdaSpace{}%
\AgdaBound{x}\AgdaSpace{}%
\AgdaOperator{\AgdaFunction{>>}}\AgdaSpace{}%
\AgdaFunction{exec}\AgdaSpace{}%
\AgdaBound{cmd}\<%
\\
\>[0]\AgdaFunction{exec}\AgdaSpace{}%
\AgdaSymbol{(}\AgdaInductiveConstructor{RECV}\AgdaSpace{}%
\AgdaBound{putx}\AgdaSpace{}%
\AgdaBound{cmd}\AgdaSymbol{)}\AgdaSpace{}%
\AgdaSymbol{=}\AgdaSpace{}%
\AgdaFunction{ask}\AgdaSpace{}%
\AgdaOperator{\AgdaField{>>=}}\AgdaSpace{}%
\AgdaField{liftIO}\AgdaSpace{}%
\AgdaOperator{\AgdaFunction{∘}}\AgdaSpace{}%
\AgdaPostulate{primRecv}\AgdaSpace{}%
\AgdaOperator{\AgdaField{>>=}}\AgdaSpace{}%
\AgdaBound{putx}\AgdaSpace{}%
\AgdaOperator{\AgdaFunction{>>}}\AgdaSpace{}%
\AgdaFunction{exec}\AgdaSpace{}%
\AgdaBound{cmd}\<%
\\
\>[0]\AgdaFunction{exec}\AgdaSpace{}%
\AgdaSymbol{(}\AgdaInductiveConstructor{SELECT}\AgdaSpace{}%
\AgdaBound{i}\AgdaSpace{}%
\AgdaBound{cmd}\AgdaSymbol{)}%
\>[21]\AgdaSymbol{=}\AgdaSpace{}%
\AgdaFunction{ask}\AgdaSpace{}%
\AgdaOperator{\AgdaField{>>=}}\AgdaSpace{}%
\AgdaField{liftIO}\AgdaSpace{}%
\AgdaOperator{\AgdaFunction{∘}}\AgdaSpace{}%
\AgdaPostulate{primSend}\AgdaSpace{}%
\AgdaBound{i}\AgdaSpace{}%
\AgdaOperator{\AgdaFunction{>>}}\AgdaSpace{}%
\AgdaFunction{exec}\AgdaSpace{}%
\AgdaBound{cmd}\<%
\\
\>[0]\AgdaFunction{exec}\AgdaSpace{}%
\AgdaSymbol{(}\AgdaInductiveConstructor{CHOICE}\AgdaSpace{}%
\AgdaBound{f-cmd}\AgdaSymbol{)}%
\>[21]\AgdaSymbol{=}\AgdaSpace{}%
\AgdaFunction{ask}\AgdaSpace{}%
\AgdaOperator{\AgdaField{>>=}}\AgdaSpace{}%
\AgdaField{liftIO}\AgdaSpace{}%
\AgdaOperator{\AgdaFunction{∘}}\AgdaSpace{}%
\AgdaPostulate{primRecv}\AgdaSpace{}%
\AgdaOperator{\AgdaField{>>=}}\AgdaSpace{}%
\AgdaSymbol{λ}\AgdaSpace{}%
\AgdaBound{x}\AgdaSpace{}%
\AgdaSymbol{→}\AgdaSpace{}%
\AgdaFunction{exec}\AgdaSpace{}%
\AgdaSymbol{(}\AgdaBound{f-cmd}\AgdaSpace{}%
\AgdaBound{x}\AgdaSymbol{)}\<%
\end{code}}
\newcommand\mstAcceptor{%
\begin{code}%
\>[0]\AgdaKeyword{record}\AgdaSpace{}%
\AgdaRecord{Accepting}\AgdaSpace{}%
\AgdaBound{A}\AgdaSpace{}%
\AgdaBound{s}\AgdaSpace{}%
\AgdaSymbol{:}\AgdaSpace{}%
\AgdaPrimitive{Set₂}\AgdaSpace{}%
\AgdaKeyword{where}\<%
\\
\>[0][@{}l@{\AgdaIndent{0}}]%
\>[2]\AgdaKeyword{constructor}\AgdaSpace{}%
\AgdaInductiveConstructor{ACC}\<%
\\
\>[2]\AgdaKeyword{field}\AgdaSpace{}%
\AgdaField{pgm}\AgdaSpace{}%
\AgdaSymbol{:}\AgdaSpace{}%
\AgdaDatatype{Cmd}\AgdaSpace{}%
\AgdaBound{A}\AgdaSpace{}%
\AgdaBound{s}\<%
\\
\\[\AgdaEmptyExtraSkip]%
\>[0]\AgdaFunction{acceptor}\AgdaSpace{}%
\AgdaSymbol{:}\AgdaSpace{}%
\AgdaRecord{Accepting}\AgdaSpace{}%
\AgdaGeneralizable{A}\AgdaSpace{}%
\AgdaGeneralizable{s}\AgdaSpace{}%
\AgdaSymbol{→}\AgdaSpace{}%
\AgdaGeneralizable{A}\AgdaSpace{}%
\AgdaSymbol{→}\AgdaSpace{}%
\AgdaDatatype{IO}\AgdaSpace{}%
\AgdaGeneralizable{A}\<%
\\
\>[0]\AgdaFunction{acceptor}\AgdaSpace{}%
\AgdaSymbol{(}\AgdaInductiveConstructor{ACC}\AgdaSpace{}%
\AgdaBound{pgm}\AgdaSymbol{)}\AgdaSpace{}%
\AgdaBound{a}\AgdaSpace{}%
\AgdaSymbol{=}\AgdaSpace{}%
\AgdaKeyword{do}\<%
\\
\>[0][@{}l@{\AgdaIndent{0}}]%
\>[2]\AgdaBound{ch}\AgdaSpace{}%
\AgdaOperator{\AgdaField{←}}\AgdaSpace{}%
\AgdaPostulate{primAccept}\<%
\\
\>[2]\AgdaOperator{\AgdaInductiveConstructor{⟨}}\AgdaSpace{}%
\AgdaBound{final}\AgdaSpace{}%
\AgdaOperator{\AgdaInductiveConstructor{,}}\AgdaSpace{}%
\AgdaSymbol{\AgdaUnderscore{}}\AgdaSpace{}%
\AgdaOperator{\AgdaInductiveConstructor{⟩}}\AgdaSpace{}%
\AgdaOperator{\AgdaField{←}}\AgdaSpace{}%
\AgdaField{runReaderT}\AgdaSpace{}%
\AgdaSymbol{(}\AgdaField{runStateT}\AgdaSpace{}%
\AgdaSymbol{(}\AgdaFunction{exec}\AgdaSpace{}%
\AgdaBound{pgm}\AgdaSymbol{)}\AgdaSpace{}%
\AgdaBound{a}\AgdaSymbol{)}\AgdaSpace{}%
\AgdaBound{ch}\<%
\\
\>[2]\AgdaFunction{pure}\AgdaSpace{}%
\AgdaBound{final}\<%
\end{code}}

%% file: latex/ST-indexed-contextfree.tex
\begin{code}[hide]%
\>[0]\AgdaSymbol{\{-\#}\AgdaSpace{}%
\AgdaKeyword{OPTIONS}\AgdaSpace{}%
\AgdaPragma{--guardedness}\AgdaSpace{}%
\AgdaSymbol{\#-\}}\AgdaSpace{}%
\AgdaComment{\{-\ required\ for\ IO\ -\}}\<%
\\
\>[0]\AgdaKeyword{module}\AgdaSpace{}%
\AgdaModule{ST-indexed-contextfree}\AgdaSpace{}%
\AgdaKeyword{where}\<%
\\
\\[\AgdaEmptyExtraSkip]%
\>[0]\AgdaKeyword{open}\AgdaSpace{}%
\AgdaKeyword{import}\AgdaSpace{}%
\AgdaModule{Data.Bool}\AgdaSpace{}%
\AgdaKeyword{using}\AgdaSpace{}%
\AgdaSymbol{(}\AgdaDatatype{Bool}\AgdaSymbol{;}\AgdaSpace{}%
\AgdaInductiveConstructor{true}\AgdaSymbol{;}\AgdaSpace{}%
\AgdaInductiveConstructor{false}\AgdaSymbol{;}\AgdaSpace{}%
\AgdaOperator{\AgdaFunction{if\AgdaUnderscore{}then\AgdaUnderscore{}else\AgdaUnderscore{}}}\AgdaSymbol{)}\<%
\\
\>[0]\AgdaKeyword{open}\AgdaSpace{}%
\AgdaKeyword{import}\AgdaSpace{}%
\AgdaModule{Data.Fin}\AgdaSpace{}%
\AgdaKeyword{using}\AgdaSpace{}%
\AgdaSymbol{(}\AgdaDatatype{Fin}\AgdaSymbol{;}\AgdaSpace{}%
\AgdaInductiveConstructor{zero}\AgdaSymbol{;}\AgdaSpace{}%
\AgdaInductiveConstructor{suc}\AgdaSymbol{;}\AgdaSpace{}%
\AgdaFunction{toℕ}\AgdaSymbol{;}\AgdaSpace{}%
\AgdaFunction{fromℕ}\AgdaSymbol{;}\AgdaSpace{}%
\AgdaFunction{opposite}\AgdaSymbol{)}\<%
\\
\>[0]\AgdaKeyword{open}\AgdaSpace{}%
\AgdaKeyword{import}\AgdaSpace{}%
\AgdaModule{Data.Fin.Properties}\AgdaSpace{}%
\AgdaKeyword{using}\AgdaSpace{}%
\AgdaSymbol{(}\AgdaFunction{toℕ-fromℕ}\AgdaSymbol{;}\AgdaSpace{}%
\AgdaFunction{toℕ-inject₁}\AgdaSymbol{)}\<%
\\
\>[0]\AgdaKeyword{open}\AgdaSpace{}%
\AgdaKeyword{import}\AgdaSpace{}%
\AgdaModule{Data.Nat}\AgdaSpace{}%
\AgdaKeyword{using}\AgdaSpace{}%
\AgdaSymbol{(}\AgdaDatatype{ℕ}\AgdaSymbol{;}\AgdaSpace{}%
\AgdaInductiveConstructor{zero}\AgdaSymbol{;}\AgdaSpace{}%
\AgdaInductiveConstructor{suc}\AgdaSymbol{)}\<%
\\
\>[0]\AgdaKeyword{open}\AgdaSpace{}%
\AgdaKeyword{import}\AgdaSpace{}%
\AgdaModule{Data.Integer}\AgdaSpace{}%
\AgdaKeyword{using}\AgdaSpace{}%
\AgdaSymbol{(}\AgdaDatatype{ℤ}\AgdaSymbol{;}\AgdaSpace{}%
\AgdaOperator{\AgdaFunction{\AgdaUnderscore{}+\AgdaUnderscore{}}}\AgdaSymbol{;}\AgdaSpace{}%
\AgdaFunction{0ℤ}\AgdaSymbol{;}\AgdaSpace{}%
\AgdaOperator{\AgdaFunction{-\AgdaUnderscore{}}}\AgdaSymbol{)}\<%
\\
\>[0]\AgdaKeyword{open}\AgdaSpace{}%
\AgdaKeyword{import}\AgdaSpace{}%
\AgdaModule{Data.Product}\AgdaSpace{}%
\AgdaKeyword{using}\AgdaSpace{}%
\AgdaSymbol{(}\AgdaOperator{\AgdaFunction{\AgdaUnderscore{}×\AgdaUnderscore{}}}\AgdaSymbol{;}\AgdaSpace{}%
\AgdaRecord{Σ}\AgdaSymbol{;}\AgdaSpace{}%
\AgdaField{proj₁}\AgdaSymbol{;}\AgdaSpace{}%
\AgdaField{proj₂}\AgdaSymbol{;}\AgdaSpace{}%
\AgdaFunction{∃-syntax}\AgdaSymbol{;}\AgdaSpace{}%
\AgdaOperator{\AgdaFunction{<\AgdaUnderscore{},\AgdaUnderscore{}>}}\AgdaSymbol{)}\AgdaSpace{}%
\AgdaKeyword{renaming}\AgdaSpace{}%
\AgdaSymbol{(}\AgdaOperator{\AgdaInductiveConstructor{\AgdaUnderscore{},\AgdaUnderscore{}}}\AgdaSpace{}%
\AgdaSymbol{to}\AgdaSpace{}%
\AgdaOperator{\AgdaInductiveConstructor{⟨\AgdaUnderscore{},\AgdaUnderscore{}⟩}}\AgdaSymbol{)}\<%
\\
\>[0]\AgdaKeyword{open}\AgdaSpace{}%
\AgdaKeyword{import}\AgdaSpace{}%
\AgdaModule{Data.Sum}\AgdaSpace{}%
\AgdaKeyword{using}\AgdaSpace{}%
\AgdaSymbol{(}\AgdaOperator{\AgdaDatatype{\AgdaUnderscore{}⊎\AgdaUnderscore{}}}\AgdaSymbol{;}\AgdaSpace{}%
\AgdaInductiveConstructor{inj₁}\AgdaSymbol{;}\AgdaSpace{}%
\AgdaInductiveConstructor{inj₂}\AgdaSymbol{)}\<%
\\
\>[0]\AgdaKeyword{open}\AgdaSpace{}%
\AgdaKeyword{import}\AgdaSpace{}%
\AgdaModule{Data.Vec}\AgdaSpace{}%
\AgdaKeyword{using}\AgdaSpace{}%
\AgdaSymbol{(}\AgdaDatatype{Vec}\AgdaSymbol{;}\AgdaSpace{}%
\AgdaInductiveConstructor{[]}\AgdaSymbol{;}\AgdaSpace{}%
\AgdaOperator{\AgdaInductiveConstructor{\AgdaUnderscore{}∷\AgdaUnderscore{}}}\AgdaSymbol{;}\AgdaSpace{}%
\AgdaFunction{lookup}\AgdaSymbol{;}\AgdaSpace{}%
\AgdaFunction{take}\AgdaSymbol{;}\AgdaSpace{}%
\AgdaFunction{drop}\AgdaSymbol{)}\<%
\\
\\[\AgdaEmptyExtraSkip]%
\>[0]\AgdaKeyword{open}\AgdaSpace{}%
\AgdaKeyword{import}\AgdaSpace{}%
\AgdaModule{Effect.Functor}\<%
\\
\\[\AgdaEmptyExtraSkip]%
\>[0]\AgdaKeyword{open}\AgdaSpace{}%
\AgdaKeyword{import}\AgdaSpace{}%
\AgdaModule{Effect.Monad}\<%
\\
\>[0]\AgdaKeyword{open}\AgdaSpace{}%
\AgdaKeyword{import}\AgdaSpace{}%
\AgdaModule{Effect.Monad.State}\<%
\\
\>[0]\AgdaKeyword{open}\AgdaSpace{}%
\AgdaKeyword{import}\AgdaSpace{}%
\AgdaModule{Effect.Monad.Reader.Transformer}\AgdaSpace{}%
\AgdaSymbol{as}\AgdaSpace{}%
\AgdaModule{Reader}\<%
\\
\>[0]\AgdaKeyword{open}\AgdaSpace{}%
\AgdaKeyword{import}\AgdaSpace{}%
\AgdaModule{Effect.Monad.State.Transformer}\AgdaSpace{}%
\AgdaSymbol{as}\AgdaSpace{}%
\AgdaModule{State}\<%
\\
\>[0]\AgdaKeyword{open}\AgdaSpace{}%
\AgdaKeyword{import}\AgdaSpace{}%
\AgdaModule{Effect.Monad.IO}\<%
\\
\\[\AgdaEmptyExtraSkip]%
\>[0]\AgdaKeyword{open}\AgdaSpace{}%
\AgdaKeyword{import}\AgdaSpace{}%
\AgdaModule{Effect.Monad.Reader.Instances}\<%
\\
\>[0]\AgdaKeyword{open}\AgdaSpace{}%
\AgdaKeyword{import}\AgdaSpace{}%
\AgdaModule{Effect.Monad.State.Instances}\<%
\\
\>[0]\AgdaKeyword{open}\AgdaSpace{}%
\AgdaKeyword{import}\AgdaSpace{}%
\AgdaModule{Effect.Monad.Identity.Instances}\<%
\\
\>[0]\AgdaKeyword{open}\AgdaSpace{}%
\AgdaKeyword{import}\AgdaSpace{}%
\AgdaModule{Effect.Monad.IO.Instances}\<%
\\
\>[0]\AgdaKeyword{open}\AgdaSpace{}%
\AgdaKeyword{import}\AgdaSpace{}%
\AgdaModule{IO.Instances}\<%
\\
\\[\AgdaEmptyExtraSkip]%
\>[0]\AgdaKeyword{open}\AgdaSpace{}%
\AgdaModule{RawMonad}\AgdaSpace{}%
\AgdaSymbol{\{\{...\}\}}\<%
\\
\>[0]\AgdaKeyword{open}\AgdaSpace{}%
\AgdaModule{RawMonadState}\AgdaSpace{}%
\AgdaSymbol{\{\{...\}\}}\<%
\\
\>[0]\AgdaKeyword{open}\AgdaSpace{}%
\AgdaModule{RawMonadReader}\AgdaSpace{}%
\AgdaSymbol{\{\{...\}\}}\<%
\\
\>[0]\AgdaKeyword{open}\AgdaSpace{}%
\AgdaModule{RawMonadIO}\AgdaSpace{}%
\AgdaSymbol{\{\{...\}\}}\<%
\\
\>[0]\AgdaKeyword{open}\AgdaSpace{}%
\AgdaModule{RawFunctor}\AgdaSpace{}%
\AgdaSymbol{\{\{...\}\}}\<%
\\
\\[\AgdaEmptyExtraSkip]%
\>[0]\AgdaKeyword{open}\AgdaSpace{}%
\AgdaKeyword{import}\AgdaSpace{}%
\AgdaModule{Function.Base}\AgdaSpace{}%
\AgdaKeyword{using}\AgdaSpace{}%
\AgdaSymbol{(}\AgdaOperator{\AgdaFunction{case\AgdaUnderscore{}of\AgdaUnderscore{}}}\AgdaSymbol{;}\AgdaSpace{}%
\AgdaOperator{\AgdaFunction{\AgdaUnderscore{}∘\AgdaUnderscore{}}}\AgdaSymbol{;}\AgdaSpace{}%
\AgdaOperator{\AgdaFunction{\AgdaUnderscore{}∘′\AgdaUnderscore{}}}\AgdaSymbol{;}\AgdaSpace{}%
\AgdaFunction{const}\AgdaSymbol{;}\AgdaSpace{}%
\AgdaFunction{constᵣ}\AgdaSymbol{;}\AgdaSpace{}%
\AgdaFunction{id}\AgdaSymbol{;}\AgdaSpace{}%
\AgdaOperator{\AgdaFunction{\AgdaUnderscore{}\$\AgdaUnderscore{}}}\AgdaSymbol{)}\<%
\\
\>[0]\AgdaKeyword{open}\AgdaSpace{}%
\AgdaKeyword{import}\AgdaSpace{}%
\AgdaModule{Relation.Binary.PropositionalEquality}\AgdaSpace{}%
\AgdaKeyword{using}\AgdaSpace{}%
\AgdaSymbol{(}\AgdaOperator{\AgdaDatatype{\AgdaUnderscore{}≡\AgdaUnderscore{}}}\AgdaSymbol{;}\AgdaSpace{}%
\AgdaInductiveConstructor{refl}\AgdaSymbol{;}\AgdaSpace{}%
\AgdaFunction{subst}\AgdaSymbol{;}\AgdaSpace{}%
\AgdaFunction{sym}\AgdaSymbol{;}\AgdaSpace{}%
\AgdaFunction{cong}\AgdaSymbol{;}\AgdaSpace{}%
\AgdaFunction{cong₂}\AgdaSymbol{;}\AgdaSpace{}%
\AgdaFunction{trans}\AgdaSymbol{;}\AgdaSpace{}%
\AgdaKeyword{module}\AgdaSpace{}%
\AgdaModule{≡-Reasoning}\AgdaSymbol{)}\<%
\\
\\[\AgdaEmptyExtraSkip]%
\>[0]\AgdaKeyword{open}\AgdaSpace{}%
\AgdaKeyword{import}\AgdaSpace{}%
\AgdaModule{IO.Base}\AgdaSpace{}%
\AgdaKeyword{using}\AgdaSpace{}%
\AgdaSymbol{(}\AgdaDatatype{IO}\AgdaSymbol{)}\<%
\\
\\[\AgdaEmptyExtraSkip]%
\>[0]\AgdaKeyword{open}\AgdaSpace{}%
\AgdaKeyword{import}\AgdaSpace{}%
\AgdaModule{Channels}\<%
\\
\\[\AgdaEmptyExtraSkip]%
\>[0]\AgdaKeyword{open}\AgdaSpace{}%
\AgdaKeyword{import}\AgdaSpace{}%
\AgdaModule{Utils}\<%
\\
\\[\AgdaEmptyExtraSkip]%
\>[0]\AgdaKeyword{pattern}\AgdaSpace{}%
\AgdaOperator{\AgdaInductiveConstructor{[\AgdaUnderscore{}]}}\AgdaSpace{}%
\AgdaBound{x}\AgdaSpace{}%
\AgdaSymbol{=}\AgdaSpace{}%
\AgdaBound{x}\AgdaSpace{}%
\AgdaOperator{\AgdaInductiveConstructor{∷}}\AgdaSpace{}%
\AgdaInductiveConstructor{[]}\<%
\\
\>[0]\AgdaKeyword{pattern}\AgdaSpace{}%
\AgdaOperator{\AgdaInductiveConstructor{[\AgdaUnderscore{},\AgdaUnderscore{}]}}\AgdaSpace{}%
\AgdaBound{x}\AgdaSpace{}%
\AgdaBound{y}\AgdaSpace{}%
\AgdaSymbol{=}\AgdaSpace{}%
\AgdaBound{x}\AgdaSpace{}%
\AgdaOperator{\AgdaInductiveConstructor{∷}}\AgdaSpace{}%
\AgdaBound{y}\AgdaSpace{}%
\AgdaOperator{\AgdaInductiveConstructor{∷}}\AgdaSpace{}%
\AgdaInductiveConstructor{[]}\<%
\\
\>[0]\AgdaKeyword{pattern}\AgdaSpace{}%
\AgdaOperator{\AgdaInductiveConstructor{[\AgdaUnderscore{},\AgdaUnderscore{},\AgdaUnderscore{}]}}\AgdaSpace{}%
\AgdaBound{x}\AgdaSpace{}%
\AgdaBound{y}\AgdaSpace{}%
\AgdaBound{z}\AgdaSpace{}%
\AgdaSymbol{=}\AgdaSpace{}%
\AgdaBound{x}\AgdaSpace{}%
\AgdaOperator{\AgdaInductiveConstructor{∷}}\AgdaSpace{}%
\AgdaBound{y}\AgdaSpace{}%
\AgdaOperator{\AgdaInductiveConstructor{∷}}\AgdaSpace{}%
\AgdaBound{z}\AgdaSpace{}%
\AgdaOperator{\AgdaInductiveConstructor{∷}}\AgdaSpace{}%
\AgdaInductiveConstructor{[]}\<%
\\
\\[\AgdaEmptyExtraSkip]%
\>[0]\AgdaKeyword{variable}\AgdaSpace{}%
\AgdaGeneralizable{n}\AgdaSpace{}%
\AgdaGeneralizable{k}\AgdaSpace{}%
\AgdaSymbol{:}\AgdaSpace{}%
\AgdaDatatype{ℕ}\<%
\\
\\[\AgdaEmptyExtraSkip]%
\>[0]\AgdaKeyword{data}\AgdaSpace{}%
\AgdaDatatype{Type}\AgdaSpace{}%
\AgdaSymbol{:}\AgdaSpace{}%
\AgdaPrimitive{Set}\AgdaSpace{}%
\AgdaKeyword{where}\<%
\\
\>[0][@{}l@{\AgdaIndent{0}}]%
\>[2]\AgdaInductiveConstructor{nat}\AgdaSpace{}%
\AgdaInductiveConstructor{int}\AgdaSpace{}%
\AgdaSymbol{:}\AgdaSpace{}%
\AgdaDatatype{Type}\<%
\end{code}
\newcommand\cstSession{%
\begin{code}%
\>[0]\AgdaKeyword{data}\AgdaSpace{}%
\AgdaDatatype{Session}\AgdaSpace{}%
\AgdaSymbol{(}\AgdaBound{n}\AgdaSpace{}%
\AgdaSymbol{:}\AgdaSpace{}%
\AgdaDatatype{ℕ}\AgdaSymbol{)}\AgdaSpace{}%
\AgdaSymbol{:}\AgdaSpace{}%
\AgdaPrimitive{Set}\AgdaSpace{}%
\AgdaKeyword{where}\<%
\\
\>[0][@{}l@{\AgdaIndent{0}}]%
\>[2]\AgdaOperator{\AgdaInductiveConstructor{⁇\AgdaUnderscore{}}}\AgdaSpace{}%
\AgdaSymbol{:}\AgdaSpace{}%
\AgdaDatatype{Type}\AgdaSpace{}%
\AgdaSymbol{→}\AgdaSpace{}%
\AgdaDatatype{Session}\AgdaSpace{}%
\AgdaBound{n}\<%
\\
\>[2]\AgdaOperator{\AgdaInductiveConstructor{‼\AgdaUnderscore{}}}\AgdaSpace{}%
\AgdaSymbol{:}\AgdaSpace{}%
\AgdaDatatype{Type}\AgdaSpace{}%
\AgdaSymbol{→}\AgdaSpace{}%
\AgdaDatatype{Session}\AgdaSpace{}%
\AgdaBound{n}\<%
\\
\>[2]\AgdaInductiveConstructor{⊕′}\AgdaSpace{}%
\AgdaSymbol{:}\AgdaSpace{}%
\AgdaSymbol{(}\AgdaBound{si}\AgdaSpace{}%
\AgdaSymbol{:}\AgdaSpace{}%
\AgdaSymbol{(}\AgdaBound{i}\AgdaSpace{}%
\AgdaSymbol{:}\AgdaSpace{}%
\AgdaDatatype{Fin}\AgdaSpace{}%
\AgdaGeneralizable{k}\AgdaSymbol{)}\AgdaSpace{}%
\AgdaSymbol{→}\AgdaSpace{}%
\AgdaDatatype{Session}\AgdaSpace{}%
\AgdaBound{n}\AgdaSymbol{)}\AgdaSpace{}%
\AgdaSymbol{→}\AgdaSpace{}%
\AgdaDatatype{Session}\AgdaSpace{}%
\AgdaBound{n}\<%
\\
\>[2]\AgdaInductiveConstructor{\&′}\AgdaSpace{}%
\AgdaSymbol{:}\AgdaSpace{}%
\AgdaSymbol{(}\AgdaBound{si}\AgdaSpace{}%
\AgdaSymbol{:}\AgdaSpace{}%
\AgdaSymbol{(}\AgdaBound{i}\AgdaSpace{}%
\AgdaSymbol{:}\AgdaSpace{}%
\AgdaDatatype{Fin}\AgdaSpace{}%
\AgdaGeneralizable{k}\AgdaSymbol{)}\AgdaSpace{}%
\AgdaSymbol{→}\AgdaSpace{}%
\AgdaDatatype{Session}\AgdaSpace{}%
\AgdaBound{n}\AgdaSymbol{)}\AgdaSpace{}%
\AgdaSymbol{→}\AgdaSpace{}%
\AgdaDatatype{Session}\AgdaSpace{}%
\AgdaBound{n}\<%
\\
\>[2]\AgdaOperator{\AgdaInductiveConstructor{\AgdaUnderscore{}⨟\AgdaUnderscore{}}}\AgdaSpace{}%
\AgdaSymbol{:}\AgdaSpace{}%
\AgdaDatatype{Session}\AgdaSpace{}%
\AgdaBound{n}\AgdaSpace{}%
\AgdaSymbol{→}\AgdaSpace{}%
\AgdaDatatype{Session}\AgdaSpace{}%
\AgdaBound{n}\AgdaSpace{}%
\AgdaSymbol{→}\AgdaSpace{}%
\AgdaDatatype{Session}\AgdaSpace{}%
\AgdaBound{n}\<%
\\
\>[2]\AgdaInductiveConstructor{skip}\AgdaSpace{}%
\AgdaSymbol{:}\AgdaSpace{}%
\AgdaDatatype{Session}\AgdaSpace{}%
\AgdaBound{n}\<%
\\
\>[2]\AgdaOperator{\AgdaInductiveConstructor{μ\AgdaUnderscore{}}}\AgdaSpace{}%
\AgdaSymbol{:}\AgdaSpace{}%
\AgdaDatatype{Session}\AgdaSpace{}%
\AgdaSymbol{(}\AgdaInductiveConstructor{suc}\AgdaSpace{}%
\AgdaBound{n}\AgdaSymbol{)}\AgdaSpace{}%
\AgdaSymbol{→}\AgdaSpace{}%
\AgdaDatatype{Session}\AgdaSpace{}%
\AgdaBound{n}\<%
\\
\>[2]\AgdaOperator{\AgdaInductiveConstructor{`\AgdaUnderscore{}}}\AgdaSpace{}%
\AgdaSymbol{:}\AgdaSpace{}%
\AgdaDatatype{Fin}\AgdaSpace{}%
\AgdaBound{n}\AgdaSpace{}%
\AgdaSymbol{→}\AgdaSpace{}%
\AgdaDatatype{Session}\AgdaSpace{}%
\AgdaBound{n}\<%
\end{code}}
\begin{code}[hide]%
\>[0]\AgdaKeyword{infixr}\AgdaSpace{}%
\AgdaNumber{30}\AgdaSpace{}%
\AgdaOperator{\AgdaFunction{\AgdaUnderscore{}⨟′\AgdaUnderscore{}}}\<%
\\
\>[0]\AgdaKeyword{infixr}\AgdaSpace{}%
\AgdaNumber{30}\AgdaSpace{}%
\AgdaOperator{\AgdaInductiveConstructor{\AgdaUnderscore{}⨟\AgdaUnderscore{}}}\AgdaSpace{}%
\AgdaOperator{\AgdaInductiveConstructor{[\AgdaUnderscore{}]\AgdaUnderscore{}⨟[\AgdaUnderscore{}]\AgdaUnderscore{}[\AgdaUnderscore{}]}}\<%
\\
\>[0]\AgdaKeyword{infixl}\AgdaSpace{}%
\AgdaNumber{40}\AgdaSpace{}%
\AgdaOperator{\AgdaInductiveConstructor{⁇\AgdaUnderscore{}}}\AgdaSpace{}%
\AgdaOperator{\AgdaInductiveConstructor{‼\AgdaUnderscore{}}}\<%
\\
\>[0]\AgdaKeyword{infixr}\AgdaSpace{}%
\AgdaNumber{50}\AgdaSpace{}%
\AgdaOperator{\AgdaInductiveConstructor{μ\AgdaUnderscore{}}}\AgdaSpace{}%
\AgdaOperator{\AgdaInductiveConstructor{`\AgdaUnderscore{}}}\<%
\\
\\[\AgdaEmptyExtraSkip]%
\>[0]\AgdaKeyword{variable}\AgdaSpace{}%
\AgdaGeneralizable{T}\AgdaSpace{}%
\AgdaSymbol{:}\AgdaSpace{}%
\AgdaDatatype{Type}\<%
\\
\>[0]\AgdaKeyword{variable}\AgdaSpace{}%
\AgdaGeneralizable{S}\AgdaSpace{}%
\AgdaGeneralizable{S₁}\AgdaSpace{}%
\AgdaGeneralizable{S₂}\AgdaSpace{}%
\AgdaSymbol{:}\AgdaSpace{}%
\AgdaDatatype{Session}\AgdaSpace{}%
\AgdaGeneralizable{n}\<%
\\
\>[0]\AgdaKeyword{variable}\AgdaSpace{}%
\AgdaGeneralizable{A₁}\AgdaSpace{}%
\AgdaGeneralizable{A₂}\AgdaSpace{}%
\AgdaGeneralizable{A′}\AgdaSpace{}%
\AgdaGeneralizable{B₁}\AgdaSpace{}%
\AgdaGeneralizable{B₂}\AgdaSpace{}%
\AgdaSymbol{:}\AgdaSpace{}%
\AgdaPrimitive{Set}\<%
\\
\\[\AgdaEmptyExtraSkip]%
\>[0]\AgdaFunction{dual}\AgdaSpace{}%
\AgdaSymbol{:}\AgdaSpace{}%
\AgdaDatatype{Session}\AgdaSpace{}%
\AgdaGeneralizable{n}\AgdaSpace{}%
\AgdaSymbol{→}\AgdaSpace{}%
\AgdaDatatype{Session}\AgdaSpace{}%
\AgdaGeneralizable{n}\<%
\\
\>[0]\AgdaFunction{dual}\AgdaSpace{}%
\AgdaSymbol{(}\AgdaOperator{\AgdaInductiveConstructor{⁇}}\AgdaSpace{}%
\AgdaBound{x}\AgdaSymbol{)}\AgdaSpace{}%
\AgdaSymbol{=}\AgdaSpace{}%
\AgdaOperator{\AgdaInductiveConstructor{‼}}\AgdaSpace{}%
\AgdaBound{x}\<%
\\
\>[0]\AgdaFunction{dual}\AgdaSpace{}%
\AgdaSymbol{(}\AgdaOperator{\AgdaInductiveConstructor{‼}}\AgdaSpace{}%
\AgdaBound{x}\AgdaSymbol{)}\AgdaSpace{}%
\AgdaSymbol{=}\AgdaSpace{}%
\AgdaOperator{\AgdaInductiveConstructor{⁇}}\AgdaSpace{}%
\AgdaBound{x}\<%
\\
\>[0]\AgdaFunction{dual}\AgdaSpace{}%
\AgdaSymbol{(}\AgdaInductiveConstructor{⊕′}\AgdaSpace{}%
\AgdaBound{Si}\AgdaSymbol{)}\AgdaSpace{}%
\AgdaSymbol{=}\AgdaSpace{}%
\AgdaInductiveConstructor{\&′}\AgdaSpace{}%
\AgdaSymbol{(}\AgdaFunction{dual}\AgdaSpace{}%
\AgdaOperator{\AgdaFunction{∘}}\AgdaSpace{}%
\AgdaBound{Si}\AgdaSymbol{)}\<%
\\
\>[0]\AgdaFunction{dual}\AgdaSpace{}%
\AgdaSymbol{(}\AgdaInductiveConstructor{\&′}\AgdaSpace{}%
\AgdaBound{Si}\AgdaSymbol{)}\AgdaSpace{}%
\AgdaSymbol{=}\AgdaSpace{}%
\AgdaInductiveConstructor{⊕′}\AgdaSpace{}%
\AgdaSymbol{(}\AgdaFunction{dual}\AgdaSpace{}%
\AgdaOperator{\AgdaFunction{∘}}\AgdaSpace{}%
\AgdaBound{Si}\AgdaSymbol{)}\<%
\\
\>[0]\AgdaFunction{dual}\AgdaSpace{}%
\AgdaSymbol{(}\AgdaBound{S₁}\AgdaSpace{}%
\AgdaOperator{\AgdaInductiveConstructor{⨟}}\AgdaSpace{}%
\AgdaBound{S₂}\AgdaSymbol{)}\AgdaSpace{}%
\AgdaSymbol{=}\AgdaSpace{}%
\AgdaFunction{dual}\AgdaSpace{}%
\AgdaBound{S₁}\AgdaSpace{}%
\AgdaOperator{\AgdaInductiveConstructor{⨟}}\AgdaSpace{}%
\AgdaFunction{dual}\AgdaSpace{}%
\AgdaBound{S₂}\<%
\\
\>[0]\AgdaFunction{dual}\AgdaSpace{}%
\AgdaInductiveConstructor{skip}\AgdaSpace{}%
\AgdaSymbol{=}\AgdaSpace{}%
\AgdaInductiveConstructor{skip}\<%
\\
\>[0]\AgdaFunction{dual}\AgdaSpace{}%
\AgdaSymbol{(}\AgdaOperator{\AgdaInductiveConstructor{μ}}\AgdaSpace{}%
\AgdaBound{S}\AgdaSymbol{)}\AgdaSpace{}%
\AgdaSymbol{=}\AgdaSpace{}%
\AgdaOperator{\AgdaInductiveConstructor{μ}}\AgdaSpace{}%
\AgdaSymbol{(}\AgdaFunction{dual}\AgdaSpace{}%
\AgdaBound{S}\AgdaSymbol{)}\<%
\\
\>[0]\AgdaFunction{dual}\AgdaSpace{}%
\AgdaSymbol{(}\AgdaOperator{\AgdaInductiveConstructor{`}}\AgdaSpace{}%
\AgdaBound{x}\AgdaSymbol{)}\AgdaSpace{}%
\AgdaSymbol{=}\AgdaSpace{}%
\AgdaOperator{\AgdaInductiveConstructor{`}}\AgdaSpace{}%
\AgdaBound{x}\<%
\\
\\[\AgdaEmptyExtraSkip]%
\>[0]\AgdaFunction{⊕}\AgdaSpace{}%
\AgdaSymbol{:}\AgdaSpace{}%
\AgdaDatatype{Vec}\AgdaSpace{}%
\AgdaSymbol{(}\AgdaDatatype{Session}\AgdaSpace{}%
\AgdaGeneralizable{n}\AgdaSymbol{)}\AgdaSpace{}%
\AgdaGeneralizable{k}\AgdaSpace{}%
\AgdaSymbol{→}\AgdaSpace{}%
\AgdaDatatype{Session}\AgdaSpace{}%
\AgdaGeneralizable{n}\<%
\\
\>[0]\AgdaFunction{⊕}\AgdaSpace{}%
\AgdaSymbol{=}\AgdaSpace{}%
\AgdaInductiveConstructor{⊕′}\AgdaSpace{}%
\AgdaOperator{\AgdaFunction{∘}}\AgdaSpace{}%
\AgdaFunction{vec→fin}\<%
\\
\\[\AgdaEmptyExtraSkip]%
\>[0]\AgdaFunction{\&}\AgdaSpace{}%
\AgdaSymbol{:}\AgdaSpace{}%
\AgdaDatatype{Vec}\AgdaSpace{}%
\AgdaSymbol{(}\AgdaDatatype{Session}\AgdaSpace{}%
\AgdaGeneralizable{n}\AgdaSymbol{)}\AgdaSpace{}%
\AgdaGeneralizable{k}\AgdaSpace{}%
\AgdaSymbol{→}\AgdaSpace{}%
\AgdaDatatype{Session}\AgdaSpace{}%
\AgdaGeneralizable{n}\<%
\\
\>[0]\AgdaFunction{\&}\AgdaSpace{}%
\AgdaSymbol{=}\AgdaSpace{}%
\AgdaInductiveConstructor{\&′}\AgdaSpace{}%
\AgdaOperator{\AgdaFunction{∘}}\AgdaSpace{}%
\AgdaFunction{vec→fin}\<%
\\
\\[\AgdaEmptyExtraSkip]%
\\[\AgdaEmptyExtraSkip]%
\>[0]\AgdaOperator{\AgdaFunction{T⟦\AgdaUnderscore{}⟧}}\AgdaSpace{}%
\AgdaSymbol{:}\AgdaSpace{}%
\AgdaDatatype{Type}\AgdaSpace{}%
\AgdaSymbol{→}\AgdaSpace{}%
\AgdaPrimitive{Set}\<%
\\
\>[0]\AgdaOperator{\AgdaFunction{T⟦}}\AgdaSpace{}%
\AgdaInductiveConstructor{nat}\AgdaSpace{}%
\AgdaOperator{\AgdaFunction{⟧}}\AgdaSpace{}%
\AgdaSymbol{=}\AgdaSpace{}%
\AgdaDatatype{ℕ}\<%
\\
\>[0]\AgdaOperator{\AgdaFunction{T⟦}}\AgdaSpace{}%
\AgdaInductiveConstructor{int}\AgdaSpace{}%
\AgdaOperator{\AgdaFunction{⟧}}\AgdaSpace{}%
\AgdaSymbol{=}\AgdaSpace{}%
\AgdaDatatype{ℤ}\<%
\\
\>[0]\<%
\end{code}
\newcommand\cstCmd{%
\begin{code}%
\>[0]\AgdaKeyword{variable}\AgdaSpace{}%
\AgdaGeneralizable{V}\AgdaSpace{}%
\AgdaGeneralizable{W}\AgdaSpace{}%
\AgdaSymbol{:}\AgdaSpace{}%
\AgdaDatatype{Vec}\AgdaSpace{}%
\AgdaPrimitive{Set}\AgdaSpace{}%
\AgdaGeneralizable{n}\<%
\\
\\[\AgdaEmptyExtraSkip]%
\>[0]\AgdaKeyword{data}\AgdaSpace{}%
\AgdaDatatype{Cmd}%
\>[12]\AgdaSymbol{:}\AgdaSpace{}%
\AgdaPrimitive{Set}\AgdaSpace{}%
\AgdaSymbol{→}\AgdaSpace{}%
\AgdaPrimitive{Set}\AgdaSpace{}%
\AgdaSymbol{→}\AgdaSpace{}%
\AgdaDatatype{Vec}\AgdaSpace{}%
\AgdaPrimitive{Set}\AgdaSpace{}%
\AgdaGeneralizable{n}\AgdaSpace{}%
\AgdaSymbol{→}\AgdaSpace{}%
\AgdaDatatype{Vec}\AgdaSpace{}%
\AgdaPrimitive{Set}\AgdaSpace{}%
\AgdaGeneralizable{n}\AgdaSpace{}%
\AgdaSymbol{→}\AgdaSpace{}%
\AgdaDatatype{Session}\AgdaSpace{}%
\AgdaGeneralizable{n}\AgdaSpace{}%
\AgdaSymbol{→}\AgdaSpace{}%
\AgdaPrimitive{Set₁}\AgdaSpace{}%
\AgdaKeyword{where}\<%
\\
\>[0][@{}l@{\AgdaIndent{0}}]%
\>[2]\AgdaInductiveConstructor{SKIP}%
\>[12]\AgdaSymbol{:}\AgdaSpace{}%
\AgdaSymbol{(}\AgdaGeneralizable{A}\AgdaSpace{}%
\AgdaSymbol{→}\AgdaSpace{}%
\AgdaGeneralizable{B}\AgdaSymbol{)}\AgdaSpace{}%
\AgdaSymbol{→}\AgdaSpace{}%
\AgdaDatatype{Cmd}\AgdaSpace{}%
\AgdaGeneralizable{A}\AgdaSpace{}%
\AgdaGeneralizable{B}\AgdaSpace{}%
\AgdaGeneralizable{V}\AgdaSpace{}%
\AgdaGeneralizable{W}\AgdaSpace{}%
\AgdaInductiveConstructor{skip}\<%
\\
\>[2]\AgdaInductiveConstructor{SEND}%
\>[12]\AgdaSymbol{:}\AgdaSpace{}%
\AgdaSymbol{(}\AgdaGeneralizable{A}\AgdaSpace{}%
\AgdaSymbol{→}\AgdaSpace{}%
\AgdaGeneralizable{B}\AgdaSpace{}%
\AgdaOperator{\AgdaFunction{×}}\AgdaSpace{}%
\AgdaOperator{\AgdaFunction{T⟦}}\AgdaSpace{}%
\AgdaGeneralizable{T}\AgdaSpace{}%
\AgdaOperator{\AgdaFunction{⟧}}\AgdaSymbol{)}\AgdaSpace{}%
\AgdaSymbol{→}\AgdaSpace{}%
\AgdaDatatype{Cmd}\AgdaSpace{}%
\AgdaGeneralizable{A}\AgdaSpace{}%
\AgdaGeneralizable{B}\AgdaSpace{}%
\AgdaGeneralizable{V}\AgdaSpace{}%
\AgdaGeneralizable{W}\AgdaSpace{}%
\AgdaSymbol{(}\AgdaOperator{\AgdaInductiveConstructor{‼}}\AgdaSpace{}%
\AgdaGeneralizable{T}\AgdaSymbol{)}\<%
\\
\>[2]\AgdaInductiveConstructor{RECV}%
\>[12]\AgdaSymbol{:}\AgdaSpace{}%
\AgdaSymbol{(}\AgdaOperator{\AgdaFunction{T⟦}}\AgdaSpace{}%
\AgdaGeneralizable{T}\AgdaSpace{}%
\AgdaOperator{\AgdaFunction{⟧}}\AgdaSpace{}%
\AgdaSymbol{→}\AgdaSpace{}%
\AgdaGeneralizable{A}\AgdaSpace{}%
\AgdaSymbol{→}\AgdaSpace{}%
\AgdaGeneralizable{B}\AgdaSymbol{)}\AgdaSpace{}%
\AgdaSymbol{→}\AgdaSpace{}%
\AgdaDatatype{Cmd}\AgdaSpace{}%
\AgdaGeneralizable{A}\AgdaSpace{}%
\AgdaGeneralizable{B}\AgdaSpace{}%
\AgdaGeneralizable{V}\AgdaSpace{}%
\AgdaGeneralizable{W}\AgdaSpace{}%
\AgdaSymbol{(}\AgdaOperator{\AgdaInductiveConstructor{⁇}}\AgdaSpace{}%
\AgdaGeneralizable{T}\AgdaSymbol{)}\<%
\\
\>[2]\AgdaInductiveConstructor{SELECT}%
\>[12]\AgdaSymbol{:}\AgdaSpace{}%
\AgdaSymbol{∀\{}\AgdaBound{Si}\AgdaSymbol{\}}\AgdaSpace{}%
\AgdaSymbol{\{}\AgdaBound{F}\AgdaSpace{}%
\AgdaSymbol{:}\AgdaSpace{}%
\AgdaDatatype{Fin}\AgdaSpace{}%
\AgdaGeneralizable{k}\AgdaSpace{}%
\AgdaSymbol{→}\AgdaSpace{}%
\AgdaPrimitive{Set}\AgdaSymbol{\}}\AgdaSpace{}%
\AgdaSymbol{→}\AgdaSpace{}%
\AgdaSymbol{(}\AgdaGeneralizable{A}\AgdaSpace{}%
\AgdaSymbol{→}\AgdaSpace{}%
\AgdaRecord{Σ}\AgdaSpace{}%
\AgdaSymbol{(}\AgdaDatatype{Fin}\AgdaSpace{}%
\AgdaGeneralizable{k}\AgdaSymbol{)}\AgdaSpace{}%
\AgdaBound{F}\AgdaSymbol{)}\<%
\\
\>[12]\AgdaSymbol{→}\AgdaSpace{}%
\AgdaSymbol{((}\AgdaBound{i}\AgdaSpace{}%
\AgdaSymbol{:}\AgdaSpace{}%
\AgdaDatatype{Fin}\AgdaSpace{}%
\AgdaGeneralizable{k}\AgdaSymbol{)}\AgdaSpace{}%
\AgdaSymbol{→}\AgdaSpace{}%
\AgdaDatatype{Cmd}\AgdaSpace{}%
\AgdaSymbol{(}\AgdaBound{F}\AgdaSpace{}%
\AgdaBound{i}\AgdaSymbol{)}\AgdaSpace{}%
\AgdaGeneralizable{B}\AgdaSpace{}%
\AgdaGeneralizable{V}\AgdaSpace{}%
\AgdaGeneralizable{W}\AgdaSpace{}%
\AgdaSymbol{(}\AgdaBound{Si}\AgdaSpace{}%
\AgdaBound{i}\AgdaSymbol{))}\<%
\\
\>[12]\AgdaSymbol{→}\AgdaSpace{}%
\AgdaDatatype{Cmd}\AgdaSpace{}%
\AgdaGeneralizable{A}\AgdaSpace{}%
\AgdaGeneralizable{B}\AgdaSpace{}%
\AgdaGeneralizable{V}\AgdaSpace{}%
\AgdaGeneralizable{W}\AgdaSpace{}%
\AgdaSymbol{(}\AgdaInductiveConstructor{⊕′}\AgdaSpace{}%
\AgdaBound{Si}\AgdaSymbol{)}\<%
\\
\>[2]\AgdaInductiveConstructor{CHOICE}%
\>[12]\AgdaSymbol{:}\AgdaSpace{}%
\AgdaSymbol{∀\{}\AgdaBound{Si}\AgdaSymbol{\}}\AgdaSpace{}%
\AgdaSymbol{→}\AgdaSpace{}%
\AgdaSymbol{((}\AgdaBound{i}\AgdaSpace{}%
\AgdaSymbol{:}\AgdaSpace{}%
\AgdaDatatype{Fin}\AgdaSpace{}%
\AgdaGeneralizable{k}\AgdaSymbol{)}\AgdaSpace{}%
\AgdaSymbol{→}\AgdaSpace{}%
\AgdaDatatype{Cmd}\AgdaSpace{}%
\AgdaGeneralizable{A}\AgdaSpace{}%
\AgdaGeneralizable{B}\AgdaSpace{}%
\AgdaGeneralizable{V}\AgdaSpace{}%
\AgdaGeneralizable{W}\AgdaSpace{}%
\AgdaSymbol{(}\AgdaBound{Si}\AgdaSpace{}%
\AgdaBound{i}\AgdaSymbol{))}\AgdaSpace{}%
\AgdaSymbol{→}\AgdaSpace{}%
\AgdaDatatype{Cmd}\AgdaSpace{}%
\AgdaGeneralizable{A}\AgdaSpace{}%
\AgdaGeneralizable{B}\AgdaSpace{}%
\AgdaGeneralizable{V}\AgdaSpace{}%
\AgdaGeneralizable{W}\AgdaSpace{}%
\AgdaSymbol{(}\AgdaInductiveConstructor{\&′}\AgdaSpace{}%
\AgdaBound{Si}\AgdaSymbol{)}\<%
\\
\>[2]\AgdaOperator{\AgdaInductiveConstructor{[\AgdaUnderscore{}]\AgdaUnderscore{}⨟[\AgdaUnderscore{}]\AgdaUnderscore{}[\AgdaUnderscore{}]}}%
\>[476I]\AgdaSymbol{:}\AgdaSpace{}%
\AgdaSymbol{(}\AgdaGeneralizable{A}\AgdaSpace{}%
\AgdaSymbol{→}\AgdaSpace{}%
\AgdaGeneralizable{A₁}\AgdaSpace{}%
\AgdaOperator{\AgdaFunction{×}}\AgdaSpace{}%
\AgdaGeneralizable{A′}\AgdaSymbol{)}\AgdaSpace{}%
\AgdaSymbol{→}\AgdaSpace{}%
\AgdaDatatype{Cmd}\AgdaSpace{}%
\AgdaGeneralizable{A₁}\AgdaSpace{}%
\AgdaGeneralizable{B₁}\AgdaSpace{}%
\AgdaGeneralizable{V}\AgdaSpace{}%
\AgdaGeneralizable{W}\AgdaSpace{}%
\AgdaGeneralizable{S₁}\<%
\\
\>[.][@{}l@{}]\<[476I]%
\>[15]\AgdaSymbol{→}\AgdaSpace{}%
\AgdaSymbol{(}\AgdaGeneralizable{A′}\AgdaSpace{}%
\AgdaSymbol{→}\AgdaSpace{}%
\AgdaGeneralizable{B₁}\AgdaSpace{}%
\AgdaSymbol{→}\AgdaSpace{}%
\AgdaGeneralizable{A₂}\AgdaSymbol{)}\AgdaSpace{}%
\AgdaSymbol{→}\AgdaSpace{}%
\AgdaDatatype{Cmd}\AgdaSpace{}%
\AgdaGeneralizable{A₂}\AgdaSpace{}%
\AgdaGeneralizable{B₂}\AgdaSpace{}%
\AgdaGeneralizable{V}\AgdaSpace{}%
\AgdaGeneralizable{W}\AgdaSpace{}%
\AgdaGeneralizable{S₂}\<%
\\
\>[15]\AgdaSymbol{→}\AgdaSpace{}%
\AgdaSymbol{(}\AgdaGeneralizable{B₁}\AgdaSpace{}%
\AgdaSymbol{→}\AgdaSpace{}%
\AgdaGeneralizable{B₂}\AgdaSpace{}%
\AgdaSymbol{→}\AgdaSpace{}%
\AgdaGeneralizable{B}\AgdaSymbol{)}\AgdaSpace{}%
\AgdaSymbol{→}\AgdaSpace{}%
\AgdaDatatype{Cmd}\AgdaSpace{}%
\AgdaGeneralizable{A}\AgdaSpace{}%
\AgdaGeneralizable{B}\AgdaSpace{}%
\AgdaGeneralizable{V}\AgdaSpace{}%
\AgdaGeneralizable{W}\AgdaSpace{}%
\AgdaSymbol{(}\AgdaGeneralizable{S₁}\AgdaSpace{}%
\AgdaOperator{\AgdaInductiveConstructor{⨟}}\AgdaSpace{}%
\AgdaGeneralizable{S₂}\AgdaSymbol{)}\<%
\\
\>[2]\AgdaInductiveConstructor{LOOP}%
\>[12]\AgdaSymbol{:}\AgdaSpace{}%
\AgdaDatatype{Cmd}\AgdaSpace{}%
\AgdaGeneralizable{A}\AgdaSpace{}%
\AgdaGeneralizable{B}\AgdaSpace{}%
\AgdaSymbol{(}\AgdaGeneralizable{A}\AgdaSpace{}%
\AgdaOperator{\AgdaInductiveConstructor{∷}}\AgdaSpace{}%
\AgdaGeneralizable{V}\AgdaSymbol{)}\AgdaSpace{}%
\AgdaSymbol{(}\AgdaGeneralizable{B}\AgdaSpace{}%
\AgdaOperator{\AgdaInductiveConstructor{∷}}\AgdaSpace{}%
\AgdaGeneralizable{W}\AgdaSymbol{)}\AgdaSpace{}%
\AgdaGeneralizable{S}\AgdaSpace{}%
\AgdaSymbol{→}\AgdaSpace{}%
\AgdaDatatype{Cmd}\AgdaSpace{}%
\AgdaGeneralizable{A}\AgdaSpace{}%
\AgdaGeneralizable{B}\AgdaSpace{}%
\AgdaGeneralizable{V}\AgdaSpace{}%
\AgdaGeneralizable{W}\AgdaSpace{}%
\AgdaSymbol{(}\AgdaOperator{\AgdaInductiveConstructor{μ}}\AgdaSpace{}%
\AgdaGeneralizable{S}\AgdaSymbol{)}\<%
\\
\>[2]\AgdaInductiveConstructor{CONTINUE}%
\>[12]\AgdaSymbol{:}\AgdaSpace{}%
\AgdaSymbol{(}\AgdaBound{i}\AgdaSpace{}%
\AgdaSymbol{:}\AgdaSpace{}%
\AgdaDatatype{Fin}\AgdaSpace{}%
\AgdaGeneralizable{n}\AgdaSymbol{)}\AgdaSpace{}%
\AgdaSymbol{→}\AgdaSpace{}%
\AgdaDatatype{Cmd}\AgdaSpace{}%
\AgdaSymbol{(}\AgdaFunction{lookup}\AgdaSpace{}%
\AgdaGeneralizable{V}\AgdaSpace{}%
\AgdaBound{i}\AgdaSymbol{)}\AgdaSpace{}%
\AgdaSymbol{(}\AgdaFunction{lookup}\AgdaSpace{}%
\AgdaGeneralizable{W}\AgdaSpace{}%
\AgdaBound{i}\AgdaSymbol{)}\AgdaSpace{}%
\AgdaGeneralizable{V}\AgdaSpace{}%
\AgdaGeneralizable{W}\AgdaSpace{}%
\AgdaSymbol{(}\AgdaOperator{\AgdaInductiveConstructor{`}}\AgdaSpace{}%
\AgdaBound{i}\AgdaSymbol{)}\<%
\end{code}}
\newcommand\cstTailComposition{%
\begin{code}%
\>[0]\AgdaOperator{\AgdaFunction{\AgdaUnderscore{}⨟′\AgdaUnderscore{}}}\AgdaSpace{}%
\AgdaSymbol{:}\AgdaSpace{}%
\AgdaDatatype{Cmd}\AgdaSpace{}%
\AgdaGeneralizable{A}\AgdaSpace{}%
\AgdaGeneralizable{B}\AgdaSpace{}%
\AgdaGeneralizable{V}\AgdaSpace{}%
\AgdaGeneralizable{W}\AgdaSpace{}%
\AgdaGeneralizable{S₁}\AgdaSpace{}%
\AgdaSymbol{→}\AgdaSpace{}%
\AgdaDatatype{Cmd}\AgdaSpace{}%
\AgdaGeneralizable{B}\AgdaSpace{}%
\AgdaGeneralizable{C}\AgdaSpace{}%
\AgdaGeneralizable{V}\AgdaSpace{}%
\AgdaGeneralizable{W}\AgdaSpace{}%
\AgdaGeneralizable{S₂}\AgdaSpace{}%
\AgdaSymbol{→}\AgdaSpace{}%
\AgdaDatatype{Cmd}\AgdaSpace{}%
\AgdaGeneralizable{A}\AgdaSpace{}%
\AgdaGeneralizable{C}\AgdaSpace{}%
\AgdaGeneralizable{V}\AgdaSpace{}%
\AgdaGeneralizable{W}\AgdaSpace{}%
\AgdaSymbol{(}\AgdaGeneralizable{S₁}\AgdaSpace{}%
\AgdaOperator{\AgdaInductiveConstructor{⨟}}\AgdaSpace{}%
\AgdaGeneralizable{S₂}\AgdaSymbol{)}\<%
\\
\>[0]\AgdaBound{cmd₁}\AgdaSpace{}%
\AgdaOperator{\AgdaFunction{⨟′}}\AgdaSpace{}%
\AgdaBound{cmd₂}\AgdaSpace{}%
\AgdaSymbol{=}\AgdaSpace{}%
\AgdaOperator{\AgdaInductiveConstructor{[}}\AgdaSpace{}%
\AgdaSymbol{(λ}\AgdaSpace{}%
\AgdaBound{x}\AgdaSpace{}%
\AgdaSymbol{→}\AgdaSpace{}%
\AgdaOperator{\AgdaInductiveConstructor{⟨}}\AgdaSpace{}%
\AgdaBound{x}\AgdaSpace{}%
\AgdaOperator{\AgdaInductiveConstructor{,}}\AgdaSpace{}%
\AgdaFunction{tt}\AgdaSpace{}%
\AgdaOperator{\AgdaInductiveConstructor{⟩}}\AgdaSymbol{)}\AgdaSpace{}%
\AgdaOperator{\AgdaInductiveConstructor{]}}\AgdaSpace{}%
\AgdaBound{cmd₁}\AgdaSpace{}%
\AgdaOperator{\AgdaInductiveConstructor{⨟[}}\AgdaSpace{}%
\AgdaFunction{constᵣ}\AgdaSpace{}%
\AgdaOperator{\AgdaInductiveConstructor{]}}\AgdaSpace{}%
\AgdaBound{cmd₂}\AgdaSpace{}%
\AgdaOperator{\AgdaInductiveConstructor{[}}\AgdaSpace{}%
\AgdaFunction{constᵣ}\AgdaSpace{}%
\AgdaOperator{\AgdaInductiveConstructor{]}}\<%
\end{code}}
\begin{code}[hide]%
\>[0]\AgdaFunction{shrink}\AgdaSpace{}%
\AgdaSymbol{:}\AgdaSpace{}%
\AgdaSymbol{∀}\AgdaSpace{}%
\AgdaSymbol{\{}\AgdaBound{n}\AgdaSymbol{\}}\AgdaSpace{}%
\AgdaSymbol{→}\AgdaSpace{}%
\AgdaDatatype{Vec}\AgdaSpace{}%
\AgdaPrimitive{Set}\AgdaSpace{}%
\AgdaBound{n}\AgdaSpace{}%
\AgdaSymbol{→}\AgdaSpace{}%
\AgdaSymbol{(}\AgdaBound{i}\AgdaSpace{}%
\AgdaSymbol{:}\AgdaSpace{}%
\AgdaDatatype{Fin}\AgdaSpace{}%
\AgdaBound{n}\AgdaSymbol{)}\AgdaSpace{}%
\AgdaSymbol{→}\AgdaSpace{}%
\AgdaDatatype{Vec}\AgdaSpace{}%
\AgdaPrimitive{Set}\AgdaSpace{}%
\AgdaSymbol{(}\AgdaInductiveConstructor{suc}\AgdaSpace{}%
\AgdaSymbol{(}\AgdaFunction{toℕ}\AgdaSpace{}%
\AgdaSymbol{(}\AgdaFunction{opposite}\AgdaSpace{}%
\AgdaBound{i}\AgdaSymbol{)))}\<%
\\
\>[0]\AgdaFunction{shrink}\AgdaSpace{}%
\AgdaSymbol{\{}\AgdaBound{n}\AgdaSymbol{\}}\AgdaSpace{}%
\AgdaBound{V}\AgdaSpace{}%
\AgdaInductiveConstructor{zero}\AgdaSpace{}%
\AgdaKeyword{rewrite}\AgdaSpace{}%
\AgdaFunction{toℕ-fromℕ}\AgdaSpace{}%
\AgdaBound{n}\AgdaSpace{}%
\AgdaSymbol{=}\AgdaSpace{}%
\AgdaBound{V}\<%
\\
\>[0]\AgdaFunction{shrink}\AgdaSpace{}%
\AgdaSymbol{\{}\AgdaInductiveConstructor{suc}\AgdaSpace{}%
\AgdaBound{n}\AgdaSymbol{\}}\AgdaSpace{}%
\AgdaSymbol{(}\AgdaBound{x}\AgdaSpace{}%
\AgdaOperator{\AgdaInductiveConstructor{∷}}\AgdaSpace{}%
\AgdaBound{V}\AgdaSymbol{)}\AgdaSpace{}%
\AgdaSymbol{(}\AgdaInductiveConstructor{suc}\AgdaSpace{}%
\AgdaBound{i}\AgdaSymbol{)}%
\>[32]\AgdaKeyword{rewrite}\AgdaSpace{}%
\AgdaFunction{toℕ-inject₁}\AgdaSpace{}%
\AgdaSymbol{(}\AgdaFunction{opposite}\AgdaSpace{}%
\AgdaBound{i}\AgdaSymbol{)}\AgdaSpace{}%
\AgdaSymbol{=}\AgdaSpace{}%
\AgdaFunction{shrink}\AgdaSpace{}%
\AgdaBound{V}\AgdaSpace{}%
\AgdaBound{i}\<%
\\
\\[\AgdaEmptyExtraSkip]%
\>[0]\AgdaFunction{CmdStore}\AgdaSpace{}%
\AgdaSymbol{:}\AgdaSpace{}%
\AgdaSymbol{∀}\AgdaSpace{}%
\AgdaBound{n}\AgdaSpace{}%
\AgdaSymbol{→}\AgdaSpace{}%
\AgdaDatatype{Vec}\AgdaSpace{}%
\AgdaPrimitive{Set}\AgdaSpace{}%
\AgdaBound{n}\AgdaSpace{}%
\AgdaSymbol{→}\AgdaSpace{}%
\AgdaDatatype{Vec}\AgdaSpace{}%
\AgdaPrimitive{Set}\AgdaSpace{}%
\AgdaBound{n}\AgdaSpace{}%
\AgdaSymbol{→}\AgdaSpace{}%
\AgdaPrimitive{Set₁}\<%
\\
\>[0]\AgdaFunction{CmdStore}\AgdaSpace{}%
\AgdaBound{n}\AgdaSpace{}%
\AgdaBound{V}\AgdaSpace{}%
\AgdaBound{W}\AgdaSpace{}%
\AgdaSymbol{=}\AgdaSpace{}%
\AgdaSymbol{(}\AgdaBound{i}\AgdaSpace{}%
\AgdaSymbol{:}\AgdaSpace{}%
\AgdaDatatype{Fin}\AgdaSpace{}%
\AgdaBound{n}\AgdaSymbol{)}\AgdaSpace{}%
\AgdaSymbol{→}\AgdaSpace{}%
\AgdaFunction{∃[}\AgdaSpace{}%
\AgdaBound{S}\AgdaSpace{}%
\AgdaFunction{]}\AgdaSpace{}%
\AgdaSymbol{(}\AgdaDatatype{Cmd}\AgdaSpace{}%
\AgdaSymbol{\{}\AgdaInductiveConstructor{suc}\AgdaSpace{}%
\AgdaSymbol{(}\AgdaFunction{toℕ}\AgdaSpace{}%
\AgdaSymbol{(}\AgdaFunction{opposite}\AgdaSpace{}%
\AgdaBound{i}\AgdaSymbol{))\}}\AgdaSpace{}%
\AgdaSymbol{(}\AgdaFunction{lookup}\AgdaSpace{}%
\AgdaBound{V}\AgdaSpace{}%
\AgdaBound{i}\AgdaSymbol{)}\AgdaSpace{}%
\AgdaSymbol{(}\AgdaFunction{lookup}\AgdaSpace{}%
\AgdaBound{W}\AgdaSpace{}%
\AgdaBound{i}\AgdaSymbol{)}\AgdaSpace{}%
\AgdaSymbol{(}\AgdaFunction{shrink}\AgdaSpace{}%
\AgdaBound{V}\AgdaSpace{}%
\AgdaBound{i}\AgdaSymbol{)}\AgdaSpace{}%
\AgdaSymbol{(}\AgdaFunction{shrink}\AgdaSpace{}%
\AgdaBound{W}\AgdaSpace{}%
\AgdaBound{i}\AgdaSymbol{)}\AgdaSpace{}%
\AgdaBound{S}\AgdaSymbol{)}\<%
\\
\\[\AgdaEmptyExtraSkip]%
\>[0]\AgdaFunction{push}\AgdaSpace{}%
\AgdaSymbol{:}\AgdaSpace{}%
\AgdaFunction{CmdStore}\AgdaSpace{}%
\AgdaGeneralizable{n}\AgdaSpace{}%
\AgdaGeneralizable{V}\AgdaSpace{}%
\AgdaGeneralizable{W}\AgdaSpace{}%
\AgdaSymbol{→}\AgdaSpace{}%
\AgdaDatatype{Cmd}\AgdaSpace{}%
\AgdaGeneralizable{A}\AgdaSpace{}%
\AgdaGeneralizable{B}\AgdaSpace{}%
\AgdaSymbol{(}\AgdaGeneralizable{A}\AgdaSpace{}%
\AgdaOperator{\AgdaInductiveConstructor{∷}}\AgdaSpace{}%
\AgdaGeneralizable{V}\AgdaSymbol{)}\AgdaSpace{}%
\AgdaSymbol{(}\AgdaGeneralizable{B}\AgdaSpace{}%
\AgdaOperator{\AgdaInductiveConstructor{∷}}\AgdaSpace{}%
\AgdaGeneralizable{W}\AgdaSymbol{)}\AgdaSpace{}%
\AgdaGeneralizable{S}\AgdaSpace{}%
\AgdaSymbol{→}\AgdaSpace{}%
\AgdaFunction{CmdStore}\AgdaSpace{}%
\AgdaSymbol{(}\AgdaInductiveConstructor{suc}\AgdaSpace{}%
\AgdaGeneralizable{n}\AgdaSymbol{)}\AgdaSpace{}%
\AgdaSymbol{(}\AgdaGeneralizable{A}\AgdaSpace{}%
\AgdaOperator{\AgdaInductiveConstructor{∷}}\AgdaSpace{}%
\AgdaGeneralizable{V}\AgdaSymbol{)}\AgdaSpace{}%
\AgdaSymbol{(}\AgdaGeneralizable{B}\AgdaSpace{}%
\AgdaOperator{\AgdaInductiveConstructor{∷}}\AgdaSpace{}%
\AgdaGeneralizable{W}\AgdaSymbol{)}\<%
\\
\>[0]\AgdaFunction{push}\AgdaSpace{}%
\AgdaSymbol{\{}\AgdaBound{n}\AgdaSymbol{\}\{}\AgdaArgument{S}\AgdaSpace{}%
\AgdaSymbol{=}\AgdaSpace{}%
\AgdaBound{S}\AgdaSymbol{\}}\AgdaSpace{}%
\AgdaBound{cms}\AgdaSpace{}%
\AgdaBound{cmd}\AgdaSpace{}%
\AgdaInductiveConstructor{zero}\AgdaSpace{}%
\AgdaKeyword{rewrite}\AgdaSpace{}%
\AgdaFunction{toℕ-fromℕ}\AgdaSpace{}%
\AgdaBound{n}\AgdaSpace{}%
\AgdaSymbol{=}\AgdaSpace{}%
\AgdaOperator{\AgdaInductiveConstructor{⟨}}\AgdaSpace{}%
\AgdaBound{S}\AgdaSpace{}%
\AgdaOperator{\AgdaInductiveConstructor{,}}\AgdaSpace{}%
\AgdaBound{cmd}\AgdaSpace{}%
\AgdaOperator{\AgdaInductiveConstructor{⟩}}\<%
\\
\>[0]\AgdaFunction{push}\AgdaSpace{}%
\AgdaBound{cms}\AgdaSpace{}%
\AgdaBound{cmd}\AgdaSpace{}%
\AgdaSymbol{(}\AgdaInductiveConstructor{suc}\AgdaSpace{}%
\AgdaBound{i}\AgdaSymbol{)}\AgdaSpace{}%
\AgdaKeyword{rewrite}\AgdaSpace{}%
\AgdaFunction{toℕ-inject₁}\AgdaSpace{}%
\AgdaSymbol{(}\AgdaFunction{opposite}\AgdaSpace{}%
\AgdaBound{i}\AgdaSymbol{)}\AgdaSpace{}%
\AgdaSymbol{=}\AgdaSpace{}%
\AgdaBound{cms}\AgdaSpace{}%
\AgdaBound{i}\<%
\\
\\[\AgdaEmptyExtraSkip]%
\>[0]\AgdaFunction{pop1}\AgdaSpace{}%
\AgdaSymbol{:}\AgdaSpace{}%
\AgdaSymbol{∀}\AgdaSpace{}%
\AgdaSymbol{\{}\AgdaBound{V}\AgdaSpace{}%
\AgdaBound{W}\AgdaSymbol{\}}\AgdaSpace{}%
\AgdaSymbol{→}\AgdaSpace{}%
\AgdaFunction{CmdStore}\AgdaSpace{}%
\AgdaSymbol{(}\AgdaInductiveConstructor{suc}\AgdaSpace{}%
\AgdaGeneralizable{n}\AgdaSymbol{)}\AgdaSpace{}%
\AgdaSymbol{(}\AgdaGeneralizable{A}\AgdaSpace{}%
\AgdaOperator{\AgdaInductiveConstructor{∷}}\AgdaSpace{}%
\AgdaBound{V}\AgdaSymbol{)}\AgdaSpace{}%
\AgdaSymbol{(}\AgdaGeneralizable{B}\AgdaSpace{}%
\AgdaOperator{\AgdaInductiveConstructor{∷}}\AgdaSpace{}%
\AgdaBound{W}\AgdaSymbol{)}\AgdaSpace{}%
\AgdaSymbol{→}\AgdaSpace{}%
\AgdaFunction{CmdStore}\AgdaSpace{}%
\AgdaGeneralizable{n}\AgdaSpace{}%
\AgdaBound{V}\AgdaSpace{}%
\AgdaBound{W}\<%
\\
\>[0]\AgdaFunction{pop1}\AgdaSpace{}%
\AgdaBound{cms}\AgdaSpace{}%
\AgdaBound{i}\AgdaSpace{}%
\AgdaKeyword{with}\AgdaSpace{}%
\AgdaBound{cms}\AgdaSpace{}%
\AgdaSymbol{(}\AgdaInductiveConstructor{suc}\AgdaSpace{}%
\AgdaBound{i}\AgdaSymbol{)}\<%
\\
\>[0]\AgdaSymbol{...}\AgdaSpace{}%
\AgdaSymbol{|}\AgdaSpace{}%
\AgdaBound{cms₁}\AgdaSpace{}%
\AgdaKeyword{rewrite}\AgdaSpace{}%
\AgdaFunction{toℕ-inject₁}\AgdaSpace{}%
\AgdaSymbol{(}\AgdaFunction{opposite}\AgdaSpace{}%
\AgdaBound{i}\AgdaSymbol{)}\AgdaSpace{}%
\AgdaSymbol{=}\AgdaSpace{}%
\AgdaBound{cms₁}\<%
\\
\\[\AgdaEmptyExtraSkip]%
\>[0]\AgdaFunction{pop}\AgdaSpace{}%
\AgdaSymbol{:}\AgdaSpace{}%
\AgdaSymbol{∀}\AgdaSpace{}%
\AgdaSymbol{\{}\AgdaBound{V}\AgdaSpace{}%
\AgdaBound{W}\AgdaSymbol{\}}\AgdaSpace{}%
\AgdaSymbol{→}\AgdaSpace{}%
\AgdaFunction{CmdStore}\AgdaSpace{}%
\AgdaSymbol{(}\AgdaInductiveConstructor{suc}\AgdaSpace{}%
\AgdaGeneralizable{n}\AgdaSymbol{)}\AgdaSpace{}%
\AgdaBound{V}\AgdaSpace{}%
\AgdaBound{W}\AgdaSpace{}%
\AgdaSymbol{→}\AgdaSpace{}%
\AgdaSymbol{(}\AgdaBound{i}\AgdaSpace{}%
\AgdaSymbol{:}\AgdaSpace{}%
\AgdaDatatype{Fin}\AgdaSpace{}%
\AgdaSymbol{(}\AgdaInductiveConstructor{suc}\AgdaSpace{}%
\AgdaGeneralizable{n}\AgdaSymbol{))}\AgdaSpace{}%
\AgdaSymbol{→}\AgdaSpace{}%
\AgdaFunction{CmdStore}\AgdaSpace{}%
\AgdaSymbol{(}\AgdaInductiveConstructor{suc}\AgdaSpace{}%
\AgdaSymbol{(}\AgdaFunction{toℕ}\AgdaSpace{}%
\AgdaSymbol{(}\AgdaFunction{opposite}\AgdaSpace{}%
\AgdaBound{i}\AgdaSymbol{)))}\AgdaSpace{}%
\AgdaSymbol{(}\AgdaFunction{shrink}\AgdaSpace{}%
\AgdaBound{V}\AgdaSpace{}%
\AgdaBound{i}\AgdaSymbol{)}\AgdaSpace{}%
\AgdaSymbol{(}\AgdaFunction{shrink}\AgdaSpace{}%
\AgdaBound{W}\AgdaSpace{}%
\AgdaBound{i}\AgdaSymbol{)}\<%
\\
\>[0]\AgdaFunction{pop}\AgdaSpace{}%
\AgdaSymbol{\{}\AgdaBound{n}\AgdaSymbol{\}}\AgdaSpace{}%
\AgdaSymbol{\{}\AgdaBound{V}\AgdaSymbol{\}}\AgdaSpace{}%
\AgdaSymbol{\{}\AgdaBound{W}\AgdaSymbol{\}}\AgdaSpace{}%
\AgdaBound{cms}\AgdaSpace{}%
\AgdaInductiveConstructor{zero}\AgdaSpace{}%
\AgdaKeyword{rewrite}\AgdaSpace{}%
\AgdaFunction{toℕ-fromℕ}\AgdaSpace{}%
\AgdaBound{n}\AgdaSpace{}%
\AgdaSymbol{=}\AgdaSpace{}%
\AgdaBound{cms}\<%
\\
\>[0]\AgdaFunction{pop}\AgdaSpace{}%
\AgdaSymbol{\{}\AgdaInductiveConstructor{suc}\AgdaSpace{}%
\AgdaBound{n}\AgdaSymbol{\}}\AgdaSpace{}%
\AgdaSymbol{\{}\AgdaBound{A}\AgdaSpace{}%
\AgdaOperator{\AgdaInductiveConstructor{∷}}\AgdaSpace{}%
\AgdaBound{V}\AgdaSymbol{\}}\AgdaSpace{}%
\AgdaSymbol{\{}\AgdaBound{B}\AgdaSpace{}%
\AgdaOperator{\AgdaInductiveConstructor{∷}}\AgdaSpace{}%
\AgdaBound{W}\AgdaSymbol{\}}\AgdaSpace{}%
\AgdaBound{cms}\AgdaSpace{}%
\AgdaSymbol{(}\AgdaInductiveConstructor{suc}\AgdaSpace{}%
\AgdaBound{i}\AgdaSymbol{)}\AgdaSpace{}%
\AgdaKeyword{rewrite}\AgdaSpace{}%
\AgdaFunction{toℕ-inject₁}\AgdaSpace{}%
\AgdaSymbol{(}\AgdaFunction{opposite}\AgdaSpace{}%
\AgdaBound{i}\AgdaSymbol{)}\AgdaSpace{}%
\AgdaSymbol{=}\AgdaSpace{}%
\AgdaFunction{pop}\AgdaSpace{}%
\AgdaSymbol{\{}\AgdaBound{n}\AgdaSymbol{\}}\AgdaSpace{}%
\AgdaSymbol{\{}\AgdaBound{V}\AgdaSymbol{\}}\AgdaSpace{}%
\AgdaSymbol{\{}\AgdaBound{W}\AgdaSymbol{\}}\AgdaSpace{}%
\AgdaSymbol{(}\AgdaFunction{pop1}\AgdaSpace{}%
\AgdaBound{cms}\AgdaSymbol{)}\AgdaSpace{}%
\AgdaBound{i}\<%
\\
\\[\AgdaEmptyExtraSkip]%
\>[0]\AgdaSymbol{\{-\#}\AgdaSpace{}%
\AgdaKeyword{TERMINATING}\AgdaSpace{}%
\AgdaSymbol{\#-\}}\AgdaSpace{}%
\AgdaComment{--\ a\ lie}\<%
\end{code}
\newcommand\cstExec{%
\begin{code}%
\>[0]\AgdaFunction{exec}\AgdaSpace{}%
\AgdaSymbol{:}\AgdaSpace{}%
\AgdaDatatype{Cmd}\AgdaSymbol{\{}\AgdaGeneralizable{n}\AgdaSymbol{\}}\AgdaSpace{}%
\AgdaGeneralizable{A}\AgdaSpace{}%
\AgdaGeneralizable{B}\AgdaSpace{}%
\AgdaGeneralizable{V}\AgdaSpace{}%
\AgdaGeneralizable{W}\AgdaSpace{}%
\AgdaGeneralizable{S}\AgdaSpace{}%
\AgdaSymbol{→}\AgdaSpace{}%
\AgdaFunction{CmdStore}\AgdaSpace{}%
\AgdaGeneralizable{n}\AgdaSpace{}%
\AgdaGeneralizable{V}\AgdaSpace{}%
\AgdaGeneralizable{W}\AgdaSpace{}%
\AgdaSymbol{→}\AgdaSpace{}%
\AgdaGeneralizable{A}\AgdaSpace{}%
\AgdaSymbol{→}\AgdaSpace{}%
\AgdaSymbol{(}\AgdaRecord{ReaderT}\AgdaSpace{}%
\AgdaPostulate{Channel}\AgdaSpace{}%
\AgdaDatatype{IO}\AgdaSymbol{)}\AgdaSpace{}%
\AgdaGeneralizable{B}\<%
\\
\>[0]\AgdaFunction{exec}\AgdaSpace{}%
\AgdaSymbol{(}\AgdaInductiveConstructor{SKIP}\AgdaSpace{}%
\AgdaBound{act}\AgdaSymbol{)}\AgdaSpace{}%
\AgdaBound{cms}\AgdaSpace{}%
\AgdaBound{a}\AgdaSpace{}%
\AgdaSymbol{=}\AgdaSpace{}%
\AgdaFunction{pure}\AgdaSpace{}%
\AgdaSymbol{(}\AgdaBound{act}\AgdaSpace{}%
\AgdaBound{a}\AgdaSymbol{)}\<%
\\
\>[0]\AgdaFunction{exec}\AgdaSpace{}%
\AgdaSymbol{(}\AgdaInductiveConstructor{SEND}\AgdaSpace{}%
\AgdaBound{getx}\AgdaSymbol{)}\AgdaSpace{}%
\AgdaBound{cms}\AgdaSpace{}%
\AgdaBound{a}\AgdaSpace{}%
\AgdaSymbol{=}\AgdaSpace{}%
\AgdaKeyword{do}\<%
\\
\>[0][@{}l@{\AgdaIndent{0}}]%
\>[2]\AgdaKeyword{let}\AgdaSpace{}%
\AgdaOperator{\AgdaInductiveConstructor{⟨}}\AgdaSpace{}%
\AgdaBound{b}\AgdaSpace{}%
\AgdaOperator{\AgdaInductiveConstructor{,}}\AgdaSpace{}%
\AgdaBound{x}\AgdaSpace{}%
\AgdaOperator{\AgdaInductiveConstructor{⟩}}\AgdaSpace{}%
\AgdaSymbol{=}\AgdaSpace{}%
\AgdaBound{getx}\AgdaSpace{}%
\AgdaBound{a}\<%
\\
\>[2]\AgdaFunction{ask}\AgdaSpace{}%
\AgdaOperator{\AgdaField{>>=}}\AgdaSpace{}%
\AgdaField{liftIO}\AgdaSpace{}%
\AgdaOperator{\AgdaFunction{∘}}\AgdaSpace{}%
\AgdaPostulate{primSend}\AgdaSpace{}%
\AgdaBound{x}\<%
\\
\>[2]\AgdaFunction{pure}\AgdaSpace{}%
\AgdaBound{b}\<%
\\
\>[0]\AgdaFunction{exec}\AgdaSpace{}%
\AgdaSymbol{(}\AgdaInductiveConstructor{RECV}\AgdaSpace{}%
\AgdaBound{putx}\AgdaSymbol{)}\AgdaSpace{}%
\AgdaBound{cms}\AgdaSpace{}%
\AgdaBound{a}\AgdaSpace{}%
\AgdaSymbol{=}\AgdaSpace{}%
\AgdaKeyword{do}\<%
\\
\>[0][@{}l@{\AgdaIndent{0}}]%
\>[2]\AgdaBound{x}\AgdaSpace{}%
\AgdaOperator{\AgdaField{←}}\AgdaSpace{}%
\AgdaFunction{ask}\AgdaSpace{}%
\AgdaOperator{\AgdaField{>>=}}\AgdaSpace{}%
\AgdaField{liftIO}\AgdaSpace{}%
\AgdaOperator{\AgdaFunction{∘}}\AgdaSpace{}%
\AgdaPostulate{primRecv}\<%
\\
\>[2]\AgdaFunction{pure}\AgdaSpace{}%
\AgdaSymbol{(}\AgdaBound{putx}\AgdaSpace{}%
\AgdaBound{x}\AgdaSpace{}%
\AgdaBound{a}\AgdaSymbol{)}\<%
\\
\>[0]\AgdaFunction{exec}\AgdaSpace{}%
\AgdaSymbol{(}\AgdaInductiveConstructor{SELECT}\AgdaSpace{}%
\AgdaBound{getx}\AgdaSpace{}%
\AgdaBound{cont}\AgdaSymbol{)}\AgdaSpace{}%
\AgdaBound{cms}\AgdaSpace{}%
\AgdaBound{a}\<%
\\
\>[0][@{}l@{\AgdaIndent{0}}]%
\>[2]\AgdaKeyword{with}\AgdaSpace{}%
\AgdaBound{getx}\AgdaSpace{}%
\AgdaBound{a}\<%
\\
\>[0]\AgdaSymbol{...}\AgdaSpace{}%
\AgdaSymbol{|}\AgdaSpace{}%
\AgdaOperator{\AgdaInductiveConstructor{⟨}}\AgdaSpace{}%
\AgdaBound{i}\AgdaSpace{}%
\AgdaOperator{\AgdaInductiveConstructor{,}}\AgdaSpace{}%
\AgdaBound{ai}\AgdaSpace{}%
\AgdaOperator{\AgdaInductiveConstructor{⟩}}\AgdaSpace{}%
\AgdaSymbol{=}\AgdaSpace{}%
\AgdaKeyword{do}\<%
\\
\>[0][@{}l@{\AgdaIndent{0}}]%
\>[2]\AgdaFunction{ask}\AgdaSpace{}%
\AgdaOperator{\AgdaField{>>=}}\AgdaSpace{}%
\AgdaField{liftIO}\AgdaSpace{}%
\AgdaOperator{\AgdaFunction{∘}}\AgdaSpace{}%
\AgdaPostulate{primSend}\AgdaSpace{}%
\AgdaBound{i}\<%
\\
\>[2]\AgdaFunction{exec}\AgdaSpace{}%
\AgdaSymbol{(}\AgdaBound{cont}\AgdaSpace{}%
\AgdaBound{i}\AgdaSymbol{)}\AgdaSpace{}%
\AgdaBound{cms}\AgdaSpace{}%
\AgdaBound{ai}\<%
\\
\>[0]\AgdaFunction{exec}\AgdaSpace{}%
\AgdaSymbol{(}\AgdaInductiveConstructor{CHOICE}\AgdaSpace{}%
\AgdaBound{cont}\AgdaSymbol{)}\AgdaSpace{}%
\AgdaBound{cms}\AgdaSpace{}%
\AgdaBound{a}\AgdaSpace{}%
\AgdaSymbol{=}\AgdaSpace{}%
\AgdaKeyword{do}\<%
\\
\>[0][@{}l@{\AgdaIndent{0}}]%
\>[2]\AgdaBound{i}\AgdaSpace{}%
\AgdaOperator{\AgdaField{←}}\AgdaSpace{}%
\AgdaFunction{ask}\AgdaSpace{}%
\AgdaOperator{\AgdaField{>>=}}\AgdaSpace{}%
\AgdaField{liftIO}\AgdaSpace{}%
\AgdaOperator{\AgdaFunction{∘}}\AgdaSpace{}%
\AgdaPostulate{primRecv}\<%
\\
\>[2]\AgdaFunction{exec}\AgdaSpace{}%
\AgdaSymbol{(}\AgdaBound{cont}\AgdaSpace{}%
\AgdaBound{i}\AgdaSymbol{)}\AgdaSpace{}%
\AgdaBound{cms}\AgdaSpace{}%
\AgdaBound{a}\<%
\\
\>[0]\AgdaFunction{exec}\AgdaSpace{}%
\AgdaOperator{\AgdaInductiveConstructor{[}}\AgdaSpace{}%
\AgdaBound{split}\AgdaSpace{}%
\AgdaOperator{\AgdaInductiveConstructor{]}}\AgdaSpace{}%
\AgdaBound{cmd₁}\AgdaSpace{}%
\AgdaOperator{\AgdaInductiveConstructor{⨟[}}\AgdaSpace{}%
\AgdaBound{cross}\AgdaSpace{}%
\AgdaOperator{\AgdaInductiveConstructor{]}}\AgdaSpace{}%
\AgdaBound{cmd₂}\AgdaSpace{}%
\AgdaOperator{\AgdaInductiveConstructor{[}}\AgdaSpace{}%
\AgdaBound{join}\AgdaSpace{}%
\AgdaOperator{\AgdaInductiveConstructor{]}}\AgdaSpace{}%
\AgdaBound{cms}\AgdaSpace{}%
\AgdaBound{a}\AgdaSpace{}%
\AgdaSymbol{=}\AgdaSpace{}%
\AgdaKeyword{do}\<%
\\
\>[0][@{}l@{\AgdaIndent{0}}]%
\>[2]\AgdaKeyword{let}\AgdaSpace{}%
\AgdaOperator{\AgdaInductiveConstructor{⟨}}\AgdaSpace{}%
\AgdaBound{a₁}\AgdaSpace{}%
\AgdaOperator{\AgdaInductiveConstructor{,}}\AgdaSpace{}%
\AgdaBound{a′}\AgdaSpace{}%
\AgdaOperator{\AgdaInductiveConstructor{⟩}}\AgdaSpace{}%
\AgdaSymbol{=}\AgdaSpace{}%
\AgdaBound{split}\AgdaSpace{}%
\AgdaBound{a}\<%
\\
\>[2]\AgdaBound{b₁}\AgdaSpace{}%
\AgdaOperator{\AgdaField{←}}\AgdaSpace{}%
\AgdaFunction{exec}\AgdaSpace{}%
\AgdaBound{cmd₁}\AgdaSpace{}%
\AgdaBound{cms}\AgdaSpace{}%
\AgdaBound{a₁}\<%
\\
\>[2]\AgdaKeyword{let}\AgdaSpace{}%
\AgdaBound{a₂}\AgdaSpace{}%
\AgdaSymbol{=}\AgdaSpace{}%
\AgdaBound{cross}\AgdaSpace{}%
\AgdaBound{a′}\AgdaSpace{}%
\AgdaBound{b₁}\<%
\\
\>[2]\AgdaBound{b₂}\AgdaSpace{}%
\AgdaOperator{\AgdaField{←}}\AgdaSpace{}%
\AgdaFunction{exec}\AgdaSpace{}%
\AgdaBound{cmd₂}\AgdaSpace{}%
\AgdaBound{cms}\AgdaSpace{}%
\AgdaBound{a₂}\<%
\\
\>[2]\AgdaFunction{pure}\AgdaSpace{}%
\AgdaSymbol{(}\AgdaBound{join}\AgdaSpace{}%
\AgdaBound{b₁}\AgdaSpace{}%
\AgdaBound{b₂}\AgdaSymbol{)}\<%
\\
\>[0]\AgdaFunction{exec}\AgdaSpace{}%
\AgdaSymbol{(}\AgdaInductiveConstructor{LOOP}\AgdaSpace{}%
\AgdaBound{cmd}\AgdaSymbol{)}\AgdaSpace{}%
\AgdaBound{cms}\AgdaSpace{}%
\AgdaBound{a}\AgdaSpace{}%
\AgdaSymbol{=}\AgdaSpace{}%
\AgdaFunction{exec}\AgdaSpace{}%
\AgdaBound{cmd}\AgdaSpace{}%
\AgdaSymbol{(}\AgdaFunction{push}\AgdaSpace{}%
\AgdaBound{cms}\AgdaSpace{}%
\AgdaBound{cmd}\AgdaSymbol{)}\AgdaSpace{}%
\AgdaBound{a}\<%
\\
\>[0]\AgdaFunction{exec}\AgdaSymbol{\{}\AgdaInductiveConstructor{suc}\AgdaSpace{}%
\AgdaBound{n}\AgdaSymbol{\}}\AgdaSpace{}%
\AgdaSymbol{(}\AgdaInductiveConstructor{CONTINUE}\AgdaSpace{}%
\AgdaBound{i}\AgdaSymbol{)}\AgdaSpace{}%
\AgdaBound{cms}\AgdaSpace{}%
\AgdaBound{a}\<%
\\
\>[0][@{}l@{\AgdaIndent{0}}]%
\>[2]\AgdaKeyword{with}\AgdaSpace{}%
\AgdaBound{cms}\AgdaSpace{}%
\AgdaBound{i}\<%
\\
\>[0]\AgdaSymbol{...}\AgdaSpace{}%
\AgdaSymbol{|}\AgdaSpace{}%
\AgdaOperator{\AgdaInductiveConstructor{⟨}}\AgdaSpace{}%
\AgdaBound{s-i}\AgdaSpace{}%
\AgdaOperator{\AgdaInductiveConstructor{,}}\AgdaSpace{}%
\AgdaBound{cmd-i}\AgdaSpace{}%
\AgdaOperator{\AgdaInductiveConstructor{⟩}}\AgdaSpace{}%
\AgdaSymbol{=}\AgdaSpace{}%
\AgdaFunction{exec}\AgdaSpace{}%
\AgdaBound{cmd-i}\AgdaSpace{}%
\AgdaSymbol{(}\AgdaFunction{pop}\AgdaSpace{}%
\AgdaBound{cms}\AgdaSpace{}%
\AgdaBound{i}\AgdaSymbol{)}\AgdaSpace{}%
\AgdaBound{a}\<%
\end{code}}
\begin{code}[hide]%
\>[0]\<%
\\
\>[0]\AgdaKeyword{record}\AgdaSpace{}%
\AgdaRecord{Accepting}\AgdaSpace{}%
\AgdaBound{A}\AgdaSpace{}%
\AgdaBound{B}\AgdaSpace{}%
\AgdaBound{S}\AgdaSpace{}%
\AgdaSymbol{:}\AgdaSpace{}%
\AgdaPrimitive{Set₁}\AgdaSpace{}%
\AgdaKeyword{where}\<%
\\
\>[0][@{}l@{\AgdaIndent{0}}]%
\>[2]\AgdaKeyword{constructor}\AgdaSpace{}%
\AgdaInductiveConstructor{ACC}\<%
\\
\>[2]\AgdaKeyword{field}\AgdaSpace{}%
\AgdaField{pgm}\AgdaSpace{}%
\AgdaSymbol{:}\AgdaSpace{}%
\AgdaDatatype{Cmd}\AgdaSpace{}%
\AgdaBound{A}\AgdaSpace{}%
\AgdaBound{B}\AgdaSpace{}%
\AgdaInductiveConstructor{[]}\AgdaSpace{}%
\AgdaInductiveConstructor{[]}\AgdaSpace{}%
\AgdaBound{S}\<%
\\
\\[\AgdaEmptyExtraSkip]%
\>[0]\AgdaFunction{acceptor}\AgdaSpace{}%
\AgdaSymbol{:}\AgdaSpace{}%
\AgdaRecord{Accepting}\AgdaSpace{}%
\AgdaGeneralizable{A}\AgdaSpace{}%
\AgdaGeneralizable{B}\AgdaSpace{}%
\AgdaGeneralizable{S}\AgdaSpace{}%
\AgdaSymbol{→}\AgdaSpace{}%
\AgdaGeneralizable{A}\AgdaSpace{}%
\AgdaSymbol{→}\AgdaSpace{}%
\AgdaDatatype{IO}\AgdaSpace{}%
\AgdaGeneralizable{B}\<%
\\
\>[0]\AgdaFunction{acceptor}\AgdaSpace{}%
\AgdaSymbol{(}\AgdaInductiveConstructor{ACC}\AgdaSpace{}%
\AgdaBound{pgm}\AgdaSymbol{)}\AgdaSpace{}%
\AgdaBound{a}\AgdaSpace{}%
\AgdaSymbol{=}\AgdaSpace{}%
\AgdaKeyword{do}\<%
\\
\>[0][@{}l@{\AgdaIndent{0}}]%
\>[2]\AgdaBound{ch}\AgdaSpace{}%
\AgdaOperator{\AgdaField{←}}\AgdaSpace{}%
\AgdaPostulate{primAccept}\<%
\\
\>[2]\AgdaBound{b}\AgdaSpace{}%
\AgdaOperator{\AgdaField{←}}\AgdaSpace{}%
\AgdaField{runReaderT}\AgdaSpace{}%
\AgdaSymbol{(}\AgdaFunction{exec}\AgdaSpace{}%
\AgdaBound{pgm}\AgdaSpace{}%
\AgdaSymbol{(λ())}\AgdaSpace{}%
\AgdaBound{a}\AgdaSymbol{)}\AgdaSpace{}%
\AgdaBound{ch}\<%
\\
\>[2]\AgdaPostulate{primClose}\AgdaSpace{}%
\AgdaBound{ch}\<%
\\
\>[2]\AgdaFunction{pure}\AgdaSpace{}%
\AgdaBound{b}\<%
\\
\\[\AgdaEmptyExtraSkip]%
\>[0]\AgdaComment{--\ examples}\<%
\end{code}
\newcommand\cstBinaryp{%
\begin{code}%
\>[0]\AgdaComment{--\ service\ protocol\ for\ a\ binary\ function}\<%
\\
\>[0]\AgdaFunction{binaryp}\AgdaSpace{}%
\AgdaSymbol{:}\AgdaSpace{}%
\AgdaDatatype{Session}\AgdaSpace{}%
\AgdaGeneralizable{n}\<%
\\
\>[0]\AgdaFunction{binaryp}\AgdaSpace{}%
\AgdaSymbol{=}\AgdaSpace{}%
\AgdaOperator{\AgdaInductiveConstructor{⁇}}\AgdaSpace{}%
\AgdaInductiveConstructor{int}\AgdaSpace{}%
\AgdaOperator{\AgdaInductiveConstructor{⨟}}\AgdaSpace{}%
\AgdaOperator{\AgdaInductiveConstructor{⁇}}\AgdaSpace{}%
\AgdaInductiveConstructor{int}\AgdaSpace{}%
\AgdaOperator{\AgdaInductiveConstructor{⨟}}\AgdaSpace{}%
\AgdaOperator{\AgdaInductiveConstructor{‼}}\AgdaSpace{}%
\AgdaInductiveConstructor{int}\<%
\\
\\[\AgdaEmptyExtraSkip]%
\>[0]\AgdaComment{--\ service\ protocol\ for\ a\ unary\ function}\<%
\\
\>[0]\AgdaFunction{unaryp}\AgdaSpace{}%
\AgdaSymbol{:}\AgdaSpace{}%
\AgdaDatatype{Session}\AgdaSpace{}%
\AgdaGeneralizable{n}\<%
\\
\>[0]\AgdaFunction{unaryp}\AgdaSpace{}%
\AgdaSymbol{=}\AgdaSpace{}%
\AgdaOperator{\AgdaInductiveConstructor{⁇}}\AgdaSpace{}%
\AgdaInductiveConstructor{int}\AgdaSpace{}%
\AgdaOperator{\AgdaInductiveConstructor{⨟}}\AgdaSpace{}%
\AgdaOperator{\AgdaInductiveConstructor{‼}}\AgdaSpace{}%
\AgdaInductiveConstructor{int}\<%
\\
\\[\AgdaEmptyExtraSkip]%
\>[0]\AgdaComment{--\ service\ protocol\ for\ choosing\ between\ a\ binary\ and\ a\ unary\ function}\<%
\\
\>[0]\AgdaFunction{arithp}\AgdaSpace{}%
\AgdaSymbol{:}\AgdaSpace{}%
\AgdaDatatype{Session}\AgdaSpace{}%
\AgdaGeneralizable{n}\<%
\\
\>[0]\AgdaFunction{arithp}\AgdaSpace{}%
\AgdaSymbol{=}\AgdaSpace{}%
\AgdaFunction{\&}\AgdaSpace{}%
\AgdaOperator{\AgdaInductiveConstructor{[}}\AgdaSpace{}%
\AgdaFunction{binaryp}\AgdaSpace{}%
\AgdaOperator{\AgdaInductiveConstructor{,}}\AgdaSpace{}%
\AgdaFunction{unaryp}\AgdaSpace{}%
\AgdaOperator{\AgdaInductiveConstructor{]}}\<%
\\
\\[\AgdaEmptyExtraSkip]%
\>[0]\AgdaComment{--\ many\ unary\ functions}\<%
\\
\>[0]\AgdaFunction{many-unaryp}\AgdaSpace{}%
\AgdaSymbol{:}\AgdaSpace{}%
\AgdaDatatype{Session}\AgdaSpace{}%
\AgdaGeneralizable{n}\<%
\\
\>[0]\AgdaFunction{many-unaryp}\AgdaSpace{}%
\AgdaSymbol{=}\AgdaSpace{}%
\AgdaOperator{\AgdaInductiveConstructor{μ}}\AgdaSpace{}%
\AgdaSymbol{(}\AgdaFunction{\&}\AgdaSpace{}%
\AgdaOperator{\AgdaInductiveConstructor{[}}\AgdaSpace{}%
\AgdaFunction{unaryp}\AgdaSpace{}%
\AgdaOperator{\AgdaInductiveConstructor{⨟}}\AgdaSpace{}%
\AgdaOperator{\AgdaInductiveConstructor{`}}\AgdaSpace{}%
\AgdaInductiveConstructor{zero}\AgdaSpace{}%
\AgdaOperator{\AgdaInductiveConstructor{,}}\AgdaSpace{}%
\AgdaInductiveConstructor{skip}\AgdaSpace{}%
\AgdaOperator{\AgdaInductiveConstructor{]}}\AgdaSymbol{)}\<%
\end{code}}
\newcommand\cstCmdExamples{%
\begin{code}%
\>[0]\AgdaFunction{addp-command}\AgdaSpace{}%
\AgdaSymbol{:}\AgdaSpace{}%
\AgdaDatatype{Cmd}\AgdaSpace{}%
\AgdaFunction{⊤}\AgdaSpace{}%
\AgdaFunction{⊤}\AgdaSpace{}%
\AgdaGeneralizable{V}\AgdaSpace{}%
\AgdaGeneralizable{W}\AgdaSpace{}%
\AgdaFunction{binaryp}\<%
\\
\>[0]\AgdaFunction{addp-command}\AgdaSpace{}%
\AgdaSymbol{=}\AgdaSpace{}%
\AgdaInductiveConstructor{RECV}\AgdaSpace{}%
\AgdaFunction{const}\AgdaSpace{}%
\AgdaOperator{\AgdaFunction{⨟′}}\AgdaSpace{}%
\AgdaInductiveConstructor{RECV}\AgdaSpace{}%
\AgdaOperator{\AgdaFunction{\AgdaUnderscore{}+\AgdaUnderscore{}}}\AgdaSpace{}%
\AgdaOperator{\AgdaFunction{⨟′}}\AgdaSpace{}%
\AgdaInductiveConstructor{SEND}\AgdaSpace{}%
\AgdaSymbol{(λ}\AgdaSpace{}%
\AgdaBound{x}\AgdaSpace{}%
\AgdaSymbol{→}\AgdaSpace{}%
\AgdaOperator{\AgdaInductiveConstructor{⟨}}\AgdaSpace{}%
\AgdaFunction{tt}\AgdaSpace{}%
\AgdaOperator{\AgdaInductiveConstructor{,}}\AgdaSpace{}%
\AgdaBound{x}\AgdaSpace{}%
\AgdaOperator{\AgdaInductiveConstructor{⟩}}\AgdaSymbol{)}\<%
\\
\\[\AgdaEmptyExtraSkip]%
\>[0]\AgdaFunction{negp-command}\AgdaSpace{}%
\AgdaSymbol{:}\AgdaSpace{}%
\AgdaDatatype{Cmd}\AgdaSpace{}%
\AgdaFunction{⊤}\AgdaSpace{}%
\AgdaFunction{⊤}\AgdaSpace{}%
\AgdaGeneralizable{V}\AgdaSpace{}%
\AgdaGeneralizable{W}\AgdaSpace{}%
\AgdaFunction{unaryp}\<%
\\
\>[0]\AgdaFunction{negp-command}\AgdaSpace{}%
\AgdaSymbol{=}\AgdaSpace{}%
\AgdaInductiveConstructor{RECV}\AgdaSpace{}%
\AgdaSymbol{(}\AgdaFunction{const}\AgdaSpace{}%
\AgdaOperator{\AgdaFunction{∘′}}\AgdaSpace{}%
\AgdaOperator{\AgdaFunction{-\AgdaUnderscore{}}}\AgdaSymbol{)}\AgdaSpace{}%
\AgdaOperator{\AgdaFunction{⨟′}}\AgdaSpace{}%
\AgdaInductiveConstructor{SEND}\AgdaSpace{}%
\AgdaSymbol{λ}\AgdaSpace{}%
\AgdaBound{x}\AgdaSpace{}%
\AgdaSymbol{→}\AgdaSpace{}%
\AgdaOperator{\AgdaInductiveConstructor{⟨}}\AgdaSpace{}%
\AgdaFunction{tt}\AgdaSpace{}%
\AgdaOperator{\AgdaInductiveConstructor{,}}\AgdaSpace{}%
\AgdaBound{x}\AgdaSpace{}%
\AgdaOperator{\AgdaInductiveConstructor{⟩}}\<%
\\
\\[\AgdaEmptyExtraSkip]%
\>[0]\AgdaFunction{arithp-command}\AgdaSpace{}%
\AgdaSymbol{:}\AgdaSpace{}%
\AgdaDatatype{Cmd}\AgdaSpace{}%
\AgdaFunction{⊤}\AgdaSpace{}%
\AgdaFunction{⊤}\AgdaSpace{}%
\AgdaGeneralizable{V}\AgdaSpace{}%
\AgdaGeneralizable{W}\AgdaSpace{}%
\AgdaFunction{arithp}\<%
\\
\>[0]\AgdaFunction{arithp-command}\AgdaSpace{}%
\AgdaSymbol{=}\AgdaSpace{}%
\AgdaInductiveConstructor{CHOICE}\AgdaSpace{}%
\AgdaSymbol{λ}\AgdaSpace{}%
\AgdaKeyword{where}\<%
\\
\>[0][@{}l@{\AgdaIndent{0}}]%
\>[2]\AgdaInductiveConstructor{zero}\AgdaSpace{}%
\AgdaSymbol{→}\AgdaSpace{}%
\AgdaFunction{addp-command}\<%
\\
\>[2]\AgdaSymbol{(}\AgdaInductiveConstructor{suc}\AgdaSpace{}%
\AgdaInductiveConstructor{zero}\AgdaSymbol{)}\AgdaSpace{}%
\AgdaSymbol{→}\AgdaSpace{}%
\AgdaFunction{negp-command}\<%
\\
\\[\AgdaEmptyExtraSkip]%
\>[0]\AgdaFunction{many-unaryp-command}\AgdaSpace{}%
\AgdaSymbol{:}\AgdaSpace{}%
\AgdaDatatype{Cmd}\AgdaSpace{}%
\AgdaDatatype{ℤ}\AgdaSpace{}%
\AgdaDatatype{ℤ}\AgdaSpace{}%
\AgdaGeneralizable{V}\AgdaSpace{}%
\AgdaGeneralizable{W}\AgdaSpace{}%
\AgdaFunction{many-unaryp}\<%
\\
\>[0]\AgdaFunction{many-unaryp-command}\AgdaSpace{}%
\AgdaSymbol{=}\AgdaSpace{}%
\AgdaInductiveConstructor{LOOP}\AgdaSpace{}%
\AgdaOperator{\AgdaFunction{\$}}\AgdaSpace{}%
\AgdaInductiveConstructor{CHOICE}\AgdaSpace{}%
\AgdaSymbol{λ}\AgdaSpace{}%
\AgdaKeyword{where}\<%
\\
\>[0][@{}l@{\AgdaIndent{0}}]%
\>[2]\AgdaInductiveConstructor{zero}\AgdaSpace{}%
\AgdaSymbol{→}\AgdaSpace{}%
\AgdaSymbol{(}\AgdaInductiveConstructor{RECV}\AgdaSpace{}%
\AgdaOperator{\AgdaFunction{\AgdaUnderscore{}+\AgdaUnderscore{}}}\AgdaSpace{}%
\AgdaOperator{\AgdaFunction{⨟′}}\AgdaSpace{}%
\AgdaInductiveConstructor{SEND}\AgdaSpace{}%
\AgdaOperator{\AgdaFunction{<}}\AgdaSpace{}%
\AgdaFunction{id}\AgdaSpace{}%
\AgdaOperator{\AgdaFunction{,}}\AgdaSpace{}%
\AgdaFunction{id}\AgdaSpace{}%
\AgdaOperator{\AgdaFunction{>}}\AgdaSymbol{)}\AgdaSpace{}%
\AgdaOperator{\AgdaFunction{⨟′}}\AgdaSpace{}%
\AgdaInductiveConstructor{CONTINUE}\AgdaSpace{}%
\AgdaInductiveConstructor{zero}\<%
\\
\>[2]\AgdaSymbol{(}\AgdaInductiveConstructor{suc}\AgdaSpace{}%
\AgdaInductiveConstructor{zero}\AgdaSymbol{)}\AgdaSpace{}%
\AgdaSymbol{→}\AgdaSpace{}%
\AgdaInductiveConstructor{SKIP}\AgdaSpace{}%
\AgdaFunction{id}\<%
\end{code}}
\newcommand\cstTreep{%
\begin{code}%
\>[0]\AgdaKeyword{data}\AgdaSpace{}%
\AgdaDatatype{IntTree}\AgdaSpace{}%
\AgdaSymbol{:}\AgdaSpace{}%
\AgdaPrimitive{Set}\AgdaSpace{}%
\AgdaKeyword{where}\<%
\\
\>[0][@{}l@{\AgdaIndent{0}}]%
\>[2]\AgdaInductiveConstructor{Leaf}%
\>[9]\AgdaSymbol{:}\AgdaSpace{}%
\AgdaDatatype{ℤ}\AgdaSpace{}%
\AgdaSymbol{→}\AgdaSpace{}%
\AgdaDatatype{IntTree}\<%
\\
\>[2]\AgdaInductiveConstructor{Branch}\AgdaSpace{}%
\AgdaSymbol{:}\AgdaSpace{}%
\AgdaDatatype{IntTree}\AgdaSpace{}%
\AgdaSymbol{→}\AgdaSpace{}%
\AgdaDatatype{IntTree}\AgdaSpace{}%
\AgdaSymbol{→}\AgdaSpace{}%
\AgdaDatatype{IntTree}\<%
\\
\\[\AgdaEmptyExtraSkip]%
\>[0]\AgdaFunction{leafp}\AgdaSpace{}%
\AgdaSymbol{:}\AgdaSpace{}%
\AgdaDatatype{Session}\AgdaSpace{}%
\AgdaGeneralizable{n}\<%
\\
\>[0]\AgdaFunction{leafp}\AgdaSpace{}%
\AgdaSymbol{=}\AgdaSpace{}%
\AgdaOperator{\AgdaInductiveConstructor{⁇}}\AgdaSpace{}%
\AgdaInductiveConstructor{int}\<%
\\
\\[\AgdaEmptyExtraSkip]%
\>[0]\AgdaFunction{branchp}\AgdaSpace{}%
\AgdaSymbol{:}\AgdaSpace{}%
\AgdaDatatype{Session}\AgdaSpace{}%
\AgdaSymbol{(}\AgdaInductiveConstructor{suc}\AgdaSpace{}%
\AgdaGeneralizable{n}\AgdaSymbol{)}\<%
\\
\>[0]\AgdaFunction{branchp}\AgdaSpace{}%
\AgdaSymbol{=}\AgdaSpace{}%
\AgdaOperator{\AgdaInductiveConstructor{`}}\AgdaSpace{}%
\AgdaInductiveConstructor{zero}\AgdaSpace{}%
\AgdaOperator{\AgdaInductiveConstructor{⨟}}\AgdaSpace{}%
\AgdaOperator{\AgdaInductiveConstructor{`}}\AgdaSpace{}%
\AgdaInductiveConstructor{zero}\<%
\\
\\[\AgdaEmptyExtraSkip]%
\>[0]\AgdaFunction{treep}\AgdaSpace{}%
\AgdaSymbol{:}\AgdaSpace{}%
\AgdaDatatype{Session}\AgdaSpace{}%
\AgdaGeneralizable{n}\<%
\\
\>[0]\AgdaFunction{treep}\AgdaSpace{}%
\AgdaSymbol{=}\AgdaSpace{}%
\AgdaOperator{\AgdaInductiveConstructor{μ}}\AgdaSpace{}%
\AgdaFunction{\&}\AgdaSpace{}%
\AgdaOperator{\AgdaInductiveConstructor{[}}\AgdaSpace{}%
\AgdaFunction{leafp}\AgdaSpace{}%
\AgdaOperator{\AgdaInductiveConstructor{,}}\AgdaSpace{}%
\AgdaFunction{branchp}\AgdaSpace{}%
\AgdaOperator{\AgdaInductiveConstructor{]}}\<%
\end{code}}
\newcommand\cstRecvTree{%
\begin{code}%
\>[0]\AgdaFunction{recvTree}\AgdaSpace{}%
\AgdaSymbol{:}\AgdaSpace{}%
\AgdaDatatype{Cmd}\AgdaSpace{}%
\AgdaFunction{⊤}\AgdaSpace{}%
\AgdaDatatype{IntTree}\AgdaSpace{}%
\AgdaGeneralizable{V}\AgdaSpace{}%
\AgdaGeneralizable{W}\AgdaSpace{}%
\AgdaFunction{treep}\<%
\\
\>[0]\AgdaFunction{recvTree}\AgdaSpace{}%
\AgdaSymbol{=}\AgdaSpace{}%
\AgdaInductiveConstructor{LOOP}\AgdaSpace{}%
\AgdaOperator{\AgdaFunction{\$}}\AgdaSpace{}%
\AgdaInductiveConstructor{CHOICE}\AgdaSpace{}%
\AgdaSymbol{λ}\AgdaSpace{}%
\AgdaKeyword{where}\<%
\\
\>[0][@{}l@{\AgdaIndent{0}}]%
\>[2]\AgdaInductiveConstructor{zero}\AgdaSpace{}%
\AgdaSymbol{→}\AgdaSpace{}%
\AgdaInductiveConstructor{RECV}\AgdaSpace{}%
\AgdaSymbol{(}\AgdaFunction{const}\AgdaSpace{}%
\AgdaOperator{\AgdaFunction{∘′}}\AgdaSpace{}%
\AgdaInductiveConstructor{Leaf}\AgdaSymbol{)}\<%
\\
\>[2]\AgdaSymbol{(}\AgdaInductiveConstructor{suc}\AgdaSpace{}%
\AgdaInductiveConstructor{zero}\AgdaSymbol{)}\AgdaSpace{}%
\AgdaSymbol{→}\AgdaSpace{}%
\AgdaOperator{\AgdaInductiveConstructor{[}}\AgdaSpace{}%
\AgdaSymbol{(λ}\AgdaSpace{}%
\AgdaBound{x}\AgdaSpace{}%
\AgdaSymbol{→}\AgdaSpace{}%
\AgdaOperator{\AgdaInductiveConstructor{⟨}}\AgdaSpace{}%
\AgdaBound{x}\AgdaSpace{}%
\AgdaOperator{\AgdaInductiveConstructor{,}}\AgdaSpace{}%
\AgdaBound{x}\AgdaSpace{}%
\AgdaOperator{\AgdaInductiveConstructor{⟩}}\AgdaSymbol{)}\AgdaSpace{}%
\AgdaOperator{\AgdaInductiveConstructor{]}}\AgdaSpace{}%
\AgdaSymbol{(}\AgdaInductiveConstructor{CONTINUE}\AgdaSpace{}%
\AgdaInductiveConstructor{zero}\AgdaSymbol{)}\AgdaSpace{}%
\AgdaOperator{\AgdaInductiveConstructor{⨟[}}\AgdaSpace{}%
\AgdaFunction{const}\AgdaSpace{}%
\AgdaOperator{\AgdaInductiveConstructor{]}}\AgdaSpace{}%
\AgdaSymbol{(}\AgdaInductiveConstructor{CONTINUE}\AgdaSpace{}%
\AgdaInductiveConstructor{zero}\AgdaSymbol{)}\AgdaSpace{}%
\AgdaOperator{\AgdaInductiveConstructor{[}}\AgdaSpace{}%
\AgdaInductiveConstructor{Branch}\AgdaSpace{}%
\AgdaOperator{\AgdaInductiveConstructor{]}}\<%
\end{code}}
\newcommand\cstSendTree{%
\begin{code}%
\>[0]\AgdaFunction{IntTreeF}\AgdaSpace{}%
\AgdaSymbol{:}\AgdaSpace{}%
\AgdaDatatype{Fin}\AgdaSpace{}%
\AgdaNumber{2}\AgdaSpace{}%
\AgdaSymbol{→}\AgdaSpace{}%
\AgdaPrimitive{Set}\<%
\\
\>[0]\AgdaFunction{IntTreeF}\AgdaSpace{}%
\AgdaInductiveConstructor{zero}\AgdaSpace{}%
\AgdaSymbol{=}\AgdaSpace{}%
\AgdaDatatype{ℤ}\<%
\\
\>[0]\AgdaFunction{IntTreeF}\AgdaSpace{}%
\AgdaSymbol{(}\AgdaInductiveConstructor{suc}\AgdaSpace{}%
\AgdaInductiveConstructor{zero}\AgdaSymbol{)}\AgdaSpace{}%
\AgdaSymbol{=}\AgdaSpace{}%
\AgdaDatatype{IntTree}\AgdaSpace{}%
\AgdaOperator{\AgdaFunction{×}}\AgdaSpace{}%
\AgdaDatatype{IntTree}\<%
\\
\\[\AgdaEmptyExtraSkip]%
\>[0]\AgdaFunction{splitTree}\AgdaSpace{}%
\AgdaSymbol{:}\AgdaSpace{}%
\AgdaDatatype{IntTree}\AgdaSpace{}%
\AgdaSymbol{→}\AgdaSpace{}%
\AgdaRecord{Σ}\AgdaSpace{}%
\AgdaSymbol{(}\AgdaDatatype{Fin}\AgdaSpace{}%
\AgdaNumber{2}\AgdaSymbol{)}\AgdaSpace{}%
\AgdaFunction{IntTreeF}\<%
\\
\>[0]\AgdaFunction{splitTree}\AgdaSpace{}%
\AgdaSymbol{(}\AgdaInductiveConstructor{Leaf}\AgdaSpace{}%
\AgdaBound{x}\AgdaSymbol{)}\AgdaSpace{}%
\AgdaSymbol{=}\AgdaSpace{}%
\AgdaOperator{\AgdaInductiveConstructor{⟨}}\AgdaSpace{}%
\AgdaInductiveConstructor{zero}\AgdaSpace{}%
\AgdaOperator{\AgdaInductiveConstructor{,}}\AgdaSpace{}%
\AgdaBound{x}\AgdaSpace{}%
\AgdaOperator{\AgdaInductiveConstructor{⟩}}\<%
\\
\>[0]\AgdaFunction{splitTree}\AgdaSpace{}%
\AgdaSymbol{(}\AgdaInductiveConstructor{Branch}\AgdaSpace{}%
\AgdaBound{t₁}\AgdaSpace{}%
\AgdaBound{t₂}\AgdaSymbol{)}\AgdaSpace{}%
\AgdaSymbol{=}\AgdaSpace{}%
\AgdaOperator{\AgdaInductiveConstructor{⟨}}\AgdaSpace{}%
\AgdaInductiveConstructor{suc}\AgdaSpace{}%
\AgdaInductiveConstructor{zero}\AgdaSpace{}%
\AgdaOperator{\AgdaInductiveConstructor{,}}\AgdaSpace{}%
\AgdaOperator{\AgdaInductiveConstructor{⟨}}\AgdaSpace{}%
\AgdaBound{t₁}\AgdaSpace{}%
\AgdaOperator{\AgdaInductiveConstructor{,}}\AgdaSpace{}%
\AgdaBound{t₂}\AgdaSpace{}%
\AgdaOperator{\AgdaInductiveConstructor{⟩}}\AgdaSpace{}%
\AgdaOperator{\AgdaInductiveConstructor{⟩}}\<%
\\
\\[\AgdaEmptyExtraSkip]%
\>[0]\AgdaFunction{sendTree}\AgdaSpace{}%
\AgdaSymbol{:}\AgdaSpace{}%
\AgdaDatatype{Cmd}\AgdaSpace{}%
\AgdaDatatype{IntTree}\AgdaSpace{}%
\AgdaFunction{⊤}\AgdaSpace{}%
\AgdaGeneralizable{V}\AgdaSpace{}%
\AgdaGeneralizable{W}\AgdaSpace{}%
\AgdaSymbol{(}\AgdaFunction{dual}\AgdaSpace{}%
\AgdaFunction{treep}\AgdaSymbol{)}\<%
\\
\>[0]\AgdaFunction{sendTree}\AgdaSpace{}%
\AgdaSymbol{=}\AgdaSpace{}%
\AgdaInductiveConstructor{LOOP}\AgdaSpace{}%
\AgdaOperator{\AgdaFunction{\$}}\AgdaSpace{}%
\AgdaInductiveConstructor{SELECT}\AgdaSpace{}%
\AgdaFunction{splitTree}\AgdaSpace{}%
\AgdaSymbol{λ}\AgdaSpace{}%
\AgdaKeyword{where}\<%
\\
\>[0][@{}l@{\AgdaIndent{0}}]%
\>[2]\AgdaInductiveConstructor{zero}\AgdaSpace{}%
\AgdaSymbol{→}\AgdaSpace{}%
\AgdaInductiveConstructor{SEND}\AgdaSpace{}%
\AgdaSymbol{(λ}\AgdaSpace{}%
\AgdaBound{z}\AgdaSpace{}%
\AgdaSymbol{→}\AgdaSpace{}%
\AgdaOperator{\AgdaInductiveConstructor{⟨}}\AgdaSpace{}%
\AgdaFunction{tt}\AgdaSpace{}%
\AgdaOperator{\AgdaInductiveConstructor{,}}\AgdaSpace{}%
\AgdaBound{z}\AgdaSpace{}%
\AgdaOperator{\AgdaInductiveConstructor{⟩}}\AgdaSymbol{)}\<%
\\
\>[2]\AgdaSymbol{(}\AgdaInductiveConstructor{suc}\AgdaSpace{}%
\AgdaInductiveConstructor{zero}\AgdaSymbol{)}\AgdaSpace{}%
\AgdaSymbol{→}\AgdaSpace{}%
\AgdaOperator{\AgdaInductiveConstructor{[}}\AgdaSpace{}%
\AgdaFunction{id}\AgdaSpace{}%
\AgdaOperator{\AgdaInductiveConstructor{]}}\AgdaSpace{}%
\AgdaSymbol{(}\AgdaInductiveConstructor{CONTINUE}\AgdaSpace{}%
\AgdaInductiveConstructor{zero}\AgdaSymbol{)}\AgdaSpace{}%
\AgdaOperator{\AgdaInductiveConstructor{⨟[}}\AgdaSpace{}%
\AgdaFunction{const}\AgdaSpace{}%
\AgdaOperator{\AgdaInductiveConstructor{]}}\AgdaSpace{}%
\AgdaSymbol{(}\AgdaInductiveConstructor{CONTINUE}\AgdaSpace{}%
\AgdaInductiveConstructor{zero}\AgdaSymbol{)}\AgdaSpace{}%
\AgdaOperator{\AgdaInductiveConstructor{[}}\AgdaSpace{}%
\AgdaFunction{const}\AgdaSpace{}%
\AgdaOperator{\AgdaInductiveConstructor{]}}\<%
\end{code}}

%% file: latex/ST-multichannel.tex
\begin{code}[hide]%
\>[0]\AgdaSymbol{\{-\#}\AgdaSpace{}%
\AgdaKeyword{OPTIONS}\AgdaSpace{}%
\AgdaPragma{--guardedness}\AgdaSpace{}%
\AgdaSymbol{\#-\}}\AgdaSpace{}%
\AgdaComment{\{-\ for\ IO\ -\}}\<%
\\
\>[0]\AgdaKeyword{module}\AgdaSpace{}%
\AgdaModule{ST-multichannel}\AgdaSpace{}%
\AgdaKeyword{where}\<%
\\
\\[\AgdaEmptyExtraSkip]%
\>[0]\AgdaKeyword{open}\AgdaSpace{}%
\AgdaKeyword{import}\AgdaSpace{}%
\AgdaModule{Data.Bool}\AgdaSpace{}%
\AgdaKeyword{using}\AgdaSpace{}%
\AgdaSymbol{(}\AgdaDatatype{Bool}\AgdaSymbol{;}\AgdaSpace{}%
\AgdaInductiveConstructor{true}\AgdaSymbol{;}\AgdaSpace{}%
\AgdaInductiveConstructor{false}\AgdaSymbol{;}\AgdaOperator{\AgdaFunction{if\AgdaUnderscore{}then\AgdaUnderscore{}else\AgdaUnderscore{}}}\AgdaSymbol{)}\<%
\\
\>[0]\AgdaKeyword{open}\AgdaSpace{}%
\AgdaKeyword{import}\AgdaSpace{}%
\AgdaModule{Data.Nat}\AgdaSpace{}%
\AgdaKeyword{using}\AgdaSpace{}%
\AgdaSymbol{(}\AgdaDatatype{ℕ}\AgdaSymbol{;}\AgdaSpace{}%
\AgdaInductiveConstructor{suc}\AgdaSymbol{;}\AgdaSpace{}%
\AgdaInductiveConstructor{zero}\AgdaSymbol{;}\AgdaSpace{}%
\AgdaOperator{\AgdaPrimitive{\AgdaUnderscore{}+\AgdaUnderscore{}}}\AgdaSymbol{)}\<%
\\
\>[0]\AgdaKeyword{open}\AgdaSpace{}%
\AgdaKeyword{import}\AgdaSpace{}%
\AgdaModule{Data.Nat.Properties}\AgdaSpace{}%
\AgdaKeyword{using}\AgdaSpace{}%
\AgdaSymbol{(}\AgdaFunction{+-suc}\AgdaSymbol{)}\<%
\\
\>[0]\AgdaKeyword{open}\AgdaSpace{}%
\AgdaKeyword{import}\AgdaSpace{}%
\AgdaModule{Data.Integer}\AgdaSpace{}%
\AgdaKeyword{using}\AgdaSpace{}%
\AgdaSymbol{(}\AgdaDatatype{ℤ}\AgdaSymbol{)}\<%
\\
\>[0]\AgdaKeyword{open}\AgdaSpace{}%
\AgdaKeyword{import}\AgdaSpace{}%
\AgdaModule{Data.Fin}\AgdaSpace{}%
\AgdaKeyword{using}\AgdaSpace{}%
\AgdaSymbol{(}\AgdaDatatype{Fin}\AgdaSymbol{;}\AgdaSpace{}%
\AgdaInductiveConstructor{suc}\AgdaSymbol{;}\AgdaSpace{}%
\AgdaInductiveConstructor{zero}\AgdaSymbol{;}\AgdaSpace{}%
\AgdaOperator{\AgdaFunction{\AgdaUnderscore{}≟\AgdaUnderscore{}}}\AgdaSymbol{)}\<%
\\
\>[0]\AgdaKeyword{open}\AgdaSpace{}%
\AgdaKeyword{import}\AgdaSpace{}%
\AgdaModule{Data.Fin.Subset}\AgdaSpace{}%
\AgdaKeyword{using}\AgdaSpace{}%
\AgdaSymbol{(}\AgdaFunction{Subset}\AgdaSymbol{)}\<%
\\
\>[0]\AgdaKeyword{open}\AgdaSpace{}%
\AgdaKeyword{import}\AgdaSpace{}%
\AgdaModule{Data.Product}\AgdaSpace{}%
\AgdaKeyword{using}\AgdaSpace{}%
\AgdaSymbol{(}\AgdaOperator{\AgdaFunction{\AgdaUnderscore{}×\AgdaUnderscore{}}}\AgdaSymbol{;}\AgdaSpace{}%
\AgdaRecord{Σ}\AgdaSymbol{;}\AgdaSpace{}%
\AgdaField{proj₁}\AgdaSymbol{;}\AgdaSpace{}%
\AgdaField{proj₂}\AgdaSymbol{)}\AgdaSpace{}%
\AgdaKeyword{renaming}\AgdaSpace{}%
\AgdaSymbol{(}\AgdaSpace{}%
\AgdaOperator{\AgdaInductiveConstructor{\AgdaUnderscore{},\AgdaUnderscore{}}}\AgdaSpace{}%
\AgdaSymbol{to}\AgdaSpace{}%
\AgdaOperator{\AgdaInductiveConstructor{⟨\AgdaUnderscore{},\AgdaUnderscore{}⟩}}\AgdaSymbol{)}\<%
\\
\>[0]\AgdaKeyword{open}\AgdaSpace{}%
\AgdaKeyword{import}\AgdaSpace{}%
\AgdaModule{Data.Sum}\AgdaSpace{}%
\AgdaKeyword{using}\AgdaSpace{}%
\AgdaSymbol{(}\AgdaOperator{\AgdaDatatype{\AgdaUnderscore{}⊎\AgdaUnderscore{}}}\AgdaSymbol{;}\AgdaSpace{}%
\AgdaInductiveConstructor{inj₁}\AgdaSymbol{;}\AgdaSpace{}%
\AgdaInductiveConstructor{inj₂}\AgdaSymbol{)}\<%
\\
\>[0]\AgdaKeyword{open}\AgdaSpace{}%
\AgdaKeyword{import}\AgdaSpace{}%
\AgdaModule{Data.Vec}\AgdaSpace{}%
\AgdaKeyword{using}\AgdaSpace{}%
\AgdaSymbol{(}\AgdaDatatype{Vec}\AgdaSymbol{;}\AgdaSpace{}%
\AgdaInductiveConstructor{[]}\AgdaSpace{}%
\AgdaSymbol{;}\AgdaSpace{}%
\AgdaOperator{\AgdaInductiveConstructor{\AgdaUnderscore{}∷\AgdaUnderscore{}}}\AgdaSymbol{;}\AgdaSpace{}%
\AgdaFunction{lookup}\AgdaSymbol{;}\AgdaSpace{}%
\AgdaFunction{remove}\AgdaSymbol{;}\AgdaSpace{}%
\AgdaFunction{updateAt}\AgdaSymbol{)}\<%
\\
\\[\AgdaEmptyExtraSkip]%
\>[0]\AgdaKeyword{open}\AgdaSpace{}%
\AgdaKeyword{import}\AgdaSpace{}%
\AgdaModule{Data.Unit}\AgdaSpace{}%
\AgdaKeyword{using}\AgdaSpace{}%
\AgdaSymbol{(}\AgdaRecord{⊤}\AgdaSymbol{;}\AgdaSpace{}%
\AgdaInductiveConstructor{tt}\AgdaSymbol{)}\<%
\\
\>[0]\AgdaKeyword{open}\AgdaSpace{}%
\AgdaKeyword{import}\AgdaSpace{}%
\AgdaModule{Data.Empty}\AgdaSpace{}%
\AgdaKeyword{using}\AgdaSpace{}%
\AgdaSymbol{(}\AgdaFunction{⊥}\AgdaSymbol{;}\AgdaSpace{}%
\AgdaFunction{⊥-elim}\AgdaSymbol{)}\<%
\\
\\[\AgdaEmptyExtraSkip]%
\>[0]\AgdaKeyword{open}\AgdaSpace{}%
\AgdaKeyword{import}\AgdaSpace{}%
\AgdaModule{Function.Base}\AgdaSpace{}%
\AgdaKeyword{using}\AgdaSpace{}%
\AgdaSymbol{(}\AgdaOperator{\AgdaFunction{case\AgdaUnderscore{}of\AgdaUnderscore{}}}\AgdaSymbol{;}\AgdaSpace{}%
\AgdaFunction{const}\AgdaSymbol{)}\<%
\\
\\[\AgdaEmptyExtraSkip]%
\>[0]\AgdaKeyword{open}\AgdaSpace{}%
\AgdaKeyword{import}\AgdaSpace{}%
\AgdaModule{Relation.Nullary}\<%
\\
\>[0][@{}l@{\AgdaIndent{0}}]%
\>[2]\AgdaKeyword{using}\AgdaSpace{}%
\AgdaSymbol{(}\AgdaOperator{\AgdaFunction{¬\AgdaUnderscore{}}}\AgdaSymbol{;}\AgdaSpace{}%
\AgdaRecord{Dec}\AgdaSymbol{;}\AgdaSpace{}%
\AgdaInductiveConstructor{yes}\AgdaSymbol{;}\AgdaSpace{}%
\AgdaInductiveConstructor{no}\AgdaSymbol{)}\<%
\\
\>[0]\AgdaKeyword{open}\AgdaSpace{}%
\AgdaKeyword{import}\AgdaSpace{}%
\AgdaModule{Relation.Binary.PropositionalEquality}\<%
\\
\>[0][@{}l@{\AgdaIndent{0}}]%
\>[2]\AgdaKeyword{using}\AgdaSpace{}%
\AgdaSymbol{(}\AgdaOperator{\AgdaDatatype{\AgdaUnderscore{}≡\AgdaUnderscore{}}}\AgdaSymbol{;}\AgdaSpace{}%
\AgdaOperator{\AgdaFunction{\AgdaUnderscore{}≢\AgdaUnderscore{}}}\AgdaSymbol{;}\AgdaSpace{}%
\AgdaInductiveConstructor{refl}\AgdaSymbol{;}\AgdaSpace{}%
\AgdaFunction{sym}\AgdaSymbol{;}\AgdaSpace{}%
\AgdaFunction{trans}\AgdaSymbol{;}\AgdaSpace{}%
\AgdaFunction{cong}\AgdaSymbol{;}\AgdaSpace{}%
\AgdaFunction{cong₂}\AgdaSymbol{;}\AgdaSpace{}%
\AgdaFunction{subst}\AgdaSymbol{;}\AgdaSpace{}%
\AgdaFunction{resp₂}\AgdaSymbol{)}\<%
\\
\\[\AgdaEmptyExtraSkip]%
\>[0]\AgdaKeyword{open}\AgdaSpace{}%
\AgdaKeyword{import}\AgdaSpace{}%
\AgdaModule{IO}\<%
\\
\\[\AgdaEmptyExtraSkip]%
\>[0]\AgdaKeyword{variable}\<%
\\
\>[0][@{}l@{\AgdaIndent{0}}]%
\>[2]\AgdaGeneralizable{m}\AgdaSpace{}%
\AgdaGeneralizable{n}\AgdaSpace{}%
\AgdaGeneralizable{o}\AgdaSpace{}%
\AgdaSymbol{:}\AgdaSpace{}%
\AgdaDatatype{ℕ}\<%
\\
\>[2]\AgdaGeneralizable{f}\AgdaSpace{}%
\AgdaSymbol{:}\AgdaSpace{}%
\AgdaDatatype{Fin}\AgdaSpace{}%
\AgdaSymbol{(}\AgdaInductiveConstructor{suc}\AgdaSpace{}%
\AgdaGeneralizable{n}\AgdaSymbol{)}\<%
\\
\>[2]\AgdaGeneralizable{A}\AgdaSpace{}%
\AgdaSymbol{:}\AgdaSpace{}%
\AgdaPrimitive{Set}\<%
\\
\\[\AgdaEmptyExtraSkip]%
\>[0]\AgdaComment{--\ splitting}\<%
\\
\\[\AgdaEmptyExtraSkip]%
\>[0]\AgdaKeyword{data}\AgdaSpace{}%
\AgdaDatatype{Split}\AgdaSpace{}%
\AgdaSymbol{:}\AgdaSpace{}%
\AgdaDatatype{ℕ}\AgdaSpace{}%
\AgdaSymbol{→}\AgdaSpace{}%
\AgdaDatatype{ℕ}\AgdaSpace{}%
\AgdaSymbol{→}\AgdaSpace{}%
\AgdaPrimitive{Set}\AgdaSpace{}%
\AgdaKeyword{where}\<%
\\
\>[0][@{}l@{\AgdaIndent{0}}]%
\>[2]\AgdaInductiveConstructor{null}\AgdaSpace{}%
\AgdaSymbol{:}\AgdaSpace{}%
\AgdaDatatype{Split}\AgdaSpace{}%
\AgdaInductiveConstructor{zero}\AgdaSpace{}%
\AgdaInductiveConstructor{zero}\<%
\\
\>[2]\AgdaInductiveConstructor{left}\AgdaSpace{}%
\AgdaSymbol{:}\AgdaSpace{}%
\AgdaDatatype{Split}\AgdaSpace{}%
\AgdaGeneralizable{m}\AgdaSpace{}%
\AgdaGeneralizable{n}\AgdaSpace{}%
\AgdaSymbol{→}\AgdaSpace{}%
\AgdaDatatype{Split}\AgdaSpace{}%
\AgdaSymbol{(}\AgdaInductiveConstructor{suc}\AgdaSpace{}%
\AgdaGeneralizable{m}\AgdaSymbol{)}\AgdaSpace{}%
\AgdaGeneralizable{n}\<%
\\
\>[2]\AgdaInductiveConstructor{right}\AgdaSpace{}%
\AgdaSymbol{:}\AgdaSpace{}%
\AgdaDatatype{Split}\AgdaSpace{}%
\AgdaGeneralizable{m}\AgdaSpace{}%
\AgdaGeneralizable{n}\AgdaSpace{}%
\AgdaSymbol{→}\AgdaSpace{}%
\AgdaDatatype{Split}\AgdaSpace{}%
\AgdaGeneralizable{m}\AgdaSpace{}%
\AgdaSymbol{(}\AgdaInductiveConstructor{suc}\AgdaSpace{}%
\AgdaGeneralizable{n}\AgdaSymbol{)}\<%
\\
\\[\AgdaEmptyExtraSkip]%
\>[0]\AgdaFunction{apply-split}\AgdaSpace{}%
\AgdaSymbol{:}\AgdaSpace{}%
\AgdaDatatype{Split}\AgdaSpace{}%
\AgdaGeneralizable{m}\AgdaSpace{}%
\AgdaGeneralizable{n}\AgdaSpace{}%
\AgdaSymbol{→}\AgdaSpace{}%
\AgdaDatatype{Vec}\AgdaSpace{}%
\AgdaGeneralizable{A}\AgdaSpace{}%
\AgdaSymbol{(}\AgdaGeneralizable{m}\AgdaSpace{}%
\AgdaOperator{\AgdaPrimitive{+}}\AgdaSpace{}%
\AgdaGeneralizable{n}\AgdaSymbol{)}\AgdaSpace{}%
\AgdaSymbol{→}\AgdaSpace{}%
\AgdaDatatype{Vec}\AgdaSpace{}%
\AgdaGeneralizable{A}\AgdaSpace{}%
\AgdaGeneralizable{m}\AgdaSpace{}%
\AgdaOperator{\AgdaFunction{×}}\AgdaSpace{}%
\AgdaDatatype{Vec}\AgdaSpace{}%
\AgdaGeneralizable{A}\AgdaSpace{}%
\AgdaGeneralizable{n}\<%
\\
\>[0]\AgdaFunction{apply-split}\AgdaSpace{}%
\AgdaInductiveConstructor{null}\AgdaSpace{}%
\AgdaInductiveConstructor{[]}\AgdaSpace{}%
\AgdaSymbol{=}\AgdaSpace{}%
\AgdaOperator{\AgdaInductiveConstructor{⟨}}\AgdaSpace{}%
\AgdaInductiveConstructor{[]}\AgdaSpace{}%
\AgdaOperator{\AgdaInductiveConstructor{,}}\AgdaSpace{}%
\AgdaInductiveConstructor{[]}\AgdaSpace{}%
\AgdaOperator{\AgdaInductiveConstructor{⟩}}\<%
\\
\>[0]\AgdaFunction{apply-split}\AgdaSpace{}%
\AgdaSymbol{(}\AgdaInductiveConstructor{left}\AgdaSpace{}%
\AgdaBound{sp}\AgdaSymbol{)}\AgdaSpace{}%
\AgdaSymbol{(}\AgdaBound{x}\AgdaSpace{}%
\AgdaOperator{\AgdaInductiveConstructor{∷}}\AgdaSpace{}%
\AgdaBound{v}\AgdaSymbol{)}\<%
\\
\>[0][@{}l@{\AgdaIndent{0}}]%
\>[2]\AgdaKeyword{with}\AgdaSpace{}%
\AgdaFunction{apply-split}\AgdaSpace{}%
\AgdaBound{sp}\AgdaSpace{}%
\AgdaBound{v}\<%
\\
\>[0]\AgdaSymbol{...}\AgdaSpace{}%
\AgdaSymbol{|}\AgdaSpace{}%
\AgdaOperator{\AgdaInductiveConstructor{⟨}}\AgdaSpace{}%
\AgdaBound{vl}\AgdaSpace{}%
\AgdaOperator{\AgdaInductiveConstructor{,}}\AgdaSpace{}%
\AgdaBound{vr}\AgdaSpace{}%
\AgdaOperator{\AgdaInductiveConstructor{⟩}}\AgdaSpace{}%
\AgdaSymbol{=}\AgdaSpace{}%
\AgdaOperator{\AgdaInductiveConstructor{⟨}}\AgdaSpace{}%
\AgdaBound{x}\AgdaSpace{}%
\AgdaOperator{\AgdaInductiveConstructor{∷}}\AgdaSpace{}%
\AgdaBound{vl}\AgdaSpace{}%
\AgdaOperator{\AgdaInductiveConstructor{,}}\AgdaSpace{}%
\AgdaBound{vr}\AgdaSpace{}%
\AgdaOperator{\AgdaInductiveConstructor{⟩}}\<%
\\
\>[0]\AgdaFunction{apply-split}\AgdaSymbol{\{}\AgdaBound{m}\AgdaSymbol{\}\{}\AgdaInductiveConstructor{suc}\AgdaSpace{}%
\AgdaBound{n}\AgdaSymbol{\}}\AgdaSpace{}%
\AgdaSymbol{(}\AgdaInductiveConstructor{right}\AgdaSpace{}%
\AgdaBound{sp}\AgdaSymbol{)}\AgdaSpace{}%
\AgdaBound{v}\<%
\\
\>[0][@{}l@{\AgdaIndent{0}}]%
\>[2]\AgdaKeyword{rewrite}\AgdaSpace{}%
\AgdaFunction{+-suc}\AgdaSpace{}%
\AgdaBound{m}\AgdaSpace{}%
\AgdaBound{n}\<%
\\
\>[2]\AgdaKeyword{with}\AgdaSpace{}%
\AgdaBound{v}\<%
\\
\>[0]\AgdaSymbol{...}\AgdaSpace{}%
\AgdaSymbol{|}\AgdaSpace{}%
\AgdaBound{x}\AgdaSpace{}%
\AgdaOperator{\AgdaInductiveConstructor{∷}}\AgdaSpace{}%
\AgdaBound{v}\<%
\\
\>[0][@{}l@{\AgdaIndent{0}}]%
\>[2]\AgdaKeyword{with}\AgdaSpace{}%
\AgdaFunction{apply-split}\AgdaSpace{}%
\AgdaBound{sp}\AgdaSpace{}%
\AgdaBound{v}\<%
\\
\>[0]\AgdaSymbol{...}\AgdaSpace{}%
\AgdaSymbol{|}\AgdaSpace{}%
\AgdaOperator{\AgdaInductiveConstructor{⟨}}\AgdaSpace{}%
\AgdaBound{vl}\AgdaSpace{}%
\AgdaOperator{\AgdaInductiveConstructor{,}}\AgdaSpace{}%
\AgdaBound{vr}\AgdaSpace{}%
\AgdaOperator{\AgdaInductiveConstructor{⟩}}\AgdaSpace{}%
\AgdaSymbol{=}\AgdaSpace{}%
\AgdaOperator{\AgdaInductiveConstructor{⟨}}\AgdaSpace{}%
\AgdaBound{vl}\AgdaSpace{}%
\AgdaOperator{\AgdaInductiveConstructor{,}}\AgdaSpace{}%
\AgdaBound{x}\AgdaSpace{}%
\AgdaOperator{\AgdaInductiveConstructor{∷}}\AgdaSpace{}%
\AgdaBound{vr}\AgdaSpace{}%
\AgdaOperator{\AgdaInductiveConstructor{⟩}}\<%
\\
\\[\AgdaEmptyExtraSkip]%
\>[0]\AgdaFunction{locate-split}\AgdaSpace{}%
\AgdaSymbol{:}\AgdaSpace{}%
\AgdaDatatype{Split}\AgdaSpace{}%
\AgdaGeneralizable{m}\AgdaSpace{}%
\AgdaGeneralizable{n}\AgdaSpace{}%
\AgdaSymbol{→}\AgdaSpace{}%
\AgdaDatatype{Fin}\AgdaSpace{}%
\AgdaSymbol{(}\AgdaGeneralizable{m}\AgdaSpace{}%
\AgdaOperator{\AgdaPrimitive{+}}\AgdaSpace{}%
\AgdaGeneralizable{n}\AgdaSymbol{)}\AgdaSpace{}%
\AgdaSymbol{→}\AgdaSpace{}%
\AgdaDatatype{Fin}\AgdaSpace{}%
\AgdaGeneralizable{m}\AgdaSpace{}%
\AgdaOperator{\AgdaDatatype{⊎}}\AgdaSpace{}%
\AgdaDatatype{Fin}\AgdaSpace{}%
\AgdaGeneralizable{n}\<%
\\
\>[0]\AgdaFunction{locate-split}\AgdaSpace{}%
\AgdaSymbol{(}\AgdaInductiveConstructor{left}\AgdaSpace{}%
\AgdaBound{sp}\AgdaSymbol{)}\AgdaSpace{}%
\AgdaInductiveConstructor{zero}\AgdaSpace{}%
\AgdaSymbol{=}\AgdaSpace{}%
\AgdaInductiveConstructor{inj₁}\AgdaSpace{}%
\AgdaInductiveConstructor{zero}\<%
\\
\>[0]\AgdaFunction{locate-split}\AgdaSpace{}%
\AgdaSymbol{(}\AgdaInductiveConstructor{left}\AgdaSpace{}%
\AgdaBound{sp}\AgdaSymbol{)}\AgdaSpace{}%
\AgdaSymbol{(}\AgdaInductiveConstructor{suc}\AgdaSpace{}%
\AgdaBound{f}\AgdaSymbol{)}\<%
\\
\>[0][@{}l@{\AgdaIndent{0}}]%
\>[2]\AgdaKeyword{with}\AgdaSpace{}%
\AgdaFunction{locate-split}\AgdaSpace{}%
\AgdaBound{sp}\AgdaSpace{}%
\AgdaBound{f}\<%
\\
\>[0]\AgdaSymbol{...}\AgdaSpace{}%
\AgdaSymbol{|}\AgdaSpace{}%
\AgdaInductiveConstructor{inj₁}\AgdaSpace{}%
\AgdaBound{x}\AgdaSpace{}%
\AgdaSymbol{=}\AgdaSpace{}%
\AgdaInductiveConstructor{inj₁}\AgdaSpace{}%
\AgdaSymbol{(}\AgdaInductiveConstructor{suc}\AgdaSpace{}%
\AgdaBound{x}\AgdaSymbol{)}\<%
\\
\>[0]\AgdaSymbol{...}\AgdaSpace{}%
\AgdaSymbol{|}\AgdaSpace{}%
\AgdaInductiveConstructor{inj₂}\AgdaSpace{}%
\AgdaBound{y}\AgdaSpace{}%
\AgdaSymbol{=}\AgdaSpace{}%
\AgdaInductiveConstructor{inj₂}\AgdaSpace{}%
\AgdaBound{y}\<%
\\
\>[0]\AgdaFunction{locate-split}\AgdaSymbol{\{}\AgdaBound{m}\AgdaSymbol{\}\{}\AgdaInductiveConstructor{suc}\AgdaSpace{}%
\AgdaBound{n}\AgdaSymbol{\}}\AgdaSpace{}%
\AgdaSymbol{(}\AgdaInductiveConstructor{right}\AgdaSpace{}%
\AgdaBound{sp}\AgdaSymbol{)}\AgdaSpace{}%
\AgdaBound{f}\<%
\\
\>[0][@{}l@{\AgdaIndent{0}}]%
\>[2]\AgdaKeyword{rewrite}\AgdaSpace{}%
\AgdaFunction{+-suc}\AgdaSpace{}%
\AgdaBound{m}\AgdaSpace{}%
\AgdaBound{n}\<%
\\
\>[2]\AgdaKeyword{with}\AgdaSpace{}%
\AgdaBound{f}\<%
\\
\>[0]\AgdaSymbol{...}\AgdaSpace{}%
\AgdaSymbol{|}\AgdaSpace{}%
\AgdaInductiveConstructor{zero}\AgdaSpace{}%
\AgdaSymbol{=}\AgdaSpace{}%
\AgdaInductiveConstructor{inj₂}\AgdaSpace{}%
\AgdaInductiveConstructor{zero}\<%
\\
\>[0]\AgdaSymbol{...}\AgdaSpace{}%
\AgdaSymbol{|}\AgdaSpace{}%
\AgdaInductiveConstructor{suc}\AgdaSpace{}%
\AgdaBound{f}\<%
\\
\>[0][@{}l@{\AgdaIndent{0}}]%
\>[2]\AgdaKeyword{with}\AgdaSpace{}%
\AgdaFunction{locate-split}\AgdaSpace{}%
\AgdaBound{sp}\AgdaSpace{}%
\AgdaBound{f}\<%
\\
\>[0]\AgdaSymbol{...}\AgdaSpace{}%
\AgdaSymbol{|}\AgdaSpace{}%
\AgdaInductiveConstructor{inj₁}\AgdaSpace{}%
\AgdaBound{x}\AgdaSpace{}%
\AgdaSymbol{=}\AgdaSpace{}%
\AgdaInductiveConstructor{inj₁}\AgdaSpace{}%
\AgdaBound{x}\<%
\\
\>[0]\AgdaSymbol{...}\AgdaSpace{}%
\AgdaSymbol{|}\AgdaSpace{}%
\AgdaInductiveConstructor{inj₂}\AgdaSpace{}%
\AgdaBound{y}\AgdaSpace{}%
\AgdaSymbol{=}\AgdaSpace{}%
\AgdaInductiveConstructor{inj₂}\AgdaSpace{}%
\AgdaSymbol{(}\AgdaInductiveConstructor{suc}\AgdaSpace{}%
\AgdaBound{y}\AgdaSymbol{)}\<%
\\
\\[\AgdaEmptyExtraSkip]%
\>[0]\AgdaComment{--\ session\ types}\<%
\\
\\[\AgdaEmptyExtraSkip]%
\>[0]\AgdaKeyword{data}\AgdaSpace{}%
\AgdaDatatype{Type}\AgdaSpace{}%
\AgdaSymbol{:}\AgdaSpace{}%
\AgdaPrimitive{Set}\AgdaSpace{}%
\AgdaKeyword{where}\<%
\\
\>[0][@{}l@{\AgdaIndent{0}}]%
\>[2]\AgdaInductiveConstructor{nat}\AgdaSpace{}%
\AgdaInductiveConstructor{int}\AgdaSpace{}%
\AgdaSymbol{:}\AgdaSpace{}%
\AgdaDatatype{Type}\<%
\end{code}
\newcommand\multiSession{%
\begin{code}%
\>[0]\AgdaKeyword{data}\AgdaSpace{}%
\AgdaDatatype{Direction}\AgdaSpace{}%
\AgdaSymbol{:}\AgdaSpace{}%
\AgdaPrimitive{Set}\AgdaSpace{}%
\AgdaKeyword{where}\<%
\\
\>[0][@{}l@{\AgdaIndent{0}}]%
\>[2]\AgdaInductiveConstructor{INP}\AgdaSpace{}%
\AgdaInductiveConstructor{OUT}\AgdaSpace{}%
\AgdaSymbol{:}\AgdaSpace{}%
\AgdaDatatype{Direction}\<%
\\
\\[\AgdaEmptyExtraSkip]%
\>[0]\AgdaKeyword{data}\AgdaSpace{}%
\AgdaDatatype{Session}\AgdaSpace{}%
\AgdaSymbol{:}\AgdaSpace{}%
\AgdaPrimitive{Set}\AgdaSpace{}%
\AgdaKeyword{where}\<%
\\
\>[0][@{}l@{\AgdaIndent{0}}]%
\>[2]\AgdaInductiveConstructor{transmit}\AgdaSpace{}%
\AgdaSymbol{:}\AgdaSpace{}%
\AgdaSymbol{(}\AgdaBound{d}\AgdaSpace{}%
\AgdaSymbol{:}\AgdaSpace{}%
\AgdaDatatype{Direction}\AgdaSymbol{)}\AgdaSpace{}%
\AgdaSymbol{→}\AgdaSpace{}%
\AgdaDatatype{Type}%
\>[39]\AgdaSymbol{→}%
\>[42]\AgdaDatatype{Session}\AgdaSpace{}%
\AgdaSymbol{→}\AgdaSpace{}%
\AgdaDatatype{Session}\<%
\\
\>[2]\AgdaInductiveConstructor{delegate}\AgdaSpace{}%
\AgdaSymbol{:}\AgdaSpace{}%
\AgdaSymbol{(}\AgdaBound{d}\AgdaSpace{}%
\AgdaSymbol{:}\AgdaSpace{}%
\AgdaDatatype{Direction}\AgdaSymbol{)}\AgdaSpace{}%
\AgdaSymbol{→}\AgdaSpace{}%
\AgdaDatatype{Session}\AgdaSpace{}%
\AgdaSymbol{→}\AgdaSpace{}%
\AgdaDatatype{Session}\AgdaSpace{}%
\AgdaSymbol{→}\AgdaSpace{}%
\AgdaDatatype{Session}\<%
\\
\>[2]\AgdaInductiveConstructor{branch}\AgdaSpace{}%
\AgdaSymbol{:}\AgdaSpace{}%
\AgdaSymbol{(}\AgdaBound{d}\AgdaSpace{}%
\AgdaSymbol{:}\AgdaSpace{}%
\AgdaDatatype{Direction}\AgdaSymbol{)}\AgdaSpace{}%
\AgdaSymbol{→}\AgdaSpace{}%
\AgdaDatatype{Session}\AgdaSpace{}%
\AgdaSymbol{→}\AgdaSpace{}%
\AgdaDatatype{Session}\AgdaSpace{}%
\AgdaSymbol{→}\AgdaSpace{}%
\AgdaDatatype{Session}\<%
\\
\>[2]\AgdaInductiveConstructor{end}\AgdaSpace{}%
\AgdaSymbol{:}\AgdaSpace{}%
\AgdaDatatype{Session}\<%
\end{code}}
\newcommand\multiMSesson{%
\begin{code}%
\>[0]\AgdaKeyword{data}\AgdaSpace{}%
\AgdaDatatype{MSession}\AgdaSpace{}%
\AgdaSymbol{:}\AgdaSpace{}%
\AgdaDatatype{ℕ}\AgdaSpace{}%
\AgdaSymbol{→}\AgdaSpace{}%
\AgdaPrimitive{Set}\<%
\\
\>[0]\AgdaKeyword{variable}\AgdaSpace{}%
\AgdaGeneralizable{M}\AgdaSpace{}%
\AgdaGeneralizable{M₁}\AgdaSpace{}%
\AgdaGeneralizable{M₂}\AgdaSpace{}%
\AgdaSymbol{:}\AgdaSpace{}%
\AgdaDatatype{MSession}\AgdaSpace{}%
\AgdaGeneralizable{n}\<%
\\
\\[\AgdaEmptyExtraSkip]%
\>[0]\AgdaFunction{Causality}\AgdaSpace{}%
\AgdaSymbol{:}\AgdaSpace{}%
\AgdaDatatype{Fin}\AgdaSpace{}%
\AgdaGeneralizable{n}\AgdaSpace{}%
\AgdaSymbol{→}\AgdaSpace{}%
\AgdaDatatype{MSession}\AgdaSpace{}%
\AgdaGeneralizable{n}\AgdaSpace{}%
\AgdaSymbol{→}\AgdaSpace{}%
\AgdaDatatype{MSession}\AgdaSpace{}%
\AgdaGeneralizable{n}\AgdaSpace{}%
\AgdaSymbol{→}\AgdaSpace{}%
\AgdaPrimitive{Set}\<%
\\
\>[0]\AgdaFunction{CheckDual0}\AgdaSpace{}%
\AgdaSymbol{:}\AgdaSpace{}%
\AgdaDatatype{MSession}\AgdaSpace{}%
\AgdaSymbol{(}\AgdaInductiveConstructor{suc}\AgdaSpace{}%
\AgdaGeneralizable{m}\AgdaSymbol{)}\AgdaSpace{}%
\AgdaSymbol{→}\AgdaSpace{}%
\AgdaDatatype{MSession}\AgdaSpace{}%
\AgdaSymbol{(}\AgdaInductiveConstructor{suc}\AgdaSpace{}%
\AgdaGeneralizable{n}\AgdaSymbol{)}\AgdaSpace{}%
\AgdaSymbol{→}\AgdaSpace{}%
\AgdaPrimitive{Set}\<%
\\
\\[\AgdaEmptyExtraSkip]%
\>[0]\AgdaKeyword{data}\AgdaSpace{}%
\AgdaDatatype{MSession}\AgdaSpace{}%
\AgdaKeyword{where}\<%
\\
\>[0][@{}l@{\AgdaIndent{0}}]%
\>[2]\AgdaInductiveConstructor{transmit}\AgdaSpace{}%
\AgdaSymbol{:}\AgdaSpace{}%
\AgdaSymbol{(}\AgdaBound{d}\AgdaSpace{}%
\AgdaSymbol{:}\AgdaSpace{}%
\AgdaDatatype{Direction}\AgdaSymbol{)}\AgdaSpace{}%
\AgdaSymbol{→}\AgdaSpace{}%
\AgdaSymbol{(}\AgdaBound{c}\AgdaSpace{}%
\AgdaSymbol{:}\AgdaSpace{}%
\AgdaDatatype{Fin}\AgdaSpace{}%
\AgdaGeneralizable{n}\AgdaSymbol{)}\AgdaSpace{}%
\AgdaSymbol{→}\AgdaSpace{}%
\AgdaSymbol{(}\AgdaBound{T}\AgdaSpace{}%
\AgdaSymbol{:}\AgdaSpace{}%
\AgdaDatatype{Type}\AgdaSymbol{)}\AgdaSpace{}%
\AgdaSymbol{→}\AgdaSpace{}%
\AgdaDatatype{MSession}\AgdaSpace{}%
\AgdaGeneralizable{n}\AgdaSpace{}%
\AgdaSymbol{→}\AgdaSpace{}%
\AgdaDatatype{MSession}\AgdaSpace{}%
\AgdaGeneralizable{n}\<%
\\
\>[2]\AgdaInductiveConstructor{branch}%
\>[11]\AgdaSymbol{:}\AgdaSpace{}%
\AgdaSymbol{(}\AgdaBound{d}\AgdaSpace{}%
\AgdaSymbol{:}\AgdaSpace{}%
\AgdaDatatype{Direction}\AgdaSymbol{)}\AgdaSpace{}%
\AgdaSymbol{→}\AgdaSpace{}%
\AgdaSymbol{(}\AgdaBound{c}\AgdaSpace{}%
\AgdaSymbol{:}\AgdaSpace{}%
\AgdaDatatype{Fin}\AgdaSpace{}%
\AgdaGeneralizable{n}\AgdaSymbol{)}\AgdaSpace{}%
\AgdaSymbol{→}\AgdaSpace{}%
\AgdaSymbol{(}\AgdaBound{M₁}\AgdaSpace{}%
\AgdaSymbol{:}\AgdaSpace{}%
\AgdaDatatype{MSession}\AgdaSpace{}%
\AgdaGeneralizable{n}\AgdaSymbol{)}\AgdaSpace{}%
\AgdaSymbol{→}\AgdaSpace{}%
\AgdaSymbol{(}\AgdaBound{M₂}\AgdaSpace{}%
\AgdaSymbol{:}\AgdaSpace{}%
\AgdaDatatype{MSession}\AgdaSpace{}%
\AgdaGeneralizable{n}\AgdaSymbol{)}\<%
\\
\>[2][@{}l@{\AgdaIndent{0}}]%
\>[4]\AgdaSymbol{→}\AgdaSpace{}%
\AgdaFunction{Causality}\AgdaSpace{}%
\AgdaBound{c}\AgdaSpace{}%
\AgdaBound{M₁}\AgdaSpace{}%
\AgdaBound{M₂}\AgdaSpace{}%
\AgdaSymbol{→}\AgdaSpace{}%
\AgdaDatatype{MSession}\AgdaSpace{}%
\AgdaGeneralizable{n}\<%
\\
\>[2]\AgdaInductiveConstructor{close}\AgdaSpace{}%
\AgdaSymbol{:}\AgdaSpace{}%
\AgdaSymbol{(}\AgdaBound{c}\AgdaSpace{}%
\AgdaSymbol{:}\AgdaSpace{}%
\AgdaDatatype{Fin}\AgdaSpace{}%
\AgdaSymbol{(}\AgdaInductiveConstructor{suc}\AgdaSpace{}%
\AgdaGeneralizable{n}\AgdaSymbol{))}\AgdaSpace{}%
\AgdaSymbol{→}\AgdaSpace{}%
\AgdaDatatype{MSession}\AgdaSpace{}%
\AgdaGeneralizable{n}\AgdaSpace{}%
\AgdaSymbol{→}\AgdaSpace{}%
\AgdaDatatype{MSession}\AgdaSpace{}%
\AgdaSymbol{(}\AgdaInductiveConstructor{suc}\AgdaSpace{}%
\AgdaGeneralizable{n}\AgdaSymbol{)}\<%
\\
\>[2]\AgdaInductiveConstructor{terminate}\AgdaSpace{}%
\AgdaSymbol{:}\AgdaSpace{}%
\AgdaDatatype{MSession}\AgdaSpace{}%
\AgdaInductiveConstructor{zero}\<%
\\
\>[2]\AgdaInductiveConstructor{connect}\AgdaSpace{}%
\AgdaSymbol{:}\AgdaSpace{}%
\AgdaDatatype{Split}\AgdaSpace{}%
\AgdaGeneralizable{m}\AgdaSpace{}%
\AgdaGeneralizable{n}\AgdaSpace{}%
\AgdaSymbol{→}\AgdaSpace{}%
\AgdaSymbol{(}\AgdaBound{M₁}\AgdaSpace{}%
\AgdaSymbol{:}\AgdaSpace{}%
\AgdaDatatype{MSession}\AgdaSpace{}%
\AgdaSymbol{(}\AgdaInductiveConstructor{suc}\AgdaSpace{}%
\AgdaGeneralizable{m}\AgdaSymbol{))}\AgdaSpace{}%
\AgdaSymbol{→}\AgdaSpace{}%
\AgdaSymbol{(}\AgdaBound{M₂}\AgdaSpace{}%
\AgdaSymbol{:}\AgdaSpace{}%
\AgdaDatatype{MSession}\AgdaSpace{}%
\AgdaSymbol{(}\AgdaInductiveConstructor{suc}\AgdaSpace{}%
\AgdaGeneralizable{n}\AgdaSymbol{))}\<%
\\
\>[2][@{}l@{\AgdaIndent{0}}]%
\>[4]\AgdaSymbol{→}\AgdaSpace{}%
\AgdaFunction{CheckDual0}\AgdaSpace{}%
\AgdaBound{M₁}\AgdaSpace{}%
\AgdaBound{M₂}\AgdaSpace{}%
\AgdaSymbol{→}\AgdaSpace{}%
\AgdaDatatype{MSession}\AgdaSpace{}%
\AgdaSymbol{(}\AgdaGeneralizable{m}\AgdaSpace{}%
\AgdaOperator{\AgdaPrimitive{+}}\AgdaSpace{}%
\AgdaGeneralizable{n}\AgdaSymbol{)}\<%
\\
\>[2]\AgdaComment{--\ assume\ new\ channel\ has\ address\ zero\ in\ both\ threads}\<%
\\
\>[2]\AgdaInductiveConstructor{delegateOUT}\AgdaSpace{}%
\AgdaSymbol{:}\AgdaSpace{}%
\AgdaSymbol{(}\AgdaBound{c}\AgdaSpace{}%
\AgdaBound{j}\AgdaSpace{}%
\AgdaSymbol{:}\AgdaSpace{}%
\AgdaDatatype{Fin}\AgdaSpace{}%
\AgdaSymbol{(}\AgdaInductiveConstructor{suc}\AgdaSpace{}%
\AgdaGeneralizable{n}\AgdaSymbol{))}\AgdaSpace{}%
\AgdaSymbol{→}\AgdaSpace{}%
\AgdaBound{c}\AgdaSpace{}%
\AgdaOperator{\AgdaFunction{≢}}\AgdaSpace{}%
\AgdaBound{j}\AgdaSpace{}%
\AgdaSymbol{→}\AgdaSpace{}%
\AgdaDatatype{Session}\AgdaSpace{}%
\AgdaSymbol{→}\AgdaSpace{}%
\AgdaDatatype{MSession}\AgdaSpace{}%
\AgdaGeneralizable{n}\AgdaSpace{}%
\AgdaSymbol{→}\AgdaSpace{}%
\AgdaDatatype{MSession}\AgdaSpace{}%
\AgdaSymbol{(}\AgdaInductiveConstructor{suc}\AgdaSpace{}%
\AgdaGeneralizable{n}\AgdaSymbol{)}\<%
\\
\>[2]\AgdaInductiveConstructor{delegateIN}%
\>[14]\AgdaSymbol{:}\AgdaSpace{}%
\AgdaSymbol{(}\AgdaBound{c}\AgdaSpace{}%
\AgdaSymbol{:}\AgdaSpace{}%
\AgdaDatatype{Fin}\AgdaSpace{}%
\AgdaGeneralizable{n}\AgdaSymbol{)}\AgdaSpace{}%
\AgdaSymbol{→}\AgdaSpace{}%
\AgdaDatatype{MSession}\AgdaSpace{}%
\AgdaSymbol{(}\AgdaInductiveConstructor{suc}\AgdaSpace{}%
\AgdaGeneralizable{n}\AgdaSymbol{)}\AgdaSpace{}%
\AgdaSymbol{→}\AgdaSpace{}%
\AgdaDatatype{MSession}\AgdaSpace{}%
\AgdaGeneralizable{n}\<%
\\
\>[2]\AgdaComment{--\ received\ channel\ has\ address\ zero\ in\ continuation}\<%
\end{code}}
\begin{code}[hide]%
\>[0]\<%
\\
\>[0]\AgdaKeyword{pattern}\AgdaSpace{}%
\AgdaInductiveConstructor{select}\AgdaSpace{}%
\AgdaBound{x}\AgdaSpace{}%
\AgdaBound{s₁}\AgdaSpace{}%
\AgdaBound{s₂}\AgdaSpace{}%
\AgdaBound{p}\AgdaSpace{}%
\AgdaSymbol{=}\AgdaSpace{}%
\AgdaInductiveConstructor{branch}\AgdaSpace{}%
\AgdaInductiveConstructor{OUT}\AgdaSpace{}%
\AgdaBound{x}\AgdaSpace{}%
\AgdaBound{s₁}\AgdaSpace{}%
\AgdaBound{s₂}\AgdaSpace{}%
\AgdaBound{p}\<%
\\
\>[0]\AgdaKeyword{pattern}\AgdaSpace{}%
\AgdaInductiveConstructor{choice}\AgdaSpace{}%
\AgdaBound{x}\AgdaSpace{}%
\AgdaBound{s₁}\AgdaSpace{}%
\AgdaBound{s₂}\AgdaSpace{}%
\AgdaBound{p}\AgdaSpace{}%
\AgdaSymbol{=}\AgdaSpace{}%
\AgdaInductiveConstructor{branch}\AgdaSpace{}%
\AgdaInductiveConstructor{INP}\AgdaSpace{}%
\AgdaBound{x}\AgdaSpace{}%
\AgdaBound{s₁}\AgdaSpace{}%
\AgdaBound{s₂}\AgdaSpace{}%
\AgdaBound{p}\<%
\\
\\[\AgdaEmptyExtraSkip]%
\>[0]\AgdaKeyword{pattern}\AgdaSpace{}%
\AgdaInductiveConstructor{recv}\AgdaSpace{}%
\AgdaBound{x}\AgdaSpace{}%
\AgdaBound{t}\AgdaSpace{}%
\AgdaBound{s}\AgdaSpace{}%
\AgdaSymbol{=}\AgdaSpace{}%
\AgdaInductiveConstructor{transmit}\AgdaSpace{}%
\AgdaInductiveConstructor{INP}\AgdaSpace{}%
\AgdaBound{x}\AgdaSpace{}%
\AgdaBound{t}\AgdaSpace{}%
\AgdaBound{s}\<%
\\
\>[0]\AgdaKeyword{pattern}\AgdaSpace{}%
\AgdaInductiveConstructor{send}\AgdaSpace{}%
\AgdaBound{x}\AgdaSpace{}%
\AgdaBound{t}\AgdaSpace{}%
\AgdaBound{s}\AgdaSpace{}%
\AgdaSymbol{=}\AgdaSpace{}%
\AgdaInductiveConstructor{transmit}\AgdaSpace{}%
\AgdaInductiveConstructor{OUT}\AgdaSpace{}%
\AgdaBound{x}\AgdaSpace{}%
\AgdaBound{t}\AgdaSpace{}%
\AgdaBound{s}\<%
\\
\\[\AgdaEmptyExtraSkip]%
\>[0]\AgdaComment{--\ adjust\ index\ f\ if\ index\ x\ is\ removed\ from\ set}\<%
\\
\\[\AgdaEmptyExtraSkip]%
\>[0]\AgdaFunction{adjust}\AgdaSpace{}%
\AgdaSymbol{:}\AgdaSpace{}%
\AgdaSymbol{(}\AgdaBound{f}\AgdaSpace{}%
\AgdaSymbol{:}\AgdaSpace{}%
\AgdaDatatype{Fin}\AgdaSpace{}%
\AgdaSymbol{(}\AgdaInductiveConstructor{suc}\AgdaSpace{}%
\AgdaGeneralizable{n}\AgdaSymbol{))}\AgdaSpace{}%
\AgdaSymbol{(}\AgdaBound{x}\AgdaSpace{}%
\AgdaSymbol{:}\AgdaSpace{}%
\AgdaDatatype{Fin}\AgdaSpace{}%
\AgdaSymbol{(}\AgdaInductiveConstructor{suc}\AgdaSpace{}%
\AgdaGeneralizable{n}\AgdaSymbol{))}\AgdaSpace{}%
\AgdaSymbol{→}\AgdaSpace{}%
\AgdaBound{f}\AgdaSpace{}%
\AgdaOperator{\AgdaFunction{≢}}\AgdaSpace{}%
\AgdaBound{x}\AgdaSpace{}%
\AgdaSymbol{→}\AgdaSpace{}%
\AgdaDatatype{Fin}\AgdaSpace{}%
\AgdaGeneralizable{n}\<%
\\
\>[0]\AgdaFunction{adjust}\AgdaSpace{}%
\AgdaInductiveConstructor{zero}\AgdaSpace{}%
\AgdaInductiveConstructor{zero}\AgdaSpace{}%
\AgdaBound{f≢x}\AgdaSpace{}%
\AgdaSymbol{=}\AgdaSpace{}%
\AgdaFunction{⊥-elim}\AgdaSpace{}%
\AgdaSymbol{(}\AgdaBound{f≢x}\AgdaSpace{}%
\AgdaInductiveConstructor{refl}\AgdaSymbol{)}\<%
\\
\>[0]\AgdaFunction{adjust}\AgdaSymbol{\{}\AgdaInductiveConstructor{suc}\AgdaSpace{}%
\AgdaBound{n}\AgdaSymbol{\}}\AgdaSpace{}%
\AgdaInductiveConstructor{zero}\AgdaSpace{}%
\AgdaSymbol{(}\AgdaInductiveConstructor{suc}\AgdaSpace{}%
\AgdaBound{x}\AgdaSymbol{)}\AgdaSpace{}%
\AgdaBound{f≢x}\AgdaSpace{}%
\AgdaSymbol{=}\AgdaSpace{}%
\AgdaInductiveConstructor{zero}\<%
\\
\>[0]\AgdaFunction{adjust}\AgdaSymbol{\{}\AgdaInductiveConstructor{suc}\AgdaSpace{}%
\AgdaBound{n}\AgdaSymbol{\}}\AgdaSpace{}%
\AgdaSymbol{(}\AgdaInductiveConstructor{suc}\AgdaSpace{}%
\AgdaBound{f}\AgdaSymbol{)}\AgdaSpace{}%
\AgdaInductiveConstructor{zero}\AgdaSpace{}%
\AgdaBound{f≢x}\AgdaSpace{}%
\AgdaSymbol{=}\AgdaSpace{}%
\AgdaBound{f}\<%
\\
\>[0]\AgdaFunction{adjust}\AgdaSymbol{\{}\AgdaInductiveConstructor{suc}\AgdaSpace{}%
\AgdaBound{n}\AgdaSymbol{\}}\AgdaSpace{}%
\AgdaSymbol{(}\AgdaInductiveConstructor{suc}\AgdaSpace{}%
\AgdaBound{f}\AgdaSymbol{)}\AgdaSpace{}%
\AgdaSymbol{(}\AgdaInductiveConstructor{suc}\AgdaSpace{}%
\AgdaBound{x}\AgdaSymbol{)}\AgdaSpace{}%
\AgdaBound{f≢x}\<%
\\
\>[0][@{}l@{\AgdaIndent{0}}]%
\>[2]\AgdaKeyword{with}\AgdaSpace{}%
\AgdaFunction{adjust}\AgdaSpace{}%
\AgdaBound{f}\AgdaSpace{}%
\AgdaBound{x}\AgdaSpace{}%
\AgdaSymbol{(λ\{}\AgdaSpace{}%
\AgdaInductiveConstructor{refl}\AgdaSpace{}%
\AgdaSymbol{→}\AgdaSpace{}%
\AgdaBound{f≢x}\AgdaSpace{}%
\AgdaInductiveConstructor{refl}\AgdaSymbol{\})}\<%
\\
\>[0]\AgdaSymbol{...}\AgdaSpace{}%
\AgdaSymbol{|}\AgdaSpace{}%
\AgdaBound{r}\AgdaSpace{}%
\AgdaSymbol{=}\AgdaSpace{}%
\AgdaInductiveConstructor{suc}\AgdaSpace{}%
\AgdaBound{r}\<%
\\
\\[\AgdaEmptyExtraSkip]%
\>[0]\AgdaComment{--\ duality}\<%
\\
\\[\AgdaEmptyExtraSkip]%
\>[0]\AgdaComment{\{-}\<%
\\
\>[0]\AgdaComment{is-dual\ :\ Session\ →\ Session\ →\ Set}\<%
\\
\>[0]\AgdaComment{is-dual\ (branch\ INP\ s₁\ s₂)\ (branch\ INP\ s₃\ s₄)\ =\ ⊥}\<%
\\
\>[0]\AgdaComment{is-dual\ (branch\ INP\ s₁\ s₂)\ (branch\ OUT\ s₃\ s₄)\ =\ is-dual\ s₁\ s₃\ ×\ is-dual\ s₂\ s₄}\<%
\\
\>[0]\AgdaComment{is-dual\ (branch\ OUT\ s₁\ s₂)\ (branch\ INP\ s₃\ s₄)\ =\ is-dual\ s₁\ s₃\ ×\ is-dual\ s₂\ s₄}\<%
\\
\>[0]\AgdaComment{is-dual\ (branch\ OUT\ s₁\ s₂)\ (branch\ OUT\ s₃\ s₄)\ =\ ⊥}\<%
\\
\>[0]\AgdaComment{is-dual\ (branch\ x\ s₁\ s₂)\ (transmit\ x₁\ x₂\ s₃)\ =\ ⊥}\<%
\\
\>[0]\AgdaComment{is-dual\ (branch\ x\ s₁\ s₂)\ end\ =\ ⊥}\<%
\\
\>[0]\AgdaComment{is-dual\ (transmit\ x\ x₁\ s₁)\ (branch\ x₂\ s₂\ s₃)\ =\ ⊥}\<%
\\
\>[0]\AgdaComment{is-dual\ (transmit\ INP\ x₁\ s₁)\ (transmit\ INP\ x₃\ s₂)\ =\ ⊥}\<%
\\
\>[0]\AgdaComment{is-dual\ (transmit\ INP\ nat\ s₁)\ (transmit\ OUT\ nat\ s₂)\ =\ is-dual\ s₁\ s₂}\<%
\\
\>[0]\AgdaComment{is-dual\ (transmit\ OUT\ nat\ s₁)\ (transmit\ INP\ nat\ s₂)\ =\ is-dual\ s₁\ s₂}\<%
\\
\>[0]\AgdaComment{is-dual\ (transmit\ OUT\ nat\ s₁)\ (transmit\ OUT\ nat\ s₂)\ =\ ⊥}\<%
\\
\>[0]\AgdaComment{is-dual\ (transmit\ x\ x₁\ s₁)\ end\ =\ ⊥}\<%
\\
\>[0]\AgdaComment{is-dual\ end\ (branch\ x\ s₂\ s₃)\ =\ ⊥}\<%
\\
\>[0]\AgdaComment{is-dual\ end\ (transmit\ x\ x₁\ s₂)\ =\ ⊥}\<%
\\
\>[0]\AgdaComment{is-dual\ end\ end\ =\ ⊤}\<%
\\
\>[0]\AgdaComment{-\}}\<%
\\
\\[\AgdaEmptyExtraSkip]%
\>[0]\AgdaFunction{dual-dir}\AgdaSpace{}%
\AgdaSymbol{:}\AgdaSpace{}%
\AgdaDatatype{Direction}\AgdaSpace{}%
\AgdaSymbol{→}\AgdaSpace{}%
\AgdaDatatype{Direction}\<%
\\
\>[0]\AgdaFunction{dual-dir}\AgdaSpace{}%
\AgdaInductiveConstructor{INP}\AgdaSpace{}%
\AgdaSymbol{=}\AgdaSpace{}%
\AgdaInductiveConstructor{OUT}\<%
\\
\>[0]\AgdaFunction{dual-dir}\AgdaSpace{}%
\AgdaInductiveConstructor{OUT}\AgdaSpace{}%
\AgdaSymbol{=}\AgdaSpace{}%
\AgdaInductiveConstructor{INP}\<%
\\
\\[\AgdaEmptyExtraSkip]%
\>[0]\AgdaFunction{dual}\AgdaSpace{}%
\AgdaSymbol{:}\AgdaSpace{}%
\AgdaDatatype{Session}\AgdaSpace{}%
\AgdaSymbol{→}\AgdaSpace{}%
\AgdaDatatype{Session}\<%
\\
\>[0]\AgdaFunction{dual}\AgdaSpace{}%
\AgdaSymbol{(}\AgdaInductiveConstructor{branch}\AgdaSpace{}%
\AgdaBound{d}\AgdaSpace{}%
\AgdaBound{s₁}\AgdaSpace{}%
\AgdaBound{s₂}\AgdaSymbol{)}\AgdaSpace{}%
\AgdaSymbol{=}\AgdaSpace{}%
\AgdaInductiveConstructor{branch}\AgdaSpace{}%
\AgdaSymbol{(}\AgdaFunction{dual-dir}\AgdaSpace{}%
\AgdaBound{d}\AgdaSymbol{)}\AgdaSpace{}%
\AgdaSymbol{(}\AgdaFunction{dual}\AgdaSpace{}%
\AgdaBound{s₁}\AgdaSymbol{)}\AgdaSpace{}%
\AgdaSymbol{(}\AgdaFunction{dual}\AgdaSpace{}%
\AgdaBound{s₂}\AgdaSymbol{)}\<%
\\
\>[0]\AgdaFunction{dual}\AgdaSpace{}%
\AgdaSymbol{(}\AgdaInductiveConstructor{transmit}\AgdaSpace{}%
\AgdaBound{d}\AgdaSpace{}%
\AgdaBound{t}\AgdaSpace{}%
\AgdaBound{s}\AgdaSymbol{)}\AgdaSpace{}%
\AgdaSymbol{=}\AgdaSpace{}%
\AgdaInductiveConstructor{transmit}\AgdaSpace{}%
\AgdaSymbol{(}\AgdaFunction{dual-dir}\AgdaSpace{}%
\AgdaBound{d}\AgdaSymbol{)}\AgdaSpace{}%
\AgdaBound{t}\AgdaSpace{}%
\AgdaSymbol{(}\AgdaFunction{dual}\AgdaSpace{}%
\AgdaBound{s}\AgdaSymbol{)}\<%
\\
\>[0]\AgdaFunction{dual}\AgdaSpace{}%
\AgdaSymbol{(}\AgdaInductiveConstructor{delegate}\AgdaSpace{}%
\AgdaBound{d}\AgdaSpace{}%
\AgdaBound{s₀}\AgdaSpace{}%
\AgdaBound{s}\AgdaSymbol{)}\AgdaSpace{}%
\AgdaSymbol{=}\AgdaSpace{}%
\AgdaInductiveConstructor{delegate}\AgdaSpace{}%
\AgdaSymbol{(}\AgdaFunction{dual-dir}\AgdaSpace{}%
\AgdaBound{d}\AgdaSymbol{)}\AgdaSpace{}%
\AgdaBound{s₀}\AgdaSpace{}%
\AgdaSymbol{(}\AgdaFunction{dual}\AgdaSpace{}%
\AgdaBound{s}\AgdaSymbol{)}\<%
\\
\>[0]\AgdaFunction{dual}\AgdaSpace{}%
\AgdaInductiveConstructor{end}\AgdaSpace{}%
\AgdaSymbol{=}\AgdaSpace{}%
\AgdaInductiveConstructor{end}\<%
\\
\\[\AgdaEmptyExtraSkip]%
\>[0]\AgdaComment{\{-}\<%
\\
\>[0]\AgdaComment{dual→is-dual\ :\ ∀\ s₁\ s₂\ →\ dual\ s₁\ ≡\ s₂\ →\ is-dual\ s₁\ s₂}\<%
\\
\>[0]\AgdaComment{dual→is-dual\ (branch\ INP\ s₁\ s₂)\ (branch\ .(dual-dir\ INP)\ .(dual\ s₁)\ .(dual\ s₂))\ refl\ =\ dual→is-dual\ s₁\ (dual\ s₁)\ refl\ ,\ dual→is-dual\ s₂\ (dual\ s₂)\ refl}\<%
\\
\>[0]\AgdaComment{dual→is-dual\ (branch\ OUT\ s₁\ s₂)\ (branch\ .(dual-dir\ OUT)\ .(dual\ s₁)\ .(dual\ s₂))\ refl\ =\ dual→is-dual\ s₁\ (dual\ s₁)\ refl\ ,\ dual→is-dual\ s₂\ (dual\ s₂)\ refl}\<%
\\
\>[0]\AgdaComment{dual→is-dual\ (branch\ d\ s₁\ s₂)\ (transmit\ d₁\ x\ s₃)\ ()}\<%
\\
\>[0]\AgdaComment{dual→is-dual\ (branch\ d\ s₁\ s₂)\ end\ ()}\<%
\\
\>[0]\AgdaComment{dual→is-dual\ (transmit\ d\ x\ s₁)\ (branch\ d₁\ s₂\ s₃)\ ()}\<%
\\
\>[0]\AgdaComment{dual→is-dual\ (transmit\ INP\ nat\ s₁)\ (transmit\ .(dual-dir\ INP)\ .nat\ .(dual\ s₁))\ refl\ =\ dual→is-dual\ s₁\ (dual\ s₁)\ refl}\<%
\\
\>[0]\AgdaComment{dual→is-dual\ (transmit\ OUT\ nat\ s₁)\ (transmit\ .(dual-dir\ OUT)\ .nat\ .(dual\ s₁))\ refl\ =\ dual→is-dual\ s₁\ (dual\ s₁)\ refl}\<%
\\
\>[0]\AgdaComment{dual→is-dual\ (transmit\ d\ x\ s₁)\ end\ ()}\<%
\\
\>[0]\AgdaComment{dual→is-dual\ end\ (branch\ d\ s₂\ s₃)\ ()}\<%
\\
\>[0]\AgdaComment{dual→is-dual\ end\ (transmit\ d\ x\ s₂)\ ()}\<%
\\
\>[0]\AgdaComment{dual→is-dual\ end\ end\ refl\ =\ tt}\<%
\\
\>[0]\<%
\\
\>[0]\AgdaComment{is-dual→dual\ :\ ∀\ s₁\ s₂\ →\ is-dual\ s₁\ s₂\ →\ dual\ s₁\ ≡\ s₂}\<%
\\
\>[0]\AgdaComment{is-dual→dual\ (branch\ INP\ s₁\ s₂)\ (branch\ OUT\ s₃\ s₄)\ (isd-s₁\ ,\ isd-s₂)}\<%
\\
\>[0]\AgdaComment{\ \ rewrite\ is-dual→dual\ s₁\ s₃\ isd-s₁}\<%
\\
\>[0]\AgdaComment{\ \ |\ \ is-dual→dual\ s₂\ s₄\ isd-s₂\ =\ refl}\<%
\\
\>[0]\AgdaComment{is-dual→dual\ (branch\ OUT\ s₁\ s₂)\ (branch\ INP\ s₃\ s₄)\ (isd-s₁\ ,\ isd-s₂)}\<%
\\
\>[0]\AgdaComment{\ \ rewrite\ is-dual→dual\ s₁\ s₃\ isd-s₁}\<%
\\
\>[0]\AgdaComment{\ \ |\ \ is-dual→dual\ s₂\ s₄\ isd-s₂\ =\ refl}\<%
\\
\>[0]\AgdaComment{is-dual→dual\ (branch\ INP\ s₁\ s₂)\ (transmit\ INP\ nat\ s₃)\ ()}\<%
\\
\>[0]\AgdaComment{is-dual→dual\ (branch\ INP\ s₁\ s₂)\ (transmit\ OUT\ nat\ s₃)\ ()}\<%
\\
\>[0]\AgdaComment{is-dual→dual\ (branch\ OUT\ s₁\ s₂)\ (transmit\ INP\ nat\ s₃)\ ()}\<%
\\
\>[0]\AgdaComment{is-dual→dual\ (branch\ OUT\ s₁\ s₂)\ (transmit\ OUT\ nat\ s₃)\ ()}\<%
\\
\>[0]\AgdaComment{is-dual→dual\ (branch\ INP\ s₁\ s₂)\ end\ ()}\<%
\\
\>[0]\AgdaComment{is-dual→dual\ (branch\ OUT\ s₁\ s₂)\ end\ ()}\<%
\\
\>[0]\AgdaComment{is-dual→dual\ (transmit\ INP\ nat\ s₁)\ (transmit\ OUT\ nat\ s₂)\ isd-s₁-s₂\ rewrite\ is-dual→dual\ s₁\ s₂\ isd-s₁-s₂\ =\ refl}\<%
\\
\>[0]\AgdaComment{is-dual→dual\ (transmit\ OUT\ nat\ s₁)\ (transmit\ INP\ nat\ s₂)\ isd-s₁-s₂\ rewrite\ is-dual→dual\ s₁\ s₂\ isd-s₁-s₂\ =\ refl}\<%
\\
\>[0]\AgdaComment{is-dual→dual\ end\ end\ isd-s₁-s₂\ =\ refl}\<%
\\
\>[0]\AgdaComment{-\}}\<%
\\
\\[\AgdaEmptyExtraSkip]%
\>[0]\AgdaComment{--\ projection}\<%
\\
\\[\AgdaEmptyExtraSkip]%
\>[0]\<%
\end{code}
\newcommand\multiProjection{%
\begin{code}%
\>[0]\AgdaFunction{project}\AgdaSpace{}%
\AgdaSymbol{:}\AgdaSpace{}%
\AgdaDatatype{Fin}\AgdaSpace{}%
\AgdaGeneralizable{n}\AgdaSpace{}%
\AgdaSymbol{→}\AgdaSpace{}%
\AgdaDatatype{MSession}\AgdaSpace{}%
\AgdaGeneralizable{n}\AgdaSpace{}%
\AgdaSymbol{→}\AgdaSpace{}%
\AgdaDatatype{Session}\<%
\\
\>[0]\AgdaFunction{project}\AgdaSpace{}%
\AgdaBound{c}\AgdaSpace{}%
\AgdaSymbol{(}\AgdaInductiveConstructor{connect}\AgdaSpace{}%
\AgdaBound{sp-c}\AgdaSpace{}%
\AgdaBound{M₁}\AgdaSpace{}%
\AgdaBound{M₂}\AgdaSpace{}%
\AgdaSymbol{\AgdaUnderscore{})}\AgdaSpace{}%
\AgdaKeyword{with}\AgdaSpace{}%
\AgdaFunction{locate-split}\AgdaSpace{}%
\AgdaBound{sp-c}\AgdaSpace{}%
\AgdaBound{c}\<%
\\
\>[0]\AgdaSymbol{...}\AgdaSpace{}%
\AgdaSymbol{|}\AgdaSpace{}%
\AgdaInductiveConstructor{inj₁}\AgdaSpace{}%
\AgdaBound{x}\AgdaSpace{}%
\AgdaSymbol{=}\AgdaSpace{}%
\AgdaFunction{project}\AgdaSpace{}%
\AgdaSymbol{(}\AgdaInductiveConstructor{suc}\AgdaSpace{}%
\AgdaBound{x}\AgdaSymbol{)}\AgdaSpace{}%
\AgdaBound{M₁}\<%
\\
\>[0]\AgdaSymbol{...}\AgdaSpace{}%
\AgdaSymbol{|}\AgdaSpace{}%
\AgdaInductiveConstructor{inj₂}\AgdaSpace{}%
\AgdaBound{y}\AgdaSpace{}%
\AgdaSymbol{=}\AgdaSpace{}%
\AgdaFunction{project}\AgdaSpace{}%
\AgdaSymbol{(}\AgdaInductiveConstructor{suc}\AgdaSpace{}%
\AgdaBound{y}\AgdaSymbol{)}\AgdaSpace{}%
\AgdaBound{M₂}\<%
\\
\>[0]\AgdaFunction{project}\AgdaSpace{}%
\AgdaBound{c}\AgdaSpace{}%
\AgdaSymbol{(}\AgdaInductiveConstructor{branch}\AgdaSpace{}%
\AgdaBound{d}\AgdaSpace{}%
\AgdaBound{x}\AgdaSpace{}%
\AgdaBound{M₁}\AgdaSpace{}%
\AgdaBound{M₂}\AgdaSpace{}%
\AgdaBound{causal}\AgdaSymbol{)}\AgdaSpace{}%
\AgdaKeyword{with}\AgdaSpace{}%
\AgdaBound{c}\AgdaSpace{}%
\AgdaOperator{\AgdaFunction{≟}}\AgdaSpace{}%
\AgdaBound{x}\<%
\\
\>[0]\AgdaSymbol{...}\AgdaSpace{}%
\AgdaSymbol{|}\AgdaSpace{}%
\AgdaInductiveConstructor{no}\AgdaSpace{}%
\AgdaBound{c≢x}\AgdaSpace{}%
\AgdaSymbol{=}\AgdaSpace{}%
\AgdaFunction{project}\AgdaSpace{}%
\AgdaBound{c}\AgdaSpace{}%
\AgdaBound{M₁}%
\>[29]\AgdaComment{--\ we\ have\ (causal\ c\ c≢x\ :\ project\ c\ M₁\ ≡\ project\ c\ M₂)}\<%
\\
\>[0]\AgdaSymbol{...}\AgdaSpace{}%
\AgdaSymbol{|}\AgdaSpace{}%
\AgdaInductiveConstructor{yes}\AgdaSpace{}%
\AgdaInductiveConstructor{refl}\AgdaSpace{}%
\AgdaSymbol{=}\AgdaSpace{}%
\AgdaInductiveConstructor{branch}\AgdaSpace{}%
\AgdaBound{d}\AgdaSpace{}%
\AgdaSymbol{(}\AgdaFunction{project}\AgdaSpace{}%
\AgdaBound{c}\AgdaSpace{}%
\AgdaBound{M₁}\AgdaSymbol{)}\AgdaSpace{}%
\AgdaSymbol{(}\AgdaFunction{project}\AgdaSpace{}%
\AgdaBound{c}\AgdaSpace{}%
\AgdaBound{M₂}\AgdaSymbol{)}\<%
\\
\>[0]\AgdaFunction{project}\AgdaSpace{}%
\AgdaBound{c}\AgdaSpace{}%
\AgdaSymbol{(}\AgdaInductiveConstructor{transmit}\AgdaSpace{}%
\AgdaBound{d}\AgdaSpace{}%
\AgdaBound{x}\AgdaSpace{}%
\AgdaBound{t}\AgdaSpace{}%
\AgdaBound{M}\AgdaSymbol{)}\AgdaSpace{}%
\AgdaKeyword{with}\AgdaSpace{}%
\AgdaBound{c}\AgdaSpace{}%
\AgdaOperator{\AgdaFunction{≟}}\AgdaSpace{}%
\AgdaBound{x}\<%
\\
\>[0]\AgdaSymbol{...}\AgdaSpace{}%
\AgdaSymbol{|}\AgdaSpace{}%
\AgdaInductiveConstructor{no}\AgdaSpace{}%
\AgdaBound{c≢x}\AgdaSpace{}%
\AgdaSymbol{=}\AgdaSpace{}%
\AgdaFunction{project}\AgdaSpace{}%
\AgdaBound{c}\AgdaSpace{}%
\AgdaBound{M}\<%
\\
\>[0]\AgdaSymbol{...}\AgdaSpace{}%
\AgdaSymbol{|}\AgdaSpace{}%
\AgdaInductiveConstructor{yes}\AgdaSpace{}%
\AgdaInductiveConstructor{refl}\AgdaSpace{}%
\AgdaSymbol{=}\AgdaSpace{}%
\AgdaInductiveConstructor{transmit}\AgdaSpace{}%
\AgdaBound{d}\AgdaSpace{}%
\AgdaBound{t}\AgdaSpace{}%
\AgdaSymbol{(}\AgdaFunction{project}\AgdaSpace{}%
\AgdaBound{c}\AgdaSpace{}%
\AgdaBound{M}\AgdaSymbol{)}\<%
\\
\>[0]\AgdaFunction{project}\AgdaSpace{}%
\AgdaSymbol{\{}\AgdaInductiveConstructor{suc}\AgdaSpace{}%
\AgdaBound{n}\AgdaSymbol{\}}\AgdaSpace{}%
\AgdaBound{c}\AgdaSpace{}%
\AgdaSymbol{(}\AgdaInductiveConstructor{close}\AgdaSpace{}%
\AgdaBound{x}\AgdaSpace{}%
\AgdaBound{M}\AgdaSymbol{)}\AgdaSpace{}%
\AgdaKeyword{with}\AgdaSpace{}%
\AgdaBound{c}\AgdaSpace{}%
\AgdaOperator{\AgdaFunction{≟}}\AgdaSpace{}%
\AgdaBound{x}\<%
\\
\>[0]\AgdaSymbol{...}\AgdaSpace{}%
\AgdaSymbol{|}\AgdaSpace{}%
\AgdaInductiveConstructor{no}\AgdaSpace{}%
\AgdaBound{c≢x}\AgdaSpace{}%
\AgdaSymbol{=}\AgdaSpace{}%
\AgdaFunction{project}\AgdaSpace{}%
\AgdaSymbol{(}\AgdaFunction{adjust}\AgdaSpace{}%
\AgdaBound{c}\AgdaSpace{}%
\AgdaBound{x}\AgdaSpace{}%
\AgdaBound{c≢x}\AgdaSymbol{)}\AgdaSpace{}%
\AgdaBound{M}\<%
\\
\>[0]\AgdaSymbol{...}\AgdaSpace{}%
\AgdaSymbol{|}\AgdaSpace{}%
\AgdaInductiveConstructor{yes}\AgdaSpace{}%
\AgdaInductiveConstructor{refl}\AgdaSpace{}%
\AgdaSymbol{=}\AgdaSpace{}%
\AgdaInductiveConstructor{end}\<%
\\
\>[0]\AgdaFunction{project}\AgdaSpace{}%
\AgdaBound{c}\AgdaSpace{}%
\AgdaSymbol{(}\AgdaInductiveConstructor{delegateOUT}\AgdaSpace{}%
\AgdaBound{x}\AgdaSpace{}%
\AgdaBound{j}\AgdaSpace{}%
\AgdaBound{x≢j}\AgdaSpace{}%
\AgdaBound{sj}\AgdaSpace{}%
\AgdaBound{M}\AgdaSymbol{)}\AgdaSpace{}%
\AgdaKeyword{with}\AgdaSpace{}%
\AgdaBound{c}\AgdaSpace{}%
\AgdaOperator{\AgdaFunction{≟}}\AgdaSpace{}%
\AgdaBound{x}\<%
\\
\>[0]\AgdaSymbol{...}\AgdaSpace{}%
\AgdaSymbol{|}\AgdaSpace{}%
\AgdaInductiveConstructor{yes}\AgdaSpace{}%
\AgdaInductiveConstructor{refl}\AgdaSpace{}%
\AgdaSymbol{=}\AgdaSpace{}%
\AgdaInductiveConstructor{delegate}\AgdaSpace{}%
\AgdaInductiveConstructor{OUT}\AgdaSpace{}%
\AgdaBound{sj}\AgdaSpace{}%
\AgdaSymbol{(}\AgdaFunction{project}\AgdaSpace{}%
\AgdaSymbol{(}\AgdaFunction{adjust}\AgdaSpace{}%
\AgdaBound{c}\AgdaSpace{}%
\AgdaBound{j}\AgdaSpace{}%
\AgdaBound{x≢j}\AgdaSymbol{)}\AgdaSpace{}%
\AgdaBound{M}\AgdaSymbol{)}\<%
\\
\>[0]\AgdaSymbol{...}\AgdaSpace{}%
\AgdaSymbol{|}\AgdaSpace{}%
\AgdaInductiveConstructor{no}\AgdaSpace{}%
\AgdaBound{c≢x}\AgdaSpace{}%
\AgdaKeyword{with}\AgdaSpace{}%
\AgdaBound{c}\AgdaSpace{}%
\AgdaOperator{\AgdaFunction{≟}}\AgdaSpace{}%
\AgdaBound{j}\<%
\\
\>[0]\AgdaSymbol{...}\AgdaSpace{}%
\AgdaSymbol{|}\AgdaSpace{}%
\AgdaInductiveConstructor{yes}\AgdaSpace{}%
\AgdaInductiveConstructor{refl}\AgdaSpace{}%
\AgdaSymbol{=}\AgdaSpace{}%
\AgdaBound{sj}\<%
\\
\>[0]\AgdaSymbol{...}\AgdaSpace{}%
\AgdaSymbol{|}\AgdaSpace{}%
\AgdaInductiveConstructor{no}\AgdaSpace{}%
\AgdaBound{c≢j}\AgdaSpace{}%
\AgdaSymbol{=}\AgdaSpace{}%
\AgdaFunction{project}\AgdaSpace{}%
\AgdaSymbol{(}\AgdaFunction{adjust}\AgdaSpace{}%
\AgdaBound{c}\AgdaSpace{}%
\AgdaBound{j}\AgdaSpace{}%
\AgdaBound{c≢j}\AgdaSymbol{)}\AgdaSpace{}%
\AgdaBound{M}\<%
\\
\>[0]\AgdaFunction{project}\AgdaSpace{}%
\AgdaBound{c}\AgdaSpace{}%
\AgdaSymbol{(}\AgdaInductiveConstructor{delegateIN}\AgdaSpace{}%
\AgdaBound{x}\AgdaSpace{}%
\AgdaBound{M}\AgdaSymbol{)}\AgdaSpace{}%
\AgdaKeyword{with}\AgdaSpace{}%
\AgdaBound{c}\AgdaSpace{}%
\AgdaOperator{\AgdaFunction{≟}}\AgdaSpace{}%
\AgdaBound{x}\<%
\\
\>[0]\AgdaSymbol{...}\AgdaSpace{}%
\AgdaSymbol{|}\AgdaSpace{}%
\AgdaInductiveConstructor{yes}\AgdaSpace{}%
\AgdaInductiveConstructor{refl}\AgdaSpace{}%
\AgdaSymbol{=}\AgdaSpace{}%
\AgdaInductiveConstructor{delegate}\AgdaSpace{}%
\AgdaInductiveConstructor{INP}\AgdaSpace{}%
\AgdaSymbol{(}\AgdaFunction{project}\AgdaSpace{}%
\AgdaInductiveConstructor{zero}\AgdaSpace{}%
\AgdaBound{M}\AgdaSymbol{)}\AgdaSpace{}%
\AgdaSymbol{(}\AgdaFunction{project}\AgdaSpace{}%
\AgdaSymbol{(}\AgdaInductiveConstructor{suc}\AgdaSpace{}%
\AgdaBound{c}\AgdaSymbol{)}\AgdaSpace{}%
\AgdaBound{M}\AgdaSymbol{)}\<%
\\
\>[0]\AgdaSymbol{...}\AgdaSpace{}%
\AgdaSymbol{|}\AgdaSpace{}%
\AgdaInductiveConstructor{no}\AgdaSpace{}%
\AgdaBound{c≢x}\AgdaSpace{}%
\AgdaSymbol{=}\AgdaSpace{}%
\AgdaFunction{project}\AgdaSpace{}%
\AgdaSymbol{(}\AgdaInductiveConstructor{suc}\AgdaSpace{}%
\AgdaBound{c}\AgdaSymbol{)}\AgdaSpace{}%
\AgdaBound{M}\<%
\\
\\[\AgdaEmptyExtraSkip]%
\>[0]\AgdaFunction{Causality}\AgdaSpace{}%
\AgdaSymbol{\{}\AgdaBound{n}\AgdaSymbol{\}}\AgdaSpace{}%
\AgdaBound{i}\AgdaSpace{}%
\AgdaBound{M₁}\AgdaSpace{}%
\AgdaBound{M₂}\AgdaSpace{}%
\AgdaSymbol{=}\AgdaSpace{}%
\AgdaSymbol{∀}\AgdaSpace{}%
\AgdaSymbol{(}\AgdaBound{j}\AgdaSpace{}%
\AgdaSymbol{:}\AgdaSpace{}%
\AgdaDatatype{Fin}\AgdaSpace{}%
\AgdaBound{n}\AgdaSymbol{)}\AgdaSpace{}%
\AgdaSymbol{→}\AgdaSpace{}%
\AgdaBound{j}\AgdaSpace{}%
\AgdaOperator{\AgdaFunction{≢}}\AgdaSpace{}%
\AgdaBound{i}\AgdaSpace{}%
\AgdaSymbol{→}\AgdaSpace{}%
\AgdaFunction{project}\AgdaSpace{}%
\AgdaBound{j}\AgdaSpace{}%
\AgdaBound{M₁}\AgdaSpace{}%
\AgdaOperator{\AgdaDatatype{≡}}\AgdaSpace{}%
\AgdaFunction{project}\AgdaSpace{}%
\AgdaBound{j}\AgdaSpace{}%
\AgdaBound{M₂}\<%
\\
\>[0]\AgdaFunction{CheckDual0}\AgdaSpace{}%
\AgdaBound{M₁}\AgdaSpace{}%
\AgdaBound{M₂}\AgdaSpace{}%
\AgdaSymbol{=}\AgdaSpace{}%
\AgdaFunction{project}\AgdaSpace{}%
\AgdaInductiveConstructor{zero}\AgdaSpace{}%
\AgdaBound{M₁}\AgdaSpace{}%
\AgdaOperator{\AgdaDatatype{≡}}\AgdaSpace{}%
\AgdaFunction{dual}\AgdaSpace{}%
\AgdaSymbol{(}\AgdaFunction{project}\AgdaSpace{}%
\AgdaInductiveConstructor{zero}\AgdaSpace{}%
\AgdaBound{M₂}\AgdaSymbol{)}\<%
\end{code}}
\begin{code}[hide]%
\>[0]\<%
\\
\>[0]\AgdaKeyword{variable}\<%
\\
\>[0][@{}l@{\AgdaIndent{0}}]%
\>[2]\AgdaGeneralizable{A′}\AgdaSpace{}%
\AgdaGeneralizable{A₁}\AgdaSpace{}%
\AgdaGeneralizable{A₂}\AgdaSpace{}%
\AgdaSymbol{:}\AgdaSpace{}%
\AgdaPrimitive{Set}\<%
\\
\>[2]\AgdaGeneralizable{T}\AgdaSpace{}%
\AgdaSymbol{:}\AgdaSpace{}%
\AgdaDatatype{Type}\<%
\\
\\[\AgdaEmptyExtraSkip]%
\>[0]\AgdaOperator{\AgdaFunction{T⟦\AgdaUnderscore{}⟧}}\AgdaSpace{}%
\AgdaSymbol{:}\AgdaSpace{}%
\AgdaDatatype{Type}\AgdaSpace{}%
\AgdaSymbol{→}\AgdaSpace{}%
\AgdaPrimitive{Set}\<%
\\
\>[0]\AgdaOperator{\AgdaFunction{T⟦}}\AgdaSpace{}%
\AgdaInductiveConstructor{nat}\AgdaSpace{}%
\AgdaOperator{\AgdaFunction{⟧}}\AgdaSpace{}%
\AgdaSymbol{=}\AgdaSpace{}%
\AgdaDatatype{ℕ}\<%
\\
\>[0]\AgdaOperator{\AgdaFunction{T⟦}}\AgdaSpace{}%
\AgdaInductiveConstructor{int}\AgdaSpace{}%
\AgdaOperator{\AgdaFunction{⟧}}\AgdaSpace{}%
\AgdaSymbol{=}\AgdaSpace{}%
\AgdaDatatype{ℤ}\<%
\end{code}
\newcommand\multiCmd{%
\begin{code}%
\>[0]\AgdaKeyword{data}\AgdaSpace{}%
\AgdaDatatype{Cmd}\AgdaSpace{}%
\AgdaSymbol{(}\AgdaBound{A}\AgdaSpace{}%
\AgdaSymbol{:}\AgdaSpace{}%
\AgdaPrimitive{Set}\AgdaSymbol{)}\AgdaSpace{}%
\AgdaSymbol{:}\AgdaSpace{}%
\AgdaSymbol{(}\AgdaBound{n}\AgdaSpace{}%
\AgdaSymbol{:}\AgdaSpace{}%
\AgdaDatatype{ℕ}\AgdaSymbol{)}\AgdaSpace{}%
\AgdaSymbol{→}\AgdaSpace{}%
\AgdaDatatype{MSession}\AgdaSpace{}%
\AgdaBound{n}\AgdaSpace{}%
\AgdaSymbol{→}\AgdaSpace{}%
\AgdaPrimitive{Set₁}\AgdaSpace{}%
\AgdaKeyword{where}\<%
\\
\>[0][@{}l@{\AgdaIndent{0}}]%
\>[2]\AgdaInductiveConstructor{CLOSE}%
\>[9]\AgdaSymbol{:}\AgdaSpace{}%
\AgdaSymbol{∀}\AgdaSpace{}%
\AgdaBound{c}\AgdaSpace{}%
\AgdaSymbol{→}\AgdaSpace{}%
\AgdaSymbol{(}\AgdaBound{A}\AgdaSpace{}%
\AgdaSymbol{→}\AgdaSpace{}%
\AgdaBound{A}\AgdaSymbol{)}\AgdaSpace{}%
\AgdaSymbol{→}\AgdaSpace{}%
\AgdaDatatype{Cmd}\AgdaSpace{}%
\AgdaBound{A}\AgdaSpace{}%
\AgdaGeneralizable{n}\AgdaSpace{}%
\AgdaGeneralizable{M}\AgdaSpace{}%
\AgdaSymbol{→}\AgdaSpace{}%
\AgdaDatatype{Cmd}\AgdaSpace{}%
\AgdaBound{A}\AgdaSpace{}%
\AgdaSymbol{(}\AgdaInductiveConstructor{suc}\AgdaSpace{}%
\AgdaGeneralizable{n}\AgdaSymbol{)}\AgdaSpace{}%
\AgdaSymbol{(}\AgdaInductiveConstructor{close}\AgdaSpace{}%
\AgdaBound{c}\AgdaSpace{}%
\AgdaGeneralizable{M}\AgdaSymbol{)}\<%
\\
\>[2]\AgdaInductiveConstructor{SEND}%
\>[9]\AgdaSymbol{:}\AgdaSpace{}%
\AgdaSymbol{∀}\AgdaSpace{}%
\AgdaBound{c}\AgdaSpace{}%
\AgdaSymbol{→}\AgdaSpace{}%
\AgdaSymbol{(}\AgdaBound{A}\AgdaSpace{}%
\AgdaSymbol{→}\AgdaSpace{}%
\AgdaOperator{\AgdaFunction{T⟦}}\AgdaSpace{}%
\AgdaGeneralizable{T}\AgdaSpace{}%
\AgdaOperator{\AgdaFunction{⟧}}\AgdaSpace{}%
\AgdaOperator{\AgdaFunction{×}}\AgdaSpace{}%
\AgdaBound{A}\AgdaSymbol{)}\AgdaSpace{}%
\AgdaSymbol{→}\AgdaSpace{}%
\AgdaDatatype{Cmd}\AgdaSpace{}%
\AgdaBound{A}\AgdaSpace{}%
\AgdaGeneralizable{n}\AgdaSpace{}%
\AgdaGeneralizable{M}\AgdaSpace{}%
\AgdaSymbol{→}\AgdaSpace{}%
\AgdaDatatype{Cmd}\AgdaSpace{}%
\AgdaBound{A}\AgdaSpace{}%
\AgdaGeneralizable{n}\AgdaSpace{}%
\AgdaSymbol{(}\AgdaInductiveConstructor{send}\AgdaSpace{}%
\AgdaBound{c}\AgdaSpace{}%
\AgdaGeneralizable{T}\AgdaSpace{}%
\AgdaGeneralizable{M}\AgdaSymbol{)}\<%
\\
\>[2]\AgdaInductiveConstructor{RECV}%
\>[9]\AgdaSymbol{:}\AgdaSpace{}%
\AgdaSymbol{∀}\AgdaSpace{}%
\AgdaBound{c}\AgdaSpace{}%
\AgdaSymbol{→}\AgdaSpace{}%
\AgdaSymbol{(}\AgdaOperator{\AgdaFunction{T⟦}}\AgdaSpace{}%
\AgdaGeneralizable{T}\AgdaSpace{}%
\AgdaOperator{\AgdaFunction{⟧}}\AgdaSpace{}%
\AgdaSymbol{→}\AgdaSpace{}%
\AgdaBound{A}\AgdaSpace{}%
\AgdaSymbol{→}\AgdaSpace{}%
\AgdaBound{A}\AgdaSymbol{)}\AgdaSpace{}%
\AgdaSymbol{→}\AgdaSpace{}%
\AgdaDatatype{Cmd}\AgdaSpace{}%
\AgdaBound{A}\AgdaSpace{}%
\AgdaGeneralizable{n}\AgdaSpace{}%
\AgdaGeneralizable{M}\AgdaSpace{}%
\AgdaSymbol{→}\AgdaSpace{}%
\AgdaDatatype{Cmd}\AgdaSpace{}%
\AgdaBound{A}\AgdaSpace{}%
\AgdaGeneralizable{n}\AgdaSpace{}%
\AgdaSymbol{(}\AgdaInductiveConstructor{recv}\AgdaSpace{}%
\AgdaBound{c}\AgdaSpace{}%
\AgdaGeneralizable{T}\AgdaSpace{}%
\AgdaGeneralizable{M}\AgdaSymbol{)}\<%
\\
\>[2]\AgdaInductiveConstructor{SELECT}\AgdaSpace{}%
\AgdaSymbol{:}\AgdaSpace{}%
\AgdaSymbol{∀}\AgdaSpace{}%
\AgdaBound{c}\AgdaSpace{}%
\AgdaSymbol{→}\AgdaSpace{}%
\AgdaSymbol{(}\AgdaBound{causal}\AgdaSpace{}%
\AgdaSymbol{:}\AgdaSpace{}%
\AgdaFunction{Causality}\AgdaSpace{}%
\AgdaBound{c}\AgdaSpace{}%
\AgdaGeneralizable{M₁}\AgdaSpace{}%
\AgdaGeneralizable{M₂}\AgdaSymbol{)}\AgdaSpace{}%
\AgdaSymbol{→}\AgdaSpace{}%
\AgdaSymbol{(}\AgdaBound{A}\AgdaSpace{}%
\AgdaSymbol{→}\AgdaSpace{}%
\AgdaDatatype{Bool}\AgdaSpace{}%
\AgdaOperator{\AgdaFunction{×}}\AgdaSpace{}%
\AgdaBound{A}\AgdaSymbol{)}\<%
\\
\>[2][@{}l@{\AgdaIndent{0}}]%
\>[4]\AgdaSymbol{→}\AgdaSpace{}%
\AgdaDatatype{Cmd}\AgdaSpace{}%
\AgdaBound{A}\AgdaSpace{}%
\AgdaGeneralizable{n}\AgdaSpace{}%
\AgdaGeneralizable{M₁}\AgdaSpace{}%
\AgdaSymbol{→}\AgdaSpace{}%
\AgdaDatatype{Cmd}\AgdaSpace{}%
\AgdaBound{A}\AgdaSpace{}%
\AgdaGeneralizable{n}\AgdaSpace{}%
\AgdaGeneralizable{M₂}\AgdaSpace{}%
\AgdaSymbol{→}\AgdaSpace{}%
\AgdaDatatype{Cmd}\AgdaSpace{}%
\AgdaBound{A}\AgdaSpace{}%
\AgdaGeneralizable{n}\AgdaSpace{}%
\AgdaSymbol{(}\AgdaInductiveConstructor{select}\AgdaSpace{}%
\AgdaBound{c}\AgdaSpace{}%
\AgdaGeneralizable{M₁}\AgdaSpace{}%
\AgdaGeneralizable{M₂}\AgdaSpace{}%
\AgdaBound{causal}\AgdaSymbol{)}\<%
\\
\>[2]\AgdaInductiveConstructor{CHOICE}\AgdaSpace{}%
\AgdaSymbol{:}\AgdaSpace{}%
\AgdaSymbol{∀}\AgdaSpace{}%
\AgdaBound{c}\AgdaSpace{}%
\AgdaSymbol{→}\AgdaSpace{}%
\AgdaSymbol{(}\AgdaBound{causal}\AgdaSpace{}%
\AgdaSymbol{:}\AgdaSpace{}%
\AgdaFunction{Causality}\AgdaSpace{}%
\AgdaBound{c}\AgdaSpace{}%
\AgdaGeneralizable{M₁}\AgdaSpace{}%
\AgdaGeneralizable{M₂}\AgdaSymbol{)}\<%
\\
\>[2][@{}l@{\AgdaIndent{0}}]%
\>[4]\AgdaSymbol{→}\AgdaSpace{}%
\AgdaDatatype{Cmd}\AgdaSpace{}%
\AgdaBound{A}\AgdaSpace{}%
\AgdaGeneralizable{n}\AgdaSpace{}%
\AgdaGeneralizable{M₁}\AgdaSpace{}%
\AgdaSymbol{→}\AgdaSpace{}%
\AgdaDatatype{Cmd}\AgdaSpace{}%
\AgdaBound{A}\AgdaSpace{}%
\AgdaGeneralizable{n}\AgdaSpace{}%
\AgdaGeneralizable{M₂}\AgdaSpace{}%
\AgdaSymbol{→}\AgdaSpace{}%
\AgdaDatatype{Cmd}\AgdaSpace{}%
\AgdaBound{A}\AgdaSpace{}%
\AgdaGeneralizable{n}\AgdaSpace{}%
\AgdaSymbol{(}\AgdaInductiveConstructor{choice}\AgdaSpace{}%
\AgdaBound{c}\AgdaSpace{}%
\AgdaGeneralizable{M₁}\AgdaSpace{}%
\AgdaGeneralizable{M₂}\AgdaSpace{}%
\AgdaBound{causal}\AgdaSymbol{)}\<%
\\
\>[2]\AgdaInductiveConstructor{CONNECT}\AgdaSpace{}%
\AgdaSymbol{:}\AgdaSpace{}%
\AgdaSymbol{∀}\AgdaSpace{}%
\AgdaSymbol{\{}\AgdaBound{M₁}\AgdaSpace{}%
\AgdaSymbol{:}\AgdaSpace{}%
\AgdaDatatype{MSession}\AgdaSpace{}%
\AgdaSymbol{(}\AgdaInductiveConstructor{suc}\AgdaSpace{}%
\AgdaGeneralizable{m}\AgdaSymbol{)\}}\AgdaSpace{}%
\AgdaSymbol{\{}\AgdaBound{M₂}\AgdaSpace{}%
\AgdaSymbol{:}\AgdaSpace{}%
\AgdaDatatype{MSession}\AgdaSpace{}%
\AgdaSymbol{(}\AgdaInductiveConstructor{suc}\AgdaSpace{}%
\AgdaGeneralizable{n}\AgdaSymbol{)\}}\AgdaSpace{}%
\AgdaSymbol{(}\AgdaBound{check}\AgdaSpace{}%
\AgdaSymbol{:}\AgdaSpace{}%
\AgdaFunction{CheckDual0}\AgdaSpace{}%
\AgdaBound{M₁}\AgdaSpace{}%
\AgdaBound{M₂}\AgdaSymbol{)}\<%
\\
\>[2][@{}l@{\AgdaIndent{0}}]%
\>[4]\AgdaSymbol{→}\AgdaSpace{}%
\AgdaSymbol{(}\AgdaBound{split}\AgdaSpace{}%
\AgdaSymbol{:}\AgdaSpace{}%
\AgdaBound{A}\AgdaSpace{}%
\AgdaSymbol{→}\AgdaSpace{}%
\AgdaBound{A}\AgdaSpace{}%
\AgdaOperator{\AgdaFunction{×}}\AgdaSpace{}%
\AgdaBound{A}\AgdaSymbol{)}\<%
\\
\>[4]\AgdaSymbol{→}\AgdaSpace{}%
\AgdaSymbol{(}\AgdaBound{sp}\AgdaSpace{}%
\AgdaSymbol{:}\AgdaSpace{}%
\AgdaDatatype{Split}\AgdaSpace{}%
\AgdaGeneralizable{m}\AgdaSpace{}%
\AgdaGeneralizable{n}\AgdaSymbol{)}\<%
\\
\>[4]\AgdaSymbol{→}\AgdaSpace{}%
\AgdaDatatype{Cmd}\AgdaSpace{}%
\AgdaBound{A}\AgdaSpace{}%
\AgdaSymbol{(}\AgdaInductiveConstructor{suc}\AgdaSpace{}%
\AgdaGeneralizable{m}\AgdaSymbol{)}\AgdaSpace{}%
\AgdaBound{M₁}\AgdaSpace{}%
\AgdaSymbol{→}\AgdaSpace{}%
\AgdaDatatype{Cmd}\AgdaSpace{}%
\AgdaBound{A}\AgdaSpace{}%
\AgdaSymbol{(}\AgdaInductiveConstructor{suc}\AgdaSpace{}%
\AgdaGeneralizable{n}\AgdaSymbol{)}\AgdaSpace{}%
\AgdaBound{M₂}\<%
\\
\>[4]\AgdaSymbol{→}\AgdaSpace{}%
\AgdaDatatype{Cmd}\AgdaSpace{}%
\AgdaBound{A}\AgdaSpace{}%
\AgdaSymbol{(}\AgdaGeneralizable{m}\AgdaSpace{}%
\AgdaOperator{\AgdaPrimitive{+}}\AgdaSpace{}%
\AgdaGeneralizable{n}\AgdaSymbol{)}\AgdaSpace{}%
\AgdaSymbol{(}\AgdaInductiveConstructor{connect}\AgdaSpace{}%
\AgdaBound{sp}\AgdaSpace{}%
\AgdaBound{M₁}\AgdaSpace{}%
\AgdaBound{M₂}\AgdaSpace{}%
\AgdaBound{check}\AgdaSymbol{)}\<%
\\
\>[2]\AgdaInductiveConstructor{SENDCH}\AgdaSpace{}%
\AgdaSymbol{:}\AgdaSpace{}%
\AgdaSymbol{∀}\AgdaSpace{}%
\AgdaSymbol{\{}\AgdaBound{sj}\AgdaSymbol{\}}\AgdaSpace{}%
\AgdaSymbol{→}\AgdaSpace{}%
\AgdaSymbol{∀}\AgdaSpace{}%
\AgdaBound{c}\AgdaSpace{}%
\AgdaBound{j}\AgdaSpace{}%
\AgdaSymbol{→}\AgdaSpace{}%
\AgdaSymbol{(}\AgdaBound{c≢j}\AgdaSpace{}%
\AgdaSymbol{:}\AgdaSpace{}%
\AgdaBound{c}\AgdaSpace{}%
\AgdaOperator{\AgdaFunction{≢}}\AgdaSpace{}%
\AgdaBound{j}\AgdaSymbol{)}\<%
\\
\>[2][@{}l@{\AgdaIndent{0}}]%
\>[4]\AgdaSymbol{→}\AgdaSpace{}%
\AgdaDatatype{Cmd}\AgdaSpace{}%
\AgdaBound{A}\AgdaSpace{}%
\AgdaGeneralizable{n}\AgdaSpace{}%
\AgdaGeneralizable{M}\AgdaSpace{}%
\AgdaSymbol{→}\AgdaSpace{}%
\AgdaDatatype{Cmd}\AgdaSpace{}%
\AgdaBound{A}\AgdaSpace{}%
\AgdaSymbol{(}\AgdaInductiveConstructor{suc}\AgdaSpace{}%
\AgdaGeneralizable{n}\AgdaSymbol{)}\AgdaSpace{}%
\AgdaSymbol{(}\AgdaInductiveConstructor{delegateOUT}\AgdaSpace{}%
\AgdaBound{c}\AgdaSpace{}%
\AgdaBound{j}\AgdaSpace{}%
\AgdaBound{c≢j}\AgdaSpace{}%
\AgdaBound{sj}\AgdaSpace{}%
\AgdaGeneralizable{M}\AgdaSymbol{)}\<%
\\
\>[2]\AgdaInductiveConstructor{RECVCH}\AgdaSpace{}%
\AgdaSymbol{:}\AgdaSpace{}%
\AgdaSymbol{∀}\AgdaSpace{}%
\AgdaBound{c}\AgdaSpace{}%
\AgdaSymbol{→}\AgdaSpace{}%
\AgdaDatatype{Cmd}\AgdaSpace{}%
\AgdaBound{A}\AgdaSpace{}%
\AgdaSymbol{(}\AgdaInductiveConstructor{suc}\AgdaSpace{}%
\AgdaGeneralizable{n}\AgdaSymbol{)}\AgdaSpace{}%
\AgdaGeneralizable{M}\AgdaSpace{}%
\AgdaSymbol{→}\AgdaSpace{}%
\AgdaDatatype{Cmd}\AgdaSpace{}%
\AgdaBound{A}\AgdaSpace{}%
\AgdaGeneralizable{n}\AgdaSpace{}%
\AgdaSymbol{(}\AgdaInductiveConstructor{delegateIN}\AgdaSpace{}%
\AgdaBound{c}\AgdaSpace{}%
\AgdaGeneralizable{M}\AgdaSymbol{)}\<%
\\
\>[2]\AgdaInductiveConstructor{END}%
\>[9]\AgdaSymbol{:}\AgdaSpace{}%
\AgdaDatatype{Cmd}\AgdaSpace{}%
\AgdaBound{A}\AgdaSpace{}%
\AgdaNumber{0}\AgdaSpace{}%
\AgdaInductiveConstructor{terminate}\<%
\end{code}}
\begin{code}[hide]%
\>[0]\AgdaKeyword{postulate}\<%
\\
\>[0][@{}l@{\AgdaIndent{0}}]%
\>[2]\AgdaPostulate{Channel}\AgdaSpace{}%
\AgdaSymbol{:}\AgdaSpace{}%
\AgdaPrimitive{Set}\<%
\\
\>[2]\AgdaPostulate{primSend}\AgdaSpace{}%
\AgdaSymbol{:}\AgdaSpace{}%
\AgdaPostulate{Channel}\AgdaSpace{}%
\AgdaSymbol{→}\AgdaSpace{}%
\AgdaGeneralizable{A}\AgdaSpace{}%
\AgdaSymbol{→}\AgdaSpace{}%
\AgdaDatatype{IO}\AgdaSpace{}%
\AgdaRecord{⊤}\<%
\\
\>[2]\AgdaPostulate{primRecv}\AgdaSpace{}%
\AgdaSymbol{:}\AgdaSpace{}%
\AgdaPostulate{Channel}\AgdaSpace{}%
\AgdaSymbol{→}\AgdaSpace{}%
\AgdaDatatype{IO}\AgdaSpace{}%
\AgdaGeneralizable{A}\<%
\\
\>[2]\AgdaPostulate{primClose}\AgdaSpace{}%
\AgdaSymbol{:}\AgdaSpace{}%
\AgdaPostulate{Channel}\AgdaSpace{}%
\AgdaSymbol{→}\AgdaSpace{}%
\AgdaDatatype{IO}\AgdaSpace{}%
\AgdaRecord{⊤}\<%
\\
\>[2]\AgdaPostulate{forkIO}%
\>[11]\AgdaSymbol{:}\AgdaSpace{}%
\AgdaDatatype{IO}\AgdaSpace{}%
\AgdaGeneralizable{A}\AgdaSpace{}%
\AgdaSymbol{→}\AgdaSpace{}%
\AgdaDatatype{IO}\AgdaSpace{}%
\AgdaRecord{⊤}\<%
\\
\>[2]\AgdaPostulate{newChan}%
\>[11]\AgdaSymbol{:}\AgdaSpace{}%
\AgdaDatatype{IO}\AgdaSpace{}%
\AgdaSymbol{(}\AgdaPostulate{Channel}\AgdaSpace{}%
\AgdaOperator{\AgdaFunction{×}}\AgdaSpace{}%
\AgdaPostulate{Channel}\AgdaSymbol{)}\<%
\end{code}
\newcommand\multiExec{%
\begin{code}%
\>[0]\AgdaFunction{exec}\AgdaSpace{}%
\AgdaSymbol{:}\AgdaSpace{}%
\AgdaDatatype{Cmd}\AgdaSpace{}%
\AgdaGeneralizable{A}\AgdaSpace{}%
\AgdaGeneralizable{n}\AgdaSpace{}%
\AgdaGeneralizable{M}\AgdaSpace{}%
\AgdaSymbol{→}\AgdaSpace{}%
\AgdaGeneralizable{A}\AgdaSpace{}%
\AgdaSymbol{→}\AgdaSpace{}%
\AgdaDatatype{Vec}\AgdaSpace{}%
\AgdaPostulate{Channel}\AgdaSpace{}%
\AgdaGeneralizable{n}\AgdaSpace{}%
\AgdaSymbol{→}\AgdaSpace{}%
\AgdaDatatype{IO}\AgdaSpace{}%
\AgdaGeneralizable{A}\<%
\\
\>[0]\AgdaFunction{exec}\AgdaSpace{}%
\AgdaSymbol{(}\AgdaInductiveConstructor{SENDCH}\AgdaSpace{}%
\AgdaBound{c}\AgdaSpace{}%
\AgdaBound{j}\AgdaSpace{}%
\AgdaBound{f≢j}\AgdaSpace{}%
\AgdaBound{cmd}\AgdaSymbol{)}\AgdaSpace{}%
\AgdaBound{state}\AgdaSpace{}%
\AgdaBound{chns}\AgdaSpace{}%
\AgdaSymbol{=}\AgdaSpace{}%
\AgdaKeyword{do}\<%
\\
\>[0][@{}l@{\AgdaIndent{0}}]%
\>[2]\AgdaPostulate{primSend}\AgdaSpace{}%
\AgdaSymbol{(}\AgdaFunction{lookup}\AgdaSpace{}%
\AgdaBound{chns}\AgdaSpace{}%
\AgdaBound{c}\AgdaSymbol{)}\AgdaSpace{}%
\AgdaSymbol{(}\AgdaFunction{lookup}\AgdaSpace{}%
\AgdaBound{chns}\AgdaSpace{}%
\AgdaBound{j}\AgdaSymbol{)}\<%
\\
\>[2]\AgdaFunction{exec}\AgdaSpace{}%
\AgdaBound{cmd}\AgdaSpace{}%
\AgdaBound{state}\AgdaSpace{}%
\AgdaSymbol{(}\AgdaFunction{remove}\AgdaSpace{}%
\AgdaBound{chns}\AgdaSpace{}%
\AgdaBound{j}\AgdaSymbol{)}\<%
\\
\>[0]\AgdaFunction{exec}\AgdaSpace{}%
\AgdaSymbol{(}\AgdaInductiveConstructor{RECVCH}\AgdaSpace{}%
\AgdaBound{c}\AgdaSpace{}%
\AgdaBound{cmd}\AgdaSymbol{)}\AgdaSpace{}%
\AgdaBound{state}\AgdaSpace{}%
\AgdaBound{chns}\AgdaSpace{}%
\AgdaSymbol{=}\AgdaSpace{}%
\AgdaKeyword{do}\<%
\\
\>[0][@{}l@{\AgdaIndent{0}}]%
\>[2]\AgdaBound{ch}\AgdaSpace{}%
\AgdaOperator{\AgdaFunction{←}}\AgdaSpace{}%
\AgdaPostulate{primRecv}\AgdaSpace{}%
\AgdaSymbol{(}\AgdaFunction{lookup}\AgdaSpace{}%
\AgdaBound{chns}\AgdaSpace{}%
\AgdaBound{c}\AgdaSymbol{)}\<%
\\
\>[2]\AgdaFunction{exec}\AgdaSpace{}%
\AgdaBound{cmd}\AgdaSpace{}%
\AgdaBound{state}\AgdaSpace{}%
\AgdaSymbol{(}\AgdaBound{ch}\AgdaSpace{}%
\AgdaOperator{\AgdaInductiveConstructor{∷}}\AgdaSpace{}%
\AgdaBound{chns}\AgdaSymbol{)}\<%
\\
\>[0]\AgdaFunction{exec}\AgdaSpace{}%
\AgdaSymbol{(}\AgdaInductiveConstructor{CONNECT}\AgdaSpace{}%
\AgdaSymbol{\AgdaUnderscore{}}\AgdaSpace{}%
\AgdaBound{split}\AgdaSpace{}%
\AgdaBound{split-ch}\AgdaSpace{}%
\AgdaBound{cmds₁}\AgdaSpace{}%
\AgdaBound{cmds₂}\AgdaSymbol{)}\AgdaSpace{}%
\AgdaBound{state}\AgdaSpace{}%
\AgdaBound{chns}\AgdaSpace{}%
\AgdaSymbol{=}\<%
\\
\>[0][@{}l@{\AgdaIndent{0}}]%
\>[2]\AgdaKeyword{let}\AgdaSpace{}%
\AgdaOperator{\AgdaInductiveConstructor{⟨}}\AgdaSpace{}%
\AgdaBound{state₁}\AgdaSpace{}%
\AgdaOperator{\AgdaInductiveConstructor{,}}\AgdaSpace{}%
\AgdaBound{state₂}\AgdaSpace{}%
\AgdaOperator{\AgdaInductiveConstructor{⟩}}\AgdaSpace{}%
\AgdaSymbol{=}\AgdaSpace{}%
\AgdaBound{split}\AgdaSpace{}%
\AgdaBound{state}\AgdaSpace{}%
\AgdaKeyword{in}\<%
\\
\>[2]\AgdaKeyword{let}\AgdaSpace{}%
\AgdaOperator{\AgdaInductiveConstructor{⟨}}\AgdaSpace{}%
\AgdaBound{chns₁}\AgdaSpace{}%
\AgdaOperator{\AgdaInductiveConstructor{,}}\AgdaSpace{}%
\AgdaBound{chns₂}\AgdaSpace{}%
\AgdaOperator{\AgdaInductiveConstructor{⟩}}\AgdaSpace{}%
\AgdaSymbol{=}\AgdaSpace{}%
\AgdaFunction{apply-split}\AgdaSpace{}%
\AgdaBound{split-ch}\AgdaSpace{}%
\AgdaBound{chns}\AgdaSpace{}%
\AgdaKeyword{in}\<%
\\
\>[2]\AgdaPostulate{newChan}\AgdaSpace{}%
\AgdaOperator{\AgdaFunction{>>=}}\AgdaSpace{}%
\AgdaSymbol{λ\{}\AgdaSpace{}%
\AgdaOperator{\AgdaInductiveConstructor{⟨}}\AgdaSpace{}%
\AgdaBound{ch₁}\AgdaSpace{}%
\AgdaOperator{\AgdaInductiveConstructor{,}}\AgdaSpace{}%
\AgdaBound{ch₂}\AgdaSpace{}%
\AgdaOperator{\AgdaInductiveConstructor{⟩}}\AgdaSpace{}%
\AgdaSymbol{→}\<%
\\
\>[2]\AgdaPostulate{forkIO}\AgdaSpace{}%
\AgdaSymbol{(}\AgdaFunction{exec}\AgdaSpace{}%
\AgdaBound{cmds₁}\AgdaSpace{}%
\AgdaBound{state₁}\AgdaSpace{}%
\AgdaSymbol{(}\AgdaBound{ch₁}\AgdaSpace{}%
\AgdaOperator{\AgdaInductiveConstructor{∷}}\AgdaSpace{}%
\AgdaBound{chns₁}\AgdaSymbol{))}\AgdaSpace{}%
\AgdaOperator{\AgdaFunction{>>}}\<%
\\
\>[2]\AgdaFunction{exec}\AgdaSpace{}%
\AgdaBound{cmds₂}\AgdaSpace{}%
\AgdaBound{state₂}\AgdaSpace{}%
\AgdaSymbol{(}\AgdaBound{ch₂}\AgdaSpace{}%
\AgdaOperator{\AgdaInductiveConstructor{∷}}\AgdaSpace{}%
\AgdaBound{chns₂}\AgdaSymbol{)}\AgdaSpace{}%
\AgdaSymbol{\}}\<%
\\
\>[0]\AgdaFunction{exec}\AgdaSpace{}%
\AgdaSymbol{(}\AgdaInductiveConstructor{CLOSE}\AgdaSpace{}%
\AgdaBound{c}\AgdaSpace{}%
\AgdaBound{gend}\AgdaSpace{}%
\AgdaBound{cmd}\AgdaSymbol{)}\AgdaSpace{}%
\AgdaBound{state}\AgdaSpace{}%
\AgdaBound{chns}\AgdaSpace{}%
\AgdaSymbol{=}\AgdaSpace{}%
\AgdaKeyword{do}\<%
\\
\>[0][@{}l@{\AgdaIndent{0}}]%
\>[2]\AgdaPostulate{primClose}\AgdaSpace{}%
\AgdaSymbol{(}\AgdaFunction{lookup}\AgdaSpace{}%
\AgdaBound{chns}\AgdaSpace{}%
\AgdaBound{c}\AgdaSymbol{)}\<%
\\
\>[2]\AgdaFunction{exec}\AgdaSpace{}%
\AgdaBound{cmd}\AgdaSpace{}%
\AgdaSymbol{(}\AgdaBound{gend}\AgdaSpace{}%
\AgdaBound{state}\AgdaSymbol{)}\AgdaSpace{}%
\AgdaSymbol{(}\AgdaFunction{remove}\AgdaSpace{}%
\AgdaBound{chns}\AgdaSpace{}%
\AgdaBound{c}\AgdaSymbol{)}\<%
\\
\>[0]\AgdaFunction{exec}\AgdaSpace{}%
\AgdaSymbol{(}\AgdaInductiveConstructor{SEND}\AgdaSpace{}%
\AgdaBound{c}\AgdaSpace{}%
\AgdaBound{getx}\AgdaSpace{}%
\AgdaBound{cmds}\AgdaSymbol{)}\AgdaSpace{}%
\AgdaBound{state}\AgdaSpace{}%
\AgdaBound{chns}\AgdaSpace{}%
\AgdaSymbol{=}\<%
\\
\>[0][@{}l@{\AgdaIndent{0}}]%
\>[2]\AgdaKeyword{let}\AgdaSpace{}%
\AgdaOperator{\AgdaInductiveConstructor{⟨}}\AgdaSpace{}%
\AgdaBound{x}\AgdaSpace{}%
\AgdaOperator{\AgdaInductiveConstructor{,}}\AgdaSpace{}%
\AgdaBound{state′}\AgdaSpace{}%
\AgdaOperator{\AgdaInductiveConstructor{⟩}}\AgdaSpace{}%
\AgdaSymbol{=}\AgdaSpace{}%
\AgdaBound{getx}\AgdaSpace{}%
\AgdaBound{state}\AgdaSpace{}%
\AgdaKeyword{in}\<%
\\
\>[2]\AgdaPostulate{primSend}\AgdaSpace{}%
\AgdaSymbol{(}\AgdaFunction{lookup}\AgdaSpace{}%
\AgdaBound{chns}\AgdaSpace{}%
\AgdaBound{c}\AgdaSymbol{)}\AgdaSpace{}%
\AgdaBound{x}\AgdaSpace{}%
\AgdaOperator{\AgdaFunction{>>}}\AgdaSpace{}%
\AgdaFunction{exec}\AgdaSpace{}%
\AgdaBound{cmds}\AgdaSpace{}%
\AgdaBound{state′}\AgdaSpace{}%
\AgdaBound{chns}\<%
\\
\>[0]\AgdaFunction{exec}\AgdaSpace{}%
\AgdaSymbol{(}\AgdaInductiveConstructor{RECV}\AgdaSpace{}%
\AgdaBound{c}\AgdaSpace{}%
\AgdaBound{putx}\AgdaSpace{}%
\AgdaBound{cmds}\AgdaSymbol{)}\AgdaSpace{}%
\AgdaBound{state}\AgdaSpace{}%
\AgdaBound{chns}\AgdaSpace{}%
\AgdaSymbol{=}\<%
\\
\>[0][@{}l@{\AgdaIndent{0}}]%
\>[2]\AgdaPostulate{primRecv}\AgdaSpace{}%
\AgdaSymbol{(}\AgdaFunction{lookup}\AgdaSpace{}%
\AgdaBound{chns}\AgdaSpace{}%
\AgdaBound{c}\AgdaSymbol{)}\AgdaSpace{}%
\AgdaOperator{\AgdaFunction{>>=}}\AgdaSpace{}%
\AgdaSymbol{λ}\AgdaSpace{}%
\AgdaBound{x}\AgdaSpace{}%
\AgdaSymbol{→}\<%
\\
\>[2]\AgdaKeyword{let}\AgdaSpace{}%
\AgdaBound{state′}\AgdaSpace{}%
\AgdaSymbol{=}\AgdaSpace{}%
\AgdaBound{putx}\AgdaSpace{}%
\AgdaBound{x}\AgdaSpace{}%
\AgdaBound{state}\AgdaSpace{}%
\AgdaKeyword{in}\<%
\\
\>[2]\AgdaFunction{exec}\AgdaSpace{}%
\AgdaBound{cmds}\AgdaSpace{}%
\AgdaBound{state′}\AgdaSpace{}%
\AgdaBound{chns}\<%
\\
\>[0]\AgdaFunction{exec}\AgdaSpace{}%
\AgdaSymbol{(}\AgdaInductiveConstructor{SELECT}\AgdaSpace{}%
\AgdaBound{c}\AgdaSpace{}%
\AgdaSymbol{\AgdaUnderscore{}}\AgdaSpace{}%
\AgdaBound{getx}\AgdaSpace{}%
\AgdaBound{cmds₁}\AgdaSpace{}%
\AgdaBound{cmds₂}\AgdaSymbol{)}\AgdaSpace{}%
\AgdaBound{state}\AgdaSpace{}%
\AgdaBound{chns}\AgdaSpace{}%
\AgdaSymbol{=}\AgdaSpace{}%
\AgdaKeyword{do}\<%
\\
\>[0][@{}l@{\AgdaIndent{0}}]%
\>[2]\AgdaKeyword{let}\AgdaSpace{}%
\AgdaOperator{\AgdaInductiveConstructor{⟨}}\AgdaSpace{}%
\AgdaBound{b}\AgdaSpace{}%
\AgdaOperator{\AgdaInductiveConstructor{,}}\AgdaSpace{}%
\AgdaBound{a}\AgdaSpace{}%
\AgdaOperator{\AgdaInductiveConstructor{⟩}}\AgdaSpace{}%
\AgdaSymbol{=}\AgdaSpace{}%
\AgdaBound{getx}\AgdaSpace{}%
\AgdaBound{state}\<%
\\
\>[2]\AgdaPostulate{primSend}\AgdaSpace{}%
\AgdaSymbol{(}\AgdaFunction{lookup}\AgdaSpace{}%
\AgdaBound{chns}\AgdaSpace{}%
\AgdaBound{c}\AgdaSymbol{)}\AgdaSpace{}%
\AgdaBound{b}\<%
\\
\>[2]\AgdaOperator{\AgdaFunction{if}}\AgdaSpace{}%
\AgdaBound{b}%
\>[1337I]\AgdaOperator{\AgdaFunction{then}}\AgdaSpace{}%
\AgdaSymbol{(}\AgdaFunction{exec}\AgdaSpace{}%
\AgdaBound{cmds₁}\AgdaSpace{}%
\AgdaBound{a}\AgdaSpace{}%
\AgdaBound{chns}\AgdaSymbol{)}\<%
\\
\>[.][@{}l@{}]\<[1337I]%
\>[7]\AgdaOperator{\AgdaFunction{else}}\AgdaSpace{}%
\AgdaSymbol{(}\AgdaFunction{exec}\AgdaSpace{}%
\AgdaBound{cmds₂}\AgdaSpace{}%
\AgdaBound{a}\AgdaSpace{}%
\AgdaBound{chns}\AgdaSymbol{)}\<%
\\
\>[0]\AgdaFunction{exec}\AgdaSpace{}%
\AgdaSymbol{(}\AgdaInductiveConstructor{CHOICE}\AgdaSpace{}%
\AgdaBound{c}\AgdaSpace{}%
\AgdaSymbol{\AgdaUnderscore{}}\AgdaSpace{}%
\AgdaBound{cmd₁}\AgdaSpace{}%
\AgdaBound{cmd₂}\AgdaSymbol{)}\AgdaSpace{}%
\AgdaBound{state}\AgdaSpace{}%
\AgdaBound{chns}\AgdaSpace{}%
\AgdaSymbol{=}\AgdaSpace{}%
\AgdaKeyword{do}\<%
\\
\>[0][@{}l@{\AgdaIndent{0}}]%
\>[2]\AgdaBound{b}\AgdaSpace{}%
\AgdaOperator{\AgdaFunction{←}}\AgdaSpace{}%
\AgdaPostulate{primRecv}\AgdaSpace{}%
\AgdaSymbol{(}\AgdaFunction{lookup}\AgdaSpace{}%
\AgdaBound{chns}\AgdaSpace{}%
\AgdaBound{c}\AgdaSymbol{)}\<%
\\
\>[2]\AgdaOperator{\AgdaFunction{if}}\AgdaSpace{}%
\AgdaBound{b}%
\>[1361I]\AgdaOperator{\AgdaFunction{then}}\AgdaSpace{}%
\AgdaFunction{exec}\AgdaSpace{}%
\AgdaBound{cmd₁}\AgdaSpace{}%
\AgdaBound{state}\AgdaSpace{}%
\AgdaBound{chns}\<%
\\
\>[.][@{}l@{}]\<[1361I]%
\>[7]\AgdaOperator{\AgdaFunction{else}}\AgdaSpace{}%
\AgdaFunction{exec}\AgdaSpace{}%
\AgdaBound{cmd₂}\AgdaSpace{}%
\AgdaBound{state}\AgdaSpace{}%
\AgdaBound{chns}\<%
\\
\>[0]\AgdaFunction{exec}\AgdaSpace{}%
\AgdaInductiveConstructor{END}\AgdaSpace{}%
\AgdaBound{state}\AgdaSpace{}%
\AgdaInductiveConstructor{[]}\AgdaSpace{}%
\AgdaSymbol{=}\AgdaSpace{}%
\AgdaKeyword{do}\<%
\\
\>[0][@{}l@{\AgdaIndent{0}}]%
\>[2]\AgdaInductiveConstructor{pure}\AgdaSpace{}%
\AgdaBound{state}\<%
\\
\\[\AgdaEmptyExtraSkip]%
\>[0]\AgdaFunction{runCmd}\AgdaSpace{}%
\AgdaSymbol{:}\AgdaSpace{}%
\AgdaDatatype{Cmd}\AgdaSpace{}%
\AgdaGeneralizable{A}\AgdaSpace{}%
\AgdaNumber{0}\AgdaSpace{}%
\AgdaGeneralizable{M}\AgdaSpace{}%
\AgdaSymbol{→}\AgdaSpace{}%
\AgdaGeneralizable{A}\AgdaSpace{}%
\AgdaSymbol{→}\AgdaSpace{}%
\AgdaDatatype{IO}\AgdaSpace{}%
\AgdaGeneralizable{A}\<%
\\
\>[0]\AgdaFunction{runCmd}\AgdaSpace{}%
\AgdaBound{cmd}\AgdaSpace{}%
\AgdaBound{init}\AgdaSpace{}%
\AgdaSymbol{=}\AgdaSpace{}%
\AgdaKeyword{do}\<%
\\
\>[0][@{}l@{\AgdaIndent{0}}]%
\>[2]\AgdaFunction{exec}\AgdaSpace{}%
\AgdaBound{cmd}\AgdaSpace{}%
\AgdaBound{init}\AgdaSpace{}%
\AgdaInductiveConstructor{[]}\<%
\end{code}}